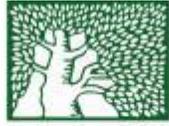

# WEIZMANN INSTITUTE OF SCIENCE

| | |
|---|---|
| **Thesis for the degree** | עבודת גמר (תזה) לתואר |
| **Doctor of Philosophy** | דוקטור לפילוסופיה |
| Submitted to the Scientific Council of the Weizmann Institute of Science Rehovot, Israel | מוגשת למועצה המדעית של מכון ויצמן למדע רחובות, ישראל |
| **By** | מאת |
| **Yevgeny Rakita** | יבגני רקיטה |
| **Between Structure and Performance in Halide Perovskites for Photovoltaic Applications: the Role of Defects** | בין מבנה לביצועים של גבישים פרובסקיטיים האלידיים עבור יישומים פוטוולטאים: תפקידם של פגמים |
| **Advisors:** | מנחים: |
| **Profs. David Cahen & Gary Hodes** | פרופ' גארי הודס ודוד כאהן |
| **September 2018** | תשרי התשע"ט |

## *Acknowledgements:*

| <u>*Who*</u> | <u>*Why*</u> |
|---|---|
| **David Cahen and Gary Hodes** | For accepting me, for guiding me, for focusing me, for history lessons, personal lessons (everyone needs it here and there) and, not less important, giving me the opportunity to make mistakes, and when I make them, guide me to the best solutions. |
| **Dr. Elena Rabinovitch, Dr. Eran Edri, Dr. Jaykrushna Das, Michael Kulbak, Dr. Nir Klein Kedem, Dr. Thomas M. Brenner, Dr. Elena Meirzadeh, Dr. Igal Levin, Arava Zohar, Dr. Davide R. Ceratti, Yonatan Orr, Natalya Weber, Hadar Kaslasi, Aya Osherov, Ayala Cohen, Galit Cohen, Omri Bar-Eli, Ayelet, Ron Tenne, Dr. Ayelet Teitelboim,** | Past and present students and postdocs I had the pleasure and luck to meet. For amazing collaborative work, support (in many aspects), assistance in scientific and technical efforts incl.: crystal growth, layer depositions, characterizations, sharing knowledge and time, and friendship. |
| **Dr. David Ehre** | For being patient with explaining the trivial (and less trivial) questions, for wonderful collaboration. David is an expert in, among other things, **pyroelectricity, impedance spectroscopy, ferroelectricity**. |
| **Dr. Sidney Cohen** | For opening his door every time I had a question on mechanical properties or just to listen for an idea. For being a guide for me in the beginning of my PhD. Sidney is an expert in, among other things, **mechanical properties and nanoindentation**, with whom I had the pleasure to work with! |
| **Dr. Michal Leskes** | For dedicating her time to do the SS-NMR experiments with me at very unconvenient times. Michal is an expert in, among other things, **Solid State NMR**, with whom I had the pleasure to work with! |
| **Prof. Dan Oron** | For the collaborative work on exposing ferroelectricity in HaPs and explaining the physics in a very approachable and way. Dan |
| | is expert in, among other things, **optics, incl. Second-Harmonic Generation**. |
| **Dr. Omer Yaffe** | For a unique collaboration that involved many hours of brain storming, enrichment, encouragement, enthusiasm, guidance and personal assistance. For keeping his door always open and always giving the feeling that there are no silly questions [even though I did not attend his course ;-)] |
| **Dr. Isai Feldman** | For teaching me the art of x-ray diffraction and the patience he had for all my questions. Isai is an expert in, among other things, **x-ray diffraction.** |
| **Dr. Ilit Lampl** | For allowing me to work independently after a very short time and being helpful with any need. Ilit is an expert in, among other things, **thermal and quantitative analysis**. |
| **Dr. Ifat Kaplan and Dr. Eugenia Klein** | For trusting me with a super-expensive equipment, guiding me through it and never stopping me whenever I tried a new idea. Both are experts in, among other things, **scanning electron microscopy (inc. EDS)**. |
| **Prof. Michael Elbaum and Dr. Vyacheslav Kalchenko** | For exposing me to new optical imaging capabilities and helping me in getting the most relevant results. Michael is an expert in, among other things, **light microscopy (reflective, confocal, fluorescent, etc.)** and Vyacheslav is an expert in, among other things, **2-photon confocal microscope** at the veterinary services. With both I had a great pleasure to work with! |
| **Dr. Aditya Sadhanala (Cambridge)** | Among other things, an expert in **Photothermal deflection spectroscopy** I had the pleasure to collaborate with! |
| **Prof. Igor Lubomirsky, Prof. Leeor Kronik, Dr. Ayelet Vilan, Dr. David Egger, Prof. Samuel Safran, , Prof. Anatoli Frenkel (Stony Brook University), Prof. Michael Toney (SSRL, Stanford), Prof. Andrew Rappe (University of Pennsylvania)** | For great discussions, time and knowledge they shared with me in formal (and less formal) personal meetings. |
| **Lilia Gofer, Benjamin Pasmantirer, Reuven Dagan and the mechanical workshop team** | For designing and making mechanically workshopped pieces with the highest accuracy and suggesting simple and brilliant solutions for difficult problems. |
| **Sharon Garusi, Alex Yoffe, Karol Bakalash,** | For support in the cleanroom and assistance with many technical issues. |
| **Ana Naamat, Ayelet Chen-Cohen and Adi Ein-Gal Bar Nahum** | For the kindest administrative assistance one can ever ask. |
| **My family** | Who supported me all these 5 years in the Weizmann institute and offering me comfort and peace of mind whenever it was required. |
| **My spouse - Rachel** | For sharing the last and stressful moments of this PhD with all the needed support, and for joining to this scientific journey with all the future unknowns on the way. |

# *List of publications*

*(All the publications are related to my PhD thesis)*

- *Authored as a primary or shared primary contributor*:

  - **Ref.** 1 - <u>Rakita, Y</u>., Cohen, S. R., Kedem, N. K., Hodes, G. & Cahen, D. Mechanical properties of $APbX_3$ (A = Cs or $CH_3NH_3$; X = I or Br) perovskite single crystals. *MRS Commun.* **5,** 623–629 (2015)

  - **Ref.** 2 - <u>Rakita, Y</u>., Kedem, N., Gupta, S., Sadhanala, A., Kalchenko, V., Böhm, M. L., Kulbak, M., Friend, R. H., Cahen, D. & Hodes, G. Low-Temperature Solution-Grown $CsPbBr_3$ Single Crystals and Their Characterization. *Cryst. Growth Des.* (2016). doi:10.1021/acs.cgd.6b00764

  - **Ref.** 3 - <u>Rakita, Y</u>., Meirzadeh, E., Bendikov, T., Kalchenko, V., Lubomirsky, I., Hodes, G., Ehre, D. & Cahen, D. $CH_3NH_3PbBr_3$ is not pyroelectric, excluding ferroelectric-enhanced photovoltaic performance. *APL Mater.* 4, 051101 (2016).

  - **Ref.** 4 - Brenner, T. M.\*, <u>Rakita, Y.</u>\*, Orr, Y., Klein, E., Feldman, I., Elbaum, M., Cahen, D. & Hodes, G. Conversion of Single Crystalline $PbI_2$ to $CH_3NH_3PbI_3$: Structural Relations and Transformation Dynamics. *Chem. Mater.* **28,** 6501–6510 (2016). (*\* equal contribution*)

  - **Ref.** 5 - <u>Rakita, Y.</u>, Bar-Elli, O., Meirzadeh, E., Kaslasi, H., Peleg, Y., Hodes, G., Lubomirsky, I., Oron, D., Ehre, D. & Cahen, D. Tetragonal $CH_3NH_3PbI_3$ is ferroelectric. *Proc. Natl. Acad. Sci.* 201702429 (2017). doi:10.1073/pnas.1702429114

  - **Ref.** 6 - <u>Rakita, Y</u>., Gupta, S., Cahen, D. & Hodes, G. Metal to Halide Perovskite (HaP): An Alternative Route to HaP Coating, Directly from Pb(0) or Sn(0) Films. *Chem. Mater.* **29,** 8620–8629 (2017).

- *Authored a secondary contributor*:

  - **Ref.** 7 - Rosenberg, J. W., Legodi, M. J., <u>Rakita, Y.,</u> Cahen, D. & Diale, M. Laplace current deep level transient spectroscopy measurements of defect states in methylammonium lead bromide single crystals. *J. Appl. Phys.* 122, 145701 (2017).

  - **Ref.** 8 - Ceratti, D. R., <u>Rakita, Y.,</u> Cremonesi, L., Tenne, R., Kalchenko, V., Elbaum, M., Oron, D., Potenza, M. A. C., Hodes, G. & Cahen, D. Self-Healing Inside $APbBr_3$ Halide Perovskite Crystals. *Adv. Mater.*(2018) doi:10.1002/adma.201706273

# Table of Contents

## *Table of Figures and Tables*

**Tables:**

# List of abbreviations

**HaP** – Halide Perovskite
**PV** – photovoltaic
**RT** – room temperature
**PL** – Photoluminescence
**XRD** – X-ray diffraction
**SEM** – Scanning Electron Microscope
**VBM** – valence band maximum
**CBM** – conduction band minimum
**DOS** – density of states
**NMR** – nuclear magnetic resonance
**AP** – acoustic phonon
**OP** – optical phonon
**ADP** – acoustic deformation potential
**ODP** – optical deformation potential
**POP** – polar optical phonos
**PZ** – piezoelectric
**THzC** – optical-pump-THz-probe conductivity
**MWC** – transient microwave conductivity
**PLQ** – photoluminescence quenching
**Chemicals**:
    **MA** – methylammonium
    **FA** – formamidinium
    **DMSO** – dimethylsulfoxide
    **DMF** – N,N – dimethyformamide
    **NMF** – N- methylformamide
    **GBL** – gamma-butyrolactone
    **MeCN** - acetonitrile
    **IPA** – iso-propanol

## *Abstract*

**Ha**lide **P**erovskite (HaP) semiconducting compounds with an ABX$_3$ stoichiometry and a (pseudo-) cubic symmetry, where the X group is a halide (I$^-$, Br$^-$ or Cl$^-$), B is a metal of the IV column (e.g., Pb or Sn) and A is a cation (organic or inorganic). My main motivation to study HaPs is their proven high outputs in converting sunlight to electricity in photovoltaic (PV) cells (>22%)[9], although, with respect to other solar-cell technologies, much simpler (solution-based) fabrication methods are required. My study focused on the fundamental structural, chemical and dielectric properties of HaPs in reference to their PV-related properties.

Since the main focus of the study relates to intrinsic properties of HaPs, single-crystals were the main model-of-choice for my experiments. Their growth and characterizations can be found in chapter 2. I do note that while film morphology, grain boundaries and different interfacing materials may be very important for device operation, these were not considered in my work.

*Below I summarize my main findings*:

(1) <u>Sections 3.2-3.3</u>: The B-X bond in the ABX$_3$ structure is found to dominate the backbone of the structure. Besides its significance in electronic and optical properties, which was already known, for the case where B is fixed to be lead (Pb) it is shown to have a dominant covalent nature with an apparent effect on its mechanical properties.[1] The A ion, which should have the right size to allow a perovskite structure, is mostly responsible for the chemical stability, particularly via the AX or AX$_3$ intermediates (see Section 4.4).

(2) <u>Section 3.4</u>: HaPs possess a mechanically soft backbone, which is reflected in their relatively *low deformation potentials*. Together with an unusual situation of an *anti*-bonding valence band maximum (suggesting shallow defect states), strong tendency to form AX$_3$ complexes with similar bandgap to that of the hosting HaP (see section 4.3) is shown. These features of bond-nature and chemistry suggest that existing defects (neutral or charged) are assumed to be readily *tolerated*, in sense that they may show no (or very low) mark on the materials (opto)electronic profile.

(3) <u>Section 3.5</u>: Being a heteropolar semiconductor, polar optical phonons (POP) are shown to be the most probable reason for scattering charges in HaPs. Comparison with a broad set of heteropolar compounds revealed that the charge mobility, which is governed by POP scattering, is *fundamentally* limited due to the 'softness' of the material. This is a *non*-trivial result, which occur due to interrelation between the deformation potential and the effective dielectric constant, $\varepsilon^*$, which is a lattice-polarizability parameter that is proportional to the mobility.

(4) <u>Chapter 4</u>: Study of HaPs' chemistry revealed highly probable recyclable degradation products (low $\Delta G_{reaction}^{degradation}$ of ~ 0.1 eV ~ 4 $k_BT$) back to the perovskite composition.[4,6] which is supported by a clear *self*-recovery of damaged APbBr$_3$ crystals (named as '*self-healing*', or *room-temperature annealing*).[8] Assuming a highly dynamic, entropy-stabilized system (supported by a careful calorimetric study done by Nagabhushana et al.)[10], '*self-healing*' should play a significant role in setting low defect densities (as found in HaPs - ~$10^{10}$ cm$^{-3}$ for single crystals). The observed defect densities in HaPs are found to fit the thermodynamic limit for formation of point defects.

(5) <u>Chapter 5</u>: The structural symmetry may allow a polar ferroelectric structure; however, despite what was claimed, its significance at room temperature in an operating PV device is questionable, mostly since it does not appear for all HaPs (MAPbI$_3$ but not MAPbBr$_3$).[3,5]

## *תקציר*

**פרובסקיטים האלידיים** (Halide Perovskites) (או בקצרה, פרובסקיטים) הינם מוליכים למחצה גבישיים בעלי מבנה (פסאודו-) קובייתי והרכב כימי של $ABX_3$, כאשר X מייצג קבוצה האלידי ($I^-$, $Br^-$, או $Cl^-$), B הינה מתכת מהטור הרביעי (כגון עופרת, Pb, ובדיל, Sn) ו-A הוא קטיון חד-ערכי (אורגני או אן-אורגני). פשטות הייצור ביחס לחומרים פוטוולטאיים קיימים, ועם זאת, היעילות הגבוהה בהמרת אנרגיית שמש לחשמל ($<22\%$)[9] מהתקנים פוטוולטאיים (photovoltaic) המורכבים מהם, מהווים למניע העיקרי שבגינו בחרתי לחקור חומרים אלו. המחקר שלי מתמקד בעיקר בתכונות המבניות, הכימיות והדיאלקטריות (dielectric) הבסיסיות של חומרים אלו, ואלו בהקשר לתכונות החשובות לחומרים פוטוולטאיים.

מכיוון שהחלק הארי במחקר מתמקד בתכונות עצמיות של פרובסקיטים, גבישים-יחידים היוו את המודל המרכזי עליו המחקר התבסס. שיטות הגידול ואפיונים של גבישים אלו מפורטות הפרק 2. אציין כי לא התייחסתי להיבטים כגון מורפולוגיה, חומרי ממשק נלווים (למשל שכבות לסלקטיביות חשמלית כמו $TiO_2$) ויחסי הגומלין של פרוספקטים עם חומרי ממשק אלו, למרות חשיבותם הלא מבוטלת של היבטים אלו בפעילות תאים פוטוולטאיים.

*עיקרי ממצאי המחקר* :

(1) <u>תת-פרקים 3.2-3.3</u> : הקשר B-X במבנה $ABX_3$ מהווה את עמוד השדרה של המבנה הפרובסקיטי. מלבד לתכונות האופטיות והאופטו-חשמליות, שהקשר ההדוק ביניהן לבין הקשר הכימי B-X היה ידוע, מצאנו כי קשר זה הוא בעל אופי קוולטי (covalent) בולט (למרות אפיונו הראשוני כחומר יוני מובהק). כמו כן, נמצא קשר הדוק בין התכונות המכאניות של החומר (בכללותו) לחוזק הקשר B-X בלבד.[1] נמצא כי הקטיון שבאתר A, שחשיבותו הקריטית במבנה בכללו היא בגודלו היוני שמאפשר מבנה פרובסקיטי, אחראי גם כל היציבות הכימית של המבנה בכללו, וזאת בעיקר בזכות תרכובות הביניים AX ו-$AX_3$ אליהם פרובסקיטים יכולים להתפרק (ראה תת-פרק 4.4).

(2) <u>תת-פרקים 3.4</u> : החוזק המכאני של פרובסקיטים נמוך, ומכיוון שעיקרו בקשר B-X, האנרגיה פוטנציאלית לעיוות (או בקצרה, ***פוטנציאל עיוות***, *deformation potential*) של קשר זה נמוכה גם כן. כמו כן, קשר זה מאופיין בפס ערכיות "אנטי-קושר" ונטייה חזקה ליצירת קומפלקסים (ראה תת-פרק 4.3), שמשמעותם רמות אנרגיה רדודות (או משולבות) ביחס לפסי הערכיות וההולכה בפרובסקיט. רמות אנרגיה רדודות ופוטנציאל עיוות נמוך מביע את היכולת של פגמים (ניטרליים או טעונים) להיות סבילים בהשפעתם על הפרופיל האופטו-חשמלי של הפרובסקיטים, ומשם על תאים פוטוולטאיים.

(3) <u>תת-פרקים 3.5</u> : מכיוון שפרובסקיטיים הם מוליכים למחצה הטרו-פולאריים (heteropolar) (לעומת סיליקון שהוא הומו-פולארי), פונונים אופטיים פולאריים (polar optical phonons), או פ.א.פ, נמצאו כמנגנון פיזור המטען (אלקטרונים או חורים) הסביר ביותר. לכן, פ.א.פ קובעים את ההולכה, או המובילויות (mobility), החשמלית בחומר. השוואת המובילויות עם זאת של הרכבים הטרו-פולאריים אחרים הראתה שככל שחומר יותר רך (או בעל פוטנציאל העיוות הנמוך),

המוביליות חשמלית של אותו חומר, *ביסודה*, נמוכה יותר. ניתן להבין תוצאה זו ע״י הקשר של ׳פוטנציאל העיוותי׳ ל־׳פורלזביליות׳ (או ׳יכולת קיטוב׳, polarizability) של החומר, כאשר האחרון מבוטא ע״י המקדם הדיאלקטרי האפקטיבי, $\varepsilon^*$ , ופרופורציונאלי למוביליות.

(4) פרק 4 : חקר של הכימיה להרכבה ופרוק של פרובסקיטים גילה שקיימים תוצרי פרוק שדורשים דורשים אנרגיה נמוכה ($\sim 0.1$ eV $\sim 4\ k_BT$), אך עם זאת, מגיבים עם עצמם במהירות ליצירת פרובסקיט.[4,6] תופעה זו תומכת ברעיון של ׳איחוי-עצמי׳ של פגמים ליצירת חומר עם ריכוז פגמים נמוך מאוד, כפי שהראנו שקורה בגבישים של $APbBr_3$.[8] בהסתמך על הרעיון שהיציבות של המבנה הפרובסקיטי מגיעה מהאנטרופיה (entropy) הגבוהה של הפרובסקיט ביחס לתוצרי הפירוק שלו (כפי שהציעה קבוצתה של Navrotsky בניסויים קלורימטריים)[10] , נראה כי ׳איחוי-עצמי׳ הינו מרכזי בהגבלת יצירת פגמים חדשים בפרובסקיטים (האלידיים). ואכן, הריכוזי של הפגמים בחומר ($\sim 10^{10}$ cm$^{-3}$) מתאים לגבול התרמודינאמי (thermodynamic) של פגמים בחומר.

(5) פרק 5 :[3,5] הסימטריה המבנית יכולה לאפשר מבנה פולארי שהוא גם פרואלקטרי (ferroelectic). למרות הוויכוח הספרותי הנרחב, מצאנו הוכחה ברורה ש־$MAPbI_3$ הוא אכן פרואלקטרי, אך עם זאת, בניגוד למה שנטען בספרות, מצאנו כי השפעת הפרואלקטריות על האופיין האופטו-חשמלי (בטמפרטורת החדר, או מעל, כפי שרוב תאי השמש חשופים אליו) הינה שנייה במחלוקת. כמו כן, הוכחנו ש־$MAPbBr_3$ *אינו* פולארי, ולכן לא יכול להיות גם פרואלקטרי. תוצאה זו מראה ש פרואלקטריות אינה תכונה יסודית בחומרים אלו, וייתכן שאינה משמעותית, גם כשקיימת.

# 1. Introduction

## 1.1. Perovskites

The perovskite structure represents a wide range of crystalline materials with the stoichiometry of *ABX$_3$*, where *A* and *B* are two different cations and *X* an anion. It was named after the Russian mineralogist Lev Perovski, who discovered the perovskite mineral CaTiO$_3$ in 1839. In its ideal form, it has a cubic structure assembled from corner-sharing [*BX$_6$*] octahedral and *A* cations positioned in cubochedral (12-fold) sites between the interconnected octahedra (see Figure 1 (a) for illustration).

Oxide-perovskites, i.e., *ABO$_3$*, have been widely studied on account of their magnetic and dielectric properties (e.g., (anti-) ferromagnetism, piezoelectricity, ferroelectricity, etc.). What makes these properties possible in these materials is symmetry breaking. In fact, most perovskites, including CaTiO$_3$, possess a lower symmetry than a perfectly cubic one (e.g., orthorhombic, tetragonal, rhombohedral, etc.).[11,12] If one uses only spherical ions, Glazer[11] showed that on the basis of possible distortions of the [*BX$_6$*] octahedron with respect to its neighboring octahedra, the perovskite structure can possess 16 different space-group symmetries (see illustrated example in Figure 1 (b)). When introducing non-spherically symmetric groups, e.g., an organic A cation such as methylammonium, CH$_3$NH$_3^+$, in (CH$_3$NH$_3$)SnBr$_3$, additional space-group symmetries are, in principle, possible. The more complete classification of perovskite structured materials symmetries was presented by Alexandrov et al.[12].

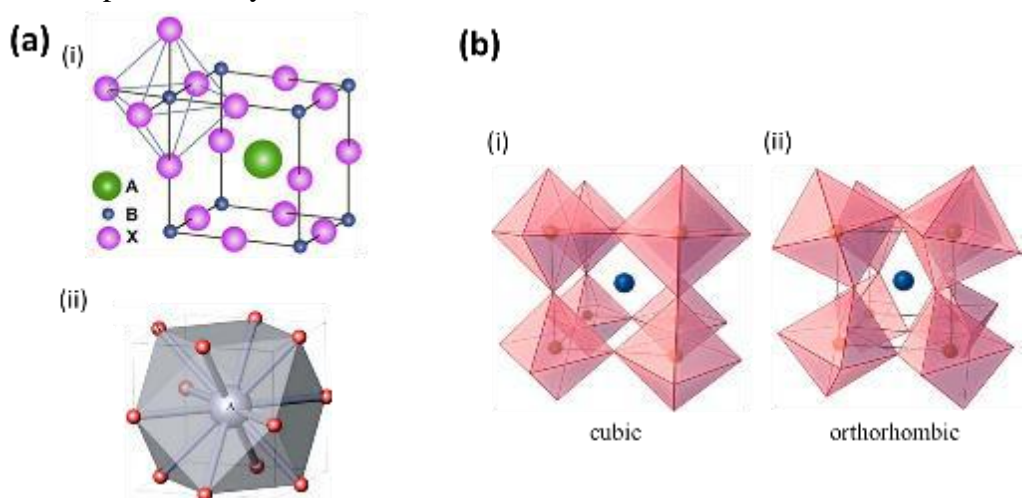

Figure 1: (a) Ball and stick models of a perovskite crystal structure where (i) is a general unit-cell view[13] and (ii) emphasizes the cubochedral 12-fold interconnectivity[14]. (b) A perovskite model that emphasizes octahedron interconnectivity in (i) a perfect (non-distorted) perovskite (cubic; space-group: $Pm\bar{3}m$) and (ii) a distorted perovskite (orthorhombic; space-group $Pnma$).

The structural distortions are usually associated with a geometrical misfit of the relative ionic dimensions. In a perfectly symmetric perovskite (i.e., cubic), the following relation must be fulfilled: $R_A + R_X = \sqrt{2}(R_B + R_X)$ ($R_A$, $R_B$ and $R_X$ are the ionic radii of 'A', 'B' and 'X'). When the ratio: $t = \frac{R_A+R_X}{\sqrt{2}(R_B+R_X)}$ (also called the Goldschmidt's *tolerance factor*)[15] deviates from $t=1$, the combination of the relevant A, B and X ions may thermodynamically favor a lower symmetry structure. Since the [$BX_6$] octahedron has an importance of its own as an individual entity in the structure, the ratio $\mu = \frac{R_B}{R_X}$ (also called *octahedral factor*) is also responsible for the favored symmetry. Empirical structure maps of $t - \mu$ show that the formation window of a perovskite is limited to $0.8 < t < 1$ and $\mu > 0.4$ for oxide-peorvskites and $0.85 < t < 1.1$ and $\mu > 0.4$ for halide-perovskites.[16,17]

Since perovskites are usually treated as ionic crystals, besides the interatomic spacing (represented by $t$ and $\mu$), the ionic charge (referring to the formal oxidation state at each lattice point) also affects the formation energy. Table 1 shows formation energies and site potentials (Madelung potentials) for perovskites with different formal oxidation states in each site.[18] The halide perovskites, [I]-[II]-(I)$_3$, have exceptionally low energy values, compared to those with divalent *X* anions. *Several experimental observations of intrinsic properties are thought to reflect these low formation energies, as will be shown in this work.*

Table 1: Electrostatic lattice energy ($E_{lattice}$) and site Madelung potentials ($V_i$) for a range of AMX$_3$ perovskite structures (using a cubic lattice with a = 6.00 Å) assuming the formal oxidation state of each ion.[18] The last row, [I]-[II]-(VII)$_3$, represents the halide perovskites.

| **Oxidation state: [cation]/(anion)** | | **E lattice [eV/Cell]** | **V$_A$ [V]** | **V$_M$ [V]** | **V$_X$ [V]** |
|---|---|---|---|---|---|
| oxide / chalcogenides | [III]-[III]-(II)$_3$ | -140.48 | -8.04 | -34.59 | 16.66 |
| | [I]-[V]-(II)$_3$ | -118.82 | -12.93 | -29.71 | 15.49 |
| | [II]-[IV]-(II)$_3$ | -106.92 | -17.81 | -24.82 | 14.33 |
| Halides | [I]-[II]-(I)$_3$ | -29.71 | -6.46 | -14.85 | 7.75 |

Two additional known symmetry-breaking mechanisms that occur in perovskites (and in many other structures) are the Jahn-Teller effect [19] and ns$^2$ lone pair steric distortion of the 14$^{th}$ row metals (i.e. Ge, Sn, Pb)[20]. The Jahn-Teller distortion might be relevant if one chooses to introduce d-block elements with high degenaracy in the d-orbitals, such as Cu$^{+2}$ or Mn$^{+2}$. Since d-orbital transition metals were out of the scope of my work, the Jahn-Teller effect is not relevant for this thesis. The $ns^2$ lone pair steric distortion may, however, play a significant role in Pb- or Sn-based halide perovskites. Since the $ns^2$ effect is more pronounced when the atomic number is smaller (Ge>Sn>Pb) (at smaller volume, due to a smaller atom, $ns^2$ delocalization increases), it is claimed to be responsible for the structural, electronic and optical differences between Pb- and Sn-based halide perovskites.[20,21] The $ns^2$ lone pair is also claimed to be responsible for pressure-related effects in Pb-based halide perovskites[22–24].

Usually, for higher temperatures, the symmetry will be higher, since thermodynamically more positional degrees of freedom become degenerate. When increasing the temperature, a common structural sequence for halide perovskites is orthorhombic to tetragonal to cubic - the highest symmetry. Additional symmetry groups may exist when the *A* cation is an asymmetric organic cation (e.g., $CH_3NH_3^+$) as shown in Figure 2.

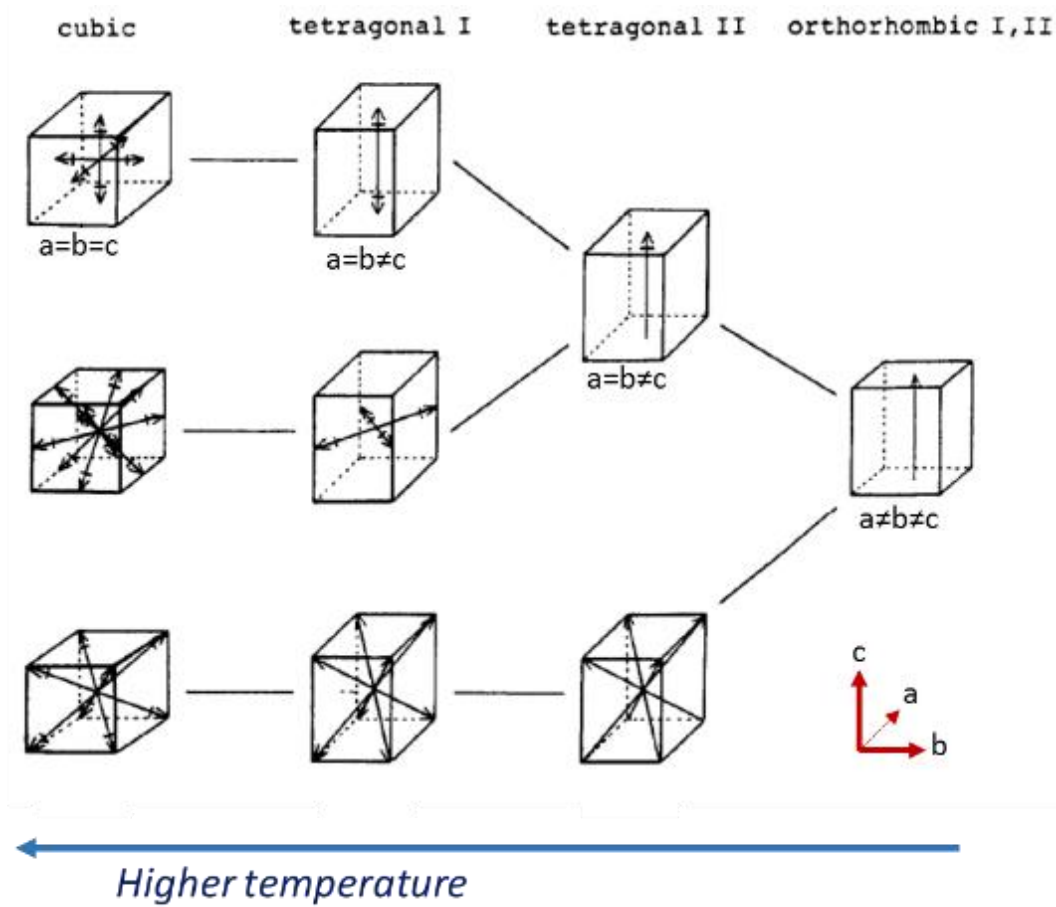

Figure 2: Typical temperature-dependent phase changes of HaPs, specifically relevant for HaPs containing a polar *A* cation, in this case $CH_3NH_3^+$. Arrows indicate the orientations of a C-N axis. ⇊: disordered around the C-N axis. ↑: ordered around the C-N axis. Shows that the type of C-N orientation depends on the B-X cage spatial symmetry, as well as a specific space-group within a general symmetry family (i.e., cubic or tetragonal). (Adopted from ref. 25)

It is important to mention so-called, lower dimensionality perovskite (or, 'perovskite-related') structures that are assembled from $[BX_6]$ octahedra with connectivity that differs from the all corner-shared one of the true perovskite structure. Such structures will usually form due to size mismatch (e.g., too large 'A' subgroup); they are, as distinguished from the 3D perovskite, 2D, 1D or 0D (via the 3D <100>, <110> or <111> orientations, respectively) 'perovskite-related' assemblies of $[BX_6]$ octahedra (see Figure 3). Due to anisotropic structural connectivity, these lower dimension 'perovskite-related' structures have a strong anisotropy in charge flow, where it is more conducting along the interconnected corner sharing $[BX_6]$, as was demonstrated by Mitzi et al.[20].

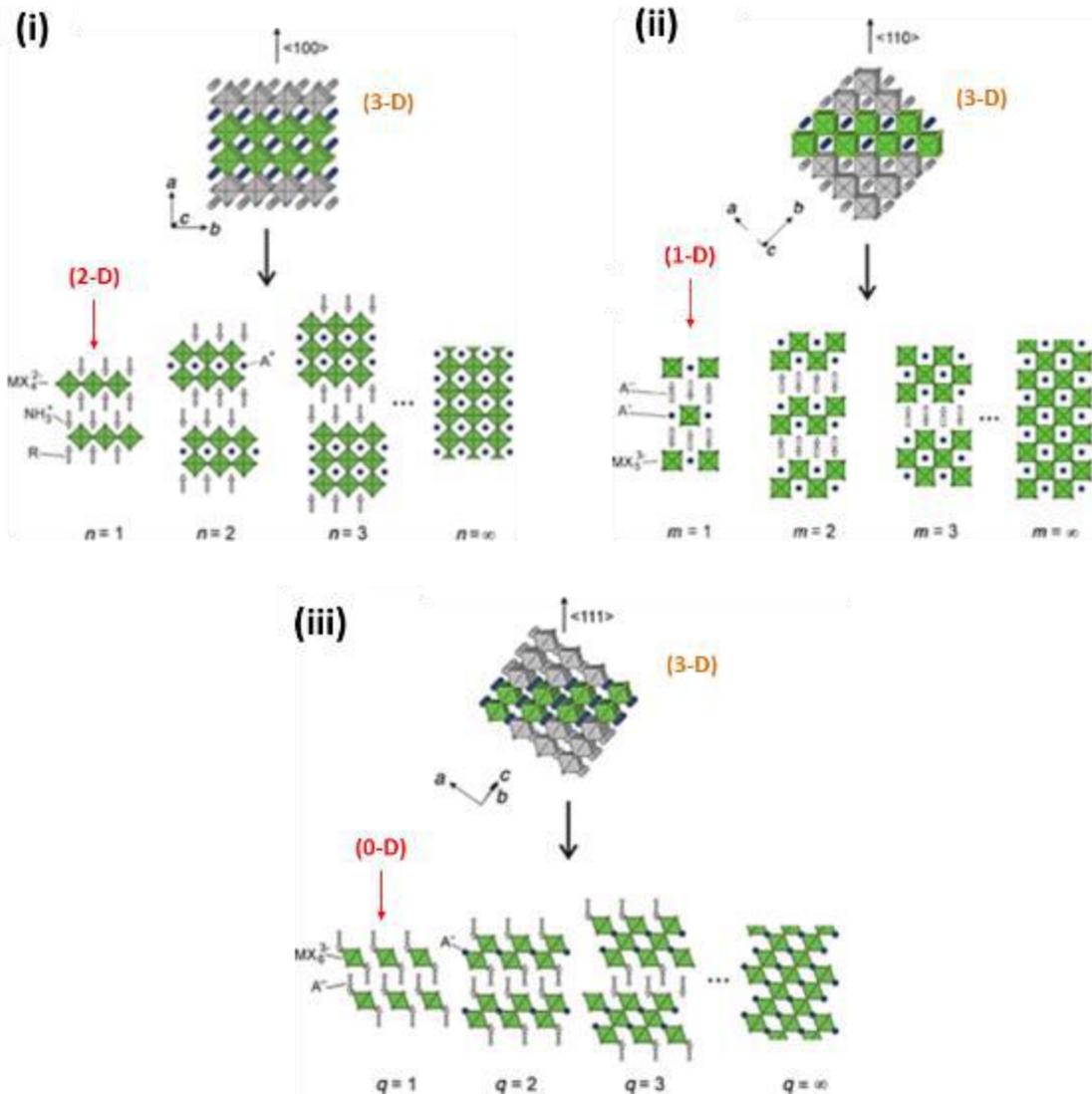

Figure 3: Schematic representation of the (i) <100>, (ii) <110> and (iii) <111> families of layered hybrid perovskites. The stoichiometry of the structures are $A'_2 A_{n-1} M_n X_{3n+1}$ for the <100> family (also known as 'Ruddlesden-Popper' structures), are $A'_2 A_m M_m X_{3m+2}$ for <110> family and $A'_2 A_{q-1} M_q X_{3q+3}$ for the <111> family. The green and gray figures represent the same 3D structure at different orientations. The green sub-section represents n=2, m=2 and q=2 that are derived from the 3D. (Adopted from ref. 26)

Regardless of the reason for the distortion and as long as we keep the same octahedral unit, it was observed [21] that the optical bandgap increases and the conductivity decreases as the symmetry decreases. Gao et al.[14] summarized that both for halide and oxide perovskites a bandgap will decrease upon:

(a)  Increase of the dimensionality (0D→3D) of the [$BX_6$] network.
(b)  Increase of the $B$–$X$–$B$ (≤180⁰) bond angle.
(c)  Decrease of the electronegativity of '$X$' anions or '$B$' cation.

The next section will introduce a more focused view on halide perovskites.

## 1.2. Halide perovskites for optoelectronics

**Ha**lide **P**erovskites, **HaP**s, have been the subject of scattered studies starting from the late 19$^{th}$ century.[27] However, it is only since the mid 1990s that they started to be explored for use in devices, especially for light-emitting devices and transistors.[26,28] Since first being demonstrated in 2009 for photovoltaic (PV) use,[29] PV cells have demonstrated unusually rapid progress and now surpass 22% light to electrical energy conversion efficiency.[9] High-energy radiation (x-, γ-rays) detectors[30,31] and strong, tunable light emitting diodes[32,33] are other technological demonstrations, showing their good optoelectronic behavior. Besides performance, what also attracted scientists (and, not less, industry) is the low material and energy input costs, needed for their fabrication. In principle, their production includes highly abundant elements and does not require temperatures much above room-temperature (RT), unlike that of other high quality optoelectronic semiconducting materials such as Si, GaAs or CdTe.[34]

For optoelectronically functional HaPs, the most common formula of the $ABX_3$ structure is when: $A$ is a large monovalent cation (e.g. methylammonium, $CH_3NH_3^+$ abbreviated to MA, formamidinium, $CH(NH_2)_2^+$ abbreviated to FA, or inorganic $Cs^+$); $B$ is usually $Pb^{2+}$ and $X$ is a halide (see Figure 4). Both because of Pb-toxicity and its similarity to Pb, Sn-based HaPs[35] (and to a much lesser extent $Ge^{2+}$-based HaPs)[36,37] are being explored. However, this usually comes at the expense of the greater tendency of the divalent cation to oxidize to the more stable tetravalent one $Sn^{4+}$ ($Ge^{4+}$), compared to $Pb^{4+}$. The reason for this increased reactivity is the reduced stability of the $ns^2$ lone pair ($n$=6, 5 or 4 for Pb, Sn or Ge) with decreasing $n$ integer. A subject of ongoing research is perovskite structures where the B cation is exchanged with two different cation with (III)/(I) oxidation states such as in $Cs_2InAgCl_6$ double-perovskites[38]. In addition, the X anion was shown to be potentially replaceable with a pseudohalide, e.g., $SCN^-$ or $BH_4^-$ [39,40].

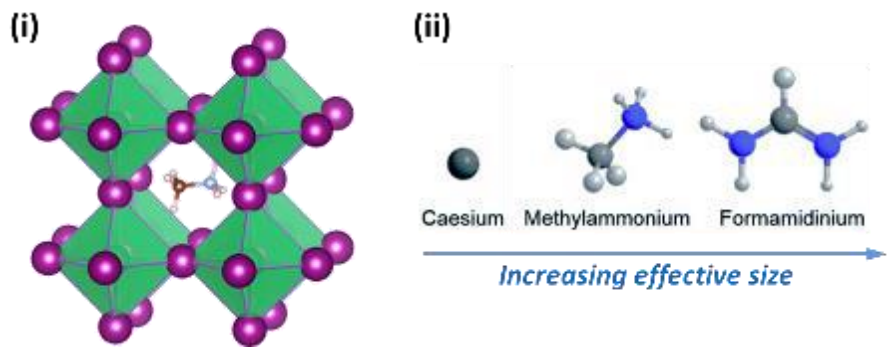

Figure 4: (i) Model of crystal structure of $CH_3NH_3PbI_3$ (= $MAPbI_3$), where the *A* group is methylammonium (MA), the *B* group is Pb and *X* is I.[41] (ii) Other commonly used cations used in optoelectronically-functional HaPs.[42]

Solid solutions made by mixing different A, B or X groups in a single layer (or crystal) can tailor properties such as optical bandgap, stability and, eventually, cell performance.[42–46] Exchanging the organic A cation (e.g., $MA^+$ or $FA^+$) with an inorganic cation (e.g., $Cs^+$) was shown to improve chemical stability of the HaP with no pronounced effect on the optoelectronic performance.[47,48] Change in the X anion is mostly used to tune the bandgap of the material for a specific application (e.g., water splitting, tandem (top stacked) solar cells, different color LEDs, etc.). Overall, the best performing (for PV) and most stable perovskites so far seem to be solid-solutions of a Pb-based HaP with FA and I as the majority A and X ions, i.e., (**FA**,Cs,MA)Pb(**I**,Br)$_3$[49–51]. *In this thesis the role of the A and X cationic groups using single crystalline HaPs is presented.*

Common ways to prepare thin HaP layers are spin- or spray-coating of a precursor solution ($MX_2$ and AX), evaporation of one or both of the precursors ($MX_2$ and/or AX) or dipping a substrate, coated with $MX_2$, in a solution, or exposing it to vapor containing AX salts.[52,53] Among the different synthesis approaches, cases where a direct reaction between an already oxidized B metal (e.g., $PbI_2$, $PbO_2$, $Pb(CH_3COO)_2$), and an AX salt were broadly explored,[4,54–56] reviewed[52,53,57] and optimized over the years. Recently, metallic Pb has been shown to convert into $MAPbI_3$ from MAI solution in IPA,[58,59] and that polyhalide ionic liquids can transform metallic Pb into iodide-based HaPs.[60] *In this work, I will show my results regarding metallic Pb (and Sn) transformation to HaP and regarding $AX_3$ polyhalide compositions that can rapidly oxidize metals to form HaPs. This unique mechanism of formation can explain 'self-healing' observations in HaPs (see chapter 4).*

## 1.3. Halide perovskites in photovoltaics

The maximum theoretical efficiencies (assuming ideal contacts without any non-radiative recombination processes) were calculated by Shockley and Queisser, (S-Q) in 1961.[61] MAPbI$_3$, which has a band gap of ~1.6 eV, can theoretically reach ~ 31% power conversion efficiency (PCE).[62] Although, in reality small HaP-based solar cells reach to 23.3% PCE in research labs, they have shown a tremendous improvement leap over the last half a decade from ~3% PCE. Si, GaAs or CdTe (the leading PV technologies nowadays) with S-Q maximal PCEs of ~31-33% PCE, practically reach 26.6%, 28.9% and 22.1% PCE, respectively,[9,63] after several decades of research. These, however, require higher energy input for their production and CdTe uses the relatively rare element tellurium.

Contact layers, morphology, composition, grain size, cell architecture, chemical stability, scalability (incl. toxicity) and reproducibility are of practical importance and, indeed over the years, have been heavily studied and, indeed, improved in HaP-based PV cells.[53,57,64,65]. In part, these developments were made possible by the superior intrinsic properties of the light-absorbing HaP films.[66,67]

For any absorber layer, the length over which a given type of charge (electrons or holes) can diffuse is the square root of the relevant diffusion coefficient-lifetime product (see Figure 5 for a broader explanation). To maximize charge extraction probability from a PV cell, the diffusion lengths should be greater than the thickness of the film, with minimal interfaces, which are usually a source for charge recombination. In other words, the higher the absorption coefficient, the thinner the film can be to absorb the light, and, thus, the smaller the charge diffusion length needs to be.

HaPs have a direct bandgap, resulting in a sharp absorption onset, and reach a high optical absorption coefficient (~ $10^5$ cm$^{-1}$).[68,69] This requires less than ~300 nm of a continuous typical Pb-based HaP layer to absorb 99% of a non-concentrated solar radiation. Photoexcited charge carrier lifetimes (~ 0.1-1 μs)[70] with sufficiently long carrier mobility (1 ~ 100 cm$^2$·V$^{-1}$·s$^{-1}$)[71] yield diffusion lengths of ~100's-1000's of nm. Moreover, many HaPs, especially MAPbI$_3$, have a very small exciton binding energy[72], a property that is tightly related to the relatively high lattice-polarizability and its static dielectric constant (~30)[73].

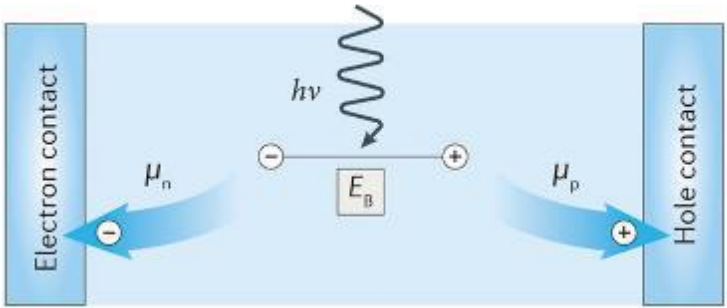 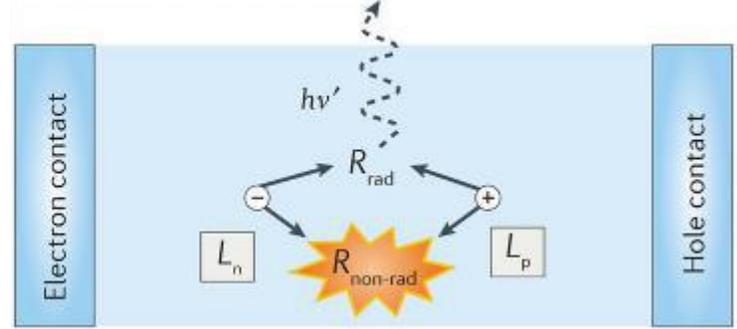

Figure 5: *General description of charge carrier generation, recombination and transport in PV semiconductors.* The carrier generation profile under optical excitation is determined by a material's absorption coefficient. Carrier generation results in free carriers and/or excitonically bound carriers, determined by the exciton binding energy ($E_B$), sample temperature and carrier density. Free carriers diffuse in the absence of an electric field with diffusion coefficients $D_{n,p}$, and in a non-zero field they will, in addition, drift with mobilities $\mu_{n,p}$; these parameters are related through the Einstein relation $\mu_{n,p} = \frac{eD_{n,p}}{k_B T}$, where $e$ is the absolute value of the electron charge, $k_B$ is the Boltzmann constant and $T$ is the temperature. The mobility and diffusion coefficient are related to the band structure of the semiconductor through the carriers' effective mass, which, in a semiclassical picture, is their apparent mass in the crystal lattice. Carrier transport is limited by scattering due to defects and/or lattice vibrations. Carriers have an average recombination lifetime $\tau$ before they recombine, during which they drift or diffuse. The lifetime is determined by the recombination rate, a result of radiative ($R_{rad}$) and non-radiative ($R_{non-rad}$) processes that depend on the carrier density, carrier trap density, dopant concentration and other factors. The carrier diffusion length $L_{n,p} = \sqrt{D_{n,p}\tau} = \sqrt{\left[\frac{k_B T}{e}\right] \cdot \mu_{n,p}\tau}$ is a measure of how far carriers migrate on average before recombining. $h\nu$ is a photon energy where: $h$-Planck's constant; $\nu$-frequency. (From Brenner, Egger et al.[66])

Overall, a ~300 nm HaP thin film, which will absorb 99% of the light, should allow all generated charges to reach the contacts and generate maximum voltage and current. What usually limits light harvesting materials to perform ideally are defects: point defects, grain boundaries, dislocations, interfaces (contact) defects, etc.. Defects can act as:

(a) Scattering centers for transporting charges. These will reduce the diffusion or drift mobility coefficients when the scattering is elastic, and increase non-radiative recombination when the scattering is inelastic.

(b) Charge trapping centers. These are the result of point defects or complexes of these that will introduce an energy level within the bandgap of the semiconductor and will mostly affect the non-radiative recombination probability.

(c) Potential barrier. Usually relate to 2D or 3D defects such as dislocations or inclusions or grain boundaries that may vary with grain size, morphology (e.g. void) or phase separation. These usually act as strong scattering centers and may limit charge flow through them or introduce surface trapping centers.

Extracting values of electronically- or optically-active defect density (or trap density) in HaPs from different experiments yielded values of ~$10^{9}$-$10^{10}$ cm$^{-3}$ for solution-grown single crystals, but ~$10^{15}$ - ~$10^{16}$ cm$^{-3}$ for polycrystalline films - *regardless of the method used for their fabrication.*[7,66,70,74]. Such very low (single crystal) to moderate (film) defect densities are usually found in (very) high-quality semiconductors, such as GaAs or Si single crystals (grown from melt, slowly cooled from high temperatures: ~$10^{9}$-$10^{11}$ cm$^{-3}$), for highly optimized polycrystalline semiconductors (polycrystalline-Si: ~$10^{13}$-$10^{14}$ cm$^{-3}$; CdTe films: ~$10^{11}$-$10^{13}$ cm$^{-3}$; Cu(In,Ga)Se$_2$ films: ~$10^{13}$ cm$^{-3}$) or single crystalline organic materials (rubrene: ~$10^{16}$ cm$^{-3}$; pentacene: ~$10^{15}$ cm$^{-3}$).[70] To maintain low defect densities in materials *other than HaPs* is rather challenging. In fact, it seems to be much easier to pass levels of $10^{16}$ cm$^{-3}$ (intentionally for p/n doping, or unintentionally) up to levels of ~$10^{21}$ cm$^{-3}$. For HaPs, however, $10^{16}$ cm$^{-3}$ seems to be an upper limit that has not yet been surpassed.

Regardless as to whether it is an advantage for optoelectronics or a disadvantage for electronics, understanding what sets that defect density limit and whether it can be broken is of fundamental interest. To answer this question, this work deals with:

- Bond nature and bond strength of HaP.
- Chemical formation and deformation paths of HaPs.
- Tolerance towards defect formation in HaPs and, once formed, possible mechanisms for their 'healing'.
- The probability that physically existing defects will have pronounced electronic or optical effects.
- Scattering mechanism: why are mobility values much lower than in similarly low defect density materials?

# 2. Crystal growth

## 2.1. General strategy for crystal growth from solution

To avoid undesired process-dependent defects (e.g., grain boundaries, kinetically-stabilized impurities or disorder due to rapidly-quenched crystallization), as can be assumed occurs when making thin films, single crystals are favored. Crystal growth of HaP single crystals is, in principle, not different from that of any other material. Therefore, knowledge regarding crystal growth can be gained from any relevant textbook.[75]

In my work, I used only solution-based crystallization and did not grow crystals from a melt of the product (although this was demonstrated to be possible for fully-inorganic HaP, e.g., $CsPbBr_3$, via the Bridgman method)[76]. A general guideline for solution-based crystal growth is a slow variation of the dissolved precursor concentration product ($Q_{sp}$=[A][B], where A and B can be $PbX_2$ and AX) and/or the solubility product ($K_{sp}$) within a supersaturated region (see Figure 6). To make it clearer, I define that below saturation $K_{sp}>Q_{sp}$, at saturation $K_{sp}=Q_{sp}$, while above saturation (also 'supersaturation') $K_{sp}<Q_{sp}$. A supersaturated state can be *metastable*, when nucleation (and sometimes even crystal growth) does not occur due to kinetic stabilization of the solution. In a metastable zone, nucleation (also called: 'secondary' nucleation) occurs only at 'activated' surfaces, which catalyze nucleation processes, such as: dust particles, impurity on the crystallization dish or a deliberate 'seed' of the desired material. By further increasing $Q_{sp}$ over $K_{sp}$, at some point (depending on the chemicals and its environmental conditions), nucleation will occur in the bulk of the solution, regardless of external catalyzing agents mentioned above. This region is *'unstable'* (sometimes called 'labile'), and such nucleation is called 'primary'.

The required slow variation of the ratio between $K_{sp}$ and $Q_{sp}$ can be achieved by:
- *slow variation of the solution temperature ($K_{sp}$ decrease).* The direction of the temperature change depends on whether the solubility is increasing or decreasing within a certain temperature range (both cases are possible).
- *slow vaporization of the solvent ($Q_{sp}$ increase)*
- *slow introduction of precursors ($Q_{sp}$ increase)*
- *slow introduction of an 'anti-solvent' (both $K_{sp}$ and $Q_{sp}$ decrease, but the $K_{sp}/Q_{sp}$ ratio decreases).*

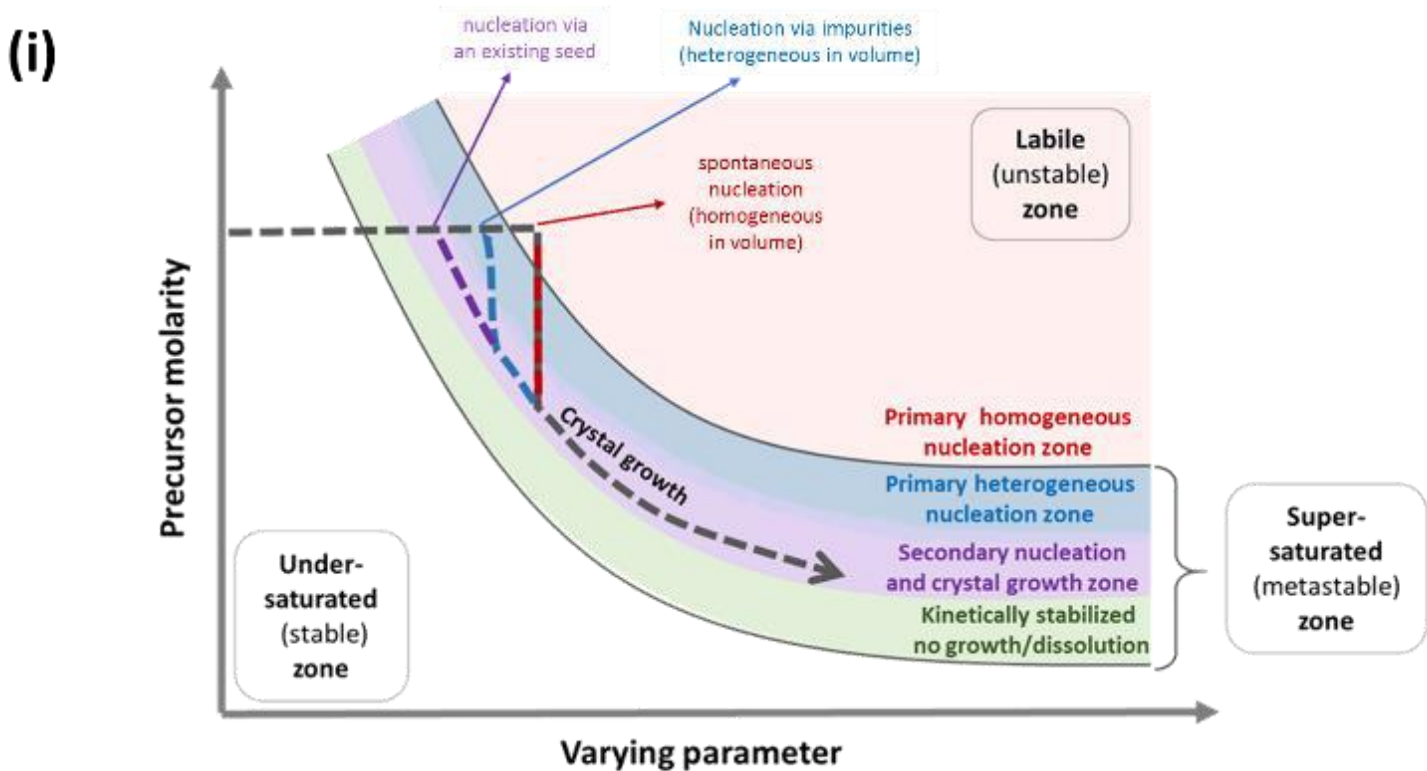

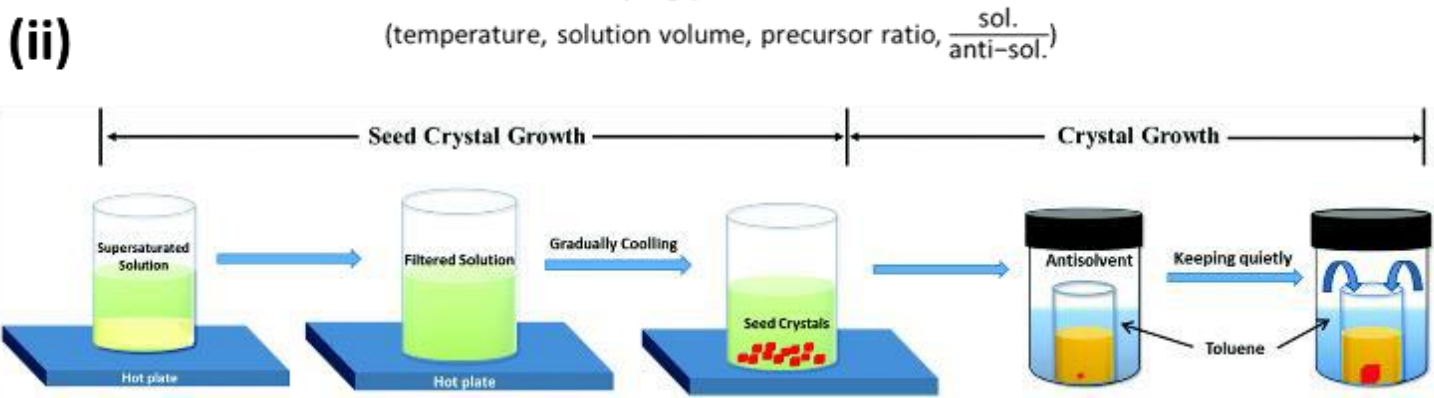

Figure 6: (i) Illustration of a typical phase diagram of precipitating solids from solution. The dotted line illustrates the change in precursor molarity inside the solution during crystallization. When the varying parameter gradually changes from the saturated (*stable*) zone towards the liable (*unstable*) zone, several nucleation steps can occur. Spontaneous and homogeneous nucleation within the volume of the solution (**red zone**) can occur when the varying parameter reaches the liable zone. Selective (and heterogeneous) nucleation can occur on top of activated surfaces, such as impurities (e.g., dust particles), which occur in a supersaturated (*metastable*) zone (**blue zone**; its exact location depends on the type of perturbation). These two processes of nucleation are called 'primary nucleation'. 'Secondary nucleation' occurs when a crystalline seed is artificially added. Regardless of the nucleation pathway, crystal growth (of existing nuclei) will occur in the supersaturated zone of the solution above the secondary nucleation limit – (**purple zone**). Dissolution of the crystal will occur within the 'under-saturated' zone. (ii) A proper crystal-growth routine:[77] preparation of a saturated solution → filtration → formation of 'seeds' via a primary heterogeneous nucleation → setting up a fresh system for crystal growth via a secondary crystallization path. In this example, the seed growth and bulk crystal growth are carried out using different crystallization techniques (CSS and VSA, respectively –see next section). However, these can be, and usually are, done via the same process.

Regarding the last point: an 'anti-solvent' is another solvent in which the solubility of the desired products is significantly lower than in the original 'solvent'. The 'anti-solvent' must be soluble in the 'solvent' and should have a higher vapor pressure than the 'solvent' in order to allow its diffusion into the crystallization solution.

The nucleation and crystal growth processes are described in Figure 6. As illustrated, prior to crystal growth, nucleation is taking place. Nucleation can be *primary*, or *secondary*, as mentioned above. Following nucleation and depletion of precursors from the solution, the crystals will keep growing at supersaturated conditions.

When avoiding homogeneous nucleation, crystal growth proceeds by formation of a stable 'seed', which may have either been in the solution previously (as illustrated in Figure 6(ii)) or formed by heterogeneous precipitation). If during growth the solution supersaturation exceeds the secondary nucleation zone towards the primary nucleation zone (homogeneous or heterogeneous), additional nucleation centers may occur – usually an undesired situation. Such a situation usually occurs by external perturbation (e.g., shaking the solution) or by a rapid change of conditions (e.g., temperature change that is too fast).

Every desired product has a window of conditions in which its formation is favored over other competing (usually undesired) compositions. Undesired compositions can be avoided by changing precursor ratio, temperature, solvent composition, or additives. Since crystallization itself is a kinetic process, addition of nuclei of the desired product to a saturated solution may favor crystallization of the desired product over competing products.

The growth mechanisms from solution can be divided into three main groups as shown in Figure 7:[78] (1) monomer-by-monomer addition; (2) crystallization via amorphous (or poorly crystalline) phase; (3) crystallization by nanocrystalline oriented attachment. Depending on the crystallization environment, there are indications that mechanism (1) and (3) are relevant for growth of HaPs.[79,80] It is worth mentioning that a recent publication showed strong indication for monomer-by-monomer addition growth, where the 'monomer' had been assigned to $[PbX_iS_j]^{2-i}$ oligomer/complex species (S is a solvent molecule).[81]

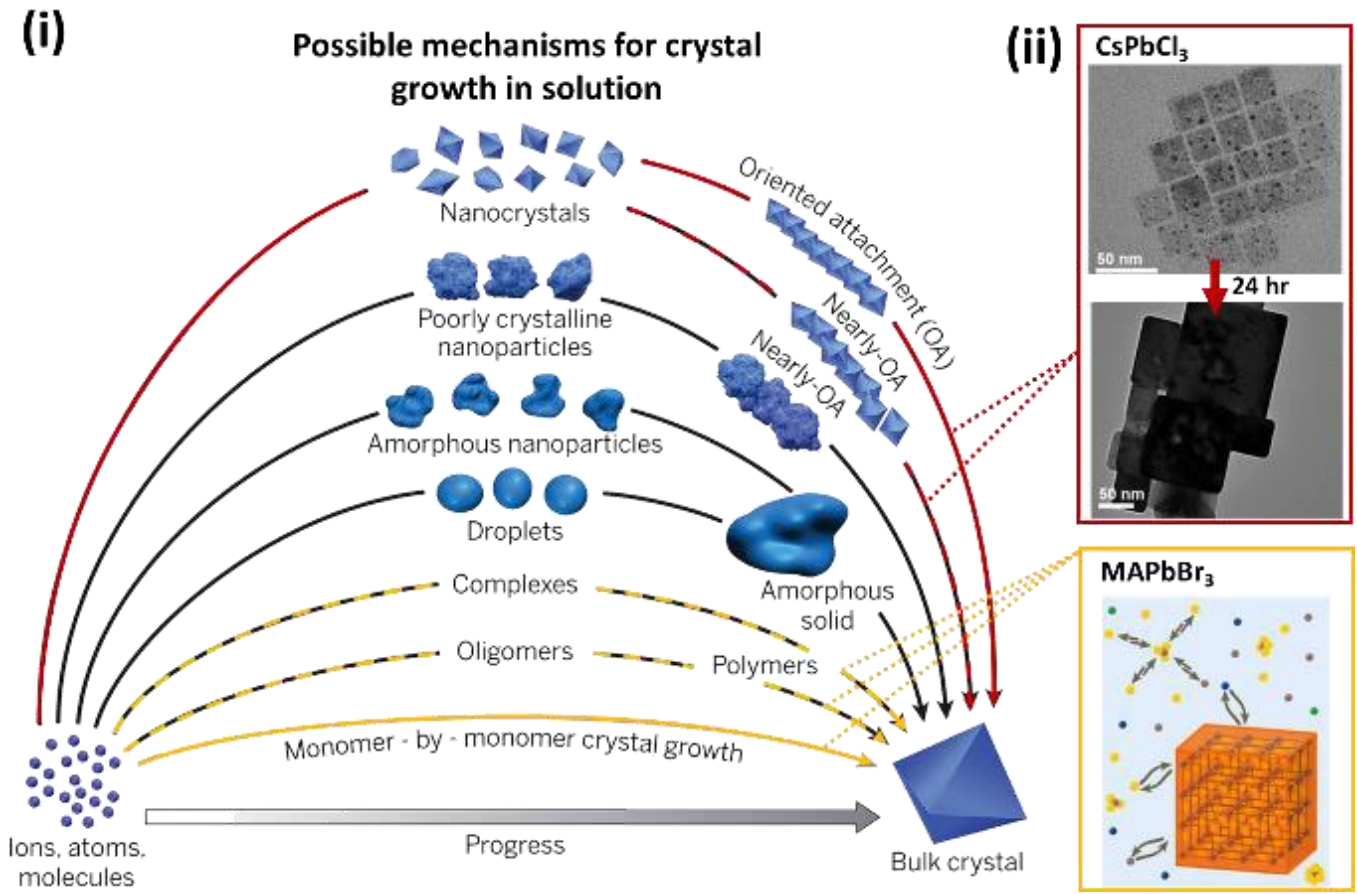

Figure 7: (i) Pathways to crystallization by (*yellow* path) monomer-by-monomer addition; *(black* path) crystallization via amorphous (or poorly crystalline) phase; (*red* path) crystallization by nanocrystalline oriented attachment.[78] (ii) (Top) indications for the *red* path in HaP: $CsPbCl_3$ nano-crystal growth that indicates growth via self-assembly of nanocrystals[80] (Bottom) indications for the *yellow* path in HaP: $MAPbX_3$ growth mechanism indicating a monomer-by monomer crystal growth when using the HRS technique. A recent publication of $MAPbBr_3$ growth mechanism strongly suggests growth via oligomers/complexes when using VSA or CSS techniques.[81] *HRS, VSA and CSS are defined in the next section)*

## 2.2. Practices in bulk crystal growth of HaPs

HaPs can be grown via one of the following methods (see also Figure 8 for illustration):

(1) <u>Vapor-Saturation of an Anti-solvent (VSA)</u> [82,70,83,2,5] : Vapors of an anti-solvents slowly diffuse into a saturated solution, where the latter is based on a solvent that is miscible with the anti-solvent but has a much lower vapor pressure. While crystal formation can occur at room temperature, heating the anti-solvent bath (to allow a higher anti-solvent residual vapor pressure) can accelerate the crystal growth. If the vapor pressure is too high, placing filter paper on top of the inner flask opening is a possible way to reduce the diffusion rate of the anti-solvent. All types of HaP single crystals can be formed with the VSA method. The grown crystals are usually of high transparency (a measure of crystal quality) and it is the method I used the most for my studies.

(2) <u>In-situ Compound Introduction (ICI)</u>[84] : N-methylformamide (NMF) in the highly acidic environment of an aqueous hydro-halide acid (HI, HBr, HCl) is gradually decomposed to methylammonium halide (MAI, MABr or MACl) and formic acid. Since a $PbX_2$ salt is already dissolved in the NMF, the solution spontaneously reaches the saturation limit and after hours to days, $MAPbX_3$ crystals start to appear. Crystals will keep growing for several days to weeks. Heat will accelerate the hydrolysis, and thus the crystal growth reaction. Crystals grown in this method are usually highly transparent; however, the grown crystals are limited to the **MA**$PbX_3$ set.

(3) <u>*Heating* of a solution when having a Retrograde-Soluble compound (HRS)</u>[2,85–87] : In aprotic-polar solvents (e.g., DMF, DMSO) and in a certain temperature range, HaPs are found to have an inverse solubility trend with temperature (i.e., solubility decreases as temperature increases). Using this, HaP single crystals can grow within minutes to hours from a gradually-heated saturated solution. Due to the rapid crystal growth, this method is widely used. However, the rapid crystallization usually results in less transparent crystals.

(4) <u>*Cooling* of a Saturated Solution (CSS)</u>[88] : When HaPs are dissolved in aqueous HX acids, HaPs can be formed when slowly cooling such saturated solutions. In practice, it was found more difficult to grow large crystals with this method, so this method may be useful when size is less of an issue.

In some cases, different additives can be used to further saturate, stabilize or dissolve remaining colloidal particles after filtration. For VSA, HX acids can be used to improve the solubility of the precursors and thus further saturate the precursor solution.[83] For HRS, formic acid is used to improve the stability of the solution and eliminate undesired colloidal particles remaining after filtration.[79] Hypophosphorous acid ($H_3PO_2$) is commonly used to stabilize HI to prevent its oxidation and formation of $I_3^-$ species in the solution.[5] In VSA and ICI, an optimum amount (usually 1-5% v/v) of $H_3PO_2$ (50% aqueous) was also found to reduce the rate of nucleation, thus resulting in fewer nucleation centers and larger crystals.

The precursor ratio also seems to be important. For $MAPbX_3$ crystals, the precursor ratio is usually 1:1 ($PbX_2$:MAX). For $CsPbX_3$, however, the precursor fraction of $PbX_2$ usually needs to be ~ 2-4 times larger than that of CsX in order to avoid precipitation of $Cs_4PbX_6$.[2,81] Further discussion on $CsPbBr_3$ crystal growth will be elaborated in section 2.4.

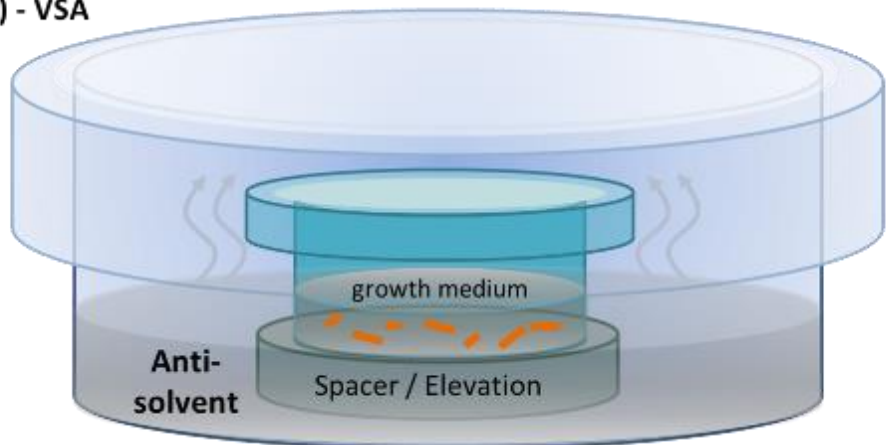

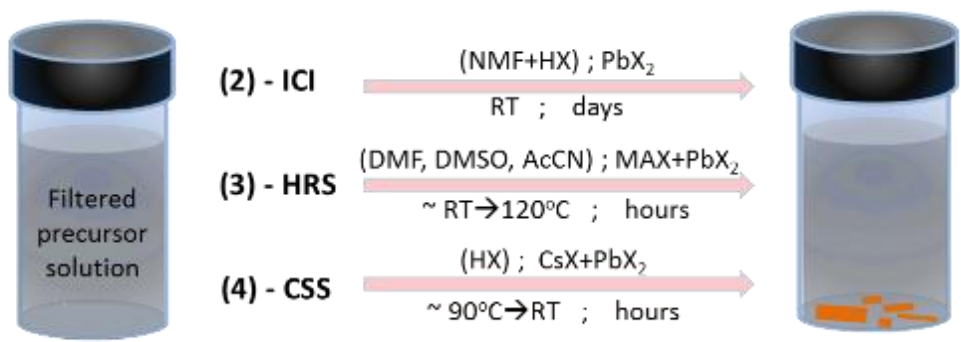

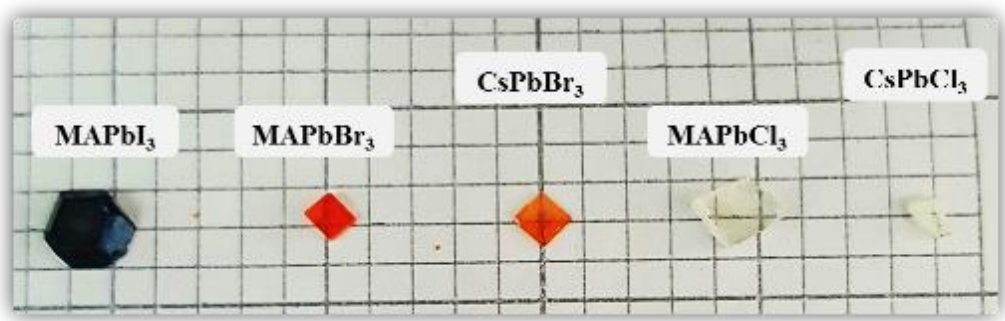

Figure 8: (i) Illustration of the commonly-used VSA setup to grow the different crystals. The filtered precursor solution is placed in a clean crystallization flask and covered with a filter paper and a glass petri dish on top to reduce anti-solvent vapor diffusion. The covered crystallization flask was then placed inside a deeper, flat-bottomed, glass dish, which contained the anti-solvent. Different architectures are possible, but a wide and flat crystallization flask has practical advantages (see main text). (ii) Illustration of other less commonly-used crystal growth methods (ICI, HRS and CSS). The general set of conditions are written above/below the arrows. (iii) Photos of HaP solution-grown crystals. All the shown crystals were grown using the VSA method (except $MAPbCl_3$ – HRS). The dimensions of the scale squares are 2x2 $mm^2$.

Beside the crystal size and quality, the crystal morphology can depend on the growth method as well as the anti-solvent in the VSA method. The MAPbCl$_3$ and MAPbBr$_3$ crystals are usually cuboidal with {100} faces. MAPbI$_3$ crystals, on the other hand, mostly result in diamond-shaped hexagonal crystals with the most developed faces being {110}, {002} and {112}. Specifically, when using the VSA method with MeCN(+HI) as a solvent[5], use of ethyl acetate as an anti-solvent will result in diamond-shaped hexagonal crystals. When the anti-solvent is diethyl ether, the MAPbI$_3$ crystals grow in a cuboidal shape with easily-distinguished crystallographic orientations (4x{110} and 2x{002}).

Regarding crystallization dishes, it is usually advised to use dishes with a wide aperture to allow an easy access to the grown crystals while they are still in the solution. Usually I used so-called 'crystallization flasks' made of Pyrex. In all the methods, except for ICI, it is important not to create strong gradients within the solution itself. A wide aperture dish may help here as well, since the solution will spread over a larger area with less height, making significant gradients less probable.

## 2.3. Characterizations of grown crystals

Crystallographic, elemental, thermal and optical profiling can be used to build a complete profile of the bulk of a newly-grown material. This will include the crystal structure, the elemental composition of the structure, the phase transitions and stability of the structure and information on optical transitions that can occur (including the bandgap ($E_g$) as well as the energetic positions of the valence band maximum (VBM) and conduction band minimum (CBM)).

If such profile is already known and what is needed is a simple identification due to first repetition attempts or changes in the growth technique, powder *x-ray diffraction* (XRD) is usually used. Specular diffraction from the natively-grown facets are useful for identifying the crystallographic orientation of an exposed facet. Where needed, the degree of structural defects (impurities or disorder) can be checked via thermal analysis or high-sensitivity elemental analysis. Degree of crystallinity ('mosaicity') or polycrystallinity can be checked via a pole-figure (or 'rocking curve' diffraction) from a characteristic facet of the grown crystal.

Based on the above, Table 2 summarizes chosen methods, available in the Weizmann institute or in Israel (with one exception – PDS), that are used or been considered for the characterization of crystals. None of these methods require electrode deposition for electrical connections, but some are destructive. More specified and advanced characterization methods (e.g., neutron diffraction) are possible, but usually require collaborations abroad and specified facilities (such as synchrotron facilities). Exact details on the techniques will be mentioned whenever relevant in the text or referenced to a specific publication.

In section 2.4 examples on some of those methods listed in Table 2 will be presented.

Table 2: The capabilities and their limitations (+ strong point; − weak point ; ± can be a strong or a weak point) for general profiling and identification of crystals. Except for PDS, the methods are limited to what is available at the Weizmann institute (WIS) or other institution in Israel if absent in the WIS (true for the date of this report). The cells colored in gray refer to 'advanced' methods and are usually complimentary to the more 'basic' ones.

| Property | Used method | Location | Limitations |
|---|---|---|---|
| **Structural (crystallographic) identity of a bulk crystal** | Powder x-ray diffraction (XRD) | WIS | ± Medium diffraction angle resolution (depends on the setup)<br>− Not sensitive to very low density of phase impurities<br>− Destructive |
| | Single crystal XRD | WIS | + Determines the exact lattice parameters<br>± Requires a truly single crystal |
| **Crystallographic orientation of an exposed (natively grown) facet** | Specular surface-XRD | WIS | ± Sensitive to the manual alignment of the crystal<br>− Requires well-defined flat surface<br>+ At high angles, can distinguish between similarly-spaced crystallographic orientations.<br>− Low diffraction spatial resolution |
| | *Electron back-scattered diffraction* (EBSD) | Bar Ilan University | + High spatial resolution (few microns)<br>− May damage the samples due to e-beam<br>− Low diffraction resolution |
| **Degree of crystallinity/disorder** | Pole-figure/Rocking curve XRD | WIS | + Measures mosaicity in a crystal<br>− Low spatial resolution |
| | Channeling of high-angle back-scattered electrons in the *scanning electron microscope* (SEM) | WIS | + High spatial resolution<br>− Not sensitive for twin surfaces<br>+ Can verify polycrystallinity<br>− Requires high-density material (e.g. Au). Not clear if for with current instrumentation, sensitivity is insufficient for HaPs |
| **Elemental analysis** *(Determination of the elemental ratio in a chosen material)* | *Energy-dispersive X-ray spectroscopy* (EDS) | WIS | + Very high spatial resolution<br>− May induce damage to the sample due to e-beam<br>− Inaccurate for low-atomic number elements (oxygen, nitrogen)<br>− Elemental sensitivity ~ 1% atomic percentage. |
| | *Wavelength-dispersive x-ray spectroscopy* (WDS) | Hebrew University | + Similar to EDS but with much higher spectral resolution.<br>+ Good for elements that have low energetic spread.<br>± Elemental sensitivity ~ 0.1 % atomic percent |
| | *Inductively coupled plasma mass spectrometry* (ICP-MS) | WIS | + Very high elemental sensitivity (ppb)<br>− Cannot detect organic compounds<br>± To see halogens an optic ICP is required (can be found at the Faculty of Agriculture) |

| | | | | |
|---|---|---|---|---|
| | | | − | Low/no spatial resolution |
| | | | − | Destructive |
| | *X-ray photoelectron spectroscopy* (XPS) | WIS | + | Accurate surface analysis |
| | | | + | Sensitive to oxidation state of the elements |
| | | | ± | Can be destructive (sample dependent) |
| | | | − | Requires UHV – will not fit volatile samples |
| | *Rutherford backscattered spectroscopy* (RBS) | Bar-Ilan University | **For high detection resolution and depth profiling** | |
| | | | + | Accurate stoichiometric ratio without a standard reference |
| | | | + | Detection limit ~ 0.001 at% (ppm) |
| | | | + | Depth profiling resolution ~5 nm |
| | | | + | Lateral resolution ~ 0.5 mm |
| | | | + | Non-destructive. |
| | *Secondary ion mass spectroscopy* (SIMS) | Tel Aviv University | + | Detection limit ~ sub-ppm – ppb |
| | | | + | Depth profiling resolution ~1 nm |
| | | | + | Lateral resolution ~ 1-10 μm |
| | | | ± | Requires ultra-high vacuum (UHV) |
| | | | − | Destructive |
| | *X-ray photoelectron spectroscopy* (XPS) | WIS | + | Accurate surface analysis |
| | | | + | Sensitive to oxidation state of the elements |
| | | | ± | Can be destructive (sample dependent) |
| | | | ± | Requires UHV – not suitable for volatile samples |
| **Thermal analysis** (Measures melting/ decomposition and phase transition temperatures) | Thermogravimetric analysis (TGA) | WIS | ± | Open chamber: results may be sensitive to atmosphere |
| | | | + | Allows analyzing volatile compounds during decomposition and its percentage in the structure (error ~1%) |
| | | | − | Destructive |
| | differential scanning calorimetry (DSC) | WIS | ± | Closed chamber*:* <ul><li>Insensitive to atmosphere.</li><li>Limited if there are volatile degradation products due to temperature.</li></ul> |
| | | | + | Highly sensitive to phase transition temperature and enthalpy |
| | | | ± | Sensitive to disorder |
| | | | − | Usually destructive |
| **Optical transitions** (bandgap/luminescing transitions) | UV-Vis-IR optical transition/ reflection | WIS | ± | An indirect technique to measure absorption |
| | | | ± | In single crystals – sensitive mostly to absorption by tail states |
| | Photothermal deflection spectroscopy (PDS) | Cambridge | + | A direct absorption measurement |
| | | | + | Sensitive to absorption onset |
| | | | + | Sensitive to in-band transitions |
| | Photoluminescence | WIS | ± | Sensitive for detecting highly-luminescent phase impurities |
| | | | + | Can be sensitive to transitions from defect states. |
| **Band alignments** (VBM, CBM and work function) | Ultraviolet photoelectron spectroscopy (UPS) | WIS | ± | Can determine both VBM and Fermi level. CBM is determined by adding up the measured optical bandgap. |
| | | | ± | Requires UHV vacuum – and a relatively conducting substrate |
| | | | ± | Highly sensitive to surface impurities. |
| | Contact potential difference (CPD) | WIS | + | Can determine the surface Fermi-level and under illumination the bulk quasi Fermi-level. |
| | | | ± | Highly sensitive to surface impurities. |

## 2.4. Growth of pure CsPbBr$_3$ crystal from solution

*This section summarizes a published work in Crystal Growth and Design and can be further explored via Ref. 2. I will share some basic characterization results as an example for application to Table 2. In further sections, where it is possible to refer to a publication in which such results are presented, basic characterization results will be omitted, while only characterization results that are critical for discussions will be presented.*

Prior to this work, it was demonstrated that single crystals of hybrid organic–inorganic HaPs (where A is MA or FA), with mm to cm dimensions, can be grown using different solution-based methods at low-temperature (around or below 100°C). All-inorganic CsPbBr$_3$ has recently gained attention as a stable HaP,[47,48] but hitherto no method was available to prepare macroscopic crystals by the low temperature solution methods (that are used to make films for photovoltaic cells). The only reported procedure of making mm-size single crystals of these materials is by Bridgman growth, where precursors are melted (above 600°C) in a quartz tube and passed through a multi-temperature zone tube furnace.[76,89]

As will be shown in section 4.4, organic elements in these compositions act as 'weak links' and decrease the stability (and thus reliability) of the entire structure, as was demonstrated in a comparison between the MAPbBr$_3$ and CsPbBr$_3$ HaPs.[48] It was also shown that changing to the fully inorganic compound (i.e. CsPbBr$_3$) does not necessarily exact a price in photovoltaic performance.[47] To be able to use single crystals to study CsPbBr$_3$ and compare it with its organic-inorganic hybrid analogs, the crystals should preferably be prepared under conditions that are comparable to those of the thin films used in device work.

Based on the phase-diagram of PbBr$_2$:CsBr compositions,[90] in addition to the 1:1 composition (CsPbBr$_3$), a 1:4 CsBr-rich Cs$_4$PbBr$_6$ or 2:1 PbBr$_2$-rich CsPb$_2$Br$_5$ compositions are expected to be formed (and are undesired). One of the major challenges was to grow exclusively the perovskite phase - completely avoiding the competing phases. Here I present two successful synthetic growth strategies to form 100% pure CsPbBr$_3$ single crystals (that later were also used for CsPbCl$_3$): VSA and HRS.[2] An important factor in this growth is that the solvent in which CsPbBr$_3$ was grown was polar-aprotic (DMSO), which allowed further growth of mixed (Cs$_x$MA$_{1-x}$)PbBr$_3$ (see section 2.5).

Crystal growth

At first, a solution of 0.4 M of $PbBr_2$:$CsBr$ with a 1:1 ratio was prepared in DMSO, followed by stirring overnight on a hotplate set at 50 °C. Then MeCN or methanol was added dropwise until an orange precipitate, which appeared with each addition, did not dissolve. Due to much lower solubility of CsBr, the final overall $PbBr_2$:$CsBr$ precursor ratio is estimated at ~1:2. It is interesting to note that the solubility of CsBr in DMSO increases when $PbBr_2$ is added to the solution – probably due to complex formation. A later publication verified this hypothesis.[81]

The crystallization setup was based on the VSA method, identical to that presented in Figure 8(i). The anti-solvent (similar to that added to the precursor solution) was MeCN or methanol. It is important to note that whenever the precursor ratio is 1:1 (for early attempts), the initial resulting crystals are highly-fluorescent, yellowish $Cs_4PbBr_6$ crystals followed by deep-orange $CsPbBr_3$ (similar to what is shown in Figure 9(ii)). We also tried $H_2O$ as an anti-solvent, but unlike MeCN and MeOH, the initial precipitating crystals are $CsPbBr_3$ and later colorless $CsPb_2Br_5$ (Figure 9(iii)). A reason for this phenomena was suggested by Liu et al.[81] showing that a crystal will grow according to the coordination number $Pb^{2+}$ atoms possess in solution. Thus $Pb^{2+}$ in DMSO will coordinate with 6 nearest neighbors in the precursor solution – similar to the $Cs_4PbBr_6$ and $CsPbBr_3$ structures; on the other hand, in $H_2O$, $Pb^{2+}$ will coordinate with 8 nearest neighbors – similarly to its coordination in the $CsPb_2Br_5$ structure.

With a different strategy (similar to that presented in Figure 8(ii)), but from the same precursor solutions, an HRS crystallization also resulted in successfully grown $CsPbBr_3$. In the HRS case, it was more difficult to isolate the $CsPbBr_3$ phase, since the $Cs_4PbBr_6$ phase precipitated and had to be filtered out (Figure 9(ii)). Overall, the VSA with MeCN as an anti-solvent resulted in the largest and the most well-shaped crystals (Figure 9(i)); the VSA crystals were therefore further characterized.

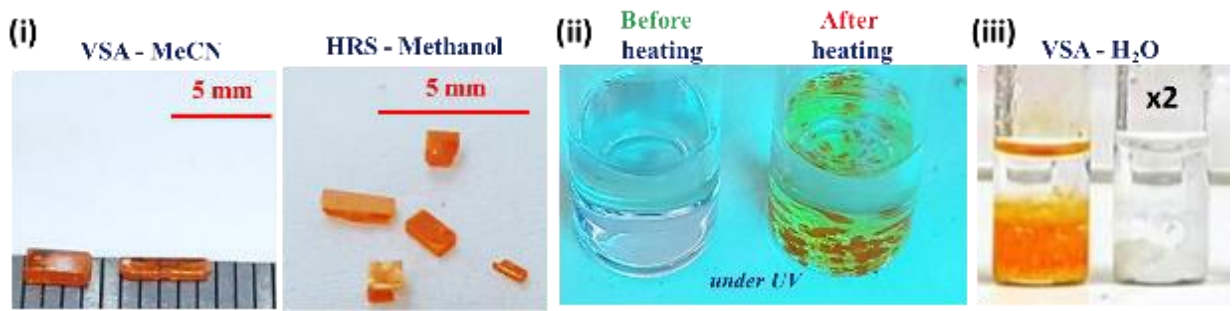

Figure 9: (i) $CsPbBr_3$ crystals grown by a (left) VSA method using MeCN-saturated solution on a 50°C hotplate and (right) HRS method using methanol additive at a heating rate of 10°C/hour up to 80°C from RT. (ii) before and after heating to 80 °C the HRS solution, showing the undesired $Cs_4PbBr_6$ fluorescent compound as a side product. (iii) Result of adding (dropwise) $H_2O$ as an anti-solvent to 0.45 M equimolar solution of CsBr and $PbBr_2$ in DMSO. An orange $CsPbBr_3$ formation followed by (double the amount of) a white $CsPb_2Br_5$ precipitate was verified by XRD.

Characterization

'Basic' characterization, following those mentioned in Table 1, will be presented in more detail than in subsequent sections. Extended details on the experimental setup can be found in the supplementary information of ref. 2.

In Figure 10 a typical structural characterization is presented. Orange crystals were pulverized and scanned in a Bragg-Brentano reflection geometry in a θ/2θ scan mode. The diffraction pattern (gathered intensity profile at different 2θ diffraction angles) was found to be highly similar to that published in the literature[76] from a melt-grown $CsPbBr_3$ at RT. The symmetry of the crystal is attributed to an orthorhombic *Pnma* space group – a common space group for a distorted perovskite structure (cf. Figure 2 in section 1.1). By aligning the natively-grown crystal surfaces parallel to the reflection plane between the x-ray source and x-ray detector (as shown in the photograph of Figure 10(i)), it was found that the natively-grown crystallographic orientations comprise two sets of {101} and one set of {020} parallel planes in a *Pnma* symmetry. Further verification of the identity of the grown single crystals was made via elemental (EDS – Figure 11) and thermal (TGA and DSC - Figure 12) analysis of the crystals, which showed that the orange crystals were indeed $CsPbBr_3$.

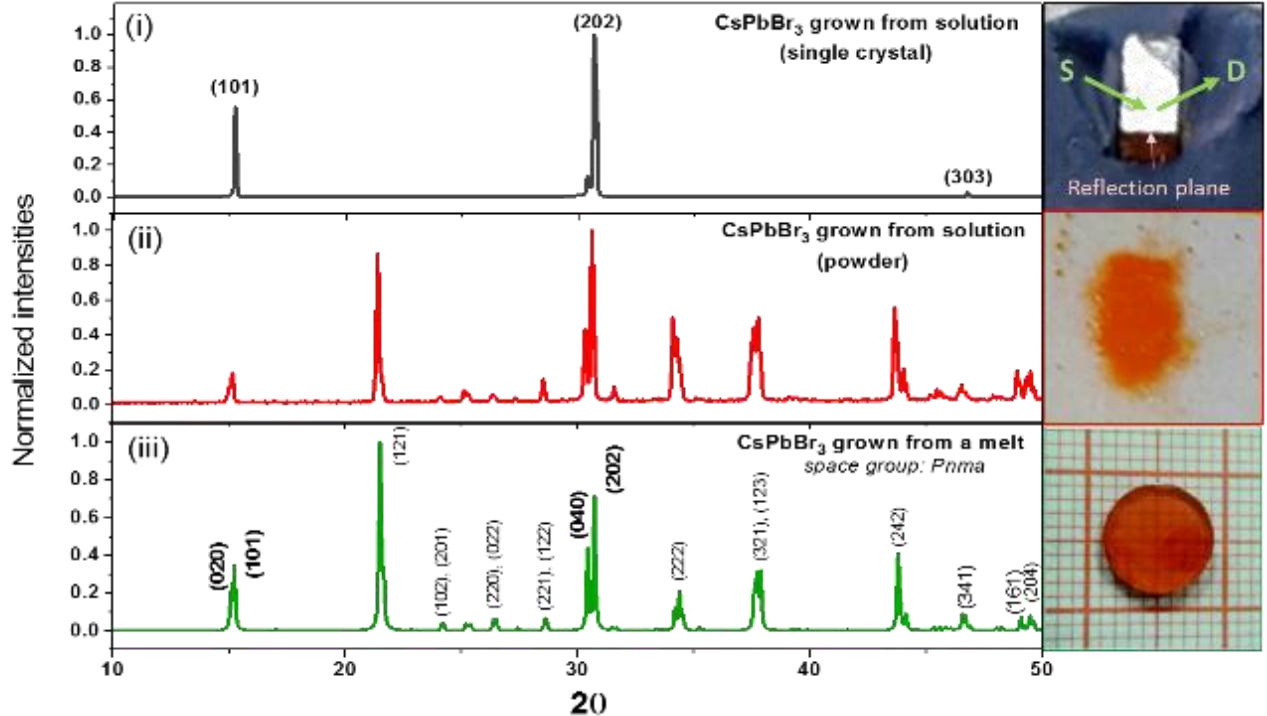

Figure 10: (i) Specular XRD pattern at RT of a $CsPbBr_3$ single crystal grown from solution (VSA/MeAC) indicating that the most developed face is a {101} face of a *Pnma* space-group. The green arrows represent the x-ray beam path from the source (S) to the detector (D). (ii) powder XRD of a pulverized solution-grown crystal; (iii) a simulated diffraction pattern based on a Crystallographic Information Framework (CIF) file gained by a complete single crystal XRD analysis taken from a thermally-grown crystal grown by Stoumpos et al [76].

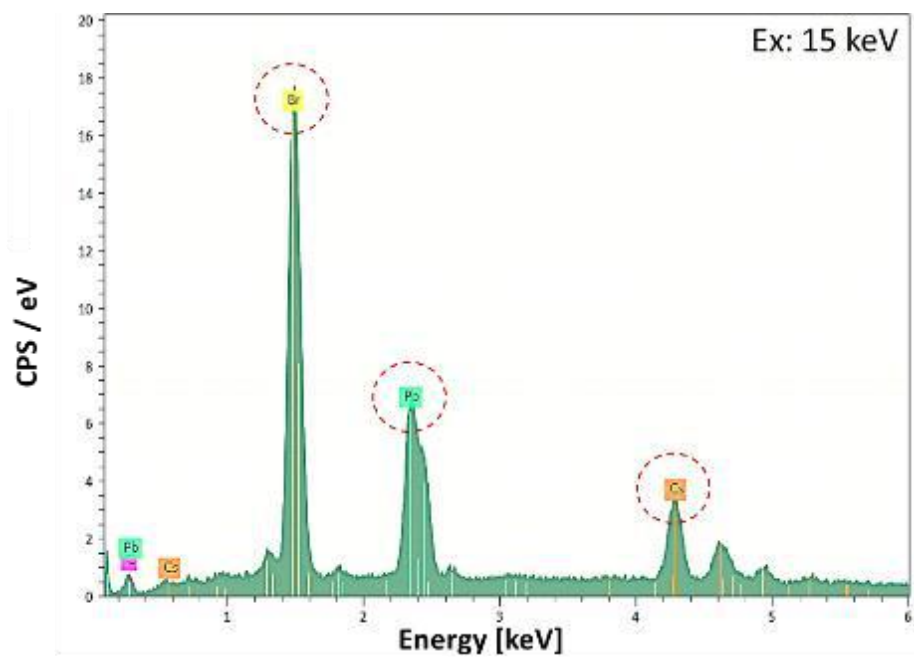

Figure 11: EDS spectrum of a CsPbBr$_3$ single crystal (excitation voltage was 15 keV). The EDS lines that were used for quantification are 1.48, 2.34 and 4.28 keV, for the Br, Pb and Cs respectively, as the most intense emission peaks for each element. The Cs:Pb:Br elemental ratio yielded (**1.00**±0.04):(**1**):(**3.4**±0.1) when taking the Pb quantity as unity. Since no beam damage is observed for CsPbBr$_3$ samples, the quantitative excess in the Br is attributed to reabsorption of Pb and Cs x-ray emissions by the 3 times more abundant Br atoms. No other elements were detected.

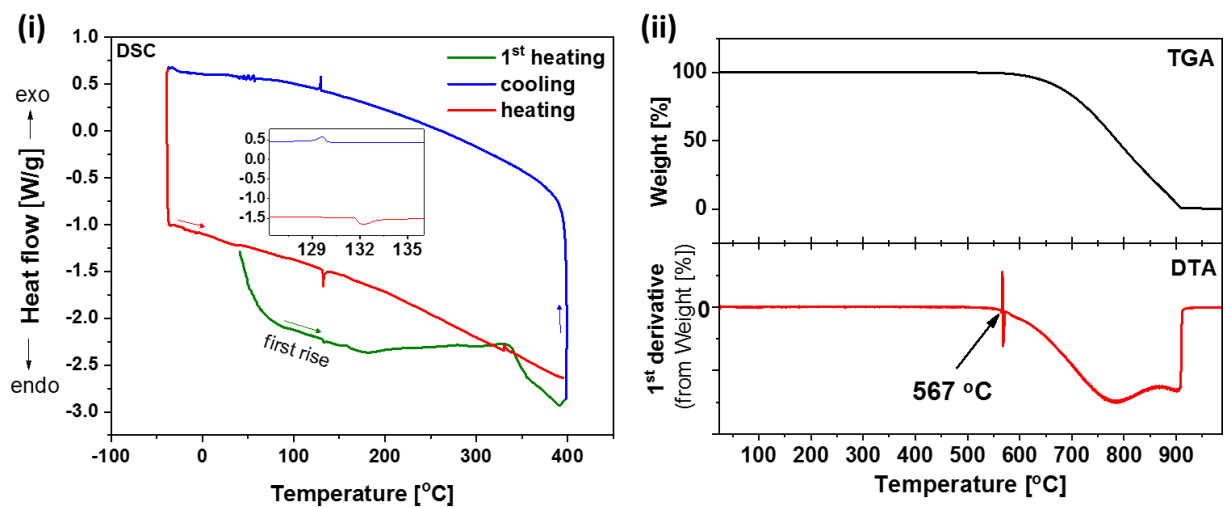

Figure 12: (i) DSC (ii) TGA and differential-TGA (DTA) of powdered samples, obtained from pulverizing solution-grown CsPbBr$_3$ crystals. The reported phase transitions of CsPbBr$_3$ are at 81°C (orthorhombic→tetragonal) and 131°C (tetragonal→cubic).[91,92] The transition at 81°C could not be detected. The transition at 131°C is clearly observed (see inset in (i)). The reported melting point at 567 °C is observed from the DTA data, which is also in agreement with melt-grown CsPbBr$_3$ crystals and spin-coated CsPbBr$_3$ thin films.[48,76] The green line in (i) is not a representative result of the material, since it is can relate to stress release, desorption of volatile species from the material or the crucibles, etc.

Besides the grown crystals, the identity of the strongly luminescent yellow side product was also investigated. We extracted the yellow, strongly fluorescent precipitate and compared it to pulverized solution-grown $CsPbBr_3$ crystals via PL and powder XRD (Figure 13). Both the side product and $CsPbBr_3$ crystals should luminesce with a green color, but clearly the total PL yield of the yellow phase is much higher (Figure 13(i)). A more quantitative analysis shows (under identical excitation and collection conditions) $10^4$ times stronger PL intensity (Figure 13(ii)) in favor of the yellow phase. By increasing the signal to noise PL from the pulverized $CsPbBr_3$ and normalizing to 1 the two spectra (Figure 13(iii)), their maxima (yellow - 516 nm $CsPbBr_3$-539 nm) are clearly distinct from each other by ~0.1 eV. It was previously reported that Pb-vacancies in a $Cs_4PbBr_6$ can cause a strong luminescence at the green region (~545 nm).[93] Powder XRD analysis of the yellow powder (Figure 13(iv)) indeed verified a $Cs_4PbBr_6$ structure (by comparing to ref. 94).

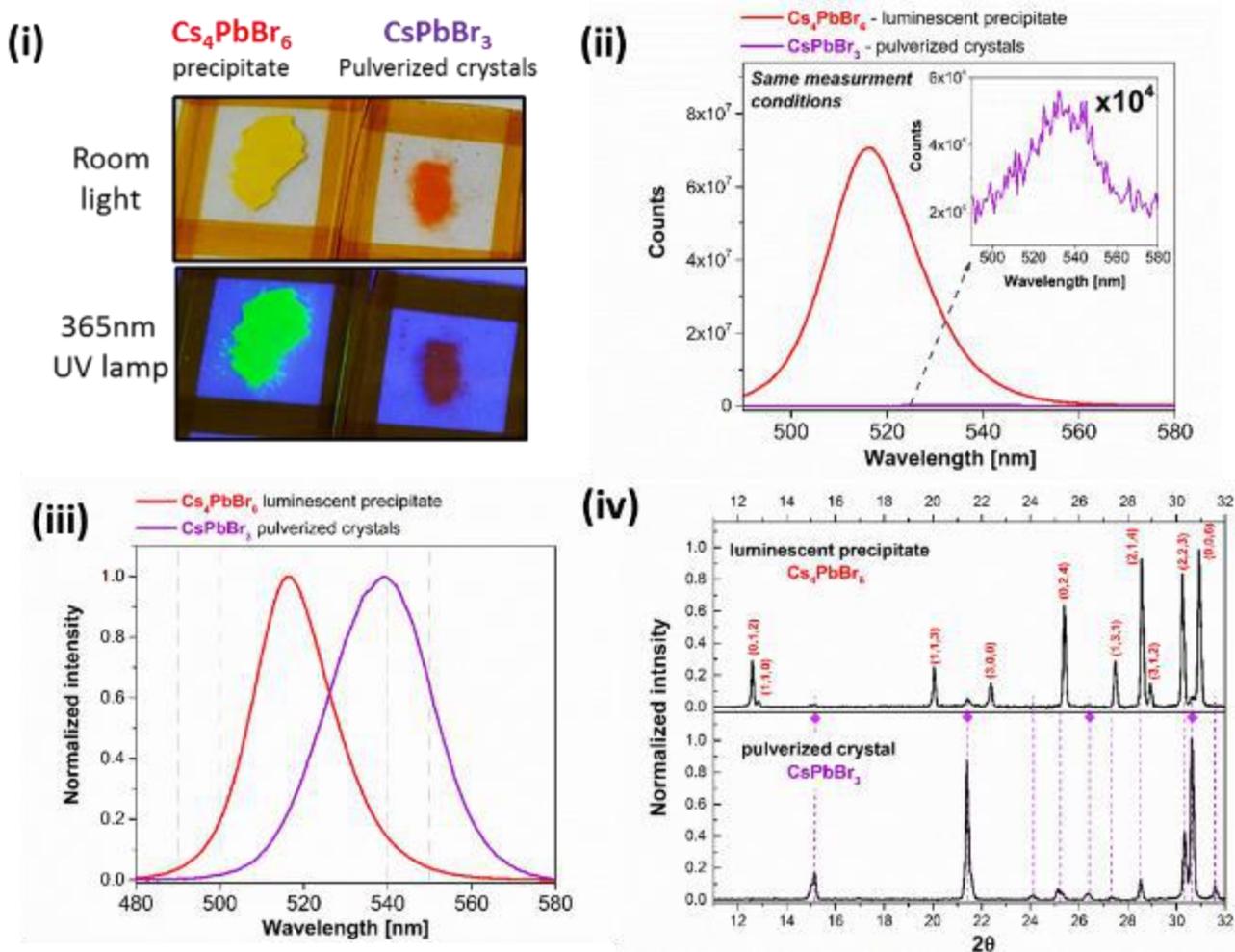

Figure 13: (i) Pictures of the extracted luminescent precipitate and pulverized $CsPbBr_3$ under room light and UV lamp. (ii) PL spectra of the two samples shown in (i) (using *identical* slit widths). (iii) Normalized PL spectrum of the two samples shown in (i) (using *different* slit width). In all three cases (i-iii), the excitation wavelength was 365 nm. (iv) Powder XRD pattern of the same powders shown in (i). The labeled red crystallographic (hkl) planes of $Cs_4PbBr_6$.[94]

PhD Thesis – Yevgeny Rakita
September, 2018

|35|

It is important to note that confined $CsPbBr_3$ quantum-dots in a CsBr matrix can show a similar effect in PL intensity and spectral blue-shift.[95] Since some observed peaks in the XRD pattern correspond to $CsPbBr_3$, both explanations for the difference in PL may hold. However, in order to get a 0.1 eV blue-shift due to quantum confinement, the $CsPbBr_3$ particles need to be at most 10 nm in diameter.[96] Based on the Scherrer equation, which allows to estimate the particle size based on the peak broadening, and a closer look on the (121) peak FWHM of the $CsPbBr_3$ in the $Cs_4PbBr_6$ pattern (@ $2\theta=21.4°$ → $\Delta(2\theta)\sim 0.2°$), the $CsPbBr_3$ particles are much larger than 10 nm. Therefore, it is more likely that the strong luminescence originates from the $Cs_4PbBr_6$ itself and not confined $CsPbBr_3$ crystals in a $Cs_4PbBr_6$ matrix.

*Dielectric, electric and optical characterizations*

To further compare between these solution grown crystal with the melt-grown $CsPbBr_3$ or other solution-grown HaPs was done by electrical (via impedance spectroscopy (IS) - Figure 14) and optical (by PDS and time-resolved PL, TRPL - Figure 15) analysis.

IS between two parallel {101} faces of a single crystal, using carbon-pasted contacts, was used to measure the bulk resistivity and relative permittivity (Figure 14). The average resistivity, $\rho$, of the solution-grown crystals was found to be $\sim 0.10 \pm 0.04$ GΩ·cm and relative permittivity $\varepsilon_r \approx 41 \pm 4$. The resistivity is an order of magnitude lower than that of crystals obtained by melt-based synthesis ($\sim$1 GΩ·cm).[76] The lower resistivity may be due to (a) unintentional doping (either by extrinsic impurities or by kinetically-frozen intrinsic defects), or (b) higher crystal quality which improves the charge mobility.

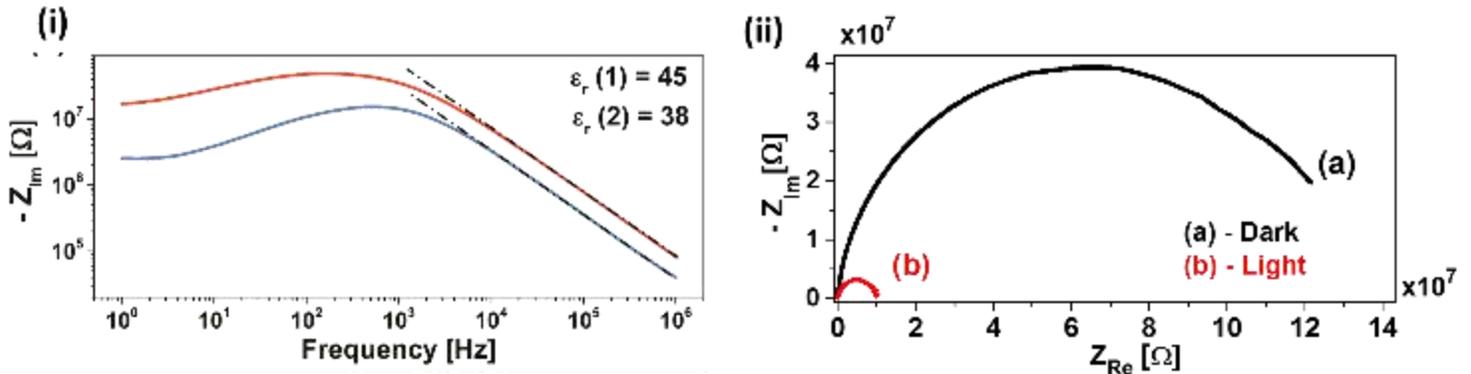

Figure 14: Impedance Spectroscopy (IS) of $CsPbBr_3$ single crystals that is used for determination of the (i) dielectric constant ($\varepsilon_r$) and (ii) resistivity ($\rho$) in the dark and under white LED illumination (5-7 mW/cm$^2$ – calibrated with a Si diode). Measurements were carried out between two {101} faces using carbon-paste electrodes with an alternating analyzing voltage of 0.1 $V_{AC}$. To find $\varepsilon_r$ we used the simple capacitor relation at high frequencies $C_{(\omega \to \infty)} = C_0 \cdot \varepsilon_r = \frac{1}{2\pi f \cdot Z_{im}}$, which can be reorganized to: $\log(Z_{im}) = -\log(f) - \log(2\pi C_0 \varepsilon_r)$. $C_0 = \varepsilon_0 \frac{A}{d}$, A is the contact area, d is the distance between the electrodes, $\varepsilon_0$ is the permittivity of vacuum, $Z_{im}$ is the imaginary impedance and f is the frequency. To find $\rho$, a Nyquist plot ($Z_{re}$ vs. $-Z_{im}$) was plotted, where the diameter of an equivalent RC circuit represents the sample resistance ($R = \frac{d}{A} \cdot \rho$).

Illumination (5-7 mW/cm2) decreases the resistivity by ~1 order of magnitude – a photo-response that is common to other HaPs and semiconductors in general.[97] We find that this effect is usually stronger for thin films than for single crystals. This is because in HaPs, including CsPbBr$_3$, the light absorption coefficient reaches ~ $10^5$ cm$^{-1}$,[98] the penetration depth is not more than 300 nm, so, except for some of the tail absorption, most of the light is absorbed in a small fraction of the total crystal thickness. Since the electrodes are across the entire bulk of the crystals and not at a single facet, this results in a much smaller photoresponse than if light were absorbed throughout the entire crystal. This is a common effect for highly-absorbing materials such as CsPbBr$_3$. A recent paper[99] comparing the PL yield and the photocurrent of CsPbBr$_3$ vs. Cs$_4$PbBr$_6$ showed a much stronger photoresponse for the less luminescent species. These two effects (PL yield and photoresponse) are related by the *exciton binding energy* ($E_B$). Upon illumination, the much lower value of $E_B$ in CsPbBr$_3$ (~30 meV) vs. Cs$_4$PbBr$_6$ (~350 meV) means that it is easier to split the exciton into a free electron and hole. Lower $E_B$ means easier exciton dissociation and, therefore, better flow of free charges.

$\varepsilon_r$, being a static (or low frequency) dielectric constant is clearly higher than the estimated optical one ($\varepsilon_{r\ (optical)}$ ~4).[100] Similarly, the optical dielectric constants of MAPbBr$_3$ and MAPbI$_3$ are found to be ~4.5 and 6.5,[101] while their GHz (microwave) dielectric constants are ~25 and 30, respectively[87].

From optical absorption data, obtained by PDS (Figure 15(i)), an Urbach energy of $E_u$=19 meV was derived from the slope of the logarithmic plot of the absorbance vs. the photon energy ($absorbance \propto \exp[\frac{E}{E_U}]$). This value is very similar to MAPbBr$_3$ single crystals (~19 meV)[102] and comparable to other values of high-quality optoelectronic materials (GaAs ~ 8 meV, crystalline-Si ~11 meV, MAPbI$_3$ ~ 15meV; CIGS ~ 25 meV)[32,68]. We also find that a thin film that was prepared by spin coating from a similar DMSO solution shows a higher Urbach energy (~36 meV; Figure 15(ii)), which implies that the crystal is of high quality. The CsPbBr$_3$ thin film fabrication procedure is reported elsewhere.[47] It is important to note that the absorption edge of the crystal shifted to lower energies and that its absorption coefficient seems to be much lower than the one from a thin film. This, however, is an artifact related to the thickness of the crystal that usually happens when comparing absorption of single macro-sized crystals with thin films (see further explanation in the caption of Fig. 15 and ref. 102).

Two different excited carrier lifetimes were found by TRPL (Figure 15(iv)): 4.4 ± 0.1 ns and 30 ± 3 ns, which are speculated to be the trap-assisted and free charge-carrier recombination-based decays, respectively.[103] Both lifetime regimes are significantly shorter than those reported for MAPbI$_3$ and MAPbBr$_3$ single crystals.[70] Other fully-inorganic photovoltaic-grade materials, CIGS and CdTe, show lifetimes of ~50 ns[104] and ~150 ns[105], respectively. In addition, a ~50 meV red-shift, which is observed in the PL decay over time (Figure 15(iii)), may indicate photon recycling[62,106,107]. An alternative reason might be diffusion and relaxation of the excited carriers to lower energy levels as the PL decays over time.

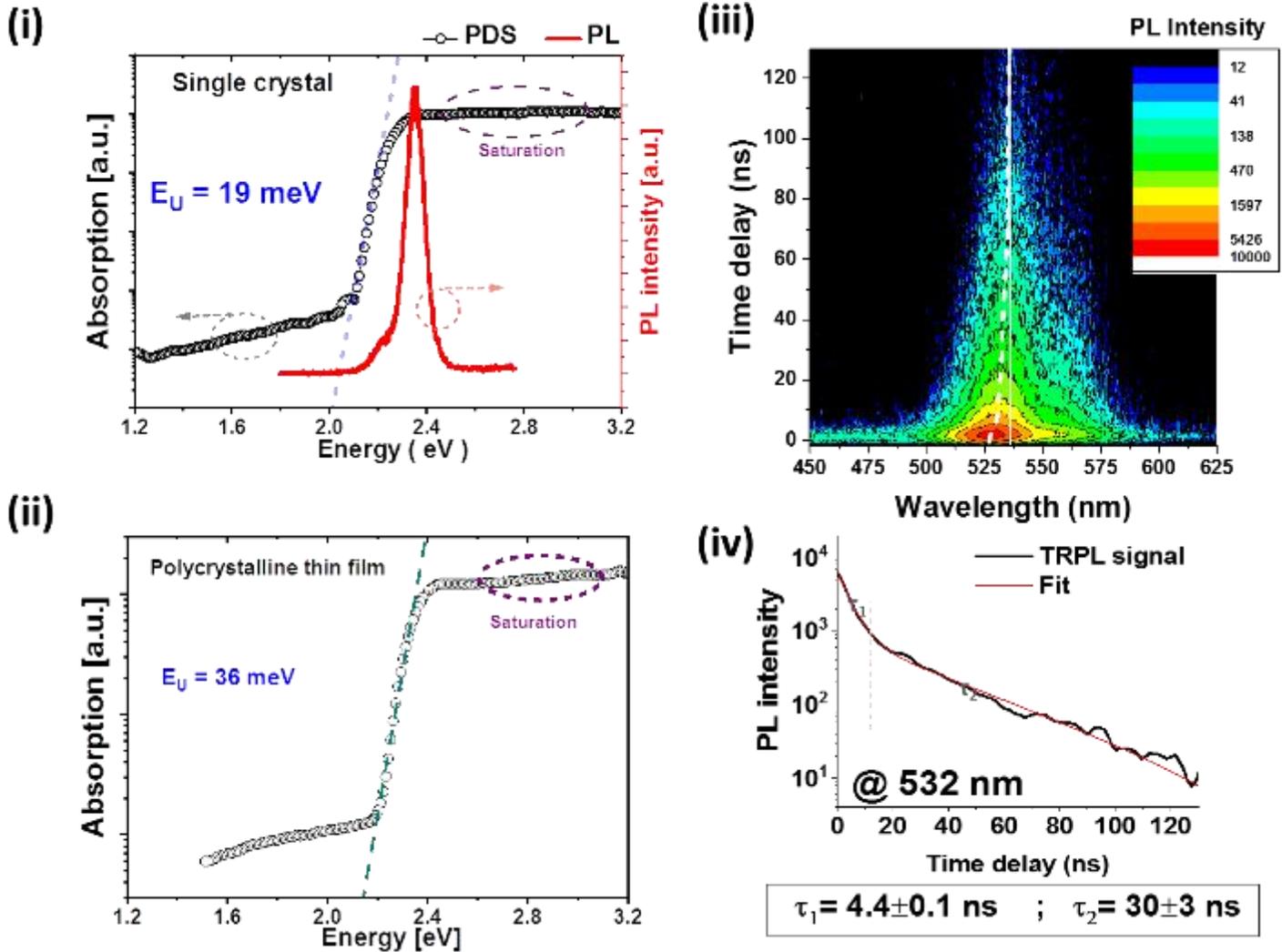

Figure 15: Optical absorbance vs. the excitation energy measured by PDS of (i) a solution grown CsPbBr$_3$ crystal and (ii) a spin-coated CsPbBr$_3$ thin film. The red plot in (i) is a (normalized) PL spectrum from the surface of a crystal. The Urbach energy (E$_U$) was derived based on $\alpha \propto \exp[E/E_U]$, which is the inverse of the dashed linear fits. (iii) A PL spectrum representation with respect to time delay and intensity. It shows a shift of ~50 meV in the maximum intensity peak over time towards longer wavelengths (see white dashed line), which can be attributed to photon recycling[62,106,107] or relaxation of the excited carriers to lower energy levels as the PL decays over time. (iv) PL intensity vs time at 532 nm from which the short and long lifetimes are extracted. $\tau_1$ and $\tau_2$ are speculated to be the trap-assisted ('monomolecular') and free-charge ('bimolecular') carrier-recombination based decays, respectively.[103]

Following the success of growing $CsPbBr_3$ from solution, further comparative analysis with HaPs having different A or X groups became possible. In the following chapters, different comparative analyses using single crystals of different $APbX_3$ compositions will be presented, including some results from mixed $(Cs_xMA_{1-x})PbBr_3$.

## 2.5. Growth of mixed crystals

Following reports that showed similar performance of solar cells based on $MAPbBr_3$ and $CsPbBr_3$ films,[47] *but* enhanced long-term stability for those made with $CsPbBr_3$[48] and enhanced stability when mixing Cs into $(FA,MA)Pb(I,Br)_3$ based films[49–51], understanding the impact of each component (A, B or X) on the overall nature of HaP crystals became of importance. It was previously shown that growth of mixed halide HaP single crystals is definitely possible, resulting in a colorful set of crystals, as shown in Figure 16.[108,109] Recently it was shown that mixing A cations can stabilize the perovskite structure. For example, in the case of $(Cs_xFA_{1-x})PbI_3$ with $0 < x < 0.1$:[46] $FA^+$ is too large to fit the lattice in a way to form $FAPbI_3$ perovskite phase and usually forms a non-perovskite phase at RT. $Cs^+$, however, is too small to form $CsPbI_3$ perovskite structure, so that their combination optimizes the preferred size and result in a 3D HaP
.

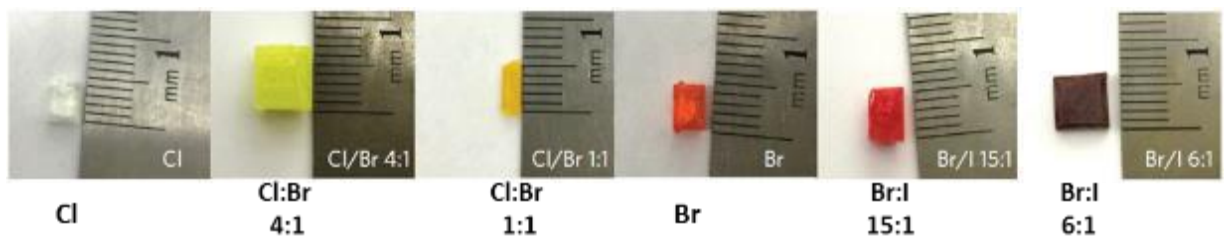

Figure 16: Photographs of single-halide and mixed-halide perovskite single crystals with different halide compositions. Image published by Y. Fang et al..[109]

Following our success in growing $CsPbBr_3$ crystals in solution, attempts to grow a complete series of $(Cs_xMA_{1-x})PbBr_3$ ($0 < x < 1$) were made in collaboration with (and led by) MSc. student Hadar Kaslasi. The complete study can be found in the Thesis work of Hadar Kaslasi. A publication on this work is in preparation.

Figure 17(i) shows such a series of $(Cs_xMA_{1-x})PbBr_3$ crystals, where the molar fraction, x, was determined via the Cs:Pb ratio from ICP-MS measurements and further verified from the weight loss of MABr, shown in TGA measurements (Figure 17(ii)). PL measurements (Figure 17(iii)) show small, but apparent, shift at the PL peak. The fact that the PL peaks are symmetric and do not split to a MA-and Cs-PbBr$_3$ spectra, suggest that the surface is a solid solution and not a heterogeneous system. Powder XRD results (Figure 17(iv)) show a gradual phase change from the higher symmetry cubic phase (MAPbBr$_3$-like), for compositions $C_1$ and $C_2$ - towards the lower-symmetry orthorhombic (CsPbBr$_3$-like) phase via a tetragonal phase - $C_3$ and $C_4$. The gradual shifts of the peaks around $2\theta \sim 30°$ and $54°$ are consistent with a solid solution nature of the mixed phase and not a heterogeneous mixed phase of CsPbBr$_3$ and MAPbBr$_3$.

It should be noted that elemental mapping of Cs, using EDS (can be found in H. Kaslasi's thesis), shows that in cases where an excess of an anti-solvent was added, a clear heterogeneity in Cs concentration may occur. CsBr has a lower solubility than MABr to begin with, which results in a faster depletion of CsBr from the solution, and requires that the anti-solvent amount will be such that the CsBr concentration will not be too different from its initial concentration. To conclude, growing *mixed* crystals require extra care to assure homogeneity.

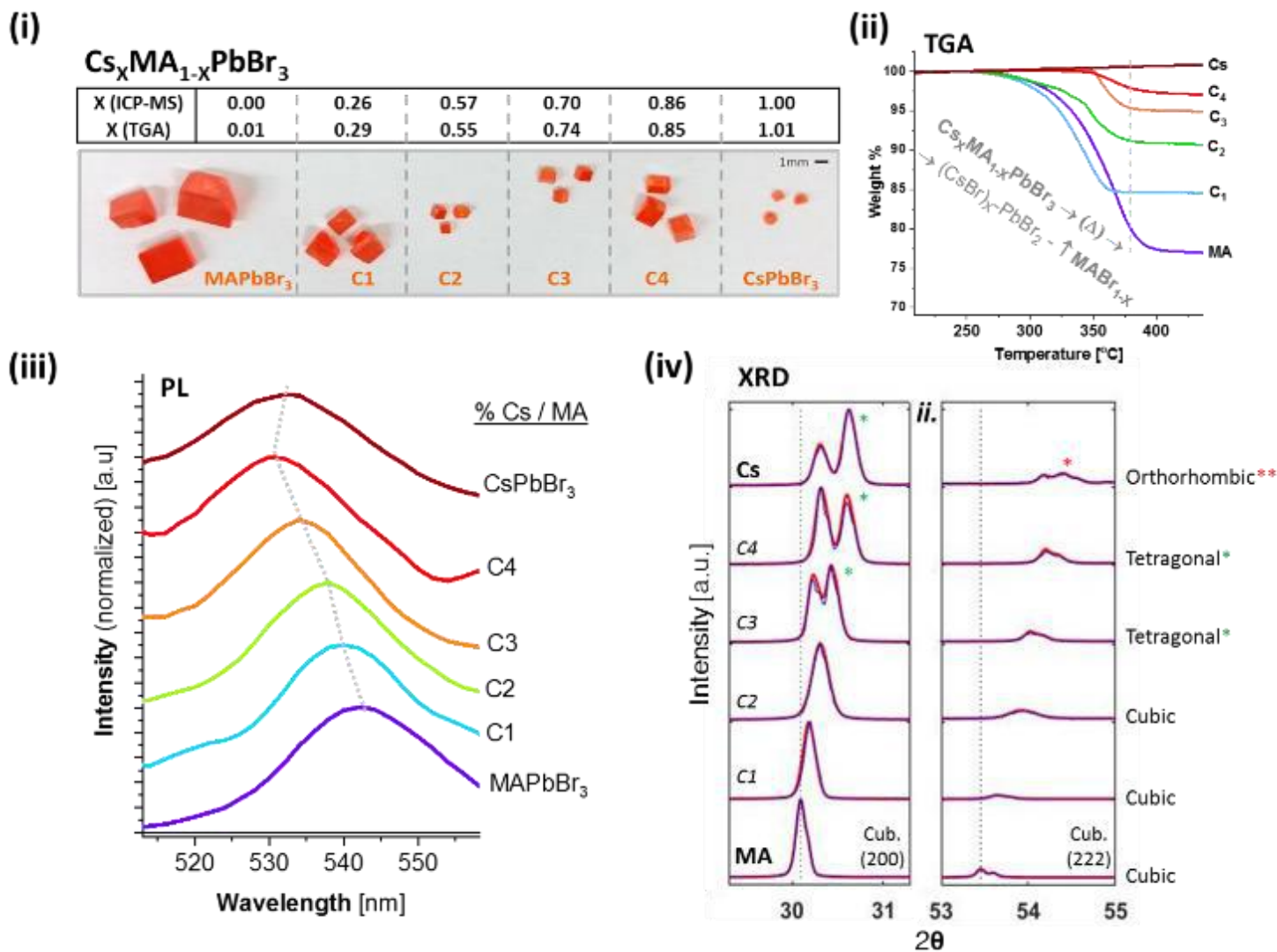

Figure 17 : (i) A photograph of pure (MA and Cs) and mixed ($C_1$-$C_4$) crystals grown by a VSA method including the result of compositional analysis. (b). A typical TGA result just after MABr evaporation and before $PbBr_2$ or CsBr evaporation. The composition analysis was calculated from the weight loss of MABr from a $(Cs_xMA_{1-x})PbBr_3$ composition. (iii) PL spectra measured from the crystal surfaces, showing small, but not negligible, change in the PL peak. The peaks neither broaden nor split, indicating that the surface is a solid solution and not heterogeneous. (iv) Powder XRD pattern (blue line) of pulverized crystals with changing Cs/Pb ratios, including the two controls ($MAPbBr_3$ and $CsPbBr_3$) and the fitted results to the most suitable space-group (red line) on top. The fittings suggest a cubic (*Pm-3m*) phase for $MAPbBr_3$ (as expected)[110], $C_1$ and $C_2$, a tetragonal (*P4/mbm* or *P4/mnc*) phase for $C_3$ and $C_4$ and an orthorhombic (*Pnma*) phase for $CsPbBr_3$ (as expected)[76]. The ranges around $2\theta$~30° and $2\theta$~54° were chosen deliberately, since they clearly show the change from cubic to tetragonal symmetry around the cubic-(200) peak at $2\theta$~30° (and its equivalents for the other symmetries) and between tetragonal to orthorhombic symmetry around the cubic-(222) at $2\theta$~54° (and its equivalents for the other symmetries). At $2\theta$~54° a *lower* shoulder (or sometimes if the signal-to-noise ratio is poor, an asymmetric broadening of the peak) is seen at slightly higher $2\theta$ angle, which is due to Cu-$K\alpha_2$ x-rays that are not filtered out from the Cu-$K\alpha_1$, which is more pronounced at higher diffraction angles (from Bragg's law).

# 3. The bond physics of HaPs

## 3.1. Theoretical concepts

The remarkable optoelectronic properties of HaPs triggered an investigation of their bond nature in an attempt to understand what makes them different from other semiconductors. As mentioned in section 1.1, the B-X bond and, to some extent, the [BX$_6$] octahedra interconnectivity in perovskites, play a crucial role in determining both optical and electronic properties. For HaPs, a nice example is the clear change in the color of a HaP crystal when replacing the halide group with a different one (Figure 16). A less significant change occurs when replacing the A group (see Figure 17(i) and (iii)), which relates both to a general bond distance and structural symmetry (i.e., [PbBr$_6$] octahedra interconnectivity) as can be deduced from the XRD pattern shown in Figure 17(iv).

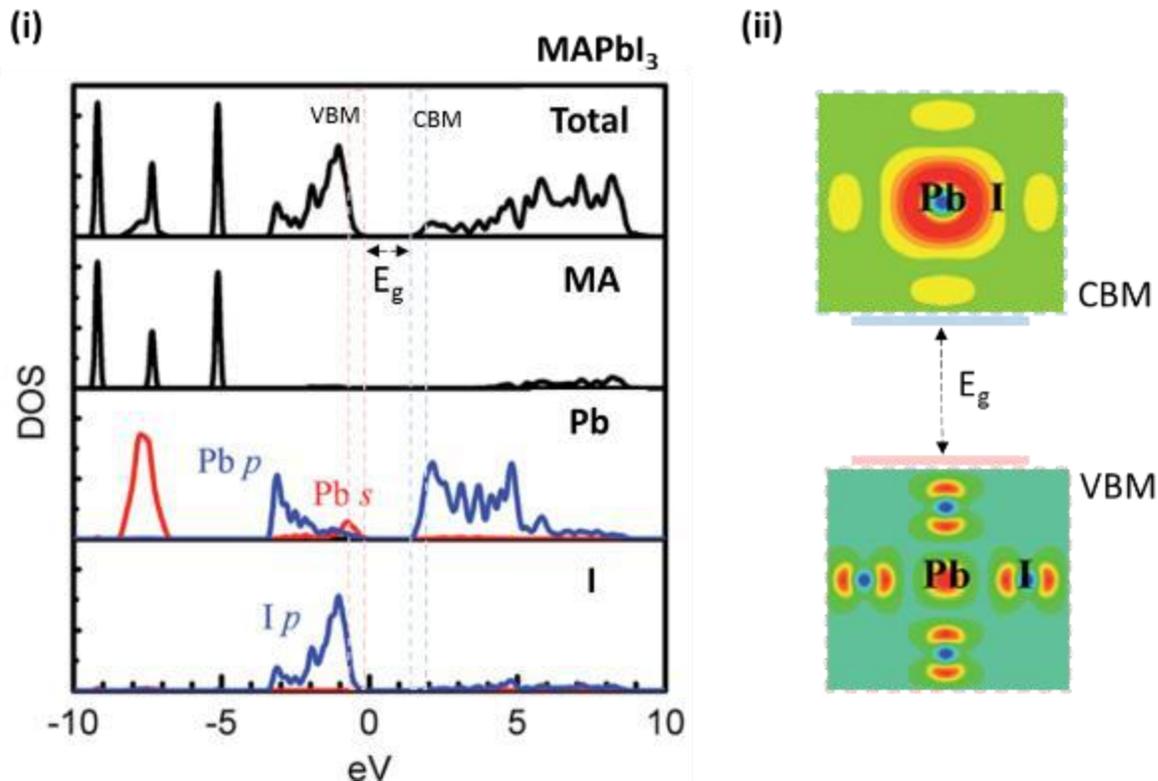

Figure 18: (i) Density of states (DOS) of MAPbI$_3$ and partial DOS of its MA, Pb and I constituents. The zero energy refers to the VBM. The A group (in this case, MA, but also true for other HaPs – see ref. 111) does not contribute states to the optical and electronical relevant states, namely the VBM and CBM. Pb($s$) and I($p$) orbitals contribute to the VBM and mostly Pb($p$) orbital contribute to the CBM. (ii) Partial charge density of the VBM and CBM. The VBM bond will be referred to as 'anti-bonding', as it shows a node in the charge density between the I and the Pb. The CBM will be referred as 'non-bonding' since the contribution from the X group is negligible compared to the Pb one. The bandgap is defined as the transition between the VBM and the CBM. Figures (i) and (ii) are borrowed from ref. 112 (cf. also section 3.4, Figure 31).

Density of states (DOS) calculations explain some of the bandgap variation with different X atoms. Figure 18 suggests that the relevant states for charge conduction and optical transitions, i.e., VBM and CBM, are composed from I and Pb orbitals, while the MA does not contribute to the VBM or CBM. The VBM shows an (unusual) 'anti-bonding' type of bond that includes Pb ($s$) and I ($p$) orbitals, while the CBM shows a 'non-bonding' type of bond, originating mostly from Pb($p$) orbitals. Figure 18 is for MAPbI$_3$ but represents many other HaPs as well (see ref. 111).

The less common situation of having an 'anti-bonding' VBM suggests that the electronic states that are associated with structural defects (such as vacancies, interstitials, surface states etc.), should be located at the VBM or CBM band edge, or even within it (see Figure 19). This gives rise to the idea of *'defect tolerance'* – structural defects that can be tolerated, i.e., do not negatively affect the main (opto)electronic properties.

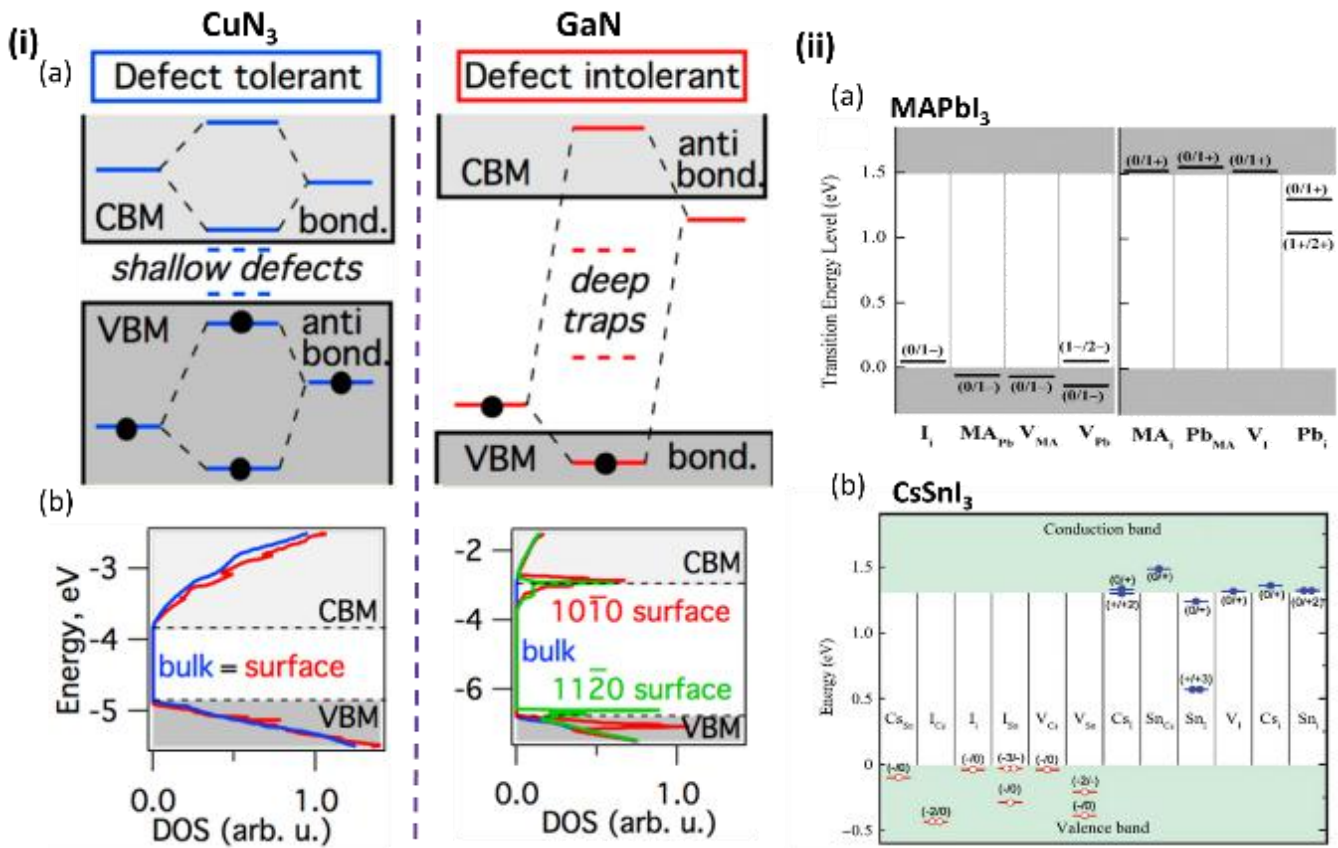

Figure 19: (i) Comparison between CuN$_3$, that possesses an *'anti-bonding'* VBM, and GaN, that possesses a *'bonding'* VBM.[113] (a) Schematic representation of defect formation levels, which were further verified by calculations (see fig. 4 in ref. 113). (b) Calculated DOS of the bulk and surface, supporting the idea that in the case of an 'anti-bonding' VBM, surface states are shallow/in-gap, and thus can be tolerated. The surface of Cu$_3$N is treated as oxygen-passivated as an energetically favorable state in ambient conditions. (ii) Calculated transition energy levels of intrinsic donors (right half) and acceptors (left half) in (a) MAPbI$_3$ [114] and (b) CsSnI$_3$ [115]. These two examples ((a) and (b)) support the idea that an 'anti-bonding' VBM in HaPs leads to shallow defect states (cf. also section 3.4, Figure 31). It should be noted that 'mid-gap' defect states in materials having an 'anti-bonding' VBM do exist, but their formation energy is significantly higher than 'shallow'/ 'in-band' defect states.[113,114]

In the following sections of this chapter, I describe experimental or experimentally-based evidence that verify and further explore the theoretical concepts introduced above. I will start with (section 3.2) measuring the average mechanical response (e.g., elastic moduli) of different Pb-based HaPs to question (experimentally) the dominance of the B-X bond in the valence band and to determine its average strength. I will continue with (section 3.3) comparing the electronic density around Pb atoms via solid state NMR to investigate the bond nature, meaning whether we should treat the interatomic bonds as covalent or ionic (or mixed). Further on (section 3.4), I use the measured elastic moduli to derive the 'deformation potential' of HaPs. Both the algebraic sign and the absolute value of the derived deformation potential will support the idea of 'defect tolerance'. Finally (section 3.5) the *intrinsic* bond nature of HaPs is correlated to an *intrinsic* PV related property, i.e., charge mobility. Based on a collection of empirically-derived constants of different polar semiconductors, I show how these are interconnected. Based on this analysis, I find that in low deformation potential materials, such as HaPs, the charge mobility is fundamentally limited to several orders of magnitude below that of more rigid polar semiconductors.

## 3.2. *Mechanical response of HaPs*

*Part of this work was published in MRS Communications and can be found in ref.* 1. *A broader set of results is presented in this section.*

As a first step in confronting theoretical concepts of what parts in a HaP structure are responsible for its VBM and, thus, its (opto)electronic properties, an investigation of its mechanical response was carried out. Based on theory,[114,116,117] the valence band (VB) possesses an anti-bonding orbital constructed from 6s Pb electrons, coupled with 3p, 4p or 5p orbitals for Cl, Br or I, respectively. In all cases the electrons of the A cation (MA or Cs) occupy orbitals with an energy that is too low to share valence electrons. Knowing this, it can be assumed that the bond stiffness should be determined by the Pb-X bond, regardless of whether the A cation is organic or inorganic. To check this point experimentally, we compared the bond stiffness (resistance to an elastic deformation) of equivalent crystallographic orientations ([100] direction in a cubic system) of different HaPs ($CsPbCl_3$, $CsPbBr_3$, $MAPbCl_3$, $MAPbBr_3$, $MAPbI_3$) using nanoindentation (Figure 20).

About the method:

Nanohardness and the indentation modulus were measured using an Agilent XP Nanoindenter. A Berkovich diamond indenter tip was loaded into the surface at a strain rate of 0.05 s$^{-1}$ to a depth of ~750 nm. The loading was done using the "Continuous Stiffness Measurement" (CSM$^{TM}$) mode [118] which gave the modulus and hardness continuously as a function of loading. The data were analyzed using standard Oliver and Pharr analysis[119]. The spacing between each indentation was at least 20 times the indentation depth (~ 800 nm). Due to the multiple indentation experiments and the low indentation diameter, the method delivers a close representation of the bulk's mechanical properties (independent of defects), meaning: Hardness (resistance to a *plastic* deformation), Young's modulus (resistance to a linear *elastic* deformation), and Creep displacement (resistance to mass flow due to a constant applied stress).

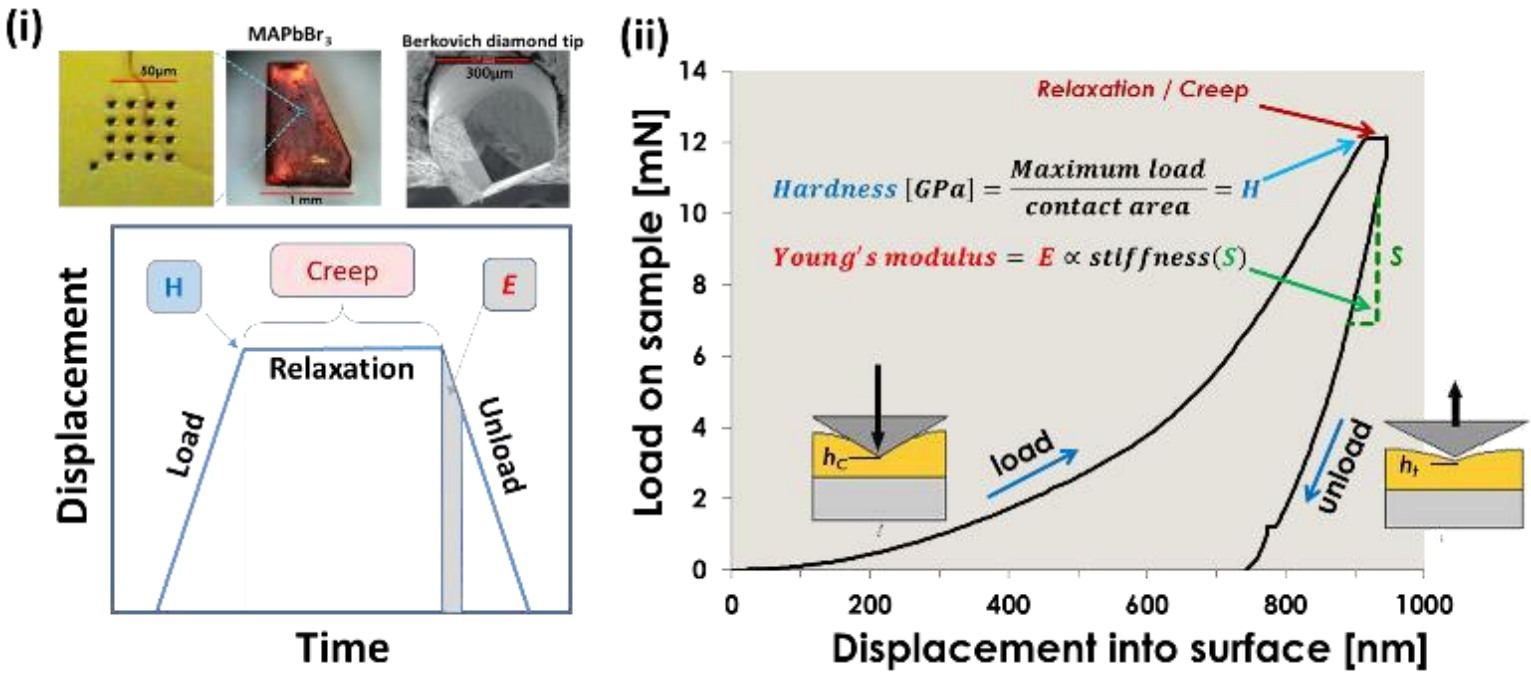

Figure 20: (i) Overview on the nanoindentation experiment: *top right* – a Berkovitch nanoindenter; *top middle* – indented MAPbBr$_3$ crystal – {100} surface; *top left* – MAPbBr$_3$ surface after indentation; *bottom* – typical displacement vs. time graph during indentation. (ii) A typical experimental result of load vs. displacement following nanoindentation. The hardness of the material can be gained from the maximum load divided by the contact area of the indenter. Young's modulus is determined from the slope of the unloading part of the experiment. Creep displacement is determined from the part where the load is kept constant.

Results and discussions:

The measured bond stiffness is represented by the longitudinal elastic modulus, *E*, called Young's modulus. By looking at the material's elasticity results shown in Figure 21 (both the comparison among different materials (i) and among HaPs themselves (ii)) it can be deduced that:

a) HaPs are soft semiconductors – significantly softer than other semiconductors that are used for photovoltaics like Si, GaAs, CdTe, CIGS. Therefore, it will be naïve to correlate directly the functional optoelectronic behavior and the bond-stiffness.

b) The bulkier the X cation, the softer the Pb-X bond - in agreement with what was deduced from theoretical calculations.[120] It also emphasizes the importance of the Pb-X bond in constructing the valence band, as discussed in section 3.1.

c) The A cation does not seem to affect the elasticity in a way that one expects from an organic group (**MA**PbBr$_3$) vs. an inorganic (**Cs**PbBr$_3$) one. The A cation has only a minor, role in determining the structural stiffness; however, the structural symmetry that the A cation imposes on the different HaPs may play some role in determining the elastic modulus.

Our values for elastic constants were reproduced in other reports by similar nanoindentation experiments[121] as well as by sound velocity measurements[122] that can be related to the Young's modulus via the Poisson ratio[123].

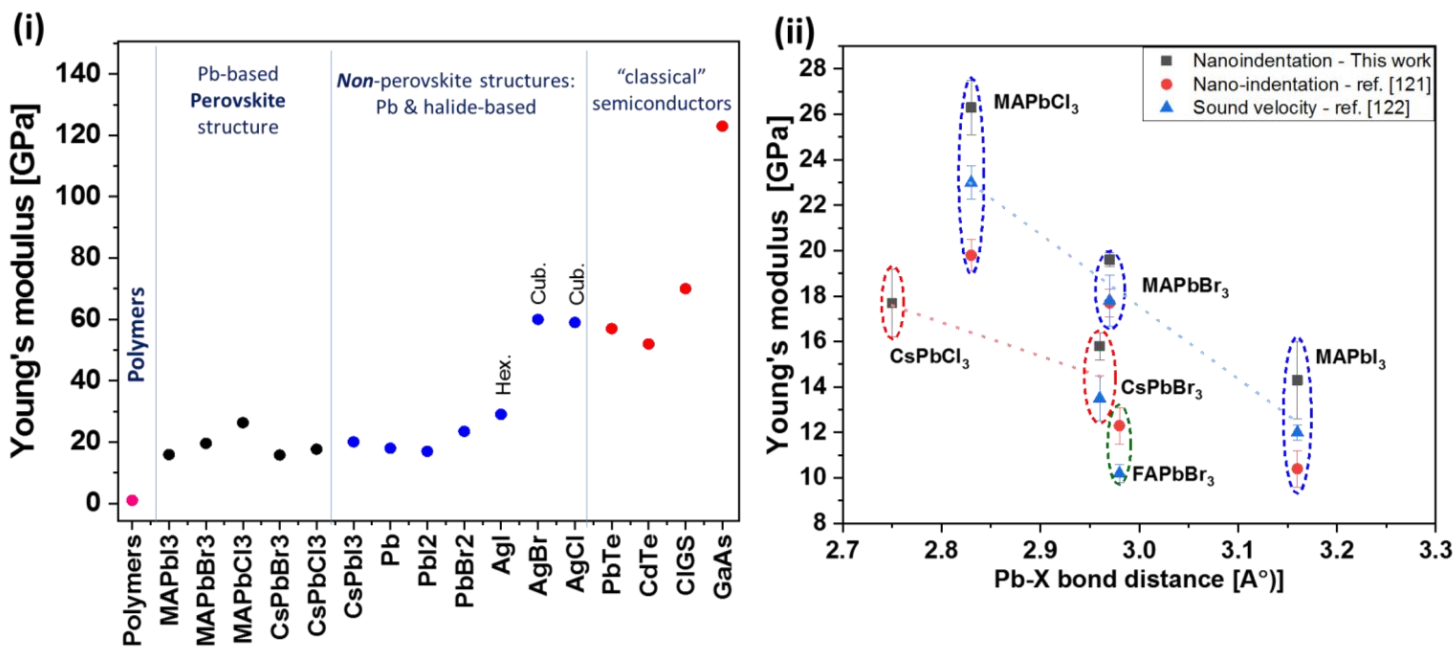

Figure 21: (i) Comparison between the Young's modulus of different semiconductors (and polymers). CsPbI$_3$ is thermodynamically unstable as a perovskite and therefore is included in the non-perovskite Pb-based compositions. (ii) Measured Young's modulus as a function of Pb-X bond length for different APbX$_3$ single crystals.

The difference in Young's modulus between Cl-, Br- and I-based HaPs can be explained by the bond length, which is inversely proportional to the bond strength (Figure 21(i)). Two *other* factors that were mentioned by Sun et al.[121] that correlate with the elastic modulus are the electronegativity and packing density. A difference in electronegativity is a known factor to increase electrostatic interaction and, since the B atom stays the same (Pb), the increase of electronegativity from I to Cl can explain the increase in bond stiffness. In section 3.3 we will see that the results of probing the electron density of the Pb atom, does *not* support the dominance of the electrostatic interactions. Increase in packing density, which for perovskites is expressed by the tolerance factor, $t$, ($t = \frac{R_A + R_X}{\sqrt{2}(R_B + R_X)}$), is known to improve resilience to elastic deformation: indeed, the elastic modulus increases from I to Cl.[121,124] The lower Young's modulus of CsPbX$_3$ with respect to that of MAPbX$_3$ can also be explained by the lower packing density of Cs-based HaPs ($t_{(CsPbBr3)}$ ~0.86 and for $t_{(MAPbBr3)}$~0.93)[17,121].

N-H---X hydrogen bonds, which can exist in MA-based, but not in the Cs-based HaPs, may be another reason for the increase in Young's modulus. However, FAPbBr$_3$, which may form hydrogen bonds and has a closer to optimum tolerance factor than both MAPbBr$_3$ and CsPbBr$_3$, has a lower elastic modulus than the other two Br-based HaPs ($E_{FAPbBr3}$ ~ 12 GPa).[122,125] Sun et al.[125], who were also surprised by that result, suggested that $t$ in FAPbBr$_3$ is actually *higher* than the optimum of $t$=1 ($t_{(FAPbBr3)}$~1.01), which then can be expressed in a lower bond stiffness. In general, for perovskites the accommodation of large cations, which leads to (too) high tolerance factors, increases the possibility of local octahedral distortions. $t$~1.01 places FAPbBr$_3$ at the edge of stability for perovskites. Although $0.85 < t < 1.1$ can sustain HaPs, $t > 1$ is a less favorable situation, where HaP tend to adopt an hexagonal or rhombohedral structure,[12] which reduces their mechanical resistance to an applied stress (cf. AgX in Figure 21(ii)). Nagabhushana et al.[10] demonstrated (both experimentally and via theoretical calculations) that HaPs with lower tolerance factors possess a lower enthalpy of formation and lower Pb-X bond strength. Another explanation that was given[125] is that, over the A cation, the more delocalized electrons in FAPbBr$_3$ reduce the N-H---X hydrogen bond strength, compared to MAPbBr$_3$. This, however, loses its relevance when comparing to CsPbBr$_3$, and the reason that CsPbBr$_3$ has a higher Young's modulus than FAPbBr$_3$ is still unclear and may relate to the 'tilt factor' between the two systems.[12] Overall, at this point, we can conclude that there should be a 'sweet-point' in the tolerance factor that will result in an optimal (highest) bond strength and, thus, structural stability.

With respect to *hardness* of HaPs, tetragonal MAPbI$_3$ has a greater hardness (0.57 ± 0.11 GPa) than cubic MAPbBr$_3$ (0.36 ± 0.01 GPa) and similar to that of the cubic MAPbCl$_3$ (0.57 ± 0.05 GPa) (note that MAPbCl$_3$ has almost twice the Young's modulus of MAPbI$_3$), despite its lower elastic modulus. Hardness (the resistance to a *plastic* deformation) is a semi-engineering parameter that depends, *like* the elastic modulus, on the overall interatomic bond stiffness but also, *unlike* the elastic modulus, on defect density (mostly cracks, dislocations and inclusions). Since a measure of hardness using nanoindentation provides information from a small area, the less frequent cracks and inclusions usually are not probed. Dislocations, however, are definitely probed and are suggested[121] to be the reason for the relatively higher hardness values of MAPbI$_3$ than expected from its elastic modulus.

To understand this point we note that distorted structural symmetry (e.g., tetragonal instead of cubic) can change the probability to form dislocations and, once formed, their slipping and entangling possibilities. In lower symmetry systems (e.g., tetragonal MAPbI$_3$), a higher dislocation density can result in their inhibited propagation, followed by (dislocation) hardening, and thus can explain the increased hardness of MAPbI$_3$ compared to MAPbBr$_3$ or MAPbCl$_3$ that have a cubic structure at RT.

Dislocation dynamics can be represented by the profile of a plastic deformation under a constant stress (i.e., creep). Time-dependent nanoindentation experiments (i.e., creep displacement) reflect the nucleation, propagation and entanglement of dislocations and, overall, represent the hardening of a material over time. Such measurements (Figure 22(i)) on cubic MAPbBr$_3$ and MAPbCl$_3$ and tetragonal MAPbI$_3$ and on a different set of HaPs done by Reyes-Martinez et al.[126] (Figure 22(ii)) showed that, indeed, plastic deformation develops similarly for similar symmetry systems, regardless of the different elastic constant. In general, lower symmetry allows less slip planes, which then results in higher probability for hardening by dislocation entanglement and an overall harder material.

To check that point, a combined *differential*-TGA analysis (Figure 23) with temperature-dependent XRD measurements were carried out on mixed (Cs$_x$MA$_{1-x}$)PbBr$_3$ crystals (by Hadar Kaslasi and Dr. Isai Feldman). The differential TGA measurements showed a clear shift (of ~ 60-70 °C) in the temperature at which MA starts to evaporate, while there is almost no difference in the temperature at which the MA evaporation rate is maximal. The difference does not correlate with Cs concentration but rather with the structural symmetry of the mixed (Cs$_x$MA$_{1-x}$)PbBr$_3$ composition at RT (cf. Figure 17(iv)). Using temperature-dependent XRD, Kaslasi and Feldman found that those mixed crystals that are not initially cubic and undergo a phase transition from orthorhombic/ tetragonal to a cubic phase at the temperature where MA start to evaporate (>260 ºC). Upon any phase transition, strain fields are commonly created. Strain-fields are a known factor for hardening and, in general, are inhibitors for mass propagation (imagine an activation barrier for mass transfer). In steel, for example, addition of C to Fe creates inclusions of a 'martensitic' phase, and result in strain-fields around the inclusions. These strain fields limit propagation of dislocations, and result in general hardening.[127] During phase transition, like happens in some $(Cs_xMA_{1-x})PbBr_3$ that are orthorhombic/tetragonal at RT, strain fields should appear and, similarly, limit mass propagation, and thus MA evaporation. This emphasizes that hardening of the material should increase its overall stability to thermal degradation or any dynamic processes that occur during degradation, as also demonstrated by creep measurements (Figure 22).

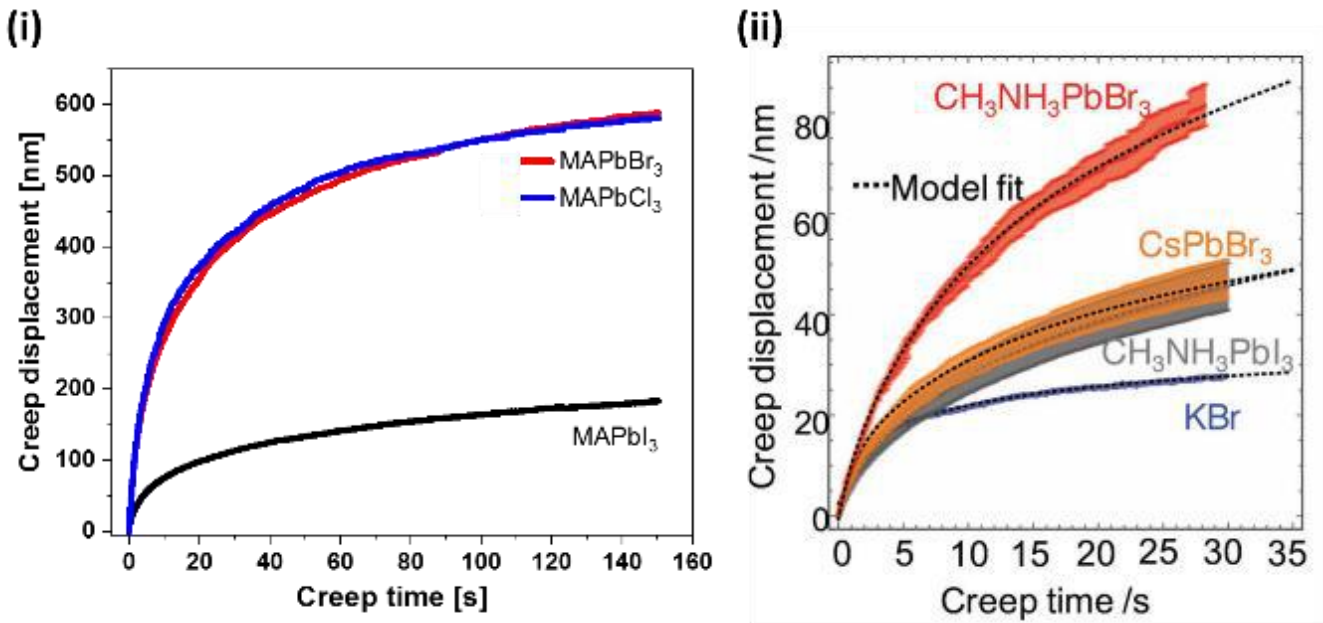

Figure 22: Time-dependent response to a constant (and identical for all measured crystals) load for measuring creep displacement comparing between: (i) two cubic systems ($MAPbBr_3$ and $MAPbCl_3$) and one tetragonal system ($MAPbI_3$); (ii) a cubic system ($MAPbBr_3$) and two lower symmetry systems ($MAPbI_3$ – tetragonal; $CsPbBr_3$ – orthorhombic). (ii) is reported by Reyes-Martinez et al.[126]. Both sets of measurements show reduction in creep displacement in lower crystallographic symmetry.

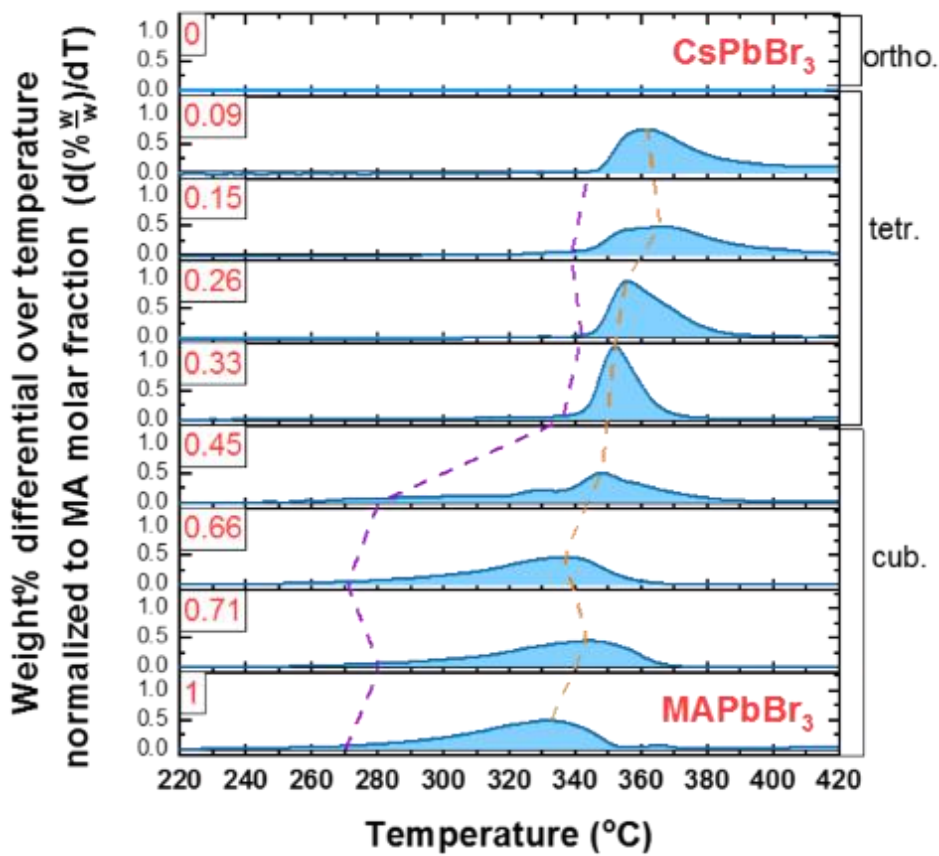

Figure 23: Differential TGA of $(Cs_xMA_{1-x})PbBr_3$ crystals, derived from TGA results, such as those shown in Figure 17(ii). The molar fraction of MA, (1-x), is given in the squares in the left corner of each rectangle. I normalized the weight losses by dividing by the molar fraction of MA in each compound, since the rate of mass loss will be different for compounds with less MA to begin with. In this way it is possible to compare the onset temperatures at which MA starts to vaporize. The dashed lines are a guide to the eye for the weight loss rate at which it starts (purple) and reaches a maximum (orange).

It can be concluded that in HaP crystals elastic deformation is clearly dependent on the Pb-X bond and, to some extent, the packing density, or *t*. Hardness, on the other hand, is strongly affected by the crystallographic symmetry. Since symmetry is strongly correlates with *t*,[12] hardness should be affected more significantly by changing the A cation than the material's Young's modulus.

## 3.3. On the bond nature of HaPs (solid state NMR results)

In previous sections it was shown that the Pb-X bond has a significant impact on the valence electrons in HaPs, as shown both theoretically and experimentally (impact on the band gap and the structural bond strength). The bond nature (i.e., how covalent or ionic), however, has not yet been determined. The importance of defining the bond nature lies in the way each individual sub-group (A, B or X) influences the material properties. For example, if the whole structure is highly ionic, the individual entities will influence the structure via coulomb interactions, with concomitant high Madelung energies, but their orbitals will be similar to their 'atomic' ones. On the other hand, when a covalent bond nature governs the structure, the electron sharing and hybridized orbital energies will dominate the material properties. Solid-state NMR (SS-NMR) on $Pb^{207}$ nuclei from different HaPs was carried out to gain insight into the Pb-X bond nature and the impact the A group has on it.

About the method:

Nuclear-Magnetic-Resonance, NMR, is a method that probes an energy gap that is formed under a strong magnetic field between two nuclear magnetic states. Although spin states of nuclei are opposite in sign, they are degenerate (equal in energy) when a magnetic field is absent (unless a material has a permanent magnetization). An external magnetic field will break their degeneracy and form an energy gap. This energy gap can be probed by an electromagnetic wave that will resonate with that gap. A several Tesla external magnet will require radio frequency electromagnetic waves. Only atoms with a halved ($\frac{1}{2}, \frac{3}{2}, \frac{5}{2}$, etc.) nuclear spin, such as $H^1$, $C^{13}$ $Br^{79}$, $Cs^{133}$, $Pb^{207}$, etc., can be probed by NMR.

Electrons that surround nuclei also have a spin that form a magnetic field. The electron spin magnetic field also interacts with the external magnetic field, which then screens the effective field from the nuclei. Changing the electron density around identical nuclei will result in a shift of the probed energy. This 'shielding' effect made by electrons changes the resonance frequency and is called a 'chemical shift'. At higher (lower) electron density the NMR frequency will be shifted 'upfield' ('downfield') and will be shown as a lower (higher) chemical shift. Therefore, NMR allows probing the electron density around a specific nucleus – in our case Pb.

Unless the local symmetry of such molecular orbitals is very high (leading to "isotropic" shift), the shielding effect will depend on the orientation of the molecular orbital with respect to the external field. Since interactions in solids are orientation-dependent, in Solid-State NMR (SS-NMR) spectroscopy, 'magic angle spinning' is required to average out this orientational dependence in order to obtain values close to the average chemical shifts. This is unnecessary in conventional NMR investigations of molecules, since rapid "molecular tumbling" averages out the chemical shift anisotropy.

In SS-NMR, besides the chemical shift, the line shape and peak width provide information about the electronic density, structural anisotropy, heterogeneity/local-disorder and dynamics of the probed nucleus. For the same five compositions used in nanoindentation experiments and shown in Figure 8(iii), i.e., MAPbI$_3$, MAPbBr$_3$, MAPbCl$_3$, CsPbBr$_3$, and CsPbCl$_3$, SS-NMR on the Pb$^{207}$ nucleus was probed as the only common element (Figure 24). The five compositions were pulverized and the powders placed in dedicated ZrO$_2$ rotors for SS-NMR, which were then sealed with a Kel-F cap. The results were acquired by direct excitation on a 11.7 T SB Bruker Avance spectrometer using a 4 mm double resonance probe at a 'magic angle spinning' frequency of 10 kHz. The chemical shift was externally referenced to solid Pb(NO$_3$)$_2$.

Results and discussion:

*These results were not published, because during the experiments, similar results, except for those of CsPbBr$_3$ and CsPbCl$_3$, were published elsewhere.*[128]

Figure 24 shows NMR peaks Pb$^{207}$ (22.1% natural abundance, spin 1/2) from the different compounds. The different chemical shifts of the Pb$^{207}$ peak represent a different electronic density around Pb for every compound. It is clearly seen that the halide has a very strong influence on the electron density of Pb, while the A cation has a much smaller influence. This implies that the halide group is clearly the dominating element in shaping the Pb-X bond and material properties related to that bond. Theory and mechanical response experiments support this result (see previous sections of this chapter).

The trend in the chemical shift with the increasing X anion size (I$^-$ > Br$^-$ > Cl$^-$) shows a decrease in the electron density around Pb. This strongly suggests that the Pb-X bond has a dominant covalent nature rather than an ionic one. This can be explained by the size of the halide: a larger X anion increases the Pb-X bond-distance (see Figure 25), which results in less electron sharing between Pb and X at the valence shell. An ionic nature would suggest that the most electronegative cation (Cl) would result in a lower electron density around Pb.

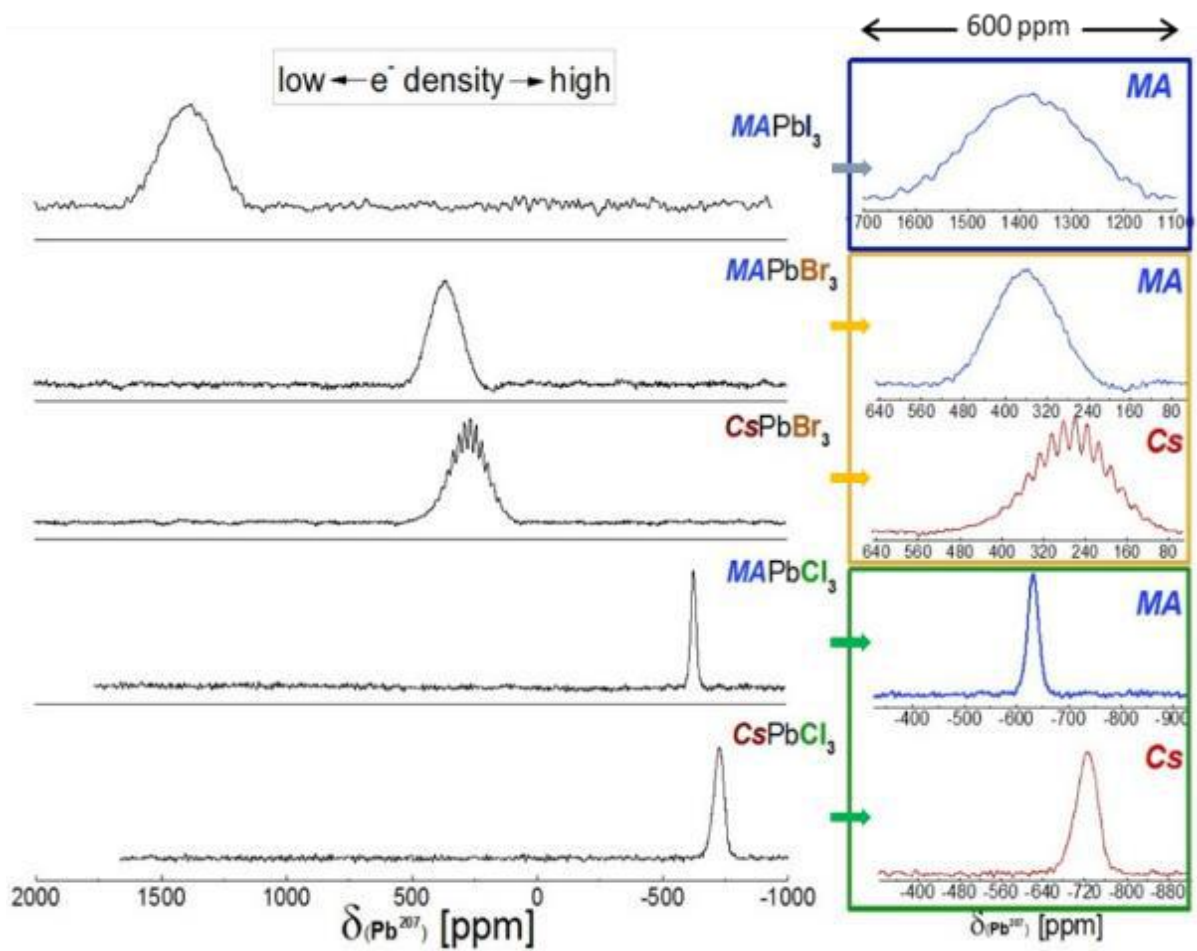

Figure 24: $^{207}$Pb SS-NMR spectra of different halide perovskites done at room temperature, acquired at a magic angle spinning frequency of 10kHz. Keeping the same angle but without spinning, the spectra looked similar. The chemical shift was externally referenced to solid Pb(NO$_3$)$_2$. The set of figures on the right are magnifications to emphasize the broadening and shift between the different samples, as well as the j-coupling splitting from the CsPbBr$_3$ sample. Cases with the same X anion share similar x-axis *values*. Cases with different A cation share similar x-axis *span*. When normalized to 100, the peak ratio between the different sub-peaks in the CsPbBr$_3$ spectrum is 94:79:63:44:28. Calculated splitting by 6 Br neighboring atoms (spin 3/2) will be: 94:79:58:37:21, while for splitting by Cs (spin 7/2, 8 nearest neighbors), the ratio will be 99:96:91:84:76.

Another important observation that support a covalent nature to the Pb-X bond is the 'j-coupling' splitting shown in CsPbBr$_3$. J-coupling is an indirect interaction between two nuclear spins that is mediated through chemical bonds connecting two spins (e.g., Pb(spin=1/2)---Br (spin=3/2)). The intensity ratio between the peaks can be derived from the Pascal triangle principle with 2nI+1 peaks (n is the number of neighboring atoms and I is the nuclear spin).[129] The expected ratio between the peaks if the split is due to six Br neighbors (n=6; I=3/2) fit very well to the actual result (see caption of Figure 24 for further details). The existence of j-coupling is, by itself, a direct proof for overlapping electrons between Pb and Br, meaning a covalent bond. An absence of similar j-coupling splitting in other HaPs (especially MAPbBr$_3$ and MAPbI$_3$) is not completely clear; however, its absence does not falsify by itself the existence of a covalent bond nature.

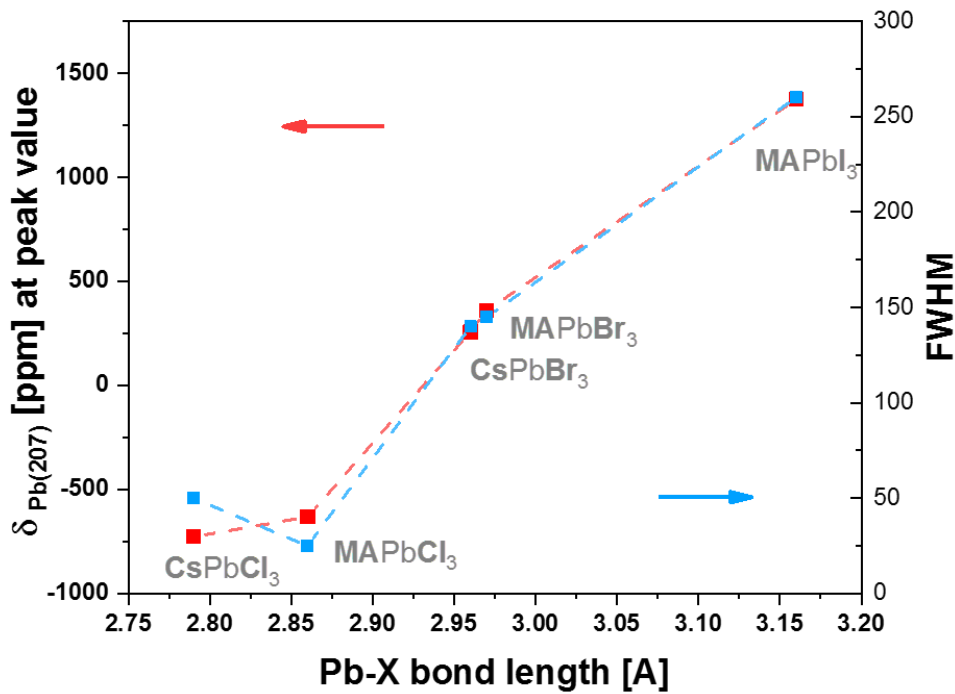

Figure 25: The chemical shift (red) and the width (blue) of the Pb$^{207}$ NMR peak as a function of the average Pb-X bond distance. The higher chemical shift represents a higher electron density shared between X and Pb atoms. A greater peak width probably represents a greater spread the Pb atom can have when the Pb-X bond loosens.

Although the effect of the A cation is small, it is still measurable. The most straightforward explanation, again, is the Pb-X bond distance, which is reduced when replacing MA by Cs (see Figure 25). Another way to explain this is by N-H---X hydrogen bonds. The halogen, which is naturally an electron acceptor, should possess more electrons when an MA group is present. Since the halides share their electrons with Pb, the Pb electron density will increase. The presence of hydrogen bonding can also be used as an argument to justify the higher elasticity modulus of MAPbBr$_3$ with respect to CsPbBr$_3$ (cf. Figure 21(ii)).

Besides the chemical shift, the peak width can provide additional information. Peak broadening usually relates (1) to an increase in a *permanent* anisotropy of the local chemical environment of the probed atom (Pb in our case) *or* (2) to a *dynamic* local structural anisotropy of the probed atom or related subunits. Since we did not observe significant change in the peak FWHM when we repeated the experiments without spinning the sample, it is reasonable to rule out the possibility of an existing permanent anisotropy of the Pb in the probed samples. Therefore, we speculate that the broadening comes from (2) - enhanced dynamics of the local structure with heavier halides in the structure. This fits the thermal parameters found from diffraction studies.[130]

Another way to explain the $Pb^{207}$-NMR chemical shifts and peak width is by considering effects from $ns^2$ electrons that can be (de)localized around the Pb atom, depending on the halide group it is in contact with (see background on $ns^2$ electrons - Section 1.1). In general, Pb complexes (also with halides) can form 'hemidirected' to 'holodirected' interconnectivity with their ligands (see Figure 26). This situation is a function of the chemical environment[131] as well as the temperature[132]. Following the ligand strength series ($I^- < Br^- < Cl^-$), stronger ligands tend to form more localized Pb-X bonding and a more hemidirected character. The meaning of a hemidirected character is that the $ns^2$ electrons are more localized, with a higher charge density around the Pb atom and a lower chemical shift, as indeed happens. The more holodirected character of the Pb-I bond, on the other hand, should result in more delocalized $ns^2$ electrons and a broader span of possible chemical environments. The temperature-dependence of the peak width in $MAPbCl_3$ is, however, not very consistent with the $ns^2$ electron delocalization, unless with increasing temperature, the $ns^2$ electrons become more localized on the Pb-X bond and not as a lone pair (i.e., $Pb^{2+}$ oxidation towards $Pb^{4+}$ may occur).

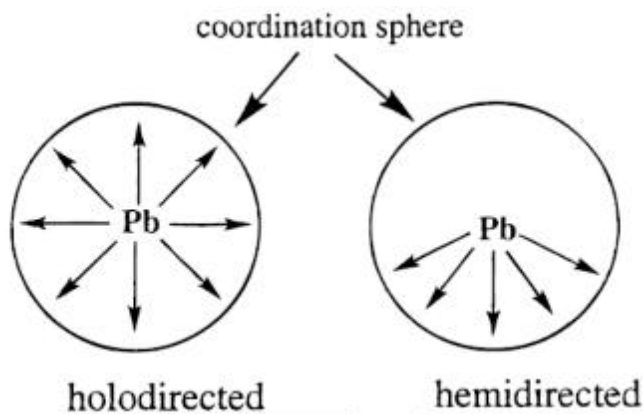

Figure 26: Schematics of a hemidirected' and a 'holodirected' coordination in Pb complexes.[131]

To emphasize the importance of ligand strength effect (and significance of the $ns^2$ electrons) over local *dynamic* disorder on the valence electrons in Pb-based HaPs, Bernard et al.[133] found by temperature-dependent $Pb^{207}$ SS-NMR of $MAPbCl_3$ [133] that the FWHM of the $Pb^{207}$-NMR peak decreases with increasing temperature. This result is quite counterintuitive if one considers enhanced dynamics at higher temperatures. The way they explain this phenomenon is by an increasing Pb-Cl spin-spin interaction with increasing temperature because of higher structural symmetry, and is expected at higher temperature (cf. Figure 2). This emphasizes that the Pb-X holodirected-type bonding character is more probable at temperature.

At this point, it can be concluded that the importance of the Pb-X bond became reinforced. A covalent nature of the bond is clearly observed. It is still not clear, however, whether individual Pb-X covalent bonds are the correct way to treat the Pb-X bond, or if the more collective (de)localization of $s^2$ electrons should be considered. In the following section, the energetic cost to deform the Pb-X bond will be considered. In addition, an 'anti-bonding'-like covalent interaction of the Pb-X bond will be further supported.

## 3.4. Defect-tolerating intrinsic features in HaPs: 'low deformation potential' and an 'anti-bonding valence band'

### 3.4.1. Background:

About 'defect tolerance':

Defects can act as scattering and/or trapping centers of electrons or holes in semiconductors. Charge transport and radiative/non-radiative recombination probabilities are, therefore, directly related to defects in semiconductors and, basically, define their (opto-)electronic quality. On the one hand, high quality optoelectronic devices, such as solar cells and LEDs, usually require low defect density to reduce non-radiative recombination, which is a major loss mechanism in these devices. For solar cells, reduction of scattering centers is also important to allow higher diffusion length (see Figure 5, section 1.3). On the other hand, for electronic devices such as diodes, transistors, etc., defects can act as electron donors (*n*-type) or electron acceptors (*p*-type). By controlling their density, spatial distribution and type (*n* or *p*), usually by deliberate extrinsic doping, one can engineer electronic devices with different properties.

As I already mentioned in the introductory chapter (section 1.3), values for defect densities in HaPs, deduced from various experiments, yielded values of ~$10^9$-$10^{10}$ cm$^{-3}$ for solution-grown single crystals. Polycrystalline films are found to have *up to* ~$10^{16}$ cm$^{-3}$ defects, which seems to be an *upper limit*. To maintain a defect density below $10^{16}$ cm$^{-3}$ in materials *other* than HaPs usually requires some effort and optimization of their fabrication: highly controlled chemical environments, proper annealing regimes, etc. For HaPs, however, $10^{16}$ cm$^{-3}$ seems to be an upper limit, *almost regardless of the method used for their fabrication*.[7,66,70,74]. HaP crystals can be grown from solution at RT in less than a few hours, usually conditions that are not at all favorable for low defect density materials. Polycrystalline films are formed in even harsher conditions, where a precursor solution is dried during spin-coating, or a co-evaporated precursor source is deposited on a cold substrate, etc. (see Figure 27). Although these methods should result in kinetically quenched products with lots of intrinsic defect states ($>>10^{16}$ cm$^{-3}$), there seems to be some factor that either limits their formation or obscures their presence.

One possible explanation for the low defect density is self-reorganization of local discontinuities (= defects) towards a more favorable thermodynamic state with less bulk defect density. In cases where defects are localized (e.g., next to surfaces or interfaces), a spread of defects into a larger bulk volume by self-diffusion may result in a local reduction of their defect density. In cases where defects are not localized, diffusion in HaPs, which is a widely-reported phenomenon[134,135] may lead to some cancelation of charged defects. Diffusion from the bulk may also remove defects by their segregation at surfaces, incl. grain boundaries.

A chemically-based strategy for defect elimination is that of *local* chemical reactions of the defects among themselves or with the host lattice. This may lead to: (1) formation of *new compounds* that are integrated with the periodicity of the hosting HaP lattice (and hardly affect the (opto)electronic properties of the hosting HaP); (2) re-formation of the HaP phase to a new phase, e.g. 2D perovskite or a non-perovskite (depends on the local stoichiometry and oxidation states of the defected area), that separates from the HaP host lattice and leave behind HaP domains with lower defect density.

These possibilities for defect elimination, as mentioned above, will be termed 'self-healing' or 'structural defect intolerance', and will be further discussed in Chapter 4.

Observation of a low defect density may suggest a highly ordered structure, but it also may reflect a situation where structural defects *do* exist, but do not perturb in a measurable manner the charge carrier dynamics. In the latter case, the cross-section for scattering or trapping of (quasi)free electrons/holes by defects is so small that it will not have an experimentally detectable optical or electronic signature (which may well depend on the detection limit of the experimental setup). Such a phenomenon is termed '*defect tolerance*' and will be discussed in this section.

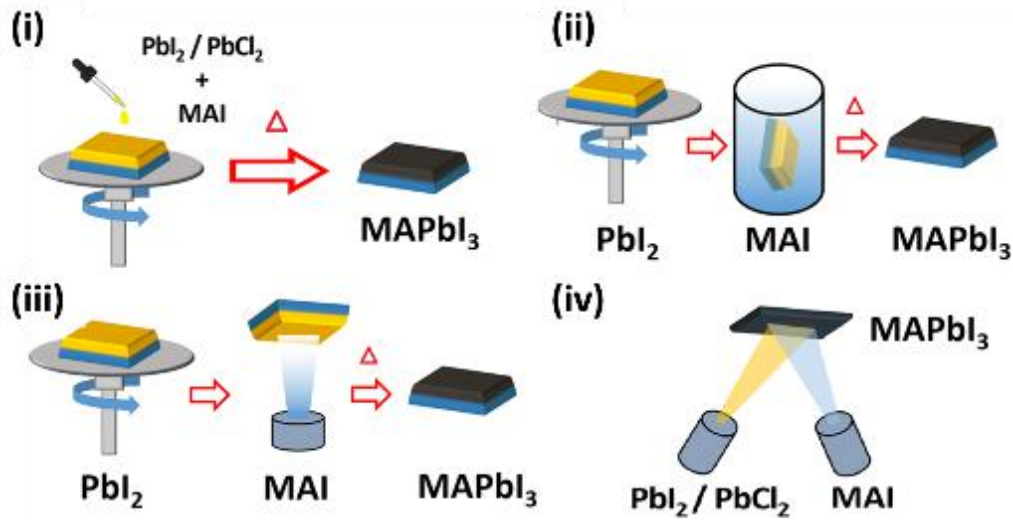

Figure 27: Deposition methods for perovskite thin films (in this example – MAPbI$_3$), including: (**i**) - single-step solution deposition by spin-coating; (**ii**) and (**iii**) - two-step deposition where the PbI$_2$ layer is first deposited by evaporation or spin coating followed by exposure to (ii) a solution containing MAI salt or (iii) MAI vapor; (iv) - thermal vapor deposition where both precursors are simultaneously co-evaporated. Adopted from ref. 34.

About 'deformation potential':

Scattering by defects occurs when a discontinuity, formed by a defect, creates an electrical energy barrier or deep valley, which then traps or scatters an electron/hole. For example, when defects are charged, Coulomb interactions between, e.g., an I$^-$ vacancy (positive) and an electron (negative) may change the electron's momentum (scatter) or bind it to the vacancy (trap). Screening/reducing the effect of localized electric fields from a defect is possible, for example, by lattice rearrangement or formation of oppositely charged defects (like Frenkel defects [vacancy and accompanying interstitial of lattice atom] or Schottky dimer defects [pair of equal but oppositely charged vacancies])[136]. Adjustment of the lattice around the defect to screen it requires additional energy (to displace lattice atoms from their [ideal lattice] equilibrium sites) that is called the *'deformation potential', **d'***.

In fact, any defect (point defect, dislocation, interface, inclusion, etc.) can form a local strain field. In the proximity of a strain field (compressive or tensile), the electrical potential landscape (e.g., a Bloch potential of a periodic lattice) distorts and forms a localized potential barrier or valley that affects charge transport.[137] The potential landscape due to a non-zero differential strain field is *proportional* to ***d'***, which means that stronger distortion from equilibrium by the strain field leads to stronger scattering/trapping.

In a perfect lattice (zero defects), thermal energy forces lattice vibrations (also called *phonons* if the vibrations are coherent). Acoustic phonons (having a wavelength greater than the lattice unit cell) or optical phonons (having a wavelength smaller than a unit cell) create periodic compressive/tensile stress fields. Such periodic lattice vibrations will induce a periodic distortion in the electrical energy landscape. For scattering by acoustic phonons (in the absence of stronger scattering mechanisms, such as ionic impurities or polar optical phonons)[137] the free charge carrier mobility will inversely relate to $(d')^2$ ($\mu \propto \frac{1}{d'^2}$). Regardless of whether acoustic phonons are relevant for scattering charge carriers in an HaP (a topic that will be broadly discussed in section 3.5), *d'* plays a role (both positive and negative) in scattering/trapping of electrons/holes in semiconductors via (avoidable) defects or (unavoidable) lattice thermal vibrations.

The ratio between the change in the bandgap, $\Delta(E_g)$, and the volumetric strain, $\frac{\Delta V}{V}$ (or $\Delta(\ln(V))$) defines *d'*. Since $\frac{\Delta V}{V}$ is related to the pressure change, $\Delta P$, $\frac{\Delta V}{V}$ via the bulk modulus, $B$, as $\Delta P = B \cdot \frac{\Delta V}{V}$, we can (assuming small changes) define *d'* as:

$$\text{Eq. 1} \qquad \boldsymbol{d'} \equiv \frac{d(E_g)}{d(\ln(V))} = -B \cdot \frac{d(E_g)}{dP}.$$

Since the volumetric thermal expansion coefficient is defined as: $\boldsymbol{\alpha_T} \equiv \frac{d(\ln(V))}{dT} = -\frac{1}{B}\frac{dP}{dT}$, we can derive *d'* also from Eq. 2:

$$\text{Eq. 2} \qquad \boldsymbol{d'} = \frac{1}{\alpha_T} \cdot \frac{d(E_g)}{dT}$$

Since the *d'* can get both positive and negative values, we will also treat the absolute value of it, |**d'**|.

### 3.4.2. Results and discussions:

MAPbI$_3$ vs. CdTe:

Figure 28 shows examples of $E_g(P)$ plots and $E_g(T)$ plots of MAPbI$_3$ and CdTe – two (pseudo)cubic semiconductors with a relatively similar bandgap. It is immediately noticed that the optical bandgap changes with an *opposite* trend between the two materials. Using Eq. 1 and Eq. 2, assisted with parameters found in ref. 1, the actual values of *d'* for the two materials are, for *MAPbI$_3$* - *d'* (Eq. 1)=**+1.2** or *d'* (Eq. 2)=**+2.6**, and for *CdTe* - *d'* (Eq. 1)= **–3.4** or *d'* (Eq. 2)= **–29.4**. Beside the opposite algebraic sign of *d'*, the absolute value of *d'* in MAPbI$_3$ – both from Eq. 1 and Eq. 2 – is smaller than that of CdTe. The origin for the discrepancy in *d'* values derived from Eq. 1 or Eq. 2 is not completely clear but, as shown in Figure 29, does not seem to be significant in a view of a broader set of materials.

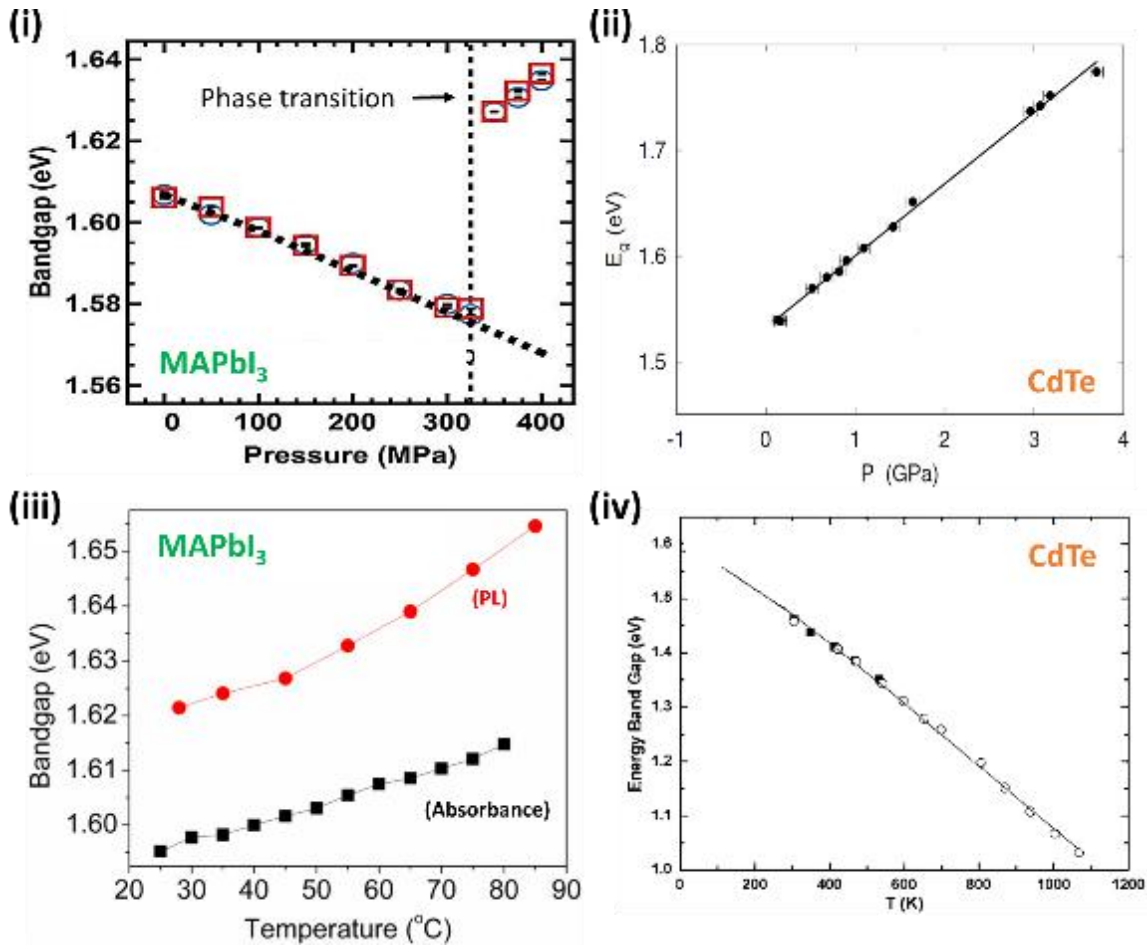

Figure 28: Bandgap, $E_g$, variation of MAPbI$_3$ and CdTe with pressure and temperature taken from literature. (i) $E_g$ (via optical absorbance) vs. pressure at 300K for (**i**) MAPbI$_3$ [138] and (**ii**) CdTe [139]. $E_g$ (via optical absorbance (or PL peak for MAPbI$_3$)) vs. temperature at atmospheric pressure for (**iii**) MAPbI$_3$ [140] and (**iv**) CdTe [141]. By using the material's parameters:[1] $B(MAPbI_3) = 13.9\ [GPa]$, $B(CdTe) = 45\ [GPa]$, $\alpha_T(MAPbI_3) = 1.3 \cdot 10^{-4} \left[\frac{1}{K}\right]$ and $\alpha_T(CdTe) = 1.8 \cdot 10^{-5} \left[\frac{1}{K}\right]$, the deformation potential of MAPbI$_3$ and CdTe was derived to be: MAPbI$_3$ - *d'* (Eq. 1)=+**1.2** or *d'* (Eq. 2)=+**2.6** ; CdTe - *d'* (Eq. 1)= **– 3.4** or *d'* (Eq. 2)= **–29.4**. In Eq.2 for MAPbI$_3$ the 'absorbance' results were used.

Figure 29 compares between *d'* values of different semiconductors having a (pseudo-)cubic perovskite, rock-salt or zinc blende or hexagonal wurtzite crystal structures (all isotropic systems with a polar interatomic bond, or heteropolar, excluding Si that is homopolar). The crystalline systems are divided to: Halides (HaPs-APbX$_3$, AgX and TlX with X=I, Br or Cl), Chalcogenides (PbX' or CdX' with X'=S, Se or Te) ; III-V (such as GaAs); Si - as an exception.

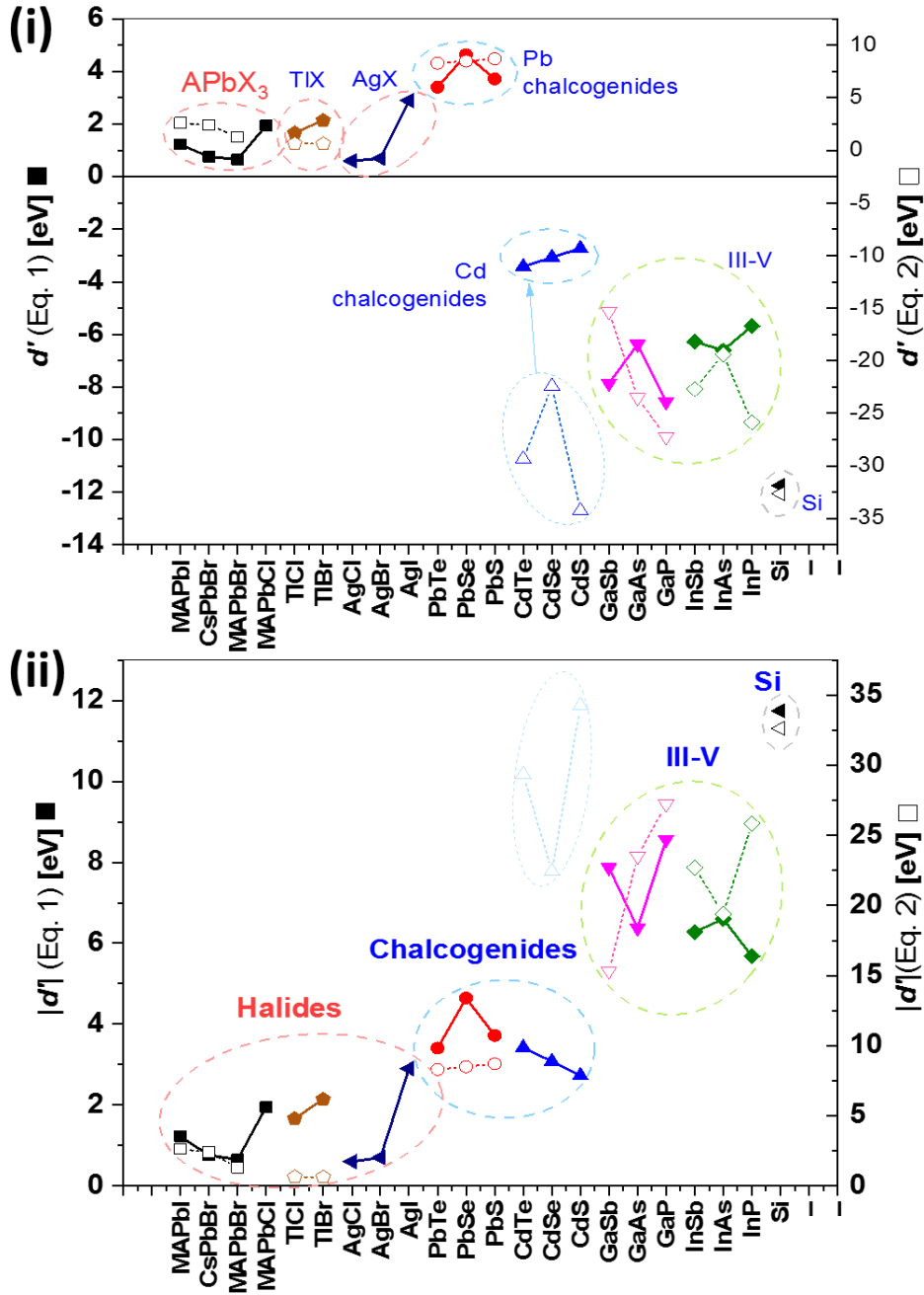

Figure 29: Deformation potentials, *d'*, of different semiconductors, derived from Eq. 1 (full symbols) and Eq. 2 (hollow symbols). All the values are based on experimental measurements[22–24,138,140–161], and for Eq. 1 are derived at T=300 K and for Eq. 2 at P=1 atm. **(i)** The algebraic *d'* value where an opposite sign in *d'* is observed between halides (HaPs, AgX, TlX) and Pb-based chalcogenides, and Cd-based chalcogenides, III-V and Si. **(ii)** same data, with |*d'*| i.e., its absolute value. Halides have the lowest absolute *d'*, then chalcogenides, then III-Vs and Si. The overall trend in |*d'*| follows the expected trend in the Madelung energy, which supports the different (partial) ionic character of the chosen heteropolar systems.

In Figure 29(i) we see two interesting features. The first is that Pb-based HaPs, Pb-based chalcogenides and other halides (AgX and TlX) possess a positive *d'* while all other semiconductors have a negative *d'*. The second (Figure 29 (ii)) is the low absolute value of *d'*, |*d'*|, of halides (including HaPs), compared to the other semiconductors. Interestingly, |*d'*| values for Pb and Cd chalcogenides are similar, while their actual *d'* values are inversed in sign.

In attempt to analyze the data in a more generalized manner (without specifying the material but its $B$, $\alpha_T$, $\frac{dE_g}{dP}$ and $\frac{dE_g}{dT}$ values) we find increasing |*d'*| with increasing $B$ and $\frac{1}{\alpha_T}$, and increasing *d'* with increasing $-\frac{dE_g}{dP}$ and $\frac{dE_g}{dT}$ (Figure 30(i) and (ii)). Except for TlX and Cd-based chalcogenides that show exceptionally low and high ratios between |*d'*|$_{(Eq. 1)}$ and |*d'*|$_{(Eq. 2)}$, respectively, the ratio between the differently-derived |*d'*| values is ~2-4 for the rest of the materials. The origin for these differences between |*d'*|$_{(Eq. 1)}$- and |*d'*|$_{(Eq. 2)}$- derived values is not yet clear. The spread is probably a result of collecting experimental data from different groups using different instrumentations and, sometimes, different methods. [22–24,138,140–161]

Figure 30(iii) and (iv) shows the interdependence between the two independent sets of literature-derived parameters: $B$ with $\frac{dE_g}{dP}$ and $\alpha_T$ with $\frac{dE_g}{dT}$. For the first set (Figure 30(iii)), we see that, in general, $\frac{dE_g}{dP}$ increases with $B$. 'Soft' semiconductors (B < ~35 GPa) tend to have a positive *d'*, which is also the case of HaPs. At an intermediate range (35 GPa < $B$ < 65 GPa) both positive and negative *d'* are observed. For 'rigid' semiconductors ($B$ > ~65 GPa), *d'* seems to be always negative.

A more obvious differentiation between a 'soft' and a 'rigid' material can be found if we plot $\frac{1}{\alpha_T}$ against $\frac{dE_g}{dT}$. 'Rigid' materials, as shown in Figure 30(iv), are such with $\alpha_T < 5·10^{-5}$ K$^{-1}$, and a narrow spread of *negative* $\frac{dE_g}{dT}$, while a 'soft' material is one with a $\alpha_T > 5·10^{-5}$ K$^{-1}$, where $\frac{dE_g}{dT}$ has a much larger spread, but *positive*.

To conclude, we define a material as '*soft*' when its bulk modulus is < ~35 GPa so that |*d'*|$_{Eq.1}$ < 2 eV, or its thermal expansion coefficient is > ~5·10$^{-5}$ K$^{-1}$, so that |*d'*|$_{Eq.2}$ < 10 eV.

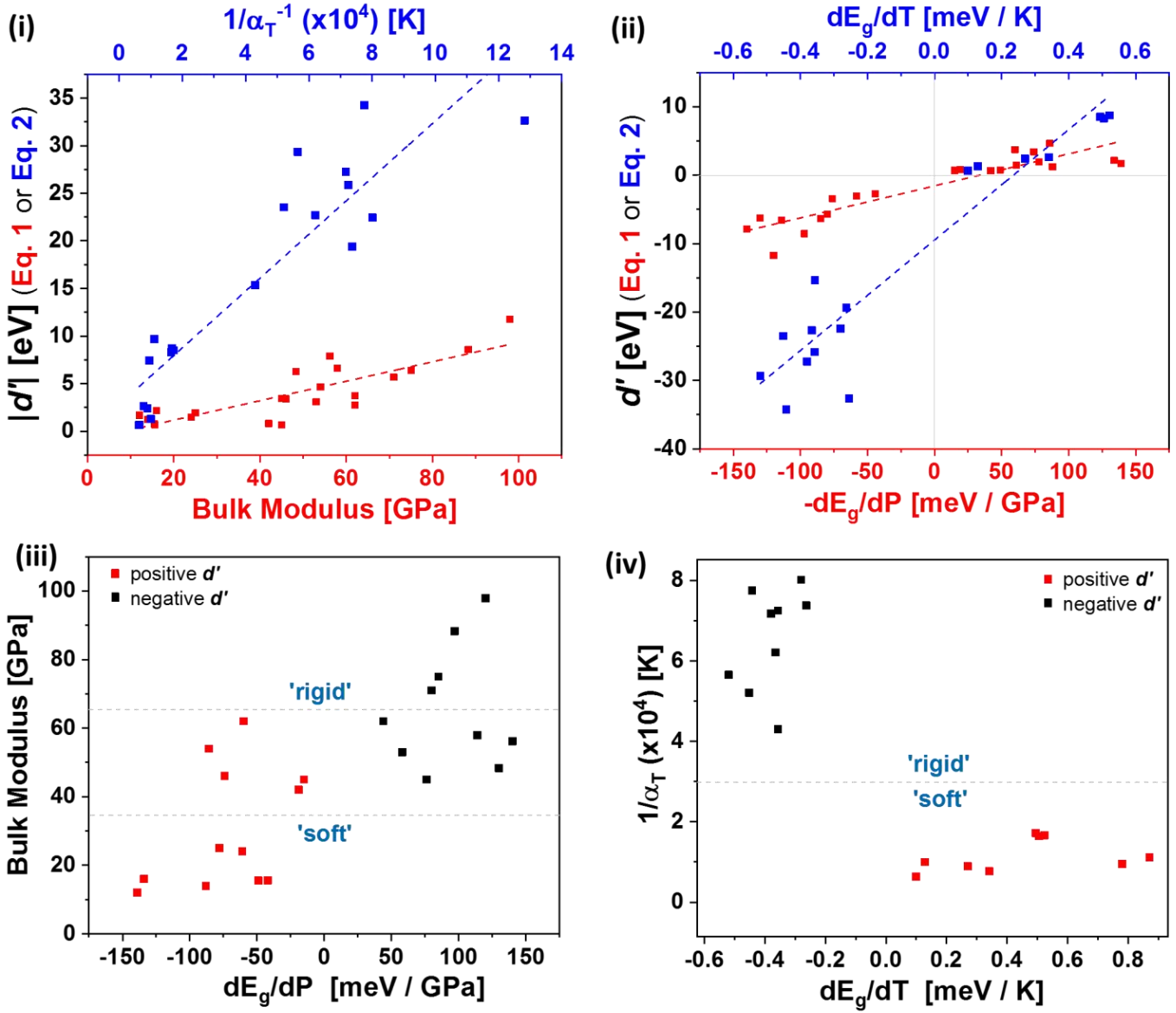

Figure 30: (i) |d'| as a function of the B and $\alpha_T$ including a linear fit (dashed lines) as a guide for the eye. The volumetric expansion coefficient, $\alpha_T$, was derived from the linear thermal expansion coefficient, assuming isotropic expansion in which $\alpha_T = \alpha_{T(lin)}$. (ii) d' (not absolute) as a function of $\frac{dE_g}{dP}$ and $\frac{dE_g}{dT}$ including a linear fit (dashed lines) as guides for the eye. (iii) Interdependence of B with $\frac{dE_g}{dP}$, showing that for 'softer' materials (lower B) $\frac{dE_g}{dP}$ becomes more negative. Recall (Eq. 1) that a negative $\frac{dE_g}{dP}$ result in a positive d'. For B~35-65 GPa (intermediate stiffness), both positive and negative d' values are found. (iv) Interdependence of $\frac{1}{\alpha_T}$ with $\frac{dE_g}{dT}$, showing that for 'softer' materials (higher $\alpha_T$) d' is negative.

### 3.4.3. Discussion:

To understand the reason for the different sign of **d'** (*positive* vs. *negative*), we look at the origin of the deformation potential. Upon a decrease in the interatomic distance (assuming no change in the symmetry group occurs), as is the case for isostatic compression or, to some extent, cooling a material, the valence and conduction bands change their energy with respect to the vacuum level, i.e., the ionization energy, IE (the energy of the VBM) and the electron affinity, EA (the energy of the CBM) change. If the EA (IE) decreases faster than the IE (EA), the bandgap will increase (decrease). Note that a decrease in EA means that CBM increases. In the remainder of this thesis I will use VBM and CBM for IE and EA, respectively.

The rate (and direction) of the change depends on the type of the interatomic orbitals. Following a model of Wei and Zunger,[150] upon compression an orbital will increase its potential energy (= energetic cost)[150] due to **(a)** an increase in the kinetic energy[i], or **(b)** a repulsive ('anti-bonding') coupling effect, due to the increased overlap between orbitals of neighboring atoms. In the case of an attractive ('bonding') coupling effect, compression will decrease the orbital potential energy. The sum of the individual effects determines the final energetic change of the specific bond. Therefore, an 'anti-bonding' orbital should always reach a higher potential energy upon compression than a 'bonding' orbital, in which the attractive orbital coupling effect may overcome the increase of the kinetic energy. Therefore, for the *usual* case, i.e., a 'bonding' VBM and an 'anti-bonding' CBM (which applies to the common, commercially important and most-studied semiconductors that have zinc blende, wurtzite or related [e.g., chalcopyrite] structures)[150], the CBM will increase more in energy than the VBM (i.e., the EA will decrease more than the IE) resulting in a *negative* **d'**. With similar reasoning, for the *unusual* case, where the VBM has an 'anti-bonding' nature with a 'non-bonding' CBM, **d'** is expected to be *positive*. These two extreme cases are illustrated in Figure 31.

The cases presented in Figure 29(i) can be divided between 'unusual' ones, with positive **d'**, which include halides (*including HaPs*) and Pb-based chalcogenides, and 'usual' semiconductors with a negative **d'**. This finding constitutes, in fact, experimental evidence for the theory-based suggestions (presented in Figure 18 and 19, section 3.1) that HaPs possess an 'anti-bonding' VBM! This empirically-based evidence for an anti-bonding VBM also strengthens the idea that HaPs, as well as other semiconductors with a positive **d'**, are expected to form defects with states that are shallow: in the gap, or in resonances inside the bands. This property, therefore, supports the idea (in the literature, cf. Figure 19 (ii)) that, intrinsically, HaPs are strongly *defect-tolerating* materials.

---

[i] The kinetic energy is proportional to $k^2$ or $1/l^2$ (for a parabolic band), where $k$ is the reciprocal lattice vector, $l$,

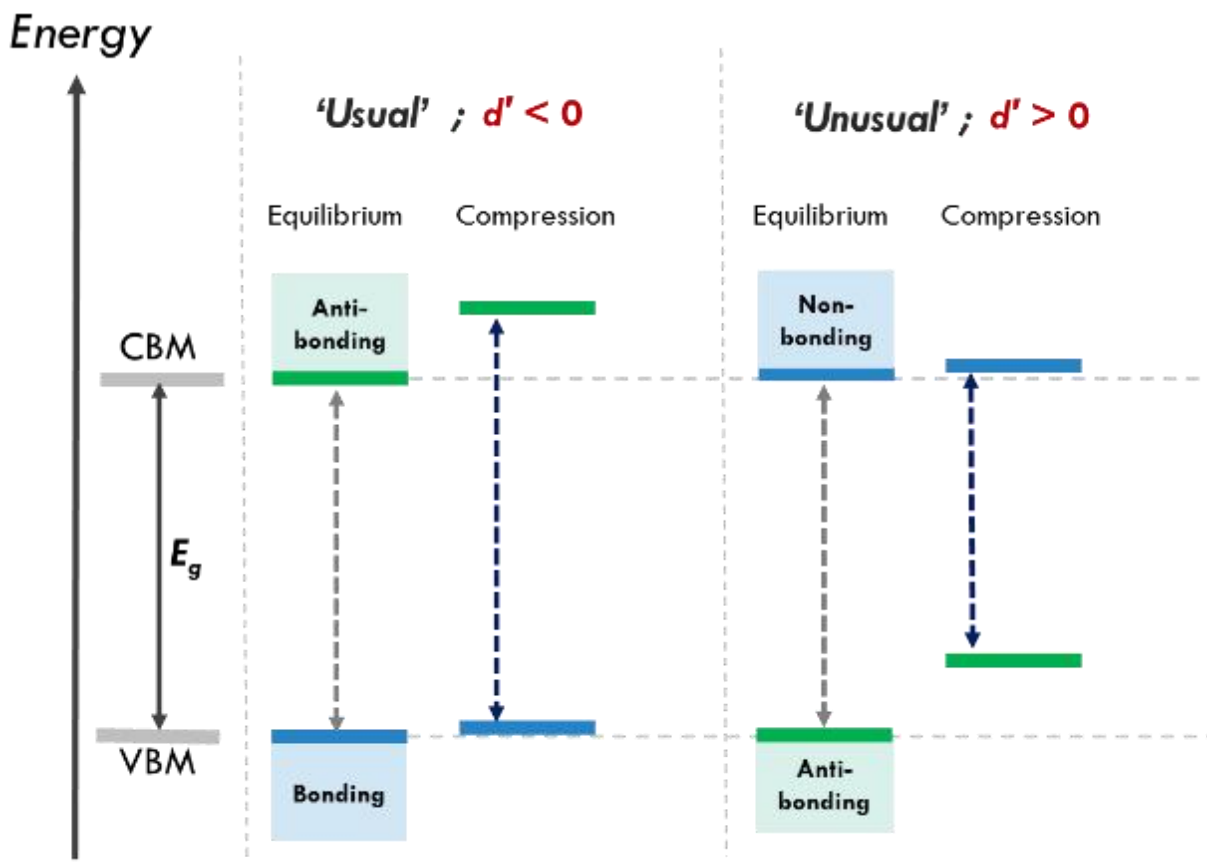

Figure 31: Schematic representations of opposite trend expected from the VBM and CBM in a semiconductor having 'bonding' VBM ('usual') or an 'anti-bonding' VBM ('unusual') due to the decrease of the interatomic bond distance (e.g. compression or cooling, before any phase transition). $d'$ is expected to be negative for the 'usual' (bonding VBM) case and positive for the 'unusual' (anti-bonding VBM) case.

To understand the importance of a low $|d'|$, as found for the 'soft' HaPs and other halides, we need to compare between the energetic landscape of a free charge carrier in a 'rigid' and a 'soft' lattice (cf. Figure 30 (iii) and (iv)) with a structural discontinuity (= defect), as shown in Figure 32. Regardless of the type of defect or the specific potential energy state that the presence of a defect introduces in an ionic crystal, it usually results in a localized extra charge (positive or negative), which needs to be neutralized ('electronic relaxation'), and a stress field, which leads to a lattice rearrangement ('ionic relaxation'). The former should be expressed in a more pronounced interaction between free charge carriers and defects: *relaxation* (or trapping of the charge) to neutralize the defect or *scattering* by a defect with the same charge sign as the free charge carrier.

The latter should be expressed in an adjacent ionic displacement that can, partially, neutralize the localized charge created by the defect, leading to a less distorted potential landscape.

Different parameters, such as the nature of the hybridized orbital, the interatomic distances that relate to the degree of orbital overlap, structural symmetry, etc., affect the energy paid for changing the interatomic distance. The less costly the displacement of adjacent ions, the higher the chances ionic relaxation will occur. Since the *ease for atomic displacement can be expressed by |d'|*, the low |*d'*| of HaPs and other halides leads to a much higher probability for ionic relaxation than in more rigid materials, and actually provides another path for '*defect tolerance*'.

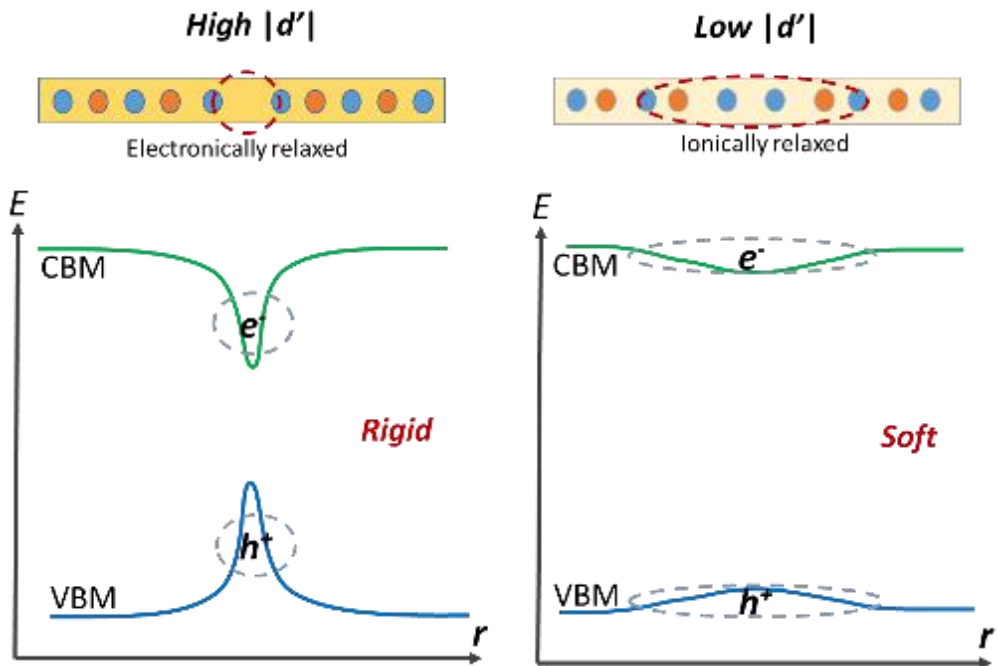

Figure 32: Illustrated examples of energy-landscape variations in space, *r*, of the CBM and VBM, due to a stress field created by a defect (e.g., vacancy, interstitial, grain-boundary, etc.) for (**left**) a rigid, high |*d'*|, system and (**right**) a soft, low |*d'*|, system. The top cartoons (correlated by *r* with the energy diagrams) represent a periodic lattice with one single vacancy (notice as missing orange ball), where the red dashed circles represent the radius of a stress field in 'rigid' (small) and 'soft' (larger) systems. The strong (weak) local energy variation in the 'rigid' (soft) case drives the 'electronic relaxation' ('ionic relaxation') of the system. In case of 'electronic relaxation' the probability for charge trapping/ scattering is high. In case of 'ionic relaxation', where the energy cost of a lattice rearrangement to reduce the energy cost of the defect is low, charge trapping/scattering due to the defect is lower.

To put this idea in a bit more quantitative perspective, using the actual values of |*d'*| (Eq. 1) for MAPbI$_3$ (1.22 eV), CdTe (3.42 eV) and GaAs (6.37 eV), the volumetric strain that is needed to exceed 1 k$_B$T (~26 meV @ RT) is (using $d' = \frac{dE}{d\ln(V)} = \frac{dE}{dV/V} \rightarrow \frac{dV}{V}[\%] = \frac{E_{(k_BT)}}{|d'|}$): 2.1% for MAPbI$_3$ ; 0.8% for CdTe ; 0.4% for GaAs. This means that 'soft' HaPs and other halides can accommodate volumetric strain that is several-fold higher than other 'rigid' materials for the same

energy cost. In fact, a volumetric strain of ~2-3%, as is the case in HaPs for energy change of $k_B T$, is a value that can cause failure in some materials.

In the following section we will try to understand the consequence of a 'soft' system in the sense of mobility of free charge carriers with respect to the more commonly modeled 'rigid' system. We will find that the ease of atomic displacement of atoms in a 'soft' lattice is actually a disadvantage when comparing charge mobility values between 'soft' and 'rigid' lattices with polar (ionic) bonds.

## 3.5. *On charge mobility in soft and heteropolar semiconductors*

### 3.5.1. Background:

The combination of properties of HaPs (e.g., high optical absorption coefficient, low effective mass, low exciton binding energy, long carrier recombination lifetimes, etc.) should (and do) allow for well-performing optoelectronic devices. However, there is one fundamental property that does not fit what is expected from the superior material properties – its carrier mobility.

When comparing mobility values of Pb-based HaPs (~1-100 cm$^2$ V$^{-1}$ s$^{-1}$) with those heteropolar semiconductors, such as GaAs or CdTe (~10$^3$-10$^5$ cm$^2$ V$^{-1}$ s$^{-1}$),[1] a significant difference is revealed. Mobility temperature dependence, which depends on the *scattering mechanism*, is found in HaPs, GaAs, CdTe and other heteropolar semiconductors to be similar and suggests scattering by polar optical phonons. Low defect density, as found for HaPs (~10$^{10}$ cm$^{-3}$ in single crystals) as well as for other high quality heteropolar semiconductors (e.g., GaAs and CdTe), make scattering by defects insignificant (at temperatures > ~100 K). Therefore, the origin of charge scattering and, thus, their mobility, probably depends on intrinsic properties of these heteropolar semiconductors. In this section I offer a detailed analysis of the different scattering mechanisms to understand what limits the mobility of HaPs and if they might reach mobility values of CdTe or GaAs.

<u>About charge mobility</u> [i]:

Charge carrier mobility, $\mu_q$, can be understood as an intrinsic property of a material, at a certain state of disorder, that represents the drift velocity [$cm/s$] of a charge carrier per unit electric field [$V/cm$], and therefore: $[\mu_q] = \left[\frac{cm^2}{V \cdot s}\right]$. The importance of the mobility value for PV devices, as explained in section 1.3, Figure 5, lies in the charge carrier diffusion length, $L$. Following Fick's law: $L = \sqrt{D\tau}$, and using the Einstein-relation for generalized diffusion: $D = \mu k_B T$, $L$ of a charged particle is $L = \sqrt{\frac{k_B T}{q} \mu_q \tau}$, with a generalized mobility $\mu = \frac{\mu_q}{q}$.

The average lifetime of carriers, electrons at the CBM and holes at the VBM, $\tau$, can be non-radiative (no ~ $E_g$ photon generated) or radiative (~ $E_g$ photon formed). The rates of all recombination paths increase with the density of excited charges (or the 'generation rate'), multiplied by a specific recombination probability that can be derived from Fermi's golden rule.[137] The overall recombination rate, $\frac{1}{\tau}$, is the sum of all the individual recombination rates, $\frac{1}{\tau_i}$.

Unlike lifetime, the mobility does not depend on the carrier density, but only on specific *scattering* rates of existing free charge carriers. The scattering rate can be derived from the same Fermi golden rule for different scattering reasons: ionized impurity, neutral impurity, grain boundary, dislocation, phonon (optical or acoustic) vibration, etc. The *average* scattering rate, $\frac{1}{\tau_s}$, is the sum of all scattering events, where $\tau_s$ is the average lifetime of carriers between two scattering events. $\tau_s$ means that an electron, which carries a charge of $q$ and an effective mass of $m^*$, on average, flows for a time of $\tau_s$ without any perturbation in momentum or energy. Thus, the average drift velocity, or the free charge carrier mobility, can be defined as:

$$\text{Eq. 3} \qquad \mu_q = \frac{q \cdot \tau_s}{m^*}$$

For each material the scattering potential, $U_S$, is what eventually defines $\tau_s$. Following Lundstrom,[137] $U_S$ depends on the specific scattering mechanism and is expressed in a matrix element of the scattering potential: $H_{k_0', k_0} \equiv \int \psi_{k_0'}(z) U_S(z) \psi_{k_0}(z) dz$, where $\psi_{k_0'}$ and $\psi_{k_0}$ are the Bloch wave functions of a charge-carrier in a periodic lattice before ($k_0$ wave vector) and after ($k_0'$ wave vector) scattering. Following Fermi's golden rule, $H_{k_0',k_0}$ affects the transition rate, $S(k_0, k_0')$, via the relation: $S(k_0, k_0') = \frac{2\pi}{\hbar} |H_{k_0',k_0}|^2 \delta(E(k_0') - E(k_0) \pm \hbar\omega)$, where $E(k_0)$ and $E(k_0')$ are the energies of the particle before and after scattering. $\hbar\omega$ is the photon energy (with $\omega$ being the photon frequency) that have been absorbed (+) or emitted (–). In cases where the momentum changes with no absolute energy change, as usually occurs for elastic scattering events, there is no photon exchange, or $\hbar\omega$ equals zero. To evaluate the scattering rate, $\frac{1}{\tau_s}$, $S(k_0, k_0')$ is summed over all final states: $\frac{1}{\tau_s} = \sum_k S(k_0, k_0')$. In cases where there is more than one scattering mechanism: $\frac{1}{\tau_{total}} = \frac{1}{\tau_{s1}} + \frac{1}{\tau_{s2}} + \cdots$ . The most significant scattering mechanisms are those that involve impurities (which at RT are usually ionized) and phonons, to which I will limit this discussion.

---

[i] (For further reading I refer to M. Lundstrom[137] [chapters 1 and 2] or S. Li[162] [chapter 8].)

Ionized impurities (*II*) set a Columbic potential, which varies with distance *r*, and is usually screened by the surroundings to a characteristic distance of $L_D \equiv \sqrt{\frac{\varepsilon_s \varepsilon_0 k_B T}{q^2 n_0}}$, so that:[i, 137]

Eq. 4 $$U_{II}(r) = \frac{q^2}{4\pi\varepsilon_s\varepsilon_0 r} e^{-\frac{r}{L_D}}$$

where $n_0$ is the doping concentration, $\varepsilon_s$ is the static dielectric constant and $\varepsilon_0$ is the vacuum permittivity. Using Eq. 4 and following the development for $\frac{1}{\tau_s}$ and then using Eq. 3:[162]

Eq. 5 $$\mu_{II} = \frac{64\sqrt{\pi}(\varepsilon_s\varepsilon_0)^2(2k_B)^{\frac{3}{2}}}{q^3\sqrt{m^*}\cdot\ln[1+A^2\cdot T^2]} \cdot \frac{T^{3/2}}{N_I} \quad ; \quad A \equiv \frac{12\pi\varepsilon_s\varepsilon_0 k_B}{q^2}\frac{1}{N_I^{1/3}}$$

where $N_I$ is the ionized impurity density. At very low temperatures, where impurities will no longer be ionized due to insufficient thermal energy, the mobility due to *neutral* impurities is:[162]

Eq. 6 $$\mu_{NI} = \frac{\pi^2 m^* q^3}{10\varepsilon_s\varepsilon_0 h^3} \cdot \frac{1}{N_N}$$

---

[i] Please note that *L* that is presented above is the electronic diffusion length, while $L_D$ is the screening length.

where $h$ is Planck's constant and $N_N$ is the density of neutral defects. In Eq. 5, the temperature dependence of the mobility is $T^{+3/2}$, while when the ionized impurity is neutralized at low temperatures, it is no longer dependent upon temperature. The temperature dependence will be further emphasized (in dark-red color), as it is going to be the most important experimental evidence from which the actual scattering mechanism is deduced.

In the case of *phonons*, which are elastic waves propagating in the lattice, one should consider optical phonons (OP) where the wavelength is of the order of a unit-cell, and acoustic phonons (AP), where the wavelength is larger than a unit cell (and, thus, with lower energy). Considering $u(r,t)$ as the wave function of the elastic phonon waves, where $u(r,t)$ describes a periodic sub-lattice atomic displacement (as in OP) and $\frac{\partial u(r,t)}{\partial r}$ describes the produced strain and a displacement of the entire lattice (as in AP), the perturbation potentials are:[137]

and

$$\text{Eq. 7} \qquad U_{OP}(r) = \mathbf{d'}_O \cdot u(r,t)$$

$$\text{Eq. 8} \qquad U_{AP}(r) = \mathbf{d'}_A \cdot \frac{\partial u(r,t)}{\partial r}$$

$\mathbf{d'}_O$ and $\mathbf{d'}_A$ are *deformation potentials* (for optical, *O*, and acoustic, *A*, phonons), which are broadly discussed in section 3.4. Optical or acoustic deformation potential scattering (abbr. ODP or ADP, respectively) can be further understood following Figure 32, where instead of an existing permanent defect, which changes the potential landscape of a free charge-carrier, OP and AP are a constant background perturbation that occur due to thermal energy. In real 3D semiconductors, OP vibrations consist of one sub-lattice moving against the other. In contrast to the simple change in volume of the unit-cell produced by longitudinal AP, optical phonon scattering is sensitive to the symmetry of the crystal. For example, selection rules forbid OP scattering of electrons at momentum *k'=(0,0,0)* (Γ point) along ⟨100⟩ directions (which includes electrons in the conduction bands of GaAs and Si). However, OP scattering does usually occur for holes at the VBM and, e.g., in Ge, also electrons at the CBM.[137] Here I will explicitly consider the generic AP scattering. Using Eq. 8 and following the development for $\frac{1}{\tau_s}$ and then using Eq. 3, electron/hole mobility, dominated by ADP scattering due to AP is:[162]

$$\text{Eq. 9} \qquad \mu_{ADP} = \frac{\sqrt{8\pi} \cdot \hbar^4 \cdot q \cdot E_Y}{3 \cdot (m^*)^{5/2} \cdot k_B^{3/2} \cdot d'^2} \cdot \mathbf{T^{-3/2}}$$

where $E_Y$ is Young's modulus. If ODP scattering is allowed, it should, in principle, give a similar temperature dependence as ADP, with different parameters. Monte-Carlo simulations of ODP in Si and Ge showed, however, that $\mu_{ODP} \sim T^{-2.1}$ to $T^{-2.4}$ [163] (but see below for $\mu_{POP}$ and $\mu_{PZ}$).

In compound semiconductors, like HaPs, GaAs or CdTe, the bond between adjacent atoms is partially ionic, where each adjacent ion will possess an effective charge $q^*$ - a fraction of the formal electronic charge that is determined by the degree of ionicity of the bond. Deformation of the lattice by phonons perturbs the dipole moment between atoms, which results in an electric field that scatters carriers. Polar OP scattering (or POP), is very strong and is typically the dominant scattering mechanism at RT, as found for GaAs and CdTe.[164–166] The perturbing potential for POP is:[137]

$$\text{Eq. 10} \qquad U_{POP}(r) = \frac{qu}{i\beta\varepsilon_0} \cdot \frac{q^*}{V_u}$$

with $V_u$ being the volume of a unit cell and $\beta$ the phonon wave vector. It is common to measure the strength of the polar interaction by the factor $\left(\frac{\varepsilon_s}{\varepsilon_\infty} - 1\right)$, which relates to $\frac{q^*}{V_u}$, as follows:[137]

$$\text{Eq. 11} \qquad \left(\frac{q^*}{V_u}\right)^2 = \frac{\varepsilon_0 \rho \omega_{LO}^2}{\varepsilon_s}\left(\frac{\varepsilon_s}{\varepsilon_\infty} - 1\right) = \varepsilon_0 \rho \omega_{LO}^2 \left(\frac{1}{\varepsilon_\infty} - \frac{1}{\varepsilon_s}\right) = \frac{\varepsilon_0 \rho \omega_{LO}^2}{\varepsilon^*}$$

with $\rho$ the density of the material, $\omega_{LO}$ the lowest *longitudinal* OP frequency and $\varepsilon_\infty$ and $\varepsilon_s$ the high frequency (optical) and static dielectric constants, respectively. Using Eq. 10 and following the development for $\frac{1}{\tau_s}$ and using Eq. 11 and Eq. 3:[162,164]

$$\text{Eq. 12} \qquad \mu_{POP} = \frac{8\pi \cdot \hbar^2}{3 \cdot q \cdot m^{*\frac{3}{2}}\sqrt{k_B}} \cdot \varepsilon_0 \varepsilon^* \cdot \frac{T^{1/2}}{\theta} \cdot \left(\exp\left(\frac{\theta}{T}\right) - 1\right)$$

where $\theta$ is the Debye temperature, which is defined as: $\theta = \frac{\hbar \cdot \omega_{LO}}{k_B}$ with $\omega_{LO}$ being the lowest longitudinal optical phonon frequency[i]. We see that, unlike other regimes, the mobility temperature dependence is exponential rather than in a power regime as for the other scattering mechanisms.

---

[i] The Debye model relates to acoustic phonons only; the highest LA phonon frequency usually agrees sufficiently well with the lowest optical LO to make it possible to use the model also with LO phonons.

AP, similar to OP, can also produce an electrostatic perturbation known as *piezoelectric scattering* (PZ), but only in a group of non-centrosymmetric semiconductors, where piezoelectricity is possible. This effect is usually weaker than POP scattering, but at low temperatures, where the number of POP phonons is small and carriers do not have sufficient energy to emit OP, PZ scattering becomes important. Its scattering potential will be:[137]

$$\text{Eq. 13} \qquad U_{PZ}(r) = \frac{q e_{PZ} u}{\varepsilon_s \varepsilon_0}$$

where $e_{PZ}$ is the material's piezoelectric constant. Using Eq. 13 and following the development for $\frac{1}{\tau_s}$ and Eq. 3:[162]

$$\text{Eq. 14} \qquad \mu_{PZ} = \frac{8\sqrt{2\pi} \cdot \hbar^2}{3 \cdot m^{*\frac{3}{2}} \sqrt{k_B}} \cdot (\varepsilon_0 \varepsilon_s)^2 \cdot \frac{2 E_Y}{e_{pz}^2} \cdot T^{-1/2}$$

It should be noted that mobility due to POP scattering is sometimes considered in the literature to have a temperature dependence of $\mu_{POP} \sim T^{-1/2}$.[167–170] This relation holds only in the limit where $T \gg \theta$, when $\exp\left(\frac{\theta}{T}\right) - 1 \approx \frac{\theta}{T}$ (using the first term of its Taylor expansion). Since $\mu_{PZ} \propto T^{-1/2}$ and $\mu_{ODP} \propto T^{-1/2}$ explicitly, this can lead to a misinterpretation. Moreover, an analysis of small ranges of mobility temperature dependence will result in an increasing $\gamma$ in $\mu_{POP} \sim T^\gamma$ $\mu_{POP} \sim T^\gamma$ with decreasing temperatures, which may be misinterpreted as a different regime, e.g., AP scattering with $\mu_{ADP} \sim T^{-3/2}$ (as also noted in refs. [171,172]).

Charge scattering can also occur from inter-valley transitions (for higher energy electrons, usually at very high temperatures)[164] or from dislocations. However, since they are less common and are specific to band-structure and dislocation density, respectively, these mechanisms are not considered.

Following Eq. 3 and the previously mentioned scattering rate summation, $\frac{1}{\tau_{total}} = \frac{1}{\tau_{s1}} + \frac{1}{\tau_{s2}} + \cdots$, the actual mobility of a real material can be estimated from:

$$\text{Eq. 15} \qquad \frac{1}{\mu_{total}} = \frac{1}{\mu_{II}} + \frac{1}{\mu_{NI}} + \frac{1}{\mu_{ADP}} + \frac{1}{\mu_{ODP}} + \frac{1}{\mu_{PZ}} + \frac{1}{\mu_{POP}}$$

where $\frac{1}{\mu_{II}}$ and $\frac{1}{\mu_{NI}}$ depend on the defect density and the temperature (ionized impurities neutralize at lower temperatures).

Usually, the regimes colored in blue are relevant at low temperatures, while the rest may be relevant at higher temperatures. Figure 33 is a good illustration of the use of Eq. 15 and the effect of impurities upon mobility.

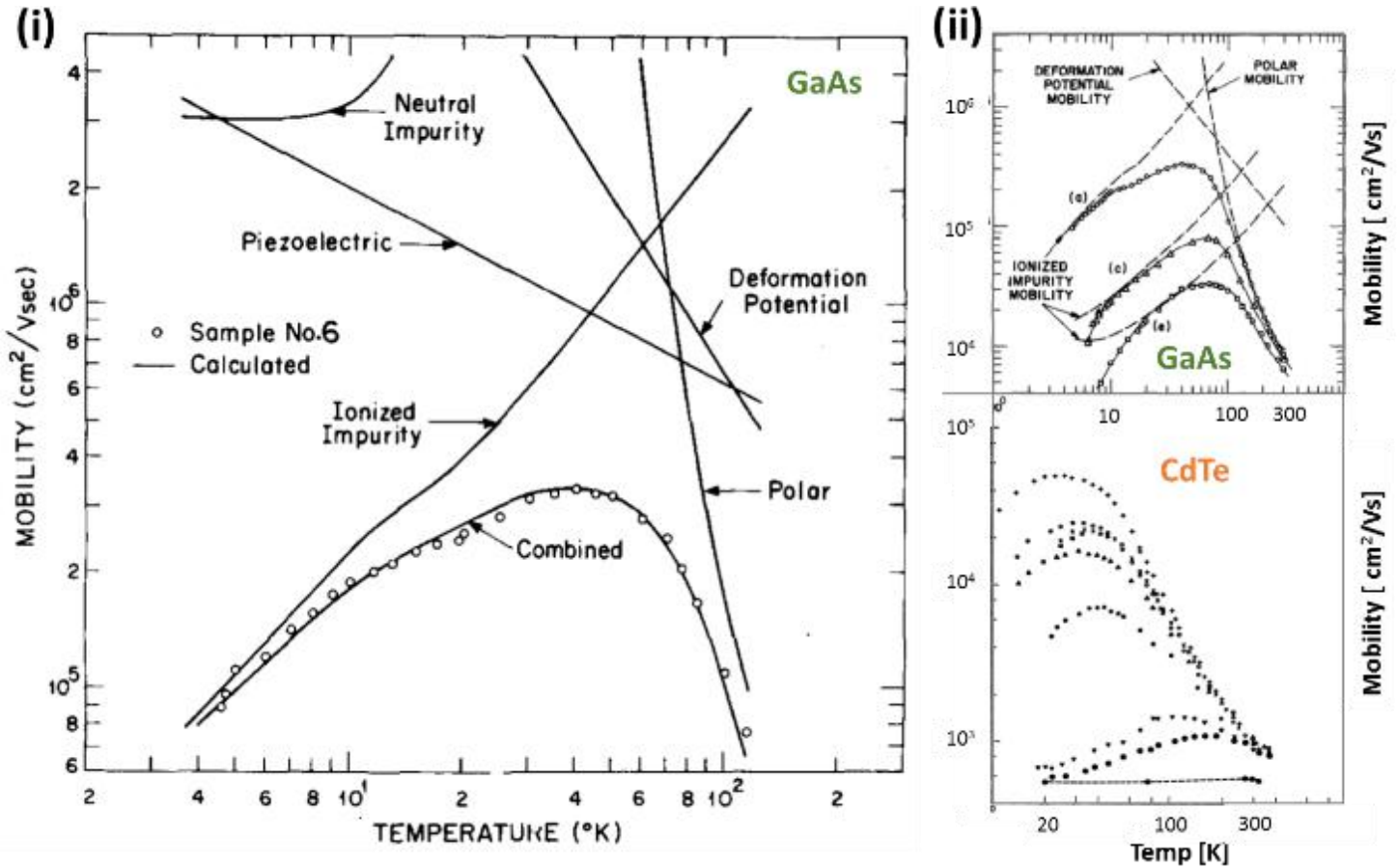

Figure 33: (i)[166] Mobility as function of temperature for GaAs (hollow circles) and simulated effective mobility (solid line following the data points) using Eq. 15 for the different scattering mechanisms (which are plotted individually). (ii) Mobility as function of temperature for GaAs[165] and CdTe[173] at different doping concentrations. The plots with the highest (lowest) maximum mobility represent lower (higher) doping densities.

Although I only analyze mobility values and did not measure them, I will touch upon this point. Both contact and non-contact probes can be used to investigate charge carrier mobilities. Methods that require electrical contacts include *Space charge limited current (SCLC)*, *Hall effect measurements and time-of-flight (TOF)*. Besides noise problems for low conductance materials and some technical preparation difficulties, these measurements tend to be influenced by the quality and type of electrical contact made. In addition, the prolonged application of low frequency or DC electrical fields to materials with non-negligible ionic conductance (such as HaPs), may cause ion migration and can potentially influence results. Furthermore, for materials with limited thermodynamic stability (see chapter 4), imposing large voltage differences across such material can lead to electrochemistry and degradation. Hall-effect measurements are usually thought of the most reliable technique. However, so far, due to the HaP's low dark conductivity, background illumination is required to reach high enough conductivity to overcome noise levels. The results of such measurements, while fine for optoelectronics, should be considered with care for the basic electronic properties of the material.

Contactless methods (meaning, no electric bias is applied) probe AC conductivity to try to avoid bias-related problems, but they are more complex to implement and analyze. These include: *optical-pump-THz-probe conductivity (THzC); transient microwave conductivity (MWC)* that avoid any contact, or *photoluminescence quenching (PLQ)*, where PL transients are monitored following the photoexcitation of a thin film that forms an interface with an electron or hole extraction layer. PLQ involves a contact, but no electric biasing. Modeling of the results relies on accurate knowledge of the charge-transfer efficiency at the interface, which is often assumed to be unity for simplicity. For further details, I refer to ref. 171 which reviews the different values and techniques that were used so far on extracting mobility values from HaPs, to which I will return in the next section.

### 3.5.2. Results and discussions:

Following the tremendous success of HaPs in optoelectronics, its charge mobility has become an issue of intense interest. As mentioned by several review papers[66,71,171], the room-temperature mobility of different Pb-based HaPs span mostly a range between 1-100 $cm^2$/V·s (see Figure 34). Despite the spread that can reach almost two orders of magnitude (cf. SCLC method), it is satisfying to find that, on average, most methods agree with each other. An exception is the PLQ method that gives lower mobility values than the other measurement techniques. It is interesting to note that there is quite a sharp difference between methods that were used for measuring single crystals (the group of electrical low frequency methods) and for thin films (mostly non-contact high frequency methods). Although values derived from single crystals were, on average, higher than those found from thin films, the different set of experimental tools reduce the significance of such a conclusion.

Besides the absolute value of the mobility, its dependence on temperature is of particular importance, since it is indicative of the scattering mechanism, as broadly explained above ("about charge mobility"). Figure 35 shows the variation of charge mobility values with temperature of MAPbI$_3$ films and MAPbBr$_3$ single crystals gathered from different sources. Similar to data at 300 K (Figure 34), we can see that the data spread is quite systematic even for different temperatures, where some additional discontinuities are present, probably due to phase transitions or other experimental reasons. Although the quality of the data is not good enough to allow a clear definition of the scattering mechanism, a clear increase in mobility with decreasing temperatures is observed. Since scattering regimes show a power relation as $\mu \propto T^\gamma$ (cf., Eq. 5, Eq. 6, Eq. 9, Eq. 14 and, if using a Taylor expansion for confined temperature regions, Eq. 12) the $\gamma$ parameter is shown for the entire data set, or for specific phase regions (see numerical values in Figure 35).

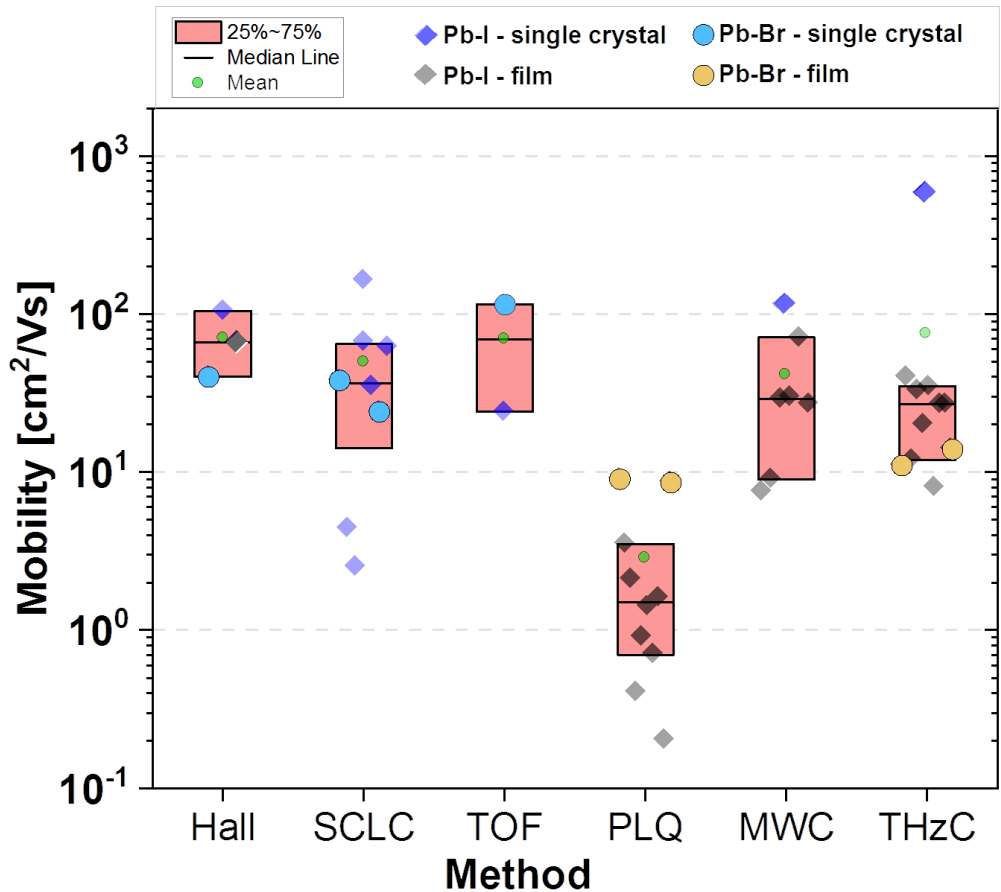

Figure 34: Spread of RT mobility values of Pb-X based (MAPbI$_3$, MAPbI$_3$(Cl), FAPbI$_3$, Cs$_{0.17}$FA$_{0.83}$PbI$_3$, MAPbBr$_3$, FAPbBr$_3$, Cs$_{0.17}$FA$_{0.83}$PbBr$_3$) HaPs. The meaning of the different data points is shown at the top of the figure. A reference for the different mobility values and an explanation for the different methods can be found in ref. 171.

For the given temperature range, scattering by impurities – ionized or neutralized – can be rejected completely, since the temperature power parameter is not negative, i.e., $\mu_{II} \propto T^{+3/2}$ and $\mu_{NI} \propto T^0$. What remains are ADP, ODP, PZ and POP scattering regimes, which all result in a negative $k$. Figure 36 show comparison to the data presented in Figure 35 with solutions of $\mu(T)$ plots (following Eq. 5 ($\mu_{II}$), Eq. 6 ($\mu_{NI}$), Eq. 9 ($\mu_{ADP}$), Eq. 12 ($\mu_{POP}$) and Eq. 14 ($\mu_{PZ}$)) using experimentally-based parameters (Table 3).

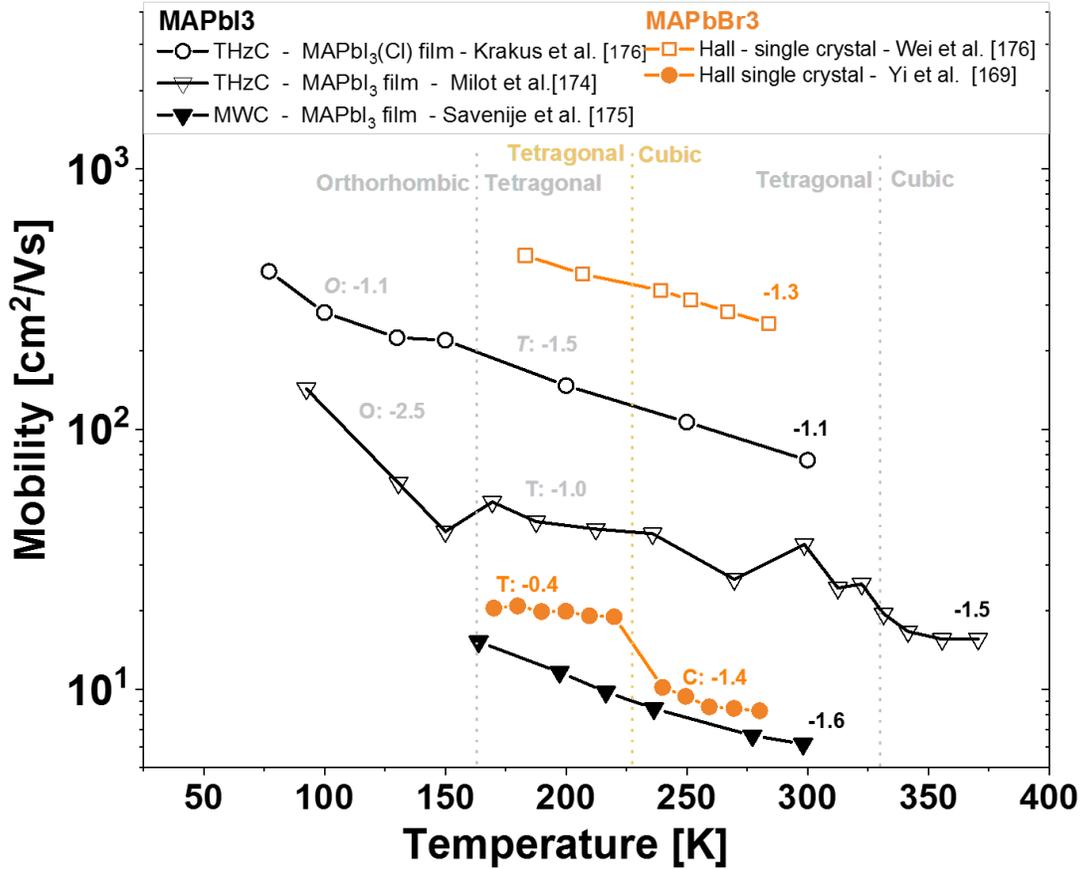

Figure 35: Mobility temperature dependence of MAPbI$_3$ (black) and MAPbBr$_3$ (orange) based on different literature sources[169,174–177]. The method and type of sample are mentioned in the legend. The mobility values for MAPbI(Cl)$_3$ in Krakus et al.[176] are derived from a plot of the scattering time, $\frac{1}{\tau_s}$, vs. temperature, using Eq. 3 with $m^* = 0.104$ [178,179]. The numerical values are the power parameters, $\gamma$, in a $\mu \propto T^\gamma$ relation of the entire range of data points. If there is a prefix of C, T or O, the fit is done only for the cubic, tetragonal or orthorhombic phases of each compound, respectively. The phase transition temperatures are shown in the figure color-coded for the two materials. It should be noted that all the presented values involve supra-bandgap illumination during measurements, which may obscure the more fundamental 'dark' mobility. The full symbols refer to mobility values that we find to be more reliable: for MAPbBr$_3$ by the Podzorov group with a track record in Hall measurements on low mobility low carrier density (organic)crystals; for MAPbI$_3$ by MWC, which is proposed by Levin et al. [180] to be more accurate than THzC for mobility measurements.

Note that $\mu_{ODP}$ simulated plots are not presented in Figure 36. The absence of an explicit analytical expression for $\mu_{ODP}$, and since Monte Carlo simulations[163] for Si and Ga show $\mu_{ODP} \sim T^{-2.1}$ to $T^{-2.4}$, attempts to derive an explicit analytical expression for were abandoned. $\mu_{PZ}$ is not plotted for MAPbBr$_3$, since it is centrosymmetric at RT (as will be further elaborated in section 5.2), and at lower temperatures no piezoelectric constants nor piezoelectricity are known. For MAPbI$_3$ $\mu_{PZ}$ is plotted twice: for the lowest 'dark' value of $e_{PZ} = 0.038 \frac{C}{m^2}$ that was reported for single crystals,[181] and for the higher 'light' value of $e_{PZ} = 0.35 \frac{C}{m^2}$, which is reported for thin films under illumination.[182]

For both MAPbI$_3$ and MAPbBr$_3$, it is clear that POP should dominate the charge carrier scattering at the reported temperatures and only at very low temperatures (< 50 K, which has still not been reported), NI scattering *may* become important. In this respect, defect densities, $N_{(I\ or\ N)}$, are assumed to be $10^{15}$ cm$^{-3}$, a commonly-reported defect density in HaP thin films. Since $\mu_{(II\ or\ NI)} \propto \frac{1}{N_{(I\ or\ N)}}$, lowering $N_{(I\ or\ N)}$ to that found in single crystals ($10^{10}$ cm$^{-3}$) will increase the $\mu_{(II\ or\ NI)}$ resulted values by five orders of magnitude, which will make NI scattering insignificant *even at lower temperatures, which has not been reported for any semiconductor till now.* [i]

Table 3: Experimentally derived material-related constants for simulations presented in Figure 36 and Figure 37. For CdTe and GaAs only POP-relevant parameters are presented. $m^*_{(MAPbI3)}$=0.104 is deduced from magnetoresistance and represents the reduced mass $\left(\frac{1}{\mu} = \frac{1}{m_e^*} + \frac{1}{m_h^*}\right)$.

| Parameter | units | MAPbI$_3$ | Ref. | MAPbBr$_3$ | Ref. | CsPbBr$_3$ | Ref. | MAPbCl$_3$ | Ref. | CdTe | GaAs | Ref. |
|---|---|---|---|---|---|---|---|---|---|---|---|---|
| $\omega_{LO}$ | cm$^{-1}$ | 133 | 170 | 166 | 170 | 135 | 172 | 225 | 170 | 168 | 296 | 183 |
| $m^*/m_e$ | --- | 0.104 | 178, 179 | 0.117 | 179 | 0.126 | 184 | 0.2 (calc.) | 170 | 0.096 | 0.066 | 183 |
| $\varepsilon_\infty$ | --- | 5.0 | 170 | 4.7 | 170 | 4.3 | 172 | 4.0 | 170 | 7.2 | 10.9 | 183 |
| $\varepsilon_S$ | --- | 33 | 170 | 32 | 170 | 29.3 | 172 | 29.8 | 170 | 10.2 | 12.9 | 183 |
| $E_Y$ | GPa | 14.2 | This work | 19.6 | This work | 15.8 | This work | 26.3 | This work | | | |
| $e_{PZ}$ | C/m$^2$ | 0.038 (dark) 0.35 (light) | 181 182 | n.a. (centro-symmetric) | --- | n.a. | --- | n.a. | --- | | | |
| $d'$ (Eq. 1) | eV | 1.22 | This work | 0.66 | This work | 0.76 | This work | 1.95 | This work | | | |
| $N_I$ (estimate) | cm$^{-3}$ | $10^{16}$ | --- | $10^{16}$ | --- | --- | --- | --- | --- | | | |
| $N_N$ (estimate) | cm$^{-3}$ | $10^{16}$ | --- | $10^{16}$ | --- | --- | --- | --- | --- | | | |

---

[i] To this point, results between 50 K - 4 K are absent. NI may become significant at these temperatures as will be shown in Figure 36.

The proximity of the power $\gamma$ parameter to -1.5 (cf. Figure 35), which corresponds to ADP, resulted in reports suggesting that ADP is the dominating scattering mechanism in HaPs. Later it was suggested that PZ scattering, due to *polar* AP, could dominate the scattering, since its absolute values were found to be lower than those resulted from ADP.[185] This is a correct statement for the lower temperatures when using the parameters mentioned in Table 3 with $e_{PZ} = 0.08 \frac{C}{m^2}$, which was reported for MAPbI$_3$ thin films under dark conditions.[182,186] Eventually, as was more recently suggested by others[167,170–172] (without the broad comparison between scattering regimes, as shown in Figure 36), POP scattering dominates charge transport in HaPs. This is actually an expected result for heteropolar semiconductors with (some) ionic character (cf., GaAs and CdTe in Figure 33),[183] and as was further emphasized in Figure 37(i).

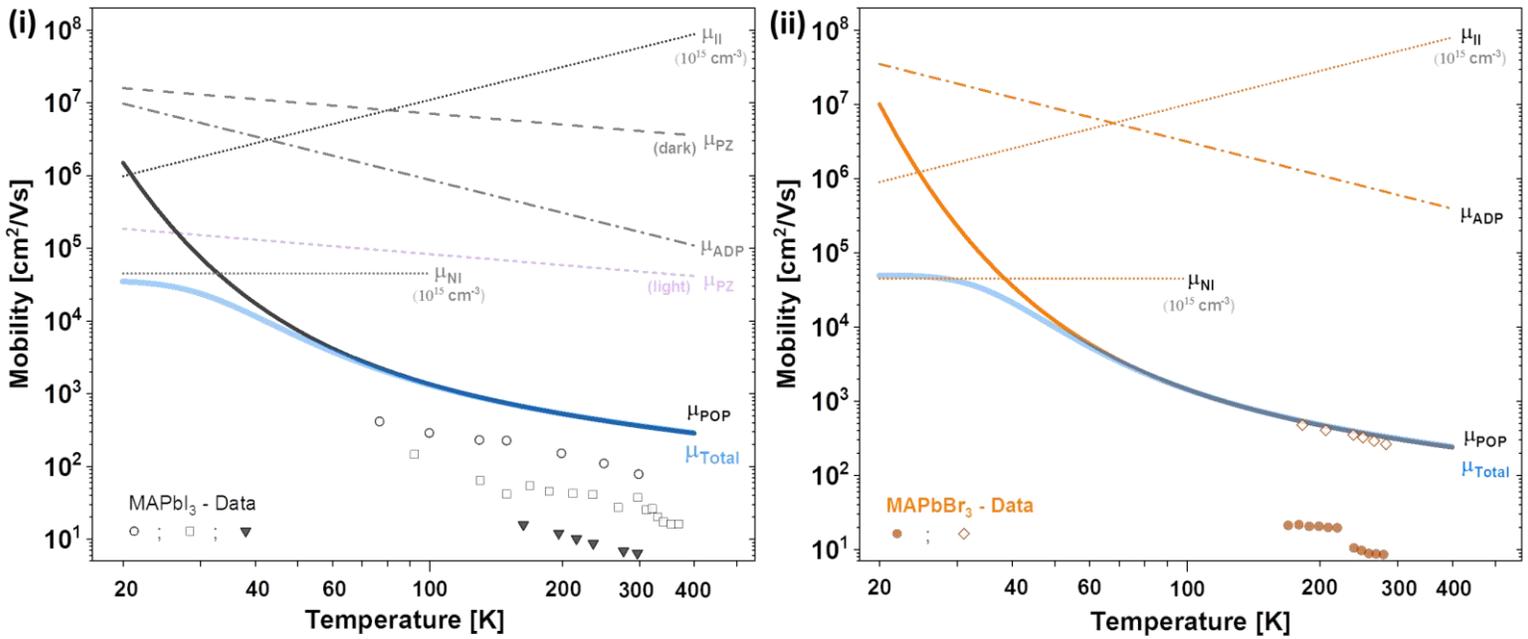

Figure 36: A log-log plot of mobility temperature dependence of (i) MAPbI$_3$ and (ii) MAPbBr$_3$ together with simulated scattering regimes of mobilities based on Eq. 5 ($\mu_{II}$), Eq. 6 ($\mu_{NI}$), Eq. 9 ($\mu_{ADP}$), Eq. 12 ($\mu_{POP}$) and Eq. 14 ($\mu_{PZ}$), using parameters taken from Table 3 . The blue curve represents the result of Eq. 15 - $\mu_{Total}$. $\mu_{Total}$ is simulated using $\mu_{PZ}$ with $e_{PZ}$ that was found for experiments under light[182]. The higher $\mu_{PZ}$ is for $e_{PZ}$ that was found for a single crystal[181]. For simulations of $\mu_{II}$ and $\mu_{NI}$, the defect density is assumed to be $10^{15}$ cm$^{-3}$ – a commonly-reported defect density in HaP thin films. Lower defect densities result in higher values of $\mu_{II}$ and $\mu_{NI}$, which become even less relevant for $\mu_{Total}$. This suggests that POP scattering is the mechanism that suites best of all other mentioned mechanisms, and determine the mobility at a wide temperature range. At low temperatures NI *may* dominate (9 meV (~100K) defects, w.r.t. band maximum/minimum, are assumed). The deviation of the data (especially the more reliable ones – full symbols – see Figure 35 ) from the simulated $\mu_{POP}$ plot suggests: (1) effective mass (*m*\*) values of the HaPs are higher than the deduced from the magneto resistance measurements (see Figure 37 and further discussion in the main text) ; (2) the pure POP model is still insufficient and it needs to be modified or it is not applicable here.

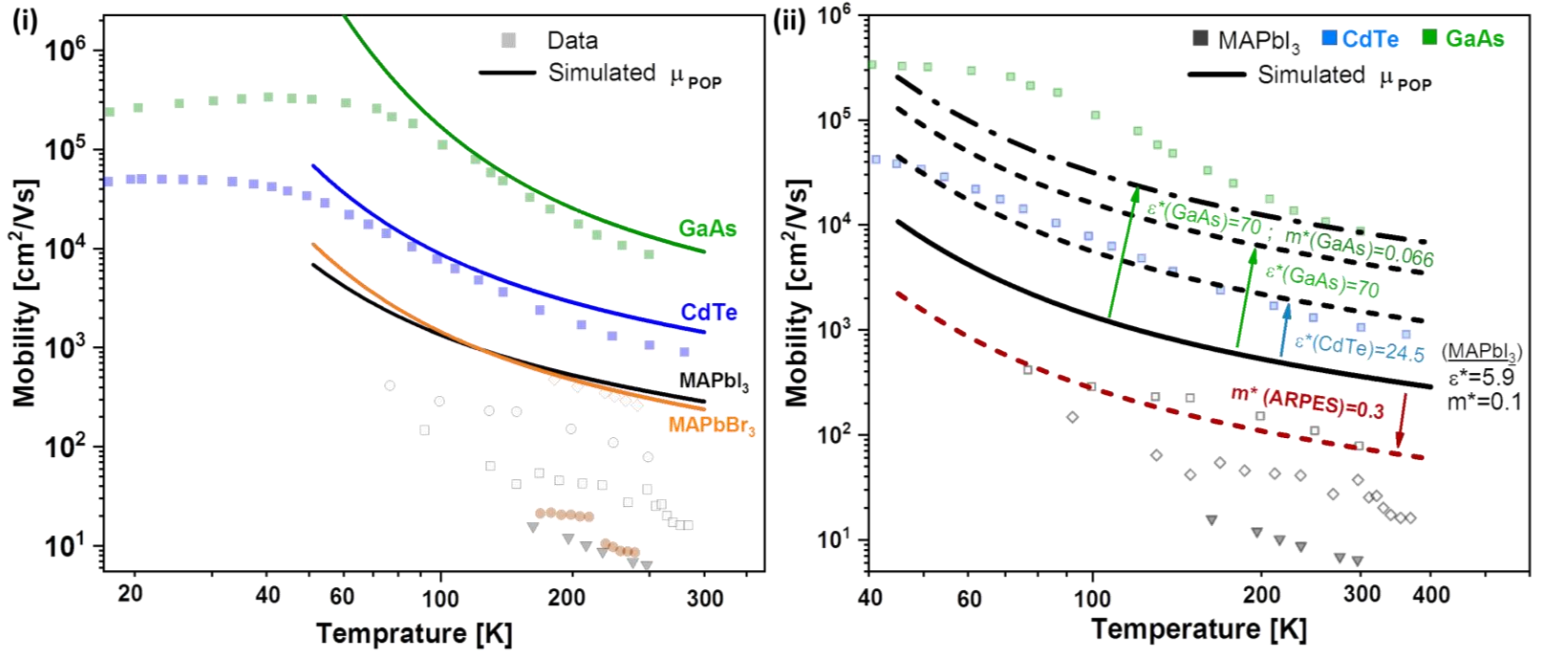

Figure 37: **(i)** Log-log plots of mobility temperature-dependence of MAPbBr$_3$, MAPbI$_3$, CdTe and GaAs together with $\mu_{POP}$ simulations based on Eq. 12 using parameters taken from Table 3. This supports the hypothesis that at higher temperature scattering regimes, by both trends and absolute values, a POP scattering regime dominates. **(ii)** Simulation of $\mu_{POP}$ using MAPbI$_3$ parameters (solid line) with higher $m^*$ values (red dashed line) or higher $\varepsilon^*$ (black dashed line). The partially varied parameter values are those of CdTe and GaAs (as indicated). This suggests that (a) the measured effective mass is incorrect, and (b) what limits the mobility of HaP to be as high as that of CdTe or GaAs is the smaller difference between $\varepsilon_\infty$ and $\varepsilon_s$, which results in higher $\varepsilon^*$ (cf. Table 3). The presented data points for CdTe and GaAs are those with the lowest defect density, presented in Figure 33(ii). The effective mass of 0.3 $m_e$ deduced by Angle-Resolved Photoemission Spectroscopy (ARPES), and represents the valence charges (hole) effective mass.

Using the same Eq. 12 for simulating $\mu_{POP}(T)$, a decent fit to the low defect density of GaAs and CdTe data is observed (see Figure 37(i)). Since POP scattering is caused by intrinsic thermal fluctuations of the lattice, regardless of the lattice's defect density, Figure 37(i) shows that the 1-2 orders of magnitude difference between HaPs and other heteropolar semiconductors is *fundamental*. It is also noticeable that HaPs exhibit only negative $\gamma$, unlike CdTe and GaAs whose mobility trend turns to from negative to positive $\gamma$ values as temperature decreases due to scattering by ionized impurities (cf. Figure 33). Nevertheless, based on simulations POP scattering results in much lower mobility values for $\mu_{POP}(T)$ than other scattering mechanisms at different temperatures, when comparing to the more 'rigid' semiconductors (compare Figure 33 with Figure 36). More generally, 'softer' heteropolar materials will require lower temperatures for other mechanisms to dominate the charge scattering compared to more 'rigid' ones (assuming low defect density). Therefore, for HaPs, measurements at much lower temperatures may be needed to see a similar 'turnover' in $\gamma$.

Two questions remain to be answered: (1) what causes the (still) ~1 order of magnitude (on average) higher predicted $\mu_{POP}(T)$ for MAPbX$_3$ HaPs (except for one set of high mobility MAPbBr$_3$ data)? ; (2) what are the fundamental properties of GaAs and CdTe that make their mobility values so much higher than those of HaPs?

For question (1), a possible answer is that the measured effective mass of $m^*$~0.11 $m_e$ is too low to be the actual effective mass of the charge carriers due to some unknown experimental misinterpretation of the magneto-transmission measurements.[178,179] The theoretically calculated $m^*$ was found to be ~0.26 $m_e$ or higher[187,188] (based on the $k$-$E$ dispersion relations). Papers that assume mobility values from, e.g., THzC measurements (unlike, e.g. Hall mobility measurements), assume an effective mass – sometimes from theory and sometimes from experiment, which is one reason for the scattered data shown in Figure 34. As noted above $m^*$ for HaPs was measured using a magneto-absorption spectroscopy technique,[178,179] while most other effective mass values are measured via cyclotron resonance or Faraday rotation techniques[183]. Without distracting from or questioning the measurement quality, these results suggest that, maybe, the models used for extracting $m^*$ from magneto-absorption for HaPs require particular adjustments.

If one wishes to make a proper comparison with the theoretical calculations from which $m^*$ is extracted (i.e., $k$-$E$ dispersion relations), Angle-Resolved Photoemission Spectroscopy (ARPES) is used to get the $k$-$E$ dispersion relations of the valence band. ARPES measurements of MAPbBr$_3$ [189] and MAPbI$_3$ [190] single crystals show that the curvature around the 'M' point of the surface Brillouin zone corresponds to effective masses of ~0.25-0.35 $m_e$. It is important to note that the VBM in HaPs lies around the 'R' point in the $k$-$E$ space (Brillouin zone), and since the ARPES measurements around the 'R' point were in the noise of the measurement, the ARPES-derived $m^*$ values may not be applicable for actual conduction. However, Yang et al., [190] assume, based on their ARPES measurements on MAPbI$_3$, that $m^*$ around the 'R' point should be as high as ~0.4 $m_e$.

Note that ARPES measurements do not tell us anything about the conduction band, but only the valence band. However, since HaPs often behave as a $p$-type material, the VBM values may be relevant after all. Overall, we see that there is some justification to assume that $m^*$~0.1 $m_e$ deduced from magneto-resistance measurements are too low for these materials, as will be also emphasized in Figure 38(iii) Figure 39(iii).

Using somewhat modified models like that of 'large polaron',[172,191] where electrons are traveling with an effectively heavier $m^*$ than their band-$m^*$ (i.e., $m^* \equiv \frac{1}{\hbar^2} \frac{\partial^2 E(k)}{\partial k^2}$) due to electrostatic interactions with a freely fluctuating highly polarizable (or soft) lattice (also known as Fröhlich coupling[192]), may result in an effectively larger $m^*$. However, the POP model used for Eq. 12 uses Fröhlich's logic, already includes coupling effects (of electrons to dipoles formed by a fluctuating lattice), so independent empirically-based verification of the 'large-polaron' model is required. It is important to mention that, according to my understanding, the lattice depolarization is independent of the electric fields induced by the electrons/holes, unlike the classical polaron model that assumes lattice polarization due to localized electric fields induced by the charges. In other words, when the magnitude of the deformation due to a static charge is smaller than the natural vibration amplitude of the system, the opposite of the classical view of a polaron, where the lattice follows the electric field of the electron holds, where the electron follows the electric fields induced by a strongly fluctuating heteropolar lattice.

Some differences may rise due to *pre-factors* that vary between different mathematically-developed POP models. For example, a factor of ~3 difference can be found between refs. [192] and [193].[i] Giving this small difference, the expression used in Eq. 12 is, therefore, an acceptable expression for heteropolar crystals, and as we saw in Figure 37(i), it results in a fair prediction of $\mu(T)$ for semiconductors with an apparently similar bond nature as that of HaPs.

Another possibility for the mismatch between the model and the measured values is that all these models assume parabolic band minima (or harmonicity), which might not be the case in a highly anharmonic system (which assumes non-parabolic bands) such as that of HaPs[194]. The importance of anharmonicity to electronic properties, and specifically mobility, is still not clear. Further theoretical efforts are required to explain experimental results that diverge the 'classical' harmonic models to dynamic and static anharmonic picture. For example, in HaPs the dynamic contribution seems to be larger than in other semiconductors,[194] which is now an ongoing research topic in the group of O. Yaffe's group. Overall, as shown in Figure 37(ii), artificially increasing the effective mass to 0.3 $m_e$ bring the simulated $\mu_{POP}(T)$ closer to the actual data of MAPbI$_3$, without changing the $\gamma$ value much.

For question (2), we know that although the 'light' electron effective mass of GaAs is about half than that of HaPs, but that of CdTe is only slightly smaller than that reported for HaPs (cf. Table 3). $\varepsilon^*$, however, which is defined as $\varepsilon^* \equiv \left(\frac{1}{\varepsilon_\infty} - \frac{1}{\varepsilon_s}\right)^{-1}$ (cf. Eq. 11), can increase quite dramatically as the difference between $\varepsilon_\infty$ and $\varepsilon_s$ decreases. In fact, at a lower bound, where $\varepsilon_s \gg \varepsilon_\infty$, $\varepsilon^* \to \varepsilon_\infty$, while at the limit where $\varepsilon_s \to \varepsilon_\infty$, $\varepsilon^* \to \infty$.[i]

---

[i] Ref. 192 is authored by Fröhlich himself, where he theoretically treat scattering by POP.

When comparing the subtraction-difference between $\varepsilon_s$ and $\varepsilon_\infty$ from Table 3, we indeed find it to be 2 to 3 for GaAs and CdTe, respectively, but ~27 for HaPs. This large difference for HaPs limits $\varepsilon^*$ to $\varepsilon_\infty$ (~5.5), while, for GaAs and CdTe it reaches 24.5 and 70.3 respectively – a much more significant difference than the one found for $m^*$.[ii] This means that changing artificially $\varepsilon_s$ values can only increase $\mu_{POP}$ but not sufficiently decrease it to the experimental values. When plotting Eq. 12 with MAPbI$_3$ parameters, except for $\varepsilon^*$, where we use that of CdTe (Figure 37(ii)), we find that indeed the plot approaches the CdTe $\mu(T)$ data. Changing the $\varepsilon^*$ parameter to that of GaAs shifts the $\mu(T)$ plot upwards, but it requires an additional reduction in $m^*$ to reach the GaAs data.

The curvature of the log-log plots over a specific temperature range is mainly dependent on the Debye temperature, $\left(\theta \equiv \frac{\hbar \cdot \omega_{LO}}{k_B}\right)$, which is similar for HaPs and CdTe, but (again) different for GaAs; therefore, it is expressed in a significant mismatch at lower temperatures.

Now we test if the different material parameters (i.e., $\varepsilon^*$; $m^*$; $\omega_{LO}$) are (in)dependent of each other. based on Eq. 12, where $\mu_{POP} \propto \varepsilon^* \cdot m^{*-1.5}$, we show in Figure 38(i) and (ii) $\varepsilon^*$ and $m^*$ with respect to $\mu_{POP}$. It is found that, indeed, $\mu_{POP}$ grows exponentially with decreasing $m^*$ and increasing $\varepsilon^*$, as one would expect. However, the power exponents are ~ -2.5 and +3.5 for $m^*$ and $\varepsilon^*$, respectively, which are much stronger than the -1.5 and +1 that one would expect. Apparently, as shown in Figure 38(iii), $m^*$ and $\varepsilon^*$ are related among themselves as $\varepsilon^* \sim \frac{1}{m^*}$. Although this relation rationalizes a power parameter of -2.5 for $\mu_{POP}(m^*)$, it still does not rationalize the +3.5 power for $\mu_{POP}(\varepsilon^*)$. Since $\mu_{POP}$ depends also on $\omega_{LO}$, an additional dependence between $\varepsilon^*$ and $\omega_{LO}$ may rationalize the very strong $\mu_{POP}(\varepsilon^*)$ dependence. And indeed, as shown in Figure 38(iv), $\varepsilon^* \propto \omega_{LO}^{(a)}$, with $a$ being ~ –0.6 for halides, ~ –1.0 for chalcogenides and ~ –1.4 for III-Vs. A

---

[i] Note that the value of $\left(\frac{1}{\varepsilon_\infty} - \frac{1}{\varepsilon_s}\right)$ exists in Marcus theory[195] for electron transfer reactions, in which the polarization of the surrounding solvent changes during electron transport. In the simplest model, the initial and final stages (before and after the charge transition) are equivalent, but the electron is moved from one site to another (like in $Fe^{3+} + e^- \to [Fe^{2+}]^\ddagger \to Fe^{3+} + e^-$). The activation energy for the charge transition is proportional to the *polarization* of the surrounding medium (usually a solvent), $\Delta e$, and the *polarizability* factor, $\left(\frac{1}{\varepsilon_\infty} - \frac{1}{\varepsilon_s}\right)$. This situation is similar to the case where an electron is changing its position in the lattice as it moves.

[ii] For the case of a polarizable liquid it is usually true that $\varepsilon_s \gg \varepsilon_\infty$ (i.e., for H$_2$O : $\varepsilon_s = 78.4$ and $\varepsilon_\infty = 2.4$). Treatment of an HaP system as a liquid-like can be found in Guo et al.[196].

relation of $\varepsilon^* \propto \omega_{LO}^{(-0.8)}$ and $\varepsilon^* \propto m^{*(-1)}$ can indeed rationalize the power parameter of +3.5 for $\mu_{POP}(m^*)$. It is important, though, to further understand the theoretical reasons behind these mutual interrelations, which should be an issue of a future work.

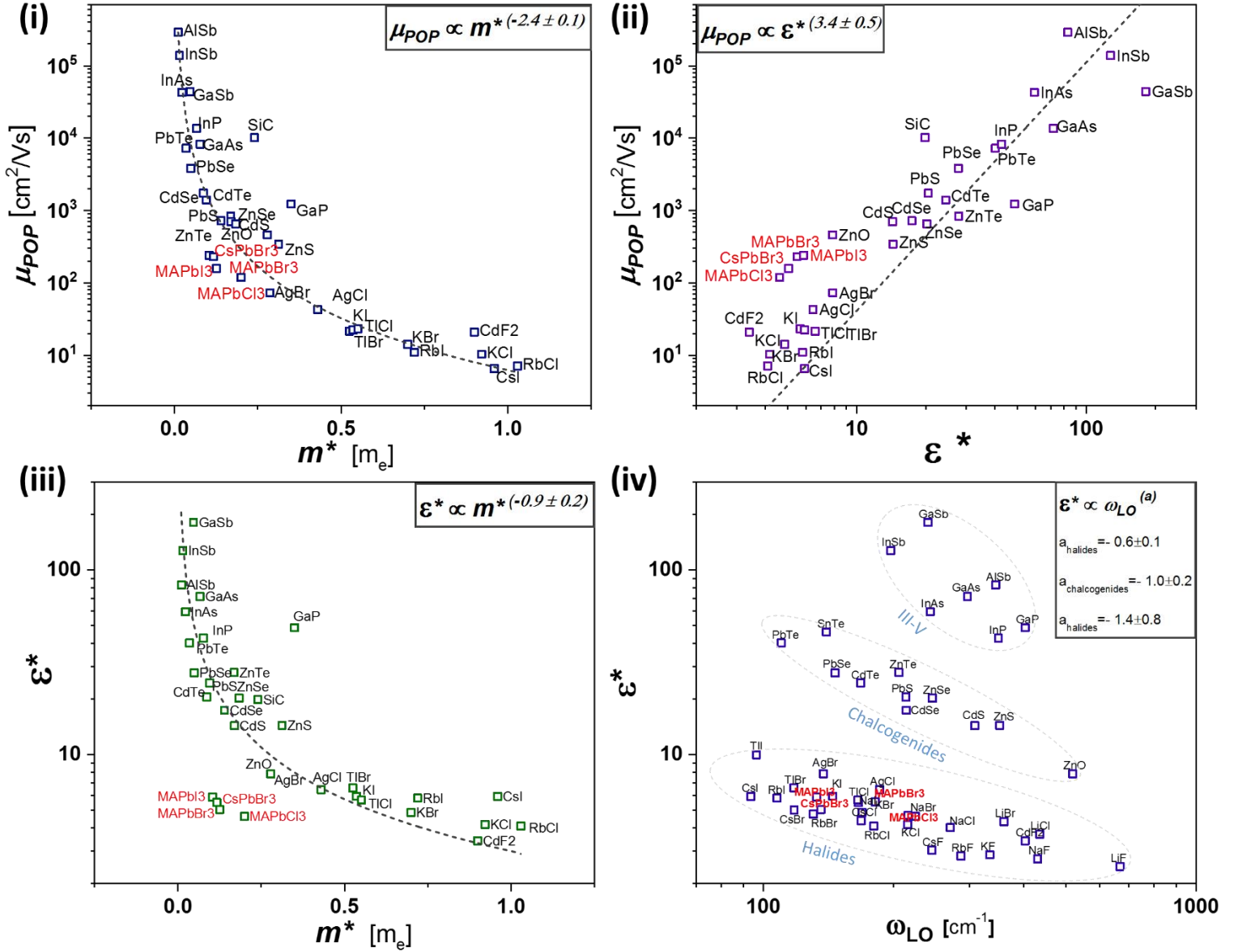

Figure 38: $\mu_{POP}$ at 300 K as a function of (**i**) $m^*$ and (**ii**) $\varepsilon^*$ of different heteropolar compounds (halides, chalcogenides and III-V's). $\mu_{POP}$ is calculated based on Eq. 12, with the empirically-based parameters from Table 3 for HaPs (colored in red) or ref. 183 for all other compositions. Power relations are found to be exceptionally high with respect to Eq. 12 $(\mu_{POP} \propto \varepsilon^* \cdot m^{*-1.5})$. (**iii**) and (**iv**) shows relations between $\varepsilon^*$ and $m^*$ or $\omega_{LO}$ that explain the exceptionally strong power relations of $\mu_{POP}$ with $m^*$ and $\varepsilon^*$ (see main text for explanation). In (**iii**), HaPs (marked in red) are shown to lie somewhat outside the general trend of other heteropolar compounds. In (**iv**), clustering of halides, chalcogenides and III-V's among themselves is observed. The spread between $m^*$ and $\omega_{LO}$ does not show any clear trend.

An interesting observation is that each subgroup is clustered in both magnitude and trends (following $\varepsilon^* \propto \omega_{LO}^{(a)}$). This clustering will not be part of further discussions. A second interesting observation, that *is* important for the sake of our discussion, is the exceptionally low $m^*$ presented in Figure 38 (iii). As mentioned above, the derived effective mass (using magneto-resistance measurements) will require further examination and verification from other methods, like cyclotron-measurements or Faraday rotation techniques.

Between $\varepsilon^*$ (or *lattice-polarizability*) and *d'*:

To understand interrelations among $\varepsilon^*$ and other composition-specific parameters, it will be useful to get a physical perspective on the meaning of $\varepsilon_\infty$ and $\varepsilon_s$. In general, $\varepsilon_i$'s can be interpreted as the screening capability of a medium of an existing electric field. $\varepsilon_\infty$ represents screening of an electric field at time scales at which atoms are static and only electrons react to an electric field. These are usually 'optical' frequencies (energies in the 'eV' range), or higher than those of atomic vibrations ('meV'), so that screening by ions can be neglected. $\varepsilon_s$ represents screening of an electric field at frequencies that are lower than those of atomic vibrations (down to DC fields), at which partially charged atoms in a lattice are capable of relaxing to a lower energy state to screen an existing electric field. As mentioned during the development of Eq. 12 ($\mu_{POP}$), a commonly-used measure for the strength of the polar interaction, or the 'screening strength' of a lattice with partially ionic charge density of $\frac{q^*}{V_n}$, is $\left(\frac{\varepsilon_s}{\varepsilon_\infty} - 1\right)$, or, if divided by $\varepsilon_s$, as $\frac{1}{\varepsilon^*}$.[137] $\frac{q^*}{V_n}$ is also known tightly related to what is called the '*lattice-polarizability*' of a crystal.

Thus, the larger $\frac{1}{\varepsilon^*}$, or the smaller $\varepsilon^*$, the more *polarizable* a system should be, with a lattice that may relax more readily. Recalling the discussion about the absolute 'deformation potential', |*d'*|, (see section 3.4, Figure 32) - the energy cost for local deformation to screen an existing electric field is directly proportional to |*d'*|. Therefore, |*d'*| should correlate with $\varepsilon^*$ in a way that for higher |*d'*| one expects higher $\varepsilon^*$. In Figure 39(i), we correlate $\varepsilon^*$ values of HaPs and other heteropolar compounds with |*d'*| (from Figure 29). Following the very strong dependence of $\mu_{POP}$ on $\varepsilon^*$ shown in Figure 38(ii), the direct implication is that $\mu_{POP}$ should strongly increase with increasing |*d'*|, as we find indeed in Figure 39(ii). Although not explicitly stated by Eq. 12, the deformation potential is also correlated to $\mu_{POP}$, but unlike ADP scattering where $\mu_{ADP} \propto |d'|^{-2}$, $\mu_{POP}$ correlates with |*d'*| with a *positive* power constant. For the limited and scattered set of data the fitting to $\mu_{POP} \propto |d'|^{(a)}$ yields $a = 3.1 \pm 0.7$. This means that the relative influence of POP (ADP) will decrease (increase) as |*d'*| increases (decreases); so, *as the system gets 'softer', the POP regime becomes more dominant*.

The effective mass, which to some extent represents the degree of hybridization towards a minimum energy state between bonded atoms, should decrease when bonds are tighter (= greater hybridization). A stiff bond means that the energy required to move an atom from its optimal momentum state (equally, its position) is high. Since the effective mass, $m^*$, is defined as the curvature of the electron energy with momentum ($m^* \equiv \frac{1}{\hbar^2}\frac{\partial^2 E(k)}{\partial k^2}$), stiffer bonds mean also a higher $m^*$. Therefore, tighter bonds usually mean *less* polarizable lattice, which means that for smaller $m^*$ a higher $\varepsilon^*$ is expected, as indeed shown in Figure 38(iii). By a similar explanation, a smaller $m^*$ should mean more rigid bonds and higher |*d'*|. This is, to some extent, true, as can be seen from Figure 39(iii). Based on the data we have and apart from a few exceptions (and HaPs are among those exceptions), 'soft' heteropolar materials mean $m^* > \sim 0.2$, and 'rigid' ones mean $m^* < \sim 0.2$.

Regarding the exception of HaPs shown in Figure 39(iii): although the divalent $B^{2+}$ metal may force stronger hybridization with halides in HaPs (than with the anions in zinc-blende or wurtzite systems like AgX and TlX), we come back to the possibility that the smaller effective masses of HaPs may result from models used to extract the value from the experiment (magnetoresistance) that are less suitable to these materials. This explanation may hold also for Figure 38(iii), and reflected in Figure 37(ii), as stated earlier, and gives rise to the possibility that for HaPs the assumptions that are applicable to zinc-blende or wurtzite or related structures do not hold. As a next step – oxide and chalcogenide perovskites need to be added up to these figures.[i]

Lastly, when the commonly used measure for the strength of polar interaction, $\left(\frac{\varepsilon_S}{\varepsilon_\infty} - 1\right)$, is plotted against the *actual* deformation potential, *d'*, (see Figure 39(iv)) a clear separation between those materials having a positive and a negative *d'* is observed. Moreover, the two properties seemed to be aligned with a relatively narrow spread, which allows one to have a decent estimate of *both the magnitude and sign* of the deformation potential, *d'*, just by knowing $\varepsilon_S$ and $\varepsilon_\infty$ (and *vice versa*). Since $\mu_{POP}$ and the $m^*$ do not align as well neither with *d'*, nor with |*d'*|, it is more difficult to get a good estimate just by knowing *d'*, and the specific values of $\varepsilon_S$ and $\varepsilon_\infty$ are required. With an attempt to get a clear separation between a 'soft' and a 'rigid' heteropolar material, it can be concluded that *a system is 'soft' when $\varepsilon_S > 2\varepsilon_\infty$*.

---

[i] We note that GaP deviates from the normal relations. GaP and GaN are known to be exceptional III-V.

To conclude this section: in case where defect density is low enough, so that impurities, whether ionized or neutralized, do not dominate free carrier scattering, *the mobility of HaPs is fundamentally smaller than that of other, more rigid, heteropolar semiconductors*. For the usually reported defect densities in HaPs, POP scattering should dictate the mobility in the material, and not ADP or PZ scattering, as suggested in the literature. It was also found that the 'softer' a heteropolar material is, the higher the chances that scattering by POP will dominate over scattering by ADP. Relatively clear relations between apparently unrelated 'fundamental' material properties, like $m^*, \varepsilon^*$ and ***d'***, were found. These allow one to estimate a fundamental property, as well as its $\mu_{POP}$, based on the other related properties.

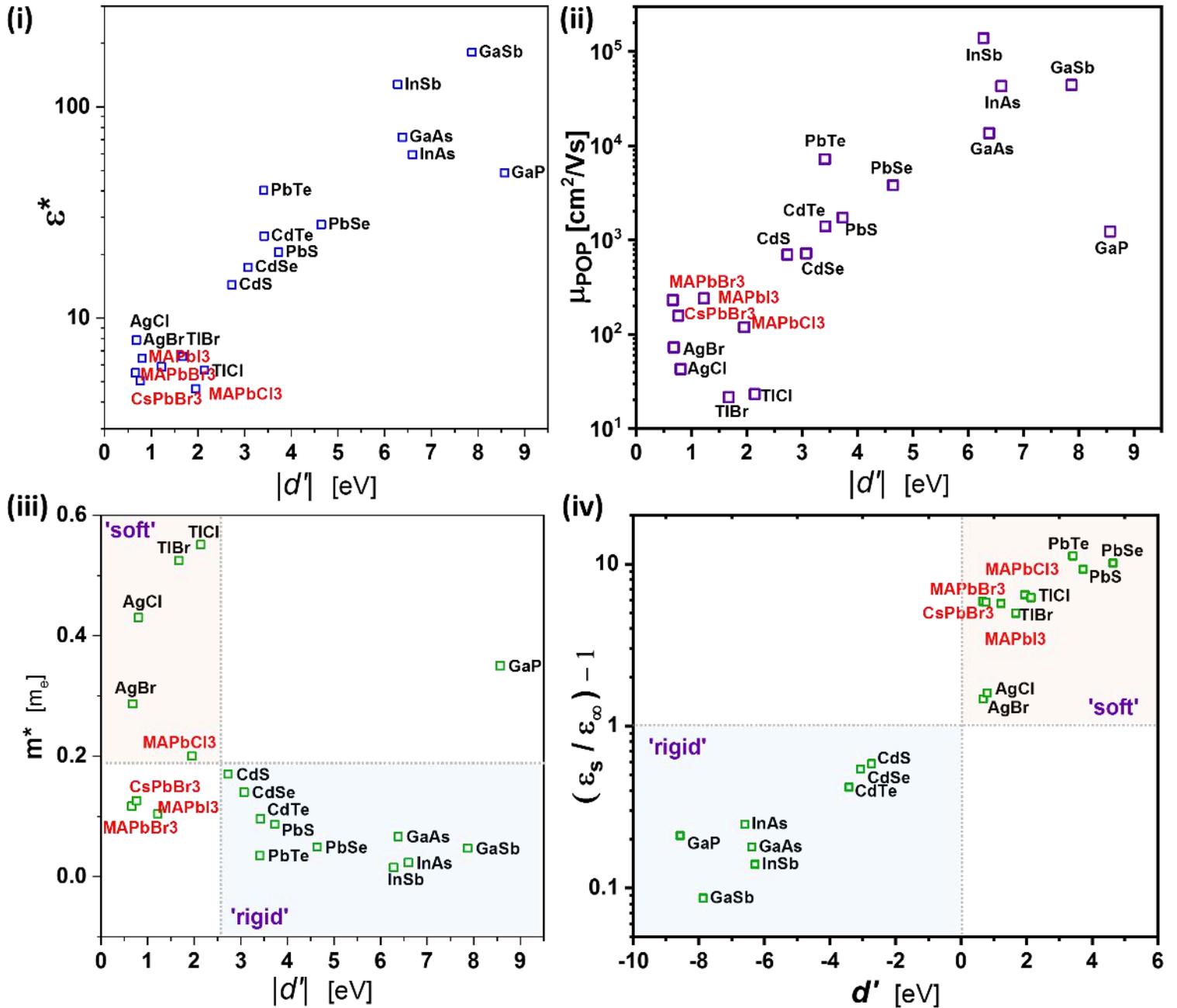

Figure 39: **(i)** $\varepsilon^*$ of HaPs (from Table 3) and other heteropolar compounds (from ref. 183) as function of the absolute magnitude of $d'$, $|d'|$ (from $dE_g/dP$), as reported in section 3.4 (Figure 29(ii)). Higher $|d'|$ results in an exponential growth of $\varepsilon^*$. **(ii)** Calculated $\mu_{POP}$ at 300K (Eq. 12) as function of $|d'|$, which increases strongly with $|d'|$ and indicates that, on average, the mobility, when limited by $\mu_{POP}$, is *fundamentally* limited by the natural 'softness' of the material. **(iii)** Correlation between $m^*$ and $|d'|$, which shows that the measured $m^*$ values of HaPs (marked in red), except that of MAPbCl$_3$ whose $m^*$ is calculated (see Table 3), are smaller than expected for similarly 'soft' compositions. **(iv)** Lattice polarizability, or $\left(\frac{\varepsilon_s}{\varepsilon_\infty} - 1\right)$ – a parameter which reflects the strength of polar interactions (see development of $\mu_{POP}$ above ~ Eq. 11) – as function of the actual $d'$, showing a nice correlation with clear separation between 'soft' and 'rigid' materials. Suggesting a predictive possibility of $\varepsilon^*$ (and thus $\mu_{POP}$) by knowing $d'$ or *vice versa*.

# 4. The thermodynamics and chemistry of HaPs

## 4.1. Theoretical concepts

As already mentioned in the introductory chapter 1 (section 1.3), defect densities in HaPs were found to be ~$10^9$-$10^{10}$ cm$^{-3}$ for solution-grown single crystals, and up to $10^{15}$-$10^{16}$ cm$^{-3}$ in thin polycrystalline films, regardless of the method used for their fabrication or measurement.[7,66,70,74]. One way to explain such low values is 'defect-tolerance', where defects do (physically) exist, but their impact is too low to be noticed. 'Low deformation potential' and an 'anti-bonding valence band' are two possible reasons for defect tolerance, as was discussed in sections 3.4 and 3.5.

In addition to tolerating defects, defect density may actually be low, despite the synthesis conditions that classically (materials engineering-wise) should result in much higher densities (e.g., RT synthesis of solution-based single crystals or kinetically-quenched deposition of polycrystalline films – see section 3.4, Figure 27). Defects in a lattice are represented as discontinuities in a periodic system (as shown in Figure 40(i)). If atoms in a defected lattice are highly mobile, there is a good chance that the defects will segregate at grain boundaries, surfaces, or form clusters in the form of inclusions (as illustrated in Figure 40). Since significant ion diffusion in HaPs has been invoked to explain experimental results,[134,135] self-reorganization of defects towards a more favorable thermodynamic state is possible.

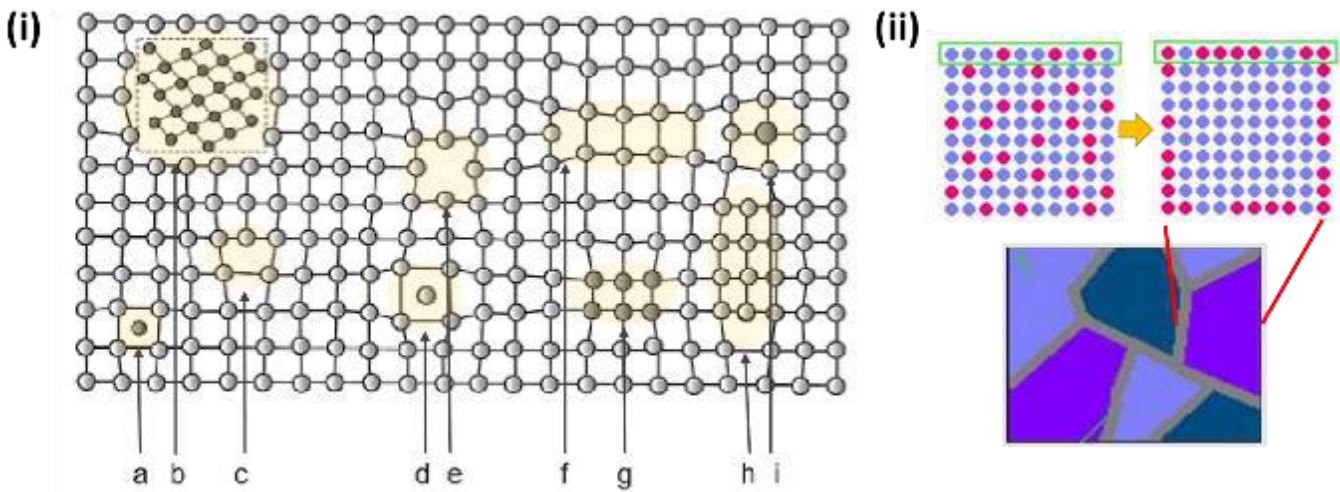

Figure 40: (i) Scheme of real crystal lattice with defects (adapted from ref. 197): *a* - interstitial impurity atom, *b* - incongruous inclusion, *c* - edge dislocation, *d* - self interstitial atom, *e* - vacancy, *f* – vacancy type dislocation loop, *g* - precipitate of impurity atoms, *h* - interstitial type dislocation loop, *i* – substitutional impurity atom. (ii) Illustration of defect segregation (adapted from ref. 198), where (top) defects, *red circles*, migrate from the bulk to the edges that result in a defect-free core with defects being segregated at the surface/interface areas, such as grain boundaries (bottom).

Thermodynamically, however, defects are *unavoidable*, since they introduce additional degrees of freedom to an ordered system that increase its entropy and stabilize it (more negative free energy of formation). The free energy includes, though, also the enthalpy. In Figure 41(i) we show the textbook scheme of the energy balance between the entropy, $\Delta S$, and enthalpy, $\Delta H$, that both increase with an increasing amount of defects. Because free energy is $\Delta G = \Delta H - T\Delta S$, it has a minimum at a non-zero defect density. Using the equations for Schottky and Frenkel defects (see Figure 41(i) and ref. 136 for further details about the meaning of these specific point defects), Figure 41(ii) shows simulated thermodynamically-equilibrated defect densities (at 300 K) with respect to the formation energy of such a defect. It shows that when the defect density is ~$10^{10}$ cm$^{-3}$ (as commonly found for solution-grown HaP singe crystals,) the formation energy is ~1.4-1.5 eV. Schottky or Frenkel defect formation energies for different, stiff materials[136] as well as oxide perovskites[199] are usually > 3 eV.

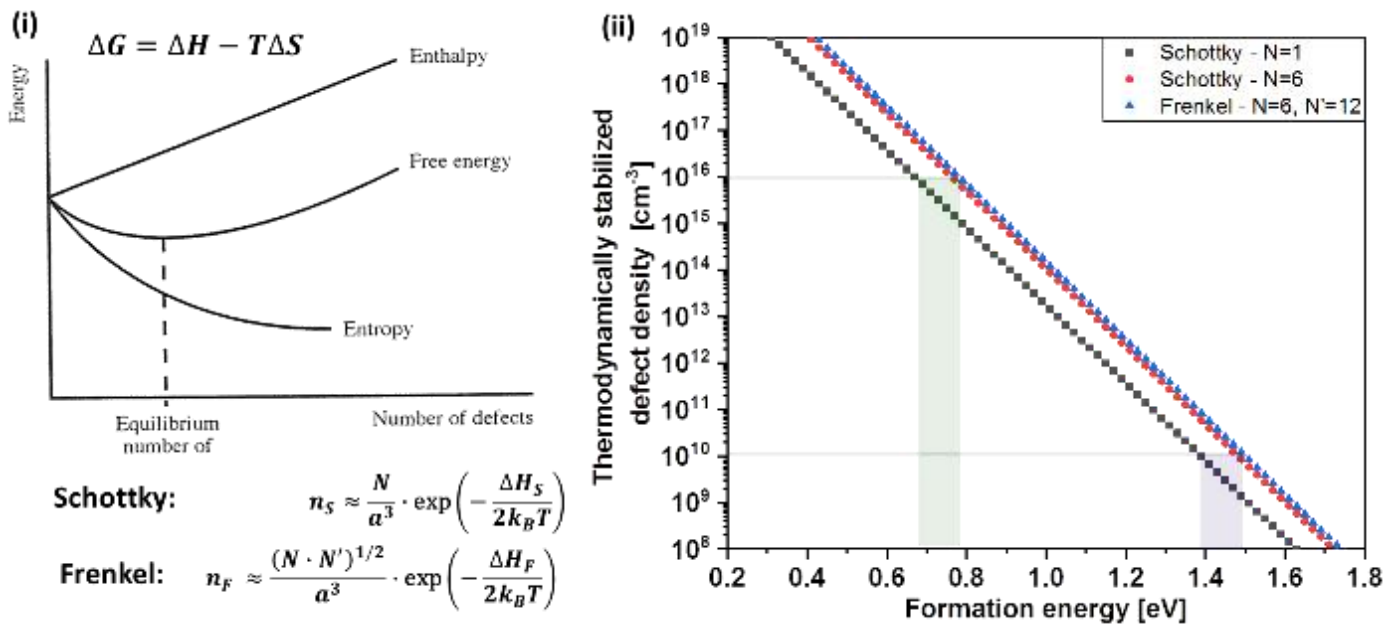

Figure 41: (**i**) A general scheme that explains the reason for a thermodynamically-stable defect density at a finite temperature (adapted from [136]). 'Enthalpy' refers to the change in enthalpy, $\Delta H$, 'entropy' to -T$\Delta S$, and 'free energy' to $\Delta G$. The equations represent equilibrium defect density, $n_i$, of two common intrinsic defects: Schottky, S, (i.e., simple lattice vacancy [pair, usually]) or Frenkel, F, (i.e., displacement of an atom into an interstitial site creating a vacancy. $a$- lattice parameter, $N$ - number of sites that may become a vacancy, $N'$ – number of interstitial sites that may accept an atom, $\Delta H_i$ – formation energy. The equations are an estimate for $n_i << N/a^3$. (**ii**) Simulated defect density from the equations in (i) for $T=300K$ and $N$ and $N'$ that are given in the legend. $N$ and $N'$ are chosen in a way that reasonably represents the perovskite system. The pale-green and purple bands are guides for the eye for the formation energies expected for defect densities that are typical for HaPs.

PhD Thesis – Yevgeny Rakita
September, 2018

|89|

Halide semiconductors, however, typically have much lower values for defect formation energies (0.7 eV for β-AgI; 1.2 eV for AgBr – both RT phases)[136], which place them on a scale of solid electrolytes, in which ion migration is a key property. In silver halides, what usually migrates is the smaller monovalent cation – $Ag^+$ – as was shown by Williams and Barr using a combined ionic and isotopic study for ion migration in AgBr and $PbBr_2$.[200] When comparing it to a divalent cationic halide salt, e.g. $PbBr_2$, it was found that the ionic conduction is driven by halide migration, while the lead is much more static (see Figure 42). The *activation energy*, $E_a$, for iodide migration, that can be derived only from ionic-conductivity experiments, was found to be ~0.31 eV, while the defect *formation energy*, $H_f$, that requires radioactive isotope tracking, was found to be ~1.6 eV. Since isotop tracing experiments are not trivial, especially since awareness of dangers, and regulations on radioactive elements became very strict, the simpler ionic conductivity experiments using impedance response as a function of temperature are more common, even though they provide only indirect evidence. Such studies on $CsPbBr_3$ and $CuPbBr_3$ revealed that the activation energy for ionic migration (which was *assumed* to be $Br^-$) is very similar to that of found via isotope tracer studies in $PbBr_2$ (Figure 42), i.e., 0.25 eV and 0.27 eV, respectively.[201,202] For $MAPbI_3$ similar values, for an *assumed* $I^-$ migration, were 0.37-0.43 eV[135,203]. Unfortunately, no isotope tracer studies have been reported for $PbI_2$, but the higher activation energy is not unreasonable.

Since halide structures, as well as HaPs, are similarly soft (as was discussed in sections 3.4 and 3.5) due to the dominant Pb-X bond, the similarity in the halide $E_a$ suggests that their $H_f$ should also be similar, i.e., ~1.6 eV. Returning to Figure 41(b), we find that $H_f$ ~1.6 eV results in a *thermodynamically equilibrated* defect density of $10^9$ cm$^{-3}$ – the measured value for HaP single crystals.[70] It is important to note that more concrete numbers for $H_f$ in HaPs (e.g., by isotopic tracing) are required to provide conclusive proof for these claims. This gives rise to

A prerequisite condition for a material to be at thermodynamic equilibrium (rather than in a kinetically-stabilized state, such as that of spatially localized extrinsic dopants in Si), is that the atoms/ions that the material is composed of are *sufficiently mobile* to allow them to reach equilibrium positions. The latter seems to be at least partially the case of HaPs. Thermodynamically equilibrated defects are present at T > 0 K. To maintain only that level (concentration) of defects, the defect formation energy ($\Delta G_{form}^{defect}$) should be unfavorable compared to the free energies of bond formation. Indeed, a positive enthalpy of formation (energetically unfavorable) from the binaries was deduced from calorimetry experiments on MAPbX$_3$ (X=I, Br) by Nagabhushana et al.[10] (the general message of these experimental results is supported by other experimental works [204–206] and theoretical calculations[207]) which means that the MAPbBr$_3$ and esp. MAPbI$_3$ are *entropically stabilized*.

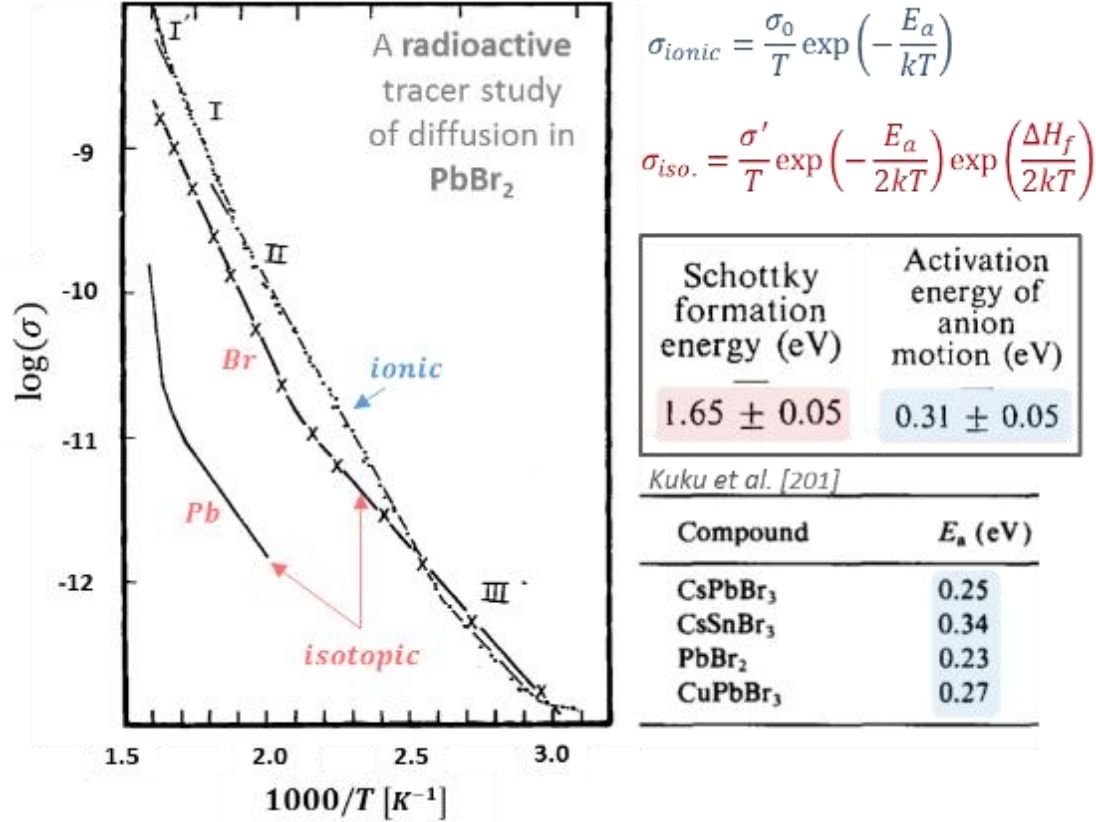

Figure 42: (Graph of) a combined study of ionic conductivity (using impedance analysis with changing temperature) with a radioactive tracer study of isotopic Pb and Br in PbBr$_2$ done by Williams et al.[200] (Table of) the activation energy, ($E_a$) and formation energy, ($H_f$) extracted from the $\sigma_{ionic}(T)$ and $\sigma_{iso.}(T)$ equations.[136] $\sigma_0$ and $\sigma'$ are exponential prefactors for the overall ionic ($\sigma_0$) or the specific isotope ($\sigma'$) conductivity. Due to the similarity between ionic and Br-isotopic conductivities, it can be deduced that the ionic conductivity in PbBr$_2$ is due to the halide. The bottom-right table gives activation energy values from similar ionic conductivity experiments (from Kuku et al.)[201] of different perovskite compounds (and PbBr$_2$). Activation energy values are assumed by the authors to be of Br$^-$ migration.

To understand the point of 'entropic stabilization' and relate it to defect formation, it is important to make the following clear:

- There is a <u>single</u> pathway that, enthalpically, is very low (few kJ/mol), or (for MAPbI$_3$) negative: [1] MAPbX$_{3(S)}$→MAX$_{(S)}$+PbX$_{2(S)}$. In these cases, what makes the HaP structure stable, *with respect to the MAX$_{(S)}$+PbX$_{2(S)}$ binaries*, is entropy – probably mostly configurational entropy. A summary of all the calorimetric studies that were done can be found in Ciccioli and Latini.[205]

- Another possible degradation pathway is:[204] [2]  $MAPbX_{3(S)} \rightarrow [MA_{(g)} + HX_{(g)}] + PbX_{2(S)}$. The reaction of breaking and vaporizing $MAX_{(s)} \rightarrow [MA_{(g)} + HX_{(g)}]$ requires a lot of heat energy (= enthalpy) (~100's kJ/mol);  for example, mixing MA with HI is very exothermic). This reaction may occur when the material is heated (or photobleached – see section 4.4), and since gaseous constituents are formed, it is a significant degradation pathway.  Reaction [1], however, does not degrade into volatile species, and therefore will not be a degradation pathway by its own.
- At any given time, there will be (Schottky/ Frenkel) point-defect concentrations that suite the enthalpy of forming such defects. This energy is large and positive (=unfavorable), as was estimated earlier (see Figure 42).
- The positive/low enthalpies w.r.t $PbX_2$ and MAX (likely) mean that the activation energy and the kinetic constant of formation/decomposition to/from this structure is small. This formation/deformation is similar to a process, usually done at high temperatures, called "*annealing*", but at RT.
- Due to the spontaneous 'annealing' at RT, defects may form/heal. If the formation/healing of the defect is faster than the interaction time of the defect with free charge – there is no point of calling these 'defect' (in the Kroeger-Vink sense), since they represent a 'steady-state' of the system.
- The fact that one can estimate defect density, means that, at least effectively, there is some electronically-relevant defect density that follows thermodynamics ($10^{10}$ cm$^{-3}$ for single crystals – see Figure 41(ii)). It means that if we do a measurement, our results imply that the system behaves *as if* it has $10^{10}$ cm$^{-3}$ static, electrically active, defects (in crystals).

I will distinguish between two types of defects: neutral defects and electrically charged, i.e., ionized ones. We saw in the previous chapter (section 3.4) that the absolute deformation potential, |*d'*| of HaPs, and other 'soft' halide-based materials, is significantly lower than that of other, more 'rigid', materials.[i] Low |*d'*| means that it is becoming more favorable energetically to have a strain-field, rather than a localized charge (as would be the case for an ionized defect). If the formation energy of an ionized defect is higher than the formation of the bonds that coordinate with it, two effects may occur:

---

[i] This is reasonable, as it scales with the free energy of formation, which is small in HaPs.

(a) The bonds surrounding this defect will break and the defect will diffuse until it will pair to an oppositely charged defect or reach a surface/interface, which naturally can bear more charged defects as it is naturally polarized.

(b) The wave function of the ionized impurity will delocalize to a volume that is larger than the first/second coordination sphere (so more bonds will be involved), and as a result it will have an effectively shallow potential well (see illustration in Figure 32 in section 3.4). It should be noted that this case is different from an ionized defect that involves an electrostatic force to distort locally the lattice, as is the case with the 'Landau polaron' picture. (There an electric charge, in Landau's case due to an ionized site, distorts a lattice from its equilibrium as a result of the electric field induced by a charge localized at a defect site).[208] Note that although a low |$d'$| is favorable for formation of a Landau polaron, in an entropically stabilized system (as deduced for HaPs)[10] a Landau polaron picture is highly unfavorable, since it will limits the number of bonding configurations and, thus the main stabilization factor - entropy.

*In polycrystalline HaPs films*, defect densities are always found (~ $10^{15}$ -$10^{16}$ cm$^{-3}$) to be higher than those in single crystals. Two possible effects potentially contribute to these higher values and both are connected to a high density of grain boundaries. First, the less- packed interfaces (e.g. grain boundaries) dictate a lower formation energy with respect to the highly-ordered and -packed bulk, and result in higher density of thermally-formed defects (cf. Figure 41(ii)). For example, transient capacitance measurements on MAPbI$_3$ films suggest that the favor of formation and migration probabilities is attributed to different types of defects when one compares surface (MA$^+$ is more likely to migrate) with bulk (I$^-$ is more likely to migrate).[135]

The second point is attributed to defect segregation at grain boundaries via ionic migration (cf. Figure 40(ii)). Assuming a sphere, where volume $V \propto R^3$ and surface $S \propto R^2$, the surface to volume ratio will be proportional to $R^{2/3}$. Replacing R with an atomic unit length, for every atom per unit volume there will be $\sim(\#atom)^{2/3}$ surface atoms. Based on a scaling approach, I plot in Figure 43 the range of possible (opto)electronically-active surface defect densities per unit volume of MAPbI$_3$ with varying grain size at different fractions of defected vs. non-defected sites. If we assume that out of all surface/interface atoms, the number of (opto)electronic relevant defects is 0.01 at%-1 at%, we find that for a MAPbI$_3$ polycrystalline film with an average grain size of ~1 μm, the density of surface defects that can be found per unit volume of such a film is ~ $10^{14}$ -$10^{16}$ cm$^{-3}$. Since a common grain size in HaP polycrystalline films is 0.1-1 micron, this scaling approach supports the experimentally derived defect density in thin films due to spontaneously formed surface defects (as 0.1 at% defect density is not such a wild idea) or as a result of defect segregation.

We can estimate, using the results of this scaling approach, that a single crystallite with radius of ~100 nm should have ~1 defect only. When we simulate a crystal of ~1 cm, we reach a defect density level of ~$10^{11}$ cm$^{-3}$ (for surface defect density of 0.01 at%); such number is already very close to the defect density of the bulk (~$10^{10}$ cm$^{-3}$) (see Figure 41).

In addition to the thermodynamic stabilization of the defect density, the overall amount of defects may not be tolerated and separation of phases may occur. This point was supported both experimentally and theoretically, as shown in Figure 44, where the MAPbI$_3$ composition is found to be stable (with respect to its binaries and other lower-dimensional HaPs, cf. Figure 3) in very narrow range of a stoichiometry. This suggests that HaPs do not support a large concentration of point defects and energetically will prefer to form alternative phases instead.

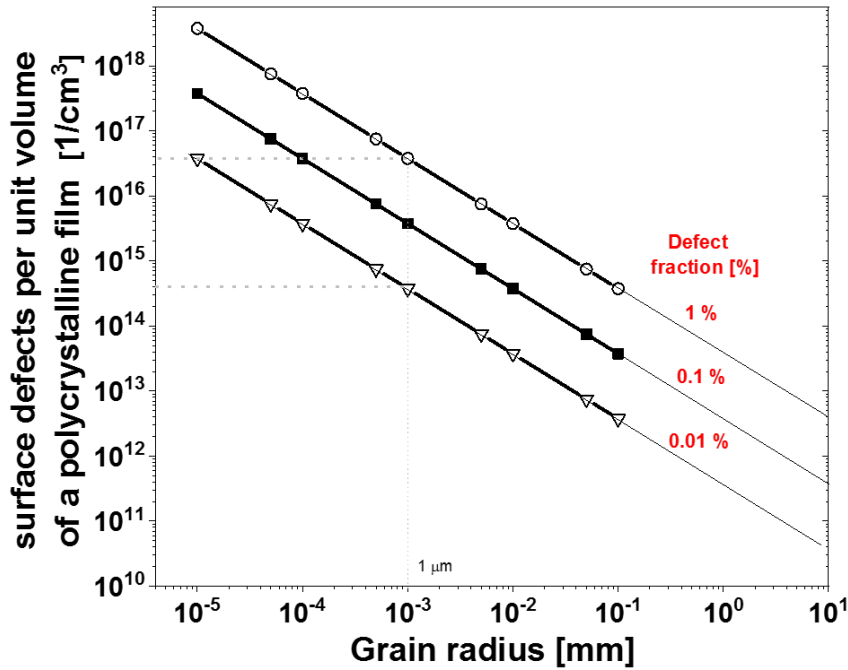

Figure 43: Surface defect density per unit volume of a polycrystalline film vs. the average grain radius of that film. The different lines assume different defect fractions per unit area of a surface/interface (marked in red). The dotted lines are guides for the eye for the commonly measured defect densities in HaPs films. The estimated results are based on a scaling approach, where the parameters were taken for MAPbI$_3$ (density,$\rho$, 4.2 gr/ml ; M$_w$=620 gr/mol). The equations that were used are: (1) $[\mathbf{defects}] = (defect\ fraction) \cdot \frac{\#S_{atoms}}{\#V_{atoms}} \cdot N_A$ ; (2) $\#S_{atoms} = (\#V_{atoms})^{2/3}$ ; (3) $\#V_{atoms} = (moles\ per\ particle) \cdot N_A$ ; (4) $(moles\ per\ particle) = \frac{\rho}{M_w} \cdot \mathbf{R^3}$. Here $N_A, \#S(V)_{atoms}$ and $R$ are Avogadro's number, number of surface (volume) atoms and the grain radius, respectively.

Other, chemically-based, strategies for defect elimination are local chemical reactions of the defects among themselves or with the host lattice. This may result in formation of new compounds that have a benign (or at least not a negative) effect on the (opto)electronic properties of the hosting HaP, or reformation of the HaP composition (true only in case of a suitable stoichiometry). For example, the poor thermodynamic stability of the HaPs makes them vulnerable to changing environmental conditions, e.g., humidity.[209–211]

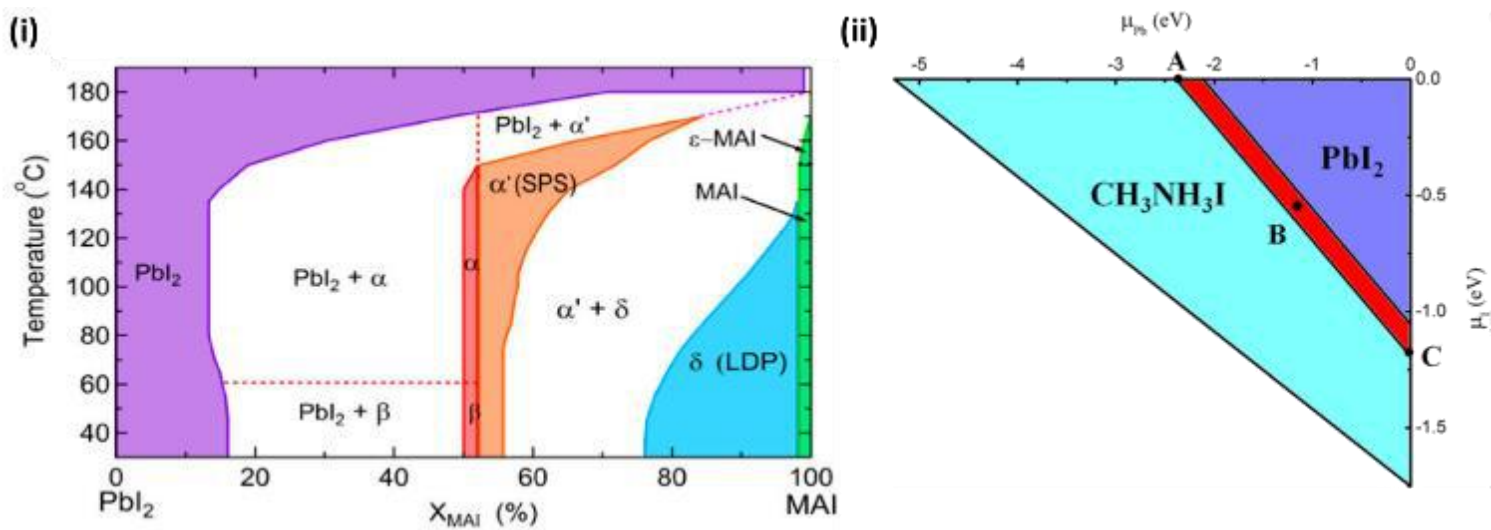

Figure 44: (i) a phase diagram of MAPbI$_3$ under different MAI ratio (x axis) at different temperatures (y-axis). Shows that the tetragonal ($\beta$) and cubic ($\alpha$) phases exist only at a very narrow precursor ratio, while for other ratios, stacked perovskite sheets ($\alpha'$-SPS), low-dimensional perovskites ($\delta - LDP$), PbI$_2$ or MAI compositions are formed (for SPS and LDP – cf. section 1.1 - Figure 3); From ref. 212  (ii) Calculated thermodynamically-stable range (red) for equilibrium growth of MAPbI$_3$ (adapted from ref. 114). Outside this region, the compound will favor decomposition to its binaries: PbI$_2$ and MAI. $\mu_i$ is the chemical potential of the constituent element referred to its most stable phase. To build this chart, the following thermodynamic (in)equalities are satisfied: (1) $\mu_{MA} + \mu_{Pb} + 3\mu_I = -5.26\ eV$; (2) $\mu_{MA} + \mu_I < \Delta H_f(MAI) = -2.87\ eV$; (3) $\mu_{Pb} + 2\mu_I = \Delta H_f(PbI_2) = -2.11\ eV$. Similarly-calculated charts are found also for other HaP compositions[213].

To conclude this section, we find that the low defect densities found in HaPs can be explained by:

- RT-'annealing' due to the combination of very low (to non) enthalpic stabilization but entropic favorability of the HaP (w.r.t its binaries - MAX and PbX$_2$). For RT annealing to occur, it is critical that activation energies for formation/decomposition will be low. Generally, low enthalpy usually implies low activation energy as well.
- High ionic mobility, which can result in defect segregation from the bulk to the surfaces/interfaces, leaving behind a thermodynamically-equilibrated, low bulk defect density.

These two phenomena may be present alone or together, and will be covered by the term '**self-healing**'. The narrow stoichiometric window in which HaPs may exist may lead to phase separation of highly-defected regions, leaving behind a 'clean' HaP structure. As long as stoichiometric and environmental conditions are carefully maintained, the resulting films/crystals should be of low defect density.

I note that it seems as if studying the thermodynamic state of HaPs *does* provide the correct picture of their performance, while kinetically stabilized states seem to be highly unlikely (short time after its fabrication), so ***questioning the role of defect (either intentional or unintentional) states may become irrelevant***. Preparation chemistry and chemical environment (e.g., at interfaces), however, should have a strong impact on the material's properties.

It is interesting that electrostatic forces in an ionic system are known to follow Madelung forces ($E_{\text{Madelung}} \propto \frac{1}{n}$ ; n = coordination number). Increasing the coordination number in an easily polarizable system automatically implies higher configurational entropy. Higher configurational entropy means an increased stability of the system. All of this can happen only if the bonds are naturally soft enough (low *d'*), where changing configurations do not cost lots of energy. For example, the delocalization of the $ns^2$ electrons becomes more energetically favorable when one moves from fluoride ('rigid' bonds) to iodide ('soft/polarizable' bonds) (hemidirected to holodirected, respectively – see Figure 26 in section 3.3).[131]

In the following sections, I present complimentary experimental evidence, which support the theoretical concepts presented here and further explores them: the chemical nature of HaP's (de)formation and defect-tolerating complex formation is presented in sections 4.2 and 4.3; the role of the A cation in the material's stability, HaP's recyclability (or 'self-healing') are presented in section 4.4.

## 4.2. Formation of MAPbI₃ from PbI₂

In this section the prototypical reaction between PbI₂ and MAI to form MAPbI₃ is discussed. We (together with Dr. Thomas M. Brenner) investigate the conversion of small, single-crystalline PbI₂ crystallites (~10-20μm in diameter) to MAPbI₃ by two commonly-used synthesis processes: reaction with MAI in solution or as a vapor. *Here I summarize what was published in ref. 4.*

The goal of this work was to reveal the transformation mechanism of the binary PbI₂ to the ternary MAPbI₃ when it is exposed to its second binary, MAI. We explored the thermodynamics and kinetics of the reaction. Our main focus was the solution-based reaction, where PbI₂ was deposited on a transparent (glass or glass coated with a conducting transparent oxide, e.g., FTO) substrate that was exposed to a solution of MAI in IPA. Results from exposure to MAI vapor (by placing the PbI₂ crystallites in front of a heated salt of MAI) are mainly used for comparison.

The first challenge was to visualize the reaction *in-situ*. Since we are working with semiconductors, photoluminescence microscopy allowed us to gain good contrast from reacted areas that converted to MAPbI₃ by filtering out any luminescence not resulting from MAPbI₃ (see Figure 45(i)) for further details on the setup). It was found (see Figure 45(ii)) that the reaction rate and the growth dynamics of MAPbI₃ strongly varies with MAI concentration. For the lower concentration range (50 mM), where the reaction is significantly slower (~1000 times), isolated nucleation events followed by growth of those individual nuclei are clearly observed. At higher concentration (100 mM), the coverage of the PbI₂ crystallites with MAPbI₃ is instantaneous and uniform. With an analogy to single crystal growth, we observe 'primary' (higher concentration) and 'secondary' (lower concentration) nucleation, followed by growth (cf. section 2.1). The fact that a secondary nucleation occurs (via more activated sites than others that result in distinctive nucleation sites) suggests the existence of a non-negligible kinetic barrier for the reaction to occur. To estimate its upper limit, we first need to find the thermodynamic threshold of that reaction.

The strong dependence of the reaction rate on MAI concentration allowed us to estimate the free energy of formation for the reaction by finding the threshold concentration at which the reaction is no longer taking place. Figure 46 shows that at RT 15 ± 5 mM MAI in IPA is the threshold concentration for the system to be at equilibrium. For the reaction PbI$_{2(s)}$ + MAI$_{(IPA)}$ → MAPbI$_{3(s)}$ we can use the formula $\Delta G^0 = k_B T \cdot \ln(K_{eq})$, where $K_{eq} = \frac{1}{[MAI]_{eq}}$, so that at *T=300K* and $[MAI]_{eq} = 15\ mM$ the free energy of the reaction is $\Delta G^0 \sim 0.11\ eV$.

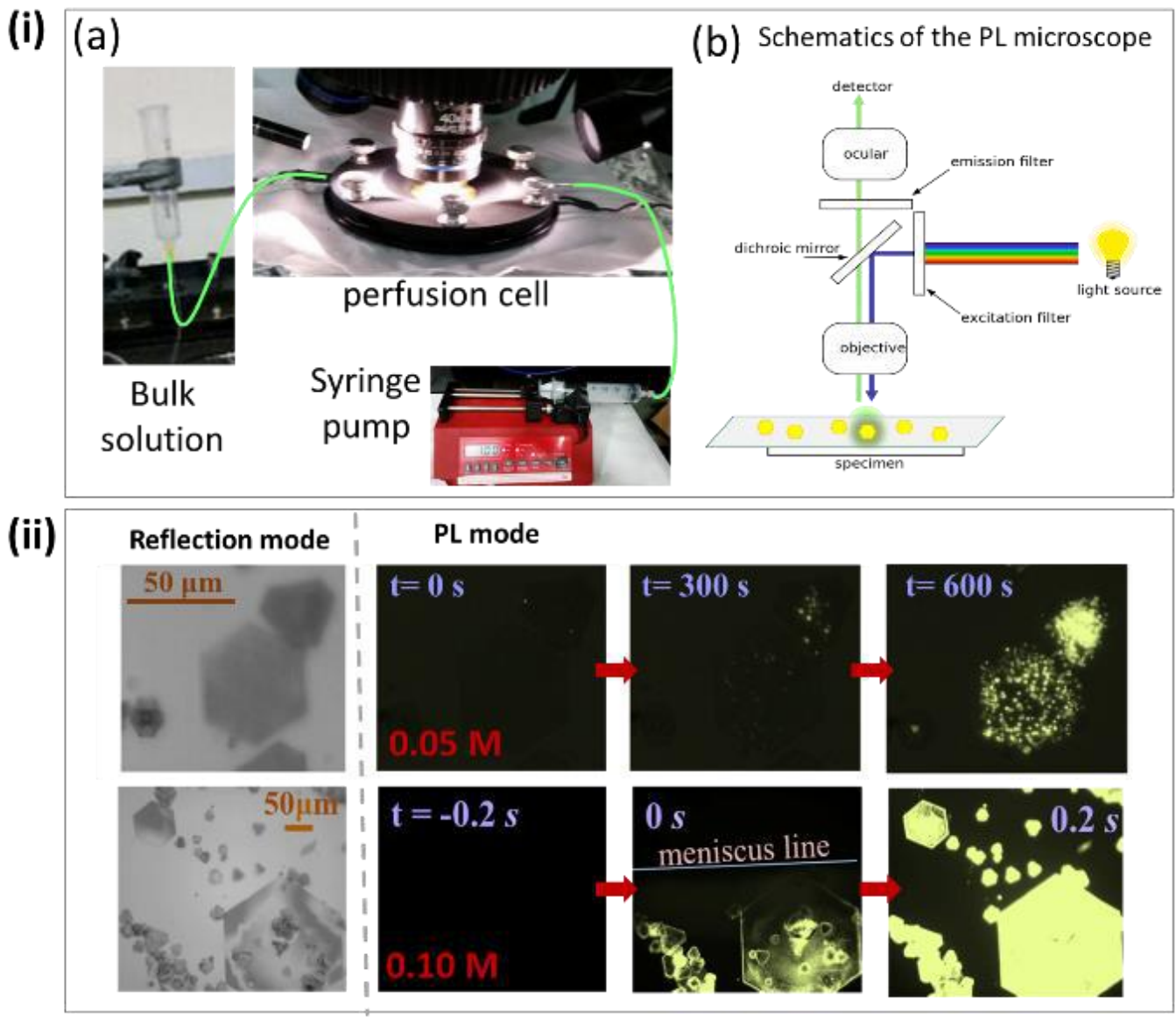

Figure 45: (i) In-situ PL microscopy measurement setup that shows (a) the syringe which contained the MAI *solution* in IPA, a *syringe-pump* to push the solution from the bulk solution towards the *perfusion cell* (green lines represent the tubing, i.e. the path of the injected solution) below the optical path in which the reaction between $PbI_2$ crystallites and MAI occurs to fromMAPbI$_3$. The perfusion cell was sealed by two glass slides with ~100 μm spacing between them, where on the one closer to the microscope, $PbI_2$ single crystallites are precipitated. A general principle of operation of the PL microscope is shown in (b), where the filter and the dichroic mirror allows blue light to shine on the sample and PL above 740 nm to be collected by the microscope, so that PL contrast is observed only when $MAPbI_3$ forms ($MAPbI_3$ luminesce is at ~780 nm). (ii) In-situ PL microscopy images of $PbI_2$ crystallites exposed to (top row) 50 mM and (bottom row) 100 mM of MAI solutions at RT. Time t=0 is the moment when the solution reached the imaged area. The reaction for 0.1 M is instantaneous and uniform, while for 0.05 M it is ~1000 times slower with highly distinct nucleation centers.

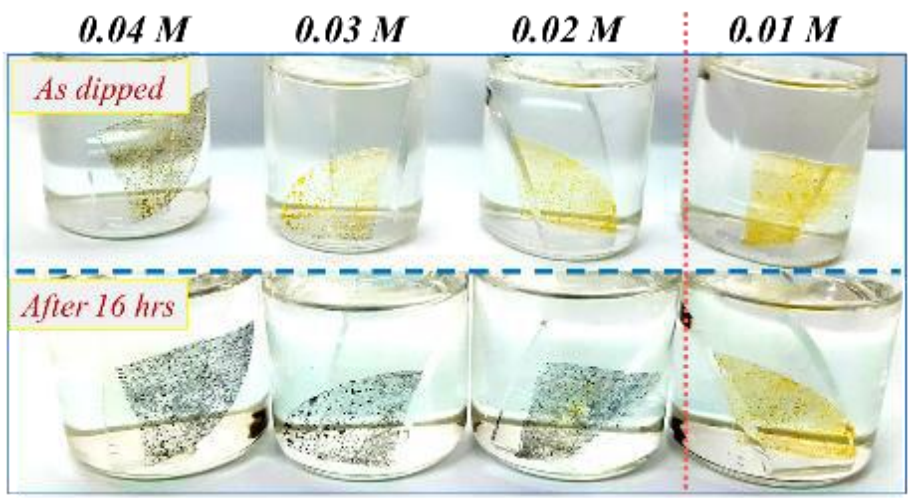

Figure 46: Photographs of PbI$_2$ single crystallites deposited on glass and exposed to different concentrations (as mentioned in each column) of MAI solution in IPA after a few seconds (top) and after 16 hr (bottom). The threshold concentration was determined to be 15 ±5 mM. The same threshold concentration was further confirmed via XRD and PL microscopy.

Regarding the kinetic barrier 100 mM is a concentration at which the system is clearly out of equilibrium and, due to the primary nucleation regime, the excess free energy is higher than the kinetic barrier. Therefore, since the system at 100 mM MAI is out of equilibrium, then we use $\Delta G = \Delta G^0 + k_B T \cdot \ln(Q)$ with $Q = \frac{1}{[MAI]}$ and $\Delta G^0 = 0.11\ eV$. We get that $\Delta G = 0.16\ eV$, which means that an upper limit for the kinetic barrier for the formation reaction (PbI$_2$+MAI→MAPbI$_3$) is < 0.05 eV, and for the decomposition reaction it is (MAPbI$_3$→PbI$_2$+MAI) ≤ 0.16 eV. Further exploration of the kinetic barriers of the formation and decomposition of MAPbI$_3$ is required to set values that are more accurate.

Although both the thermodynamic (~0.11 eV) and kinetic (≤ 0.16 eV) energies are higher than room temperature thermal energy (0.026 eV), which makes the MAPbI$_3$ stable by itself, it is still vulnerable to applied external energies, e.g., ~0.9 V of operational voltages in a solar cell, > 1eV of an incoming photon. Therefore, it might decompose upon illumination or because of migrating ions upon a strong applied electric field, as was clearly demonstrated.[214–216]. When applying an electric field together with light, the effect of degradation and ion migration is shown to be accelerated.[217] These effects will be further described in the following sections.

The next step was an *ex situ* characterization of the microscopic morphology of the reacted PbI$_2$ single crystallites. Although in both solution-based and vapor-based MAI exposure, the resulting product is MAPbI$_3$, SEM imaging (see Figure 47) and XRD analysis (see published paper)[4] reveal that the solution-reacted PbI$_2$ crystallites show a strong orientational relation between $PbI_2$ and $MAPbI_3$, while the vapor-reacted $PbI_2$ crystallites do not share a similar morphological or structural correlation. For the solution-based reaction, XRD suggested that the $PbI_2$ (001) plane becomes the $MAPbI_3$-(202) plane, which perfectly agrees with the SEM morphological images (see further explanation in Figure 47 caption). When comparing the two orientations in the two structures, it is found that in both structures the $[PbI_6]$ octahedral orientations are aligned in the same direction. This suggests that any formation of $MAPbI_3$ from $PbI_2$ or decomposition to $PbI_2$ does not require rotation of the $[PbI_6]$ octahedra.

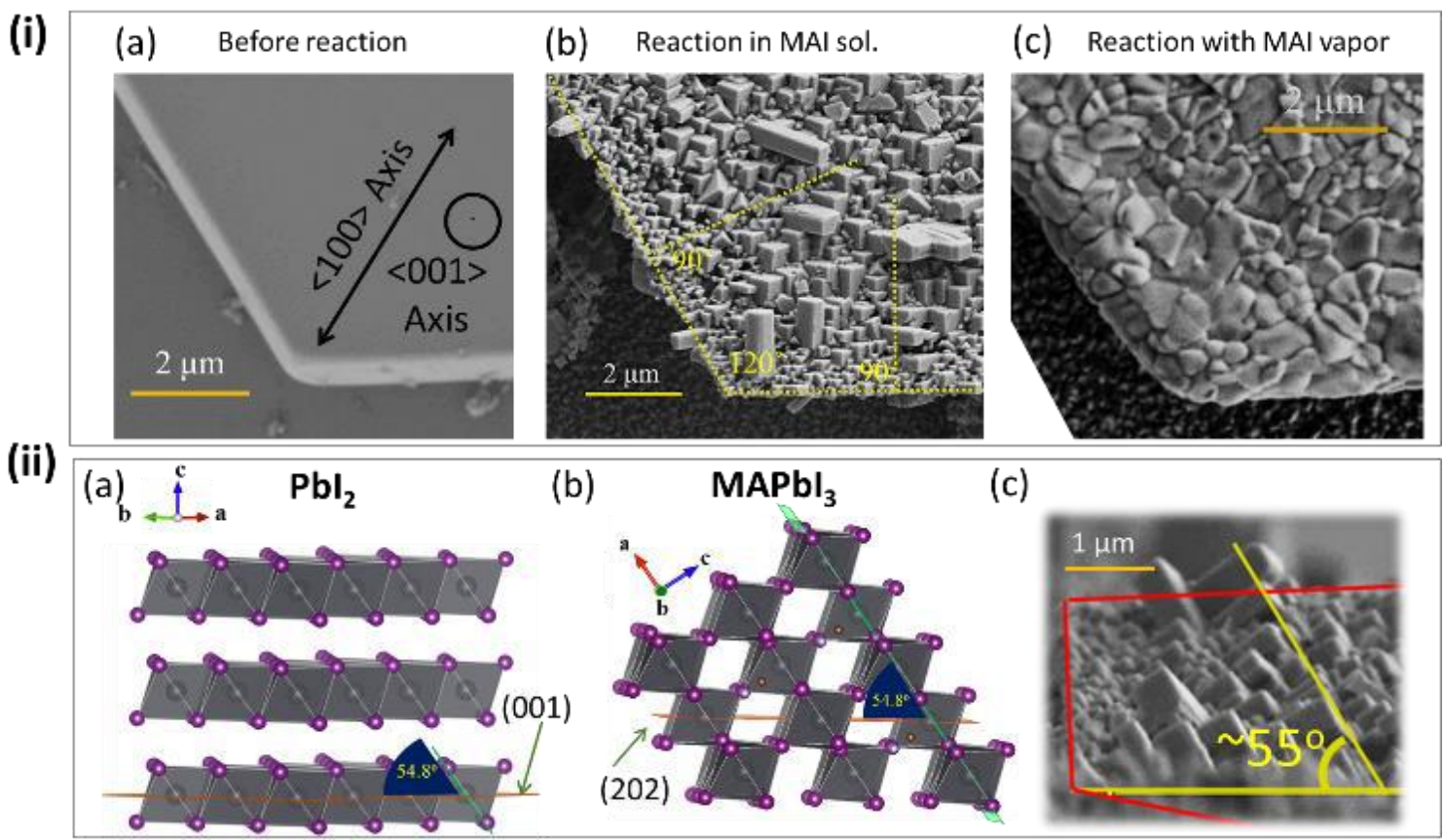

Figure 47: (i) SEM images of: (a) $PbI_2$ single crystallite precipitated on a conducting substrate from a supersaturated $PbI_2$ aqueous solution that was heated to 90 °C and slowly cooled to room temperature; (b) Solution-reacted (for 55 hr) $PbI_2$ crystallite in 0.1M MAI solution dissolved in IPA. It shows highly-oriented tetragonal crystallites of $MAPbI_3$ on the original $PbI_2$ substrate – the lines perpendicular to the hexagonal boundary of the original crystallites are guides for the eye, drawn at 90° to the hexagon edges. Preferred orientation and composition were also verified by XRD and EDS (see ref. 4); (c) MAI vapor-reacted $PbI_2$ crystallite in an enclosed chamber containing MAI salt that was heated to 140 °C. Preferred orientation was not observed, either morphologically or via XRD. (ii), (a) $PbI_2$ crystal structure viewed with the *<001>* direction pointing up the page. (b) – $MAPbI_3$ crystal structure viewed along the *<202>* direction up the page, which is the preferred orientation of the liquid-reacted polycrystalline shown in (i,b). This is further supported by (ii,c) that shows a tilt angle of ~55° of the $MAPbI_3$ crystallites with respect to the initial $PbI_2$ orientation. This same angle is the angle towards which $PbI_6$ octahedra in both $PbI_2$ and $MAPbI_3$ are tilting in (ii,a) and (ii,b).

This observation implies that the reaction pathway in solution is via a *topotactic* relation between $PbI_2$ and $MAPbI_3$. A topotactic reaction is when there is an orientation relationship between the 'parent' substrate and the 'child' product structures. If intercalation occurs (as commonly occurs in topotactic reactions), e.g. MAI into the $PbI_2$ structure, this can dramatically decrease activation barriers for formation and decomposition between structures. Intercalation reactions are common in electrochemical solid state systems[218–220], where sub-lattice unit migration (e.g., $Li^+$) is a key property.

Following nucleation, it was evident that grain growth occurs by a further dissolution−recrystallization process, apparently via $[PbI_x]^{(2-x)}$ complexes created in IPA.[221] Since we find the reaction via vapor phase produces material lacking a preferred orientation, this suggests the transformation in the vapor phase case is dominated by a deconstruction−reconstruction of the $PbI_2$ to $MAPbI_3$, which is probably due to the higher thermal energy involved in the process.

To conclude this section, it was shown that the thermodynamic barrier of $PbI_2$ transformation to $MAPbI_3$ (~0.11 eV) and the kinetic barrier for $MAPbI_3$ decomposition ($\leq 0.16$ eV) is significantly lower than the energies that this material may experience in optoelectronic devices. It is important to note that the mentioned free and kinetic energy values are relevant only for the specific system we used ($PbI_2$ conversion to $MAPbI_3$ from MAI in IPA solution). The nature of the solvent may, however, affect these values. It was also shown that one possible contribution to the low energetic values is the orientational relation between $PbI_2$ and $MAPbI_3$, which suggests a low energy path for formation (decomposition) of $MAPbI_3$ from (to) $PbI_2$ once MAI is introduced.

## 4.3. Formation of ABX₃ from metallic Pb (or Sn) using AX or AX₃

In the previous section I focused on a simple chemical addition reaction, where MAI seems to intercalate PbI$_2$ followed by growth via dissolution-reconstruction of the MAPbI$_3$ structure. In this section, it is shown that HaPs (not only MAPbI$_3$) can be formed in the same MAX solution in IPA (or other alcohols) via reduction of Pb$^0$ (or Sn$^0$). I investigate the conversion mechanism of the metal to HaPs and show that in an oxidizing environment the metal reacts with an AX salt to form an HaP. *This part is mostly a summary of what was published in ref. 6*. I add a new part (not published but partially related to a publication by Petrov et al. [60]) about oxidation with AX$_3$ species that is found to be highly corrosive and capable of forming ABX$_3$ compounds once a B metal is exposed to an AX$_3$ molten salt.

As a proof of concept, Figure 48(i) shows that when elemental Pb$^0$ film is exposed (even at RT) to a simple alcoholic (e.g., IPA) solution of an AX salt (e.g., MAI, FABr) a spontaneous reaction occurs to transform the metallic Pb$^0$ to Pb-based HaPs. Using the same concept for Sn metal or using inorganic salts (e.g., CsBr) instead of organic ones, similar results (meaning: CsPbBr$_3$, MASnI$_3$ or the perovskite-related structure of Cs$_2$SnI$_6$) are obtained. However, for these latter cases, a more reactive (corrosive) environment is required, for example: more polar solvents (e.g., methanol or ethanol), an increased acidity (e.g., added HI, trifluoroacetic acid) or increased presence of polyhalides (e.g., addition of I$_2$ to form I$_3^-$).

Since the reaction takes place in ambient atmosphere and metallic Pb is used, it is reasonable to think that an oxidized PbO$_x$ (due to atmospheric corrosion)[222] is the material that is responsible for the formation of the perovskite in the presence of AX – similar to the case of PbI$_2$.[55] Whatever PbO$_x$ is there, it was not detected on a freshly-evaporated Pb film and, as shown in Figure 48(ii), although Pb$^0$ is definitely present together with the perovskite after certain limited reaction time, its quantity decreases until it completely converts to MAPbI$_3$.

Addition of a strong acid (e.g., HI (57% aqueous) or trifluoroacetic acid (99%)) to the MAX solution in IPA results in acceleration of the reaction (Figure 48(iii)). Added base (e.g., KOH), however, impedes the transformation. An acceleration of the reaction was also observed upon increasing the concentration of AX salt (similarly to the PbI$_2$ conversion presented in the previous section). Addition of halides (e.g., I$_2$ or Br$_2$), which most likely form polyhalide species (e.g., I$_3^-$) as clearly observed by the dark-brown color of the solution, was also found to accelerate the reaction.

Analyzing the morphology of the reacted films under different reaction conditions always showed smaller crystallites when the reaction went faster. As an example, Figure 48(iv) shows smaller MAPbI$_3$ crystallites when MAI concentration is higher (other similar examples can be found in ref. 6). When using Sn instead of Pb for Sn-based perovskites (MASnI$_3$, Cs$_2$SnI$_6$), an acidic environment is especially important - much more than for the Pb-based perovskites, for which added acid is not usually necessary for the reaction to take place.

Based on our observations, we suggest the transformation mechanism to be via an initial *fast* reaction between a native oxide (e.g., PbO$_x$) and the AX reagent, followed by a *slower* oxidation of the metallic Pb$^0$ (Sn$^0$) to Pb$^{2+}$ (Sn$^{2+}$ or Sn$^{4+}$) with a dissolution-reconstruction growth of the HaP, as also occurred in the case of PbI$_2$ transformation to MAPbI$_3$ described earlier. The small resulting crystallites obtained under stronger oxidation conditions occur probably due to larger number of nucleation sites – similarly to what happened when using higher MAI concentration to form MAPbI$_3$ from PbI$_2$ using a very similar MAI solution in IPA (cf. Figure 45(ii)).

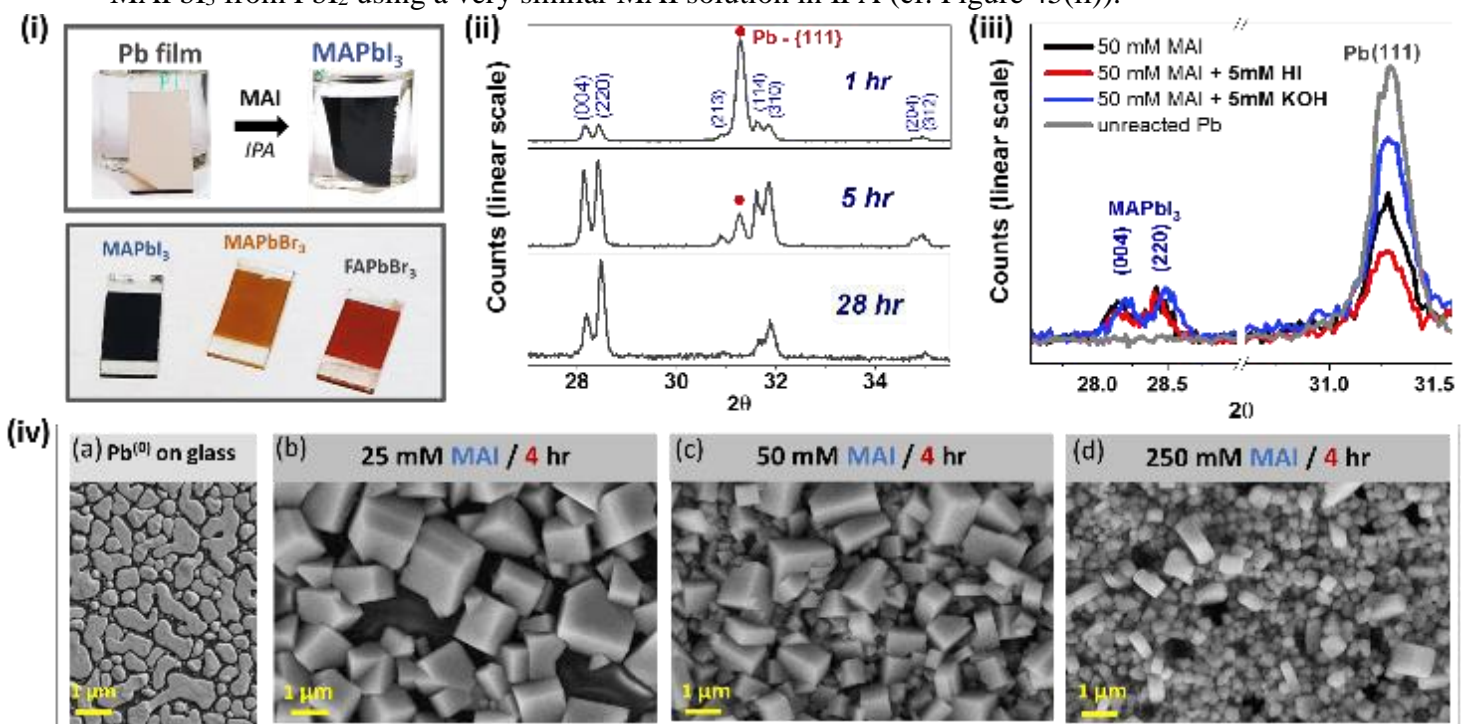

Figure 48: (i) (*top*) Pb film (~ 100 nm) evaporated on dense-TiO$_2$ /FTO/glass substrate before and after treatment with MAI (50 mM for ~ 2 hr) dissolved in IPA; (*bottom*) HaP films after treatment of similar Pb films in solutions of (from left to right) 50mM MAI, 70 mM MABr and 70 mM FABr for ~ 2 hr @ 20 ºC, 4 hr @ 50 ºC and 5 hr @ 50 ºC, respectively. (ii) XRD patterns of a reacted Pb film (deposited on a glass substrate; ~150 nm) with 50 mM MAI solution in IPA for (top) 1 hr, (middle) 5 hr and (bottom) 28 hr. The Pb peak at 2θ~31˚ indicates the completeness of the reaction if compared with Pb peak from MAPbI$_3$. (iii) XRD patterns (normalized to the MAPbI$_3$ (110) peak) of Pb films (~100 nm), deposited on FTO, that were treated with 50 mM of MAI, dissolved in IPA, showing that at lower pH the reaction rate is higher (see Pb/MAPbI$_3$ peak ratio). (iv) SEM images that compare between (a) Pb film (before reaction) and (b-d) Pb film treated at RT for 4 hours with MAI dissolved in IPA at different concentrations (shown at the top of the images). For the higher MAI concentrations, the reaction rate is faster and the resulting MAPbI$_3$ crystallites are smaller. Similar decreases of MAPbI$_3$ crystallite size is observed when decreasing pH or adding I$_2$.

We can sum up what seem to be the most probable pathways for the oxidation of Pb metal in the following equations:

(1) $Pb^0 + 2H^+ \rightarrow Pb^{2+} + H_2$

(2) $Pb^0 + X_3^- \rightarrow PbX_2 + X^-$

It should be noted that when the A group is MA or FA, the reaction is more rapid than when the inorganic Cs is used. One reason may be the native acidity of a MA or FA group that can release $H^+$ to the solution and dissolve in the alcoholic solution as, for example methylamine ($CH_3NH_2$). Moreover, it was shown that dipping Pb film in a diluted solution of HI or KI with some trifluoroacetic acid, the Pb film turned yellow (which is $PbI_2$).

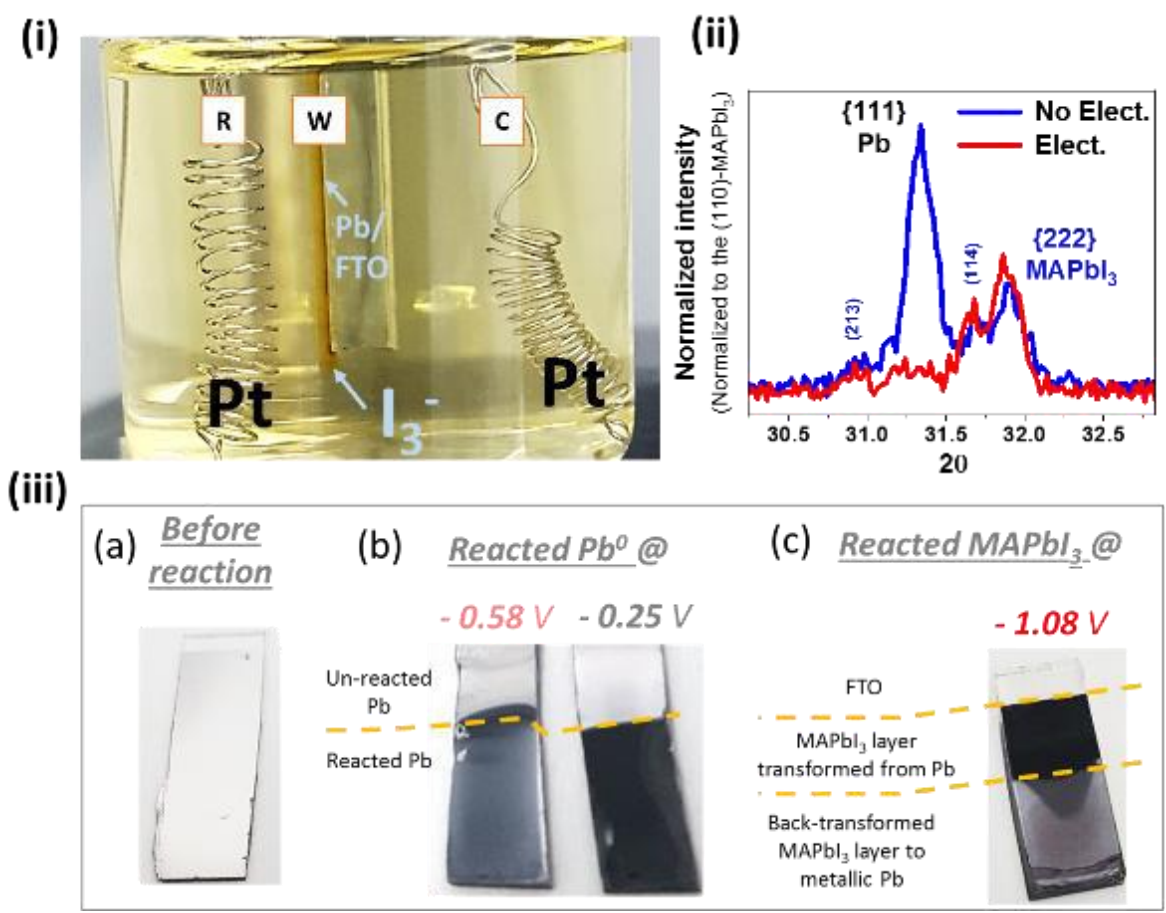

Figure 49: Electrochemically-assisted conversion of Pb (~ 100 nm on FTO) to MAPbI$_3$ in a solution of 50 mM MAI in IPA. (i) A photograph of the reaction system ~ 1 min after applying 0.75 V between the reference (R) and the working (W) electrodes. Both counter (C) and R electrodes are Pt coils. The working electrode is Pb on FTO/glass. The brown cloud next to the Pb electrode is electrochemically-generated polyiodide. (ii) XRD patterns of (red) electrochemically-assisted (1.0 V) and (blue) non-electrochemically reacted films in 50 mM MAI/IPA. The disappearing Pb-{111} peak demonstrates the acceleration in reaction rate due to the applied bias. (iii) Photographed samples of: (a) unreacted Pb film deposited on glass; (b) reacted Pb films deposited on glass, after 5 min in 50 mM MAI in IPA at (left) -0.58 V and (right) -0.25 V bias vs. SHE. (c) MAPbI$_3$ on FTO (obtained after transforming Pb), reacted in a similar solution as in (b) but at -1.08 V vs SHE. All potentials were measured vs. Ag/AgI and then related to the SHE scale.

To further support the reaction mechanism we applied a positive bias on the $Pb^0$ electrode vs. a Pt counter electrode (see Figure 49(i)). It was shown that the reaction rate was dramatically accelerated from hours to minutes as was verified via XRD (see Figure 49 (ii)). This experiment shows that Pb oxidation is indeed the rate-determining step. It is also evident that $I^-$ is being oxidized to form $I_3^-$ as seen by the brown cloud next to the working (Pb) electrode (Figure 49 (i)). The polyhalide so formed can further react with the oxidized $Pb^{2+}$ to form $PbI_2$ that can then further react with MAI. The most likely counter reaction is the reduction of $H^+$ to $H_2$, which was evident at the cathode as bubbles at very high current densities.

Applying an increasingly *negative* potential slows the conversion reaction and at -1.08 V (vs. standard hydrogen electrode (SHE)), a reacted halide perovskite film transforms back to metallic Pb (see Figure 49 (iii)). Since $Sn^{2+}$ tends to oxidize to $Sn^{4+}$ (one of the drawbacks of Sn-based perovskites), such an electrochemical reduction may also be useful for controllably reducing $Sn^{4+}$ to $Sn^{2+}$ and improve Sn-based HaPs. This concept, however, was not tested in practice.

To summarize the electrochemical reactions:

(3) *Working*: $Pb^0 \rightarrow Pb^{2+} + 2e^-$

(4) *Working*: $3I^- \rightarrow I_3^- + 2e^-$

(5) *Counter*: $2H^+ + 2e^- \rightarrow H_2$

The same effect was also demonstrated for a Br analogue, where in practice it can reduce the time for Pb conversion to $MAPbX_3$ from hours to minutes.[6]

An alternative route to $MAPbI_3$ is by exposure of $Pb^0$ to a molten salt of $MAI_3$. Similar compounds, such as $CsI_3$ and $CsBr_3$, are known and to be used as strong oxidation agents, even for noble metals like Au.[223] Recently $MAI_3$ was demonstrated to be able to etch metallic Pb to form $MAPbI_3$, as was first demonstrated by Petrov et al.[60], as well as metallic Au to form a perovskite-related structure $(MA)_2Au_2I_6$.[224] The formation of the $AX_3$ salt is usually spontaneous and proceeds as follows:

(6) $AX + X_2 \rightarrow AX_3$

In the cases of organic A cations, even at RT where the I-based precursors are solid, the reaction is rapid, as demonstrated in Figure 50(i). However, in cases where the A cation is Cs, $CsX_3$ formation is slower (a few hours) and may require elevated temperatures for the reaction to be complete. Unlike $MAI_3$, all other $AX_3$ compounds are solid at RT (see Table 4). It is worth mentioning that since $X_2$ are volatile species (especially $Br_2$), these compounds slowly release $X_2$ species, as can be clearly sensed by a strong odor from $Br_2$.

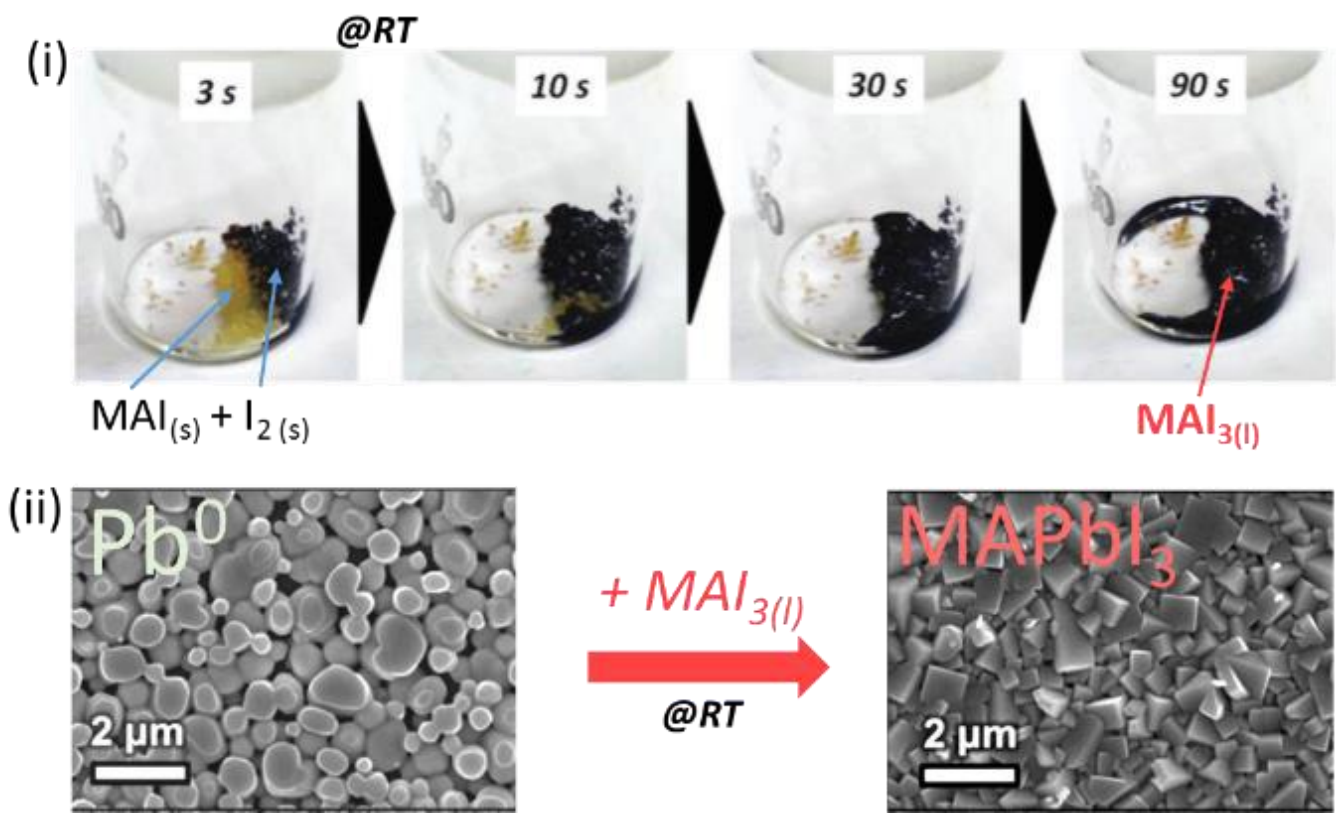

Figure 50: (i) synthesis of MAI$_3$ molten salt by simply mixing the two solids: MAI and I$_2$. The reaction at RT takes no more than few minutes. (ii) SEM images of evaporated Pb$^0$ (left) and the same film after exposing it to MAI$_3$, using a spin-coater. Pay attention that the resulting morphology is similar to that found after exposing PbI$_2$ or Pb$^0$ to MAI in IPA (cf. Figure 47(i) and Figure 48(iv)). These figures (with some graphical adaptations) are from Petrov et al.[60]

| AX$_3$ | Melting temperature [ °C] | Bandgap (indirect) [eV] |
|---|---|---|
| MAI$_3$ | 16-17 | n=2 →1.83<br>**n=0.5 →1.63** |
| MA(I,Br)$_3$ | 79 | n=2 →2.05<br>**n=0.5 →1.63-1.76** |
| MABr$_3$ | 90 | n=2 →2.73<br>**n=0.5 →2.19-2.31** |
| FAI$_3$ | 80 | n=2 →1.97<br>**n=0.5 →1.43-1.51** |
| FABr$_3$ | 65 | n=2 →2.47<br>**n=0.5 →2.05-2.12** |
| CsI$_3$ | 207 [a] | n.a. |
| CsBr$_3$ | Decomposes at 160 | **n=2 →2.14**<br>n= 0.5 →1.88 |

Table 4: AX$_3$ compounds synthesized following reaction (6), including their melting/ decomposition temperatures. The bandgap of some of the compound was determined via transmission spectroscopy and transformation to a Tauc-plot representation with n=2 (direct transition) or n=0.5 (indirect transition),[225] similarly to that presented in Figure 51. The values in **bold** are assumed to be the dominant transition based on visual color of the material.

[a] CsI$_3$ was not synthesized in-house but added from ref. 226 for completeness.

When exposing different metals (Pb, Au or even Pt) to *any* of the molten salts, the metals react with a reaction rate that increase as: Pb > Au > Pt. When exposing Pb to a controlled amount of the polyhalide salt, a perovskite (related) compound appears. At least for the case of MAPbI$_3$, the morphology of the resulting film shows well-defined MAPbI$_3$ crystallites (see Figure 50(ii)) – similar to that obtained by reacting Pb or PbI$_2$ in a MAI solution of IPA (cf. Figure 47(i) and Figure 48(iv)). The redox equations can be described as:

(7) *Red*: AX$_3$ + $2e^-$ → AX + 2X$^-$

(8) *Ox*:  Pb$^0$ → Pb$^{2+}$ + $2e^-$

An interesting point that should be noted is the similarity in the bandgap of AX$_3$ species with those of the APbX$_3$. To emphasize this point, transmission spectra (and their corresponding Tauc plots) are plotted in Figure 51(i) showing that for MAPbX$_3$ and AX$_3$, optical bandgaps are clearly similar. The main difference between the perovskite and the polyhalide compounds is the directness of the optical transition, meaning that for HaPs it is a direct transition, while for the polyhalide compounds it seems to be indirect. When accounting in the possibility of *defect hybridization*, the spontaneously occurring 'guest' AX$_3$ compounds with a similar bandgap to their 'hosting' APbX$_3$ compounds may result in (opto)electronically tolerated defects as emphasized buy Walsh and Zunger[227] (see Figure 51(ii)). To strengthen this point one should identify the actual band alignment of the 'host' and 'guest' compounds (i.e., the energetic position of the AX$_3$ complex w.r.t. the HaPs), e.g., via UPS or IPES. Such experimental evidences are still missing. Together with the 'defect-tolerating' concept that was broadly discussed in sections 3.1 and 3.4 (i.e., shallow defects due to an 'anti-bonding' VBM), this complex formation mechanism suggests *another* 'defect-tolerating' mechanism.

To validate the potential of a HaP film fabricated using a metal as a precursor, some preliminary attempts to fabricate thin films for PV cells were made. After some optimization of Pb thickness, MAX/additive concentration and temperature, the metallic Pb was transformed to HaP via dipping in a MAX solution of IPA (without electrochemical assistance;. examples for cells fabricated with an applied field can be found in the paper.[6]) Other examples (by other groups) of solar cell fabrication via Pb$^0$ transformation using MAI solution in IPA or direct from MAI$_3$ molten salt can be found in Refs. [58] and [60], respectively.

Commonly-used fabrication methods for polycrystalline films (such as co-evaporation, spin- or spray-coating) are, usually, kinetically-stabilized due to rapid quenching during fabrication; therefore, reproducibility issues may (and do) arise. Conversion of BX$_2$ or B, to ABX$_3$, as presented in these two sections, suggests a thermodynamically-driven formation of the HaP, which may, in principle, make layers that are as good as single crystals. Measuring the PL lifetime with TRPL of converted Pb films to MAPbI$_3$ and MAPbBr$_3$ (Figure 51(i)) shows a mono-molecular decay (referred to the longer time constant, $\tau_2$) that compares favorably to values found for films used for the state-of-the-art devices and single crystals.[66] This suggests that this method results in a good quality film. The main drawback (at present) is the less-controlled or favorable morphology.

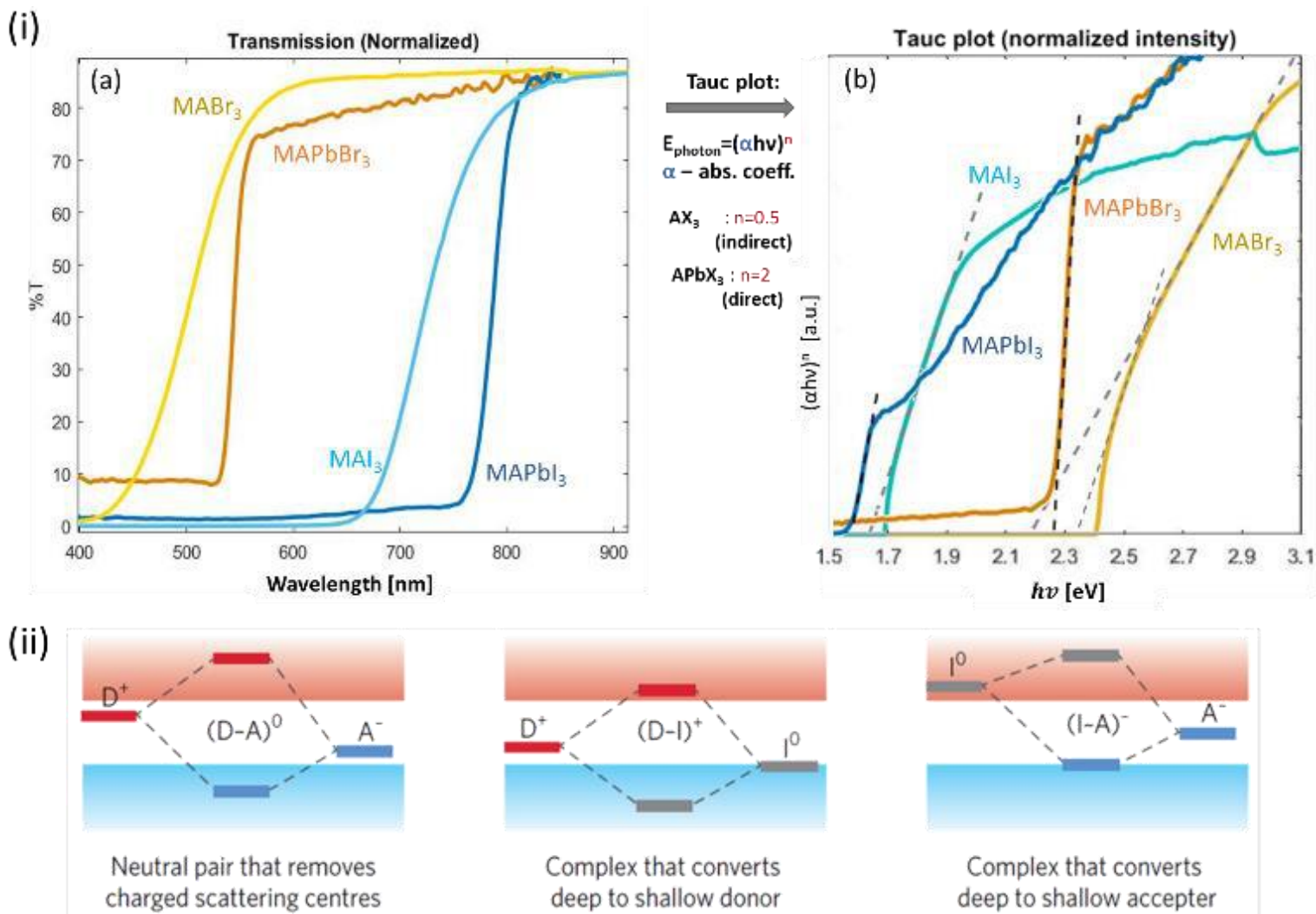

Figure 51: (i)(a) transmission spectra of MAPbX$_3$ and MAX$_3$. Based on the sharp absorption onset in MAPbX$_3$ we deduce that the optical transition is 'direct', while due to the moderate absorption onset in MAX$_3$ compounds we deduce the transition is 'indirect'. The bandgap was derived by using a Tauc plot (b), as described in ref. 225. The obtained optical bandgaps are: $E_g$(MAPbBr$_3$)= 2.26 eV ; $E_g$(MABr$_3$)= 2.19 - 2.31 eV ; $E_g$(MAPbI$_3$)= 1.57 eV ; $E_g$(MAI$_3$)= 1.63 eV. $E_g$ for MABr$_3$ is within a range since there are two linear parts so the bandgap is not clearly defined. (ii) An illustration (adopted from Walsh and Zunger[227]) representing a conceptual idea that individual defect may complex with their surroundings (e.g., other defects) to form a new type of hybridized species with energy states that will resonate with the bulk bands (VBM or CBM). This concept suggest a path for defect-tolerance since the hybridized species are (opto)electronically inactive.

Preliminary efforts to use these films to make photovoltaic cells showed respectable voltages, but very low current density with a significant hysteresis (see Figure 52(ii)). The reasonable voltages are attributed to the long lifetimes of the excited carriers. Long lifetimes mean a reduced recombination probability, which results in an increased charge extraction probability. The overall low performance is attributed to a poor morphology of the layer resulting in possible shorts or parts with thickness larger than the diffusion length in the material. The very prominent hysteresis in the I-V plots, which is a common problem in HaP-based photovoltaic cells, can have different reasons, such as ion migration, charge trapping or electrochemical reactions.[180,228,229] As shown above (Figure 49), electrochemistry is definitely a possibility in HaPs. At the same time, it seems as if upon removal of an electric bias (e.g., for solar cells, during night time), an electrochemical process can spontaneously reverse itself (similar to the reaction between $Pb^0$ and A-X), which supports possible "self-healing" (as will be further discussed *and exemplified* in the following sections).

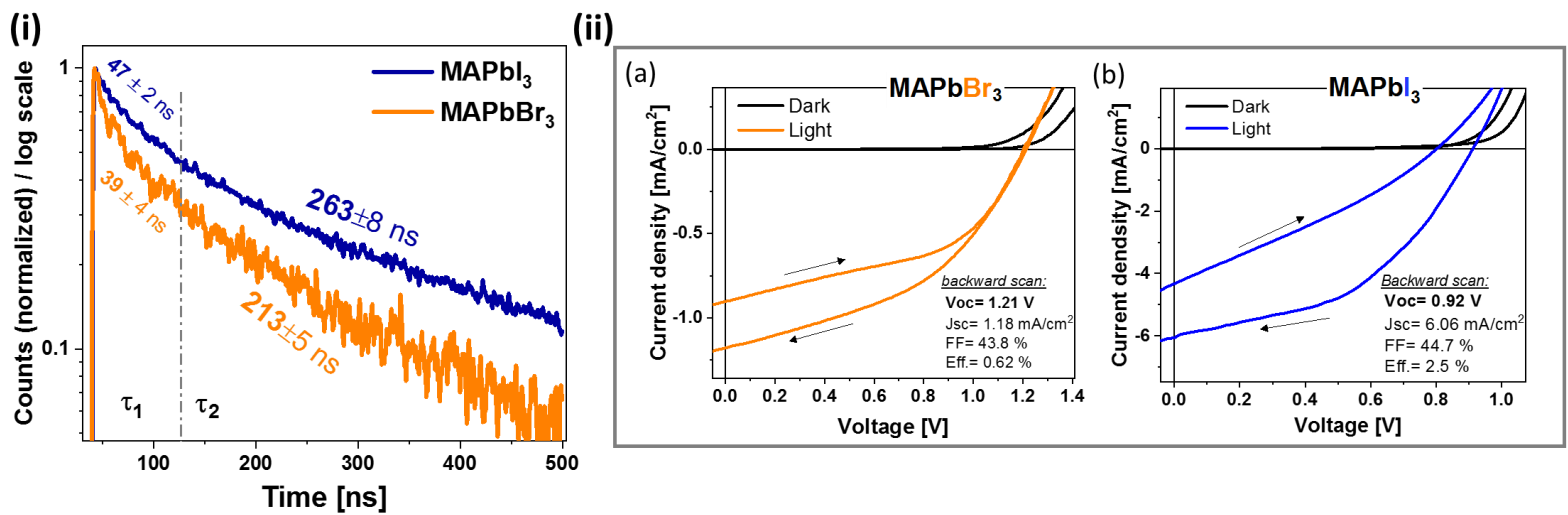

Figure 52: (i) TRPL of $MAPbI_3$ and $MAPbBr_3$ formed from reaction of Pb films (on glass) with 50 mM MAI with 1 molar% of HI (blue) and 70 mM MABr (orange) for 5 hr and 16 hr, respectively. (ii) J-V scans of: (a) $MAPbBr_3$-based solar cell (Pb film in 70 mM MABr solution in IPA reacted for ~24 hr at 50 °C with a cell architecture of FTO/d-TiO$_2$/MAPbBr$_3$/Au) ; (b) $MAPbI_3$-based solar cell (50 mM MAI + 10% (molar) $I_2$ in IPA for ~6 hr at 50°C with a cell architecture FTO/d-TiO$_2$/MAPbI$_3$/spiro-OMeTAD/Au). The black curves are obtained under dark conditions, while the colored curves are obtained under solar-simulated 1-sun (100 mW/cm$^2$). Arrows on each curve indicate the direction of the scans.

In this section, it was shown that formation of the HaP is not limited to the addition of AX salt to an already oxidized Pb but also that AX (assisted with an oxidative environment) or $AX_3$ melts are capable of reacting with metallic Pb to form HaPs. It is important to note that Pb-, Ag- and Tl-halide compounds decompose quite readily to metallic Pb, Ag or Tl and a halide.[230] This reaction is widely used in the photography industry, where a metal-halide (usually AgI) layer deposited on a film exposed to light form clusters of metallic Ag, that are then used to build the contrast of an image. The complimentary reaction is oxidation of $2X^-$ to $X_2$. Since in HaP the possible existence of AX and $X_2$ in proximity may leads to formation $AX_3$, a metallic $Pb^0$ will rapidly turn back to $APbX_3$ – a 'self recyclability' capability that together with rapid ion diffusivity encourages 'self healing'. This point will be emphasized in the following section.

## 4.4.     On degradation, self-recovery and the role of the A cation

*This part is based, in part, on ref.* 8

One of the major drawbacks of HaPs is their poor stability. They are known to change their performance and/or even structure in different environments (e.g., $N_2$, $O_2$, $H_2O$, $I_2$)[231–236], degrade upon irradiation of photons [234,237] or electrons[238] and even decompose under a moderate electric field[135,216,232]. Reviewing the chemistry of formation of HaPs in the previous sections of this chapter showed that there is (1) a high probability for ion migration, (2) low formation and activation energies between the ternary HaP and its binaries and (3) highly reactive decomposition products (i.e., $AX_3$) that can oxidize back any free metal formed in the decomposition.[i] In this section I discuss the degradation mechanism of HaPs (mostly *photon degradation* as a natural degradation path for any PV application), present evidence for partial or complete recovery and explain the role of the A cation (organic vs. inorganic) both for degradation and for recovery. For the sake of comparison, I refer only to results from $APbBr_3$ compounds (A=MA, FA or Cs), since all three Br-based compounds possess a perovskite structure at RT (unlike $FAPbI_3$ or $CsPbI_3$).

It was shown in section 3.2, Figure 23, that at high enough temperatures (>250 ºC) MABr leaves the $MAPbBr_3$ matrix, leaving behind $PbBr_2$. When replacing MA with Cs, this intermediate stage is absent and the entire $CsPbBr_3$ melts, since CsBr has a higher melting temperature than that of $CsPbBr_3$[48]. It was also postulated that when mixing Cs in MA to an extent that the symmetry of the mixed $(Cs_xMA_{1-x})PbBr_3$ is decreased (e.g., from cubic to tetragonal), the threshold temperature at which MA evaporates from the structure is increased – probably due to dislocation hardening, that may impede (MABr) mass migration.

---

[i] These *three* points suggest that 'entropic stabilization', 'self-recyclability' or 'self-healing', which are mentioned earlier, are three analogous terms that should exist in HaP with high probability.

Similar stability-related evidence are found also in other degradation experiments, where in principle, Cs-based compounds are more stable than those of MA- (or FA-)based ones. For example, under an electron beam (see Figure 53(i)), the $CsPbBr_3$ crystal surface is found to much more stable than of $MAPbBr_3$ (similar results for polycrystalline thin films are found in Kedem et al.[238]). When monitoring the degradation of an operating solar cell, the current density at maximum power point is a good indicator for degradation. Similar to thermal and electron-beam degradation, when comparing the current density (at the maximum power point) of solar cells with different absorber layers, i.e., $MAPbBr_3$ vs. $CsPbBr_3$ with identical contact layers, it is found[48] that Cs-based solar cells are more stable with very little obvious loss in performance (see Figure 53(ii)). Since the overall bond strength of these materials is similar (that of $MAPbBr_3$ is even slightly higher), these results suggest that the volatility and reactivity of the AX binary strongly affects the overall durability of the compounds to different degradation paths. In the case of degradation from an electron beam, carbonization due to the presence of an organic MA group may be an important reason for degradation.

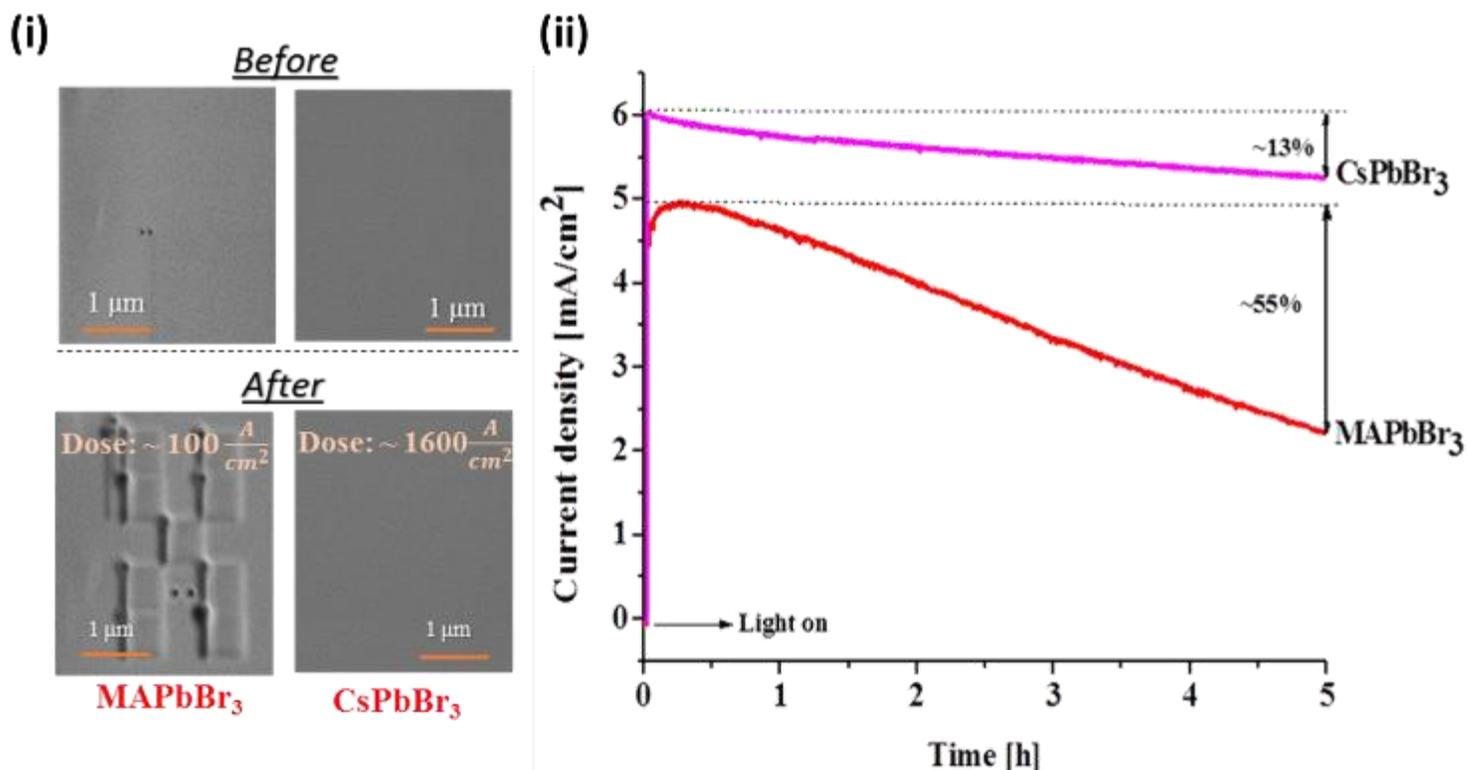

Figure 53: (i) Comparison between MAPbBr$_3$ and CsPbBr$_3$ under high electron-beam bombardment doses (imaging dose: ~4 A/cm$^2$; e-beam bleaching currents: MAPbBr$_3$ ~100 A/cm$^2$, CsPbBr$_3$ ~1600 A/cm$^2$; acceleration voltage: 3kV; detection mode: secondary electrons). (ii) Current density measured at an applied bias close to the initial maximum power point versus time under solar-simulated 1-sun (100 mW/cm$^2$) illumination for MAPbBr$_3$- and CsPbBr$_3$-based cells (taken from Kulbak et al.[48]). CsPbBr$_3$ is found to be more durable to both e$^-$-beam and 1-sun illumination.

Unlike thermal degradation and that due to an electron beam, in operating solar cell we also need to consider degradation due to photon exposure, which causes *photo-degradation,* or electric fields, which may cause local oxidation/ reduction and ion migration. These may eventually result in loss of power conversion efficiency. Since ion migration has already been much discussed in the literature, as mentioned in section 4.1, I will mainly focus on photo-degradation.

Photo-degradation has been well-known for many years in different halide-based compounds, such as AgX, TlX and PbX$_2$.[230] In fact, AgX compounds (in some cases mixed with PbX$_2$) are well-known for their use in photography as the photoactive layers (see Figure 54(i)). These metal-halide compounds are known to undergo reduction to the metal and oxidation of the halide upon illumination, where the metallic residue is known to become a contrast agent in a photographic film (which is then further developed (growth of additional Ag$^0$), fixed (removal of AgX excess), washed and dried[239]). Recently,[237] XPS surface analysis of MAPbI$_3$ film revealed that, similar to metal-halide compounds, MAPbI$_3$ undergo an accelerated degradation upon illumination (in addition to that which occurs due to the probing x-ray beam of the XPS itself) in the form of Pb$^{2+}$ reduction to metallic Pb (see Figure 54(ii)).

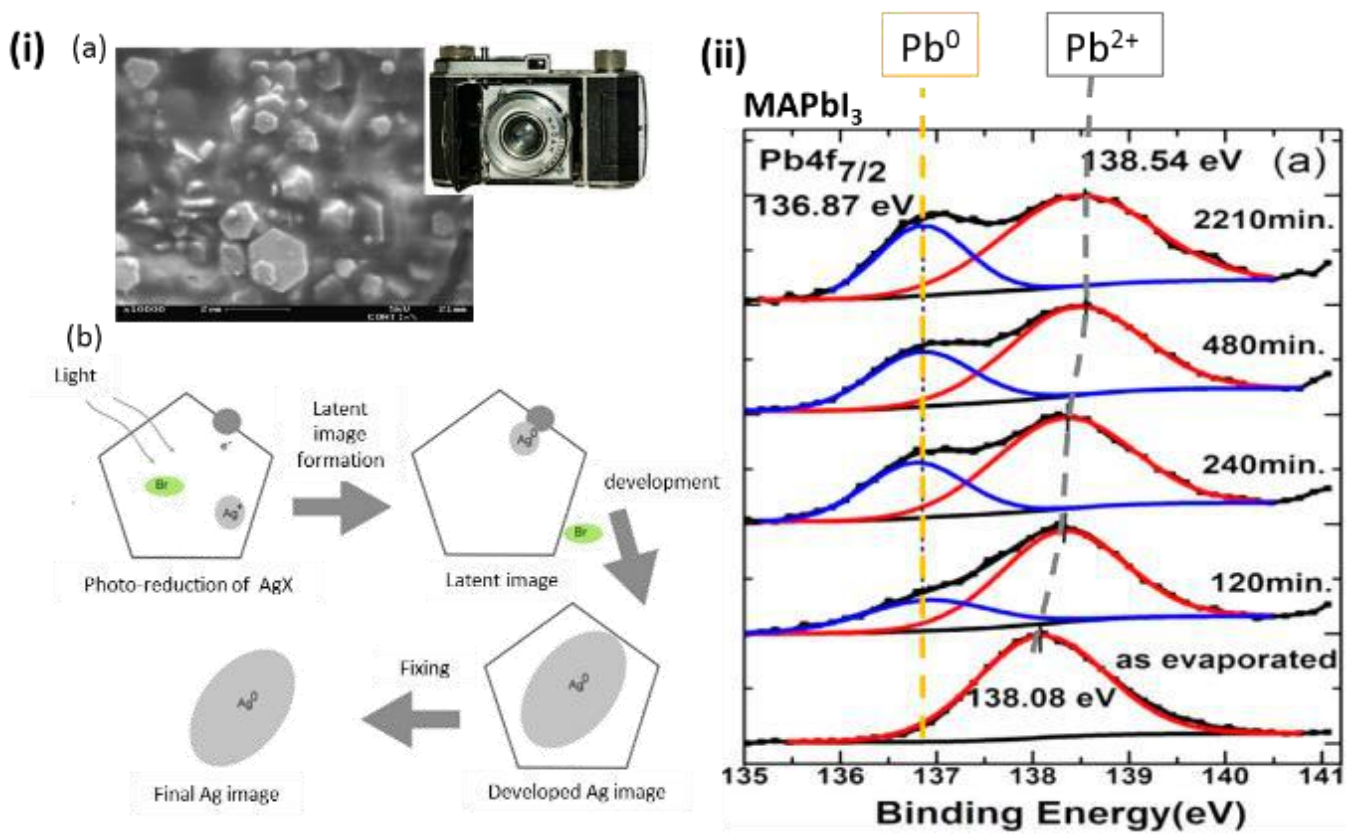

Figure 54: (i) (a) SEM image of the active surface of a photographic film.[240] (b) Part of the development process chemistry of a photographic film.[241] (ii) XPS analysis of and *in-situ* illuminated MAPbI$_3$ film showing that Pb$^0$ is formed upon illumination (adopted from Li et al.[237]).

In a collaborative work led by Dr. Davide Ceratti in our group, using confocal microscopy, where information (e.g., reflection or PL) is collected only from the focal point of the light source, further information regarding photo-degradation was collected from {100} surfaces of APbBr$_3$ single crystals. Using a pulsed photon flux with intensities much higher than those of sunlight, degradation in the absolute PL intensity from the surface (using single-photon absorption @488nm - Figure 55) or the bulk (using two-photon absorption @800 nm - Figure 56(ii)) is studied. Under high light intensities (in our case for surface degradation – $10^7$-$10^8$ mW/cm$^2$, which are ~ $10^5$-$10^6$ 'suns'), local heating and stress formation may occur[242] in addition to a pure photo-degradation. It is also possible that internal redox reactions take part in the photo-degradation, in addition to any mechanical effects that result upon strong irradiation.

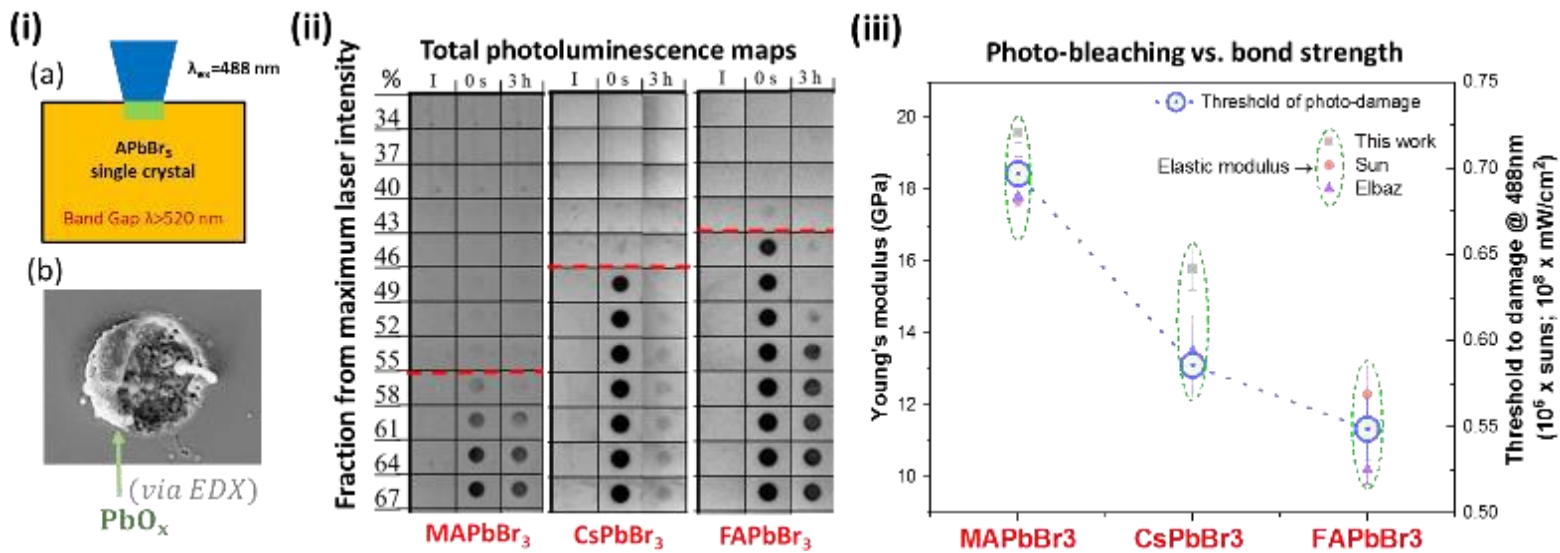

Figure 55: (i) (a) Illustration for photo-bleaching experiment done under a confocal optical microscope in which {100} facet of APbBr$_3$ single crystals is bombarded with a 488 nm pulsed laser. The bleaching area (10 μm) is >> than the laser spot size (~ 0.5 μm). The maximum bleaching intensity was 1.2 x10$^8$ mW/cm$^2$ (b) SEM image of CsPbBr$_3$ after bleaching at maximum intensity., EDS analysis (and comparing with MAPbBr$_3$ and FAPbBr$_3$) showed that photo-bleaching led to the formation of Pb$^0$, which, in the case of CsPbBr$_3$, is then oxidized in air to PbO$_x$. At half the power, such damage was not observed. (ii) A stack of PL total intensity images done via the experiment described in (i,a) at different bleaching intensities. Each row represents a different area to which a different bleaching power was applied (percentage from the maximum laser power is mentioned let to the images). Each row represents the initial (I) stage (before bleaching), just after photo-bleaching (0s) and after 3 hours in air (3h). The horizontal red-dashed lines are the thresholds at which a clear observation of PL degradation (with respect to the unbleached areas) just after photo-bleaching is observed. (iii) Comparison between the threshold to damage found in (ii) [units of "10$^6$·suns", i.e., 10$^8$ mW/cm$^2$], and the bond stiffness that is represented as the measured Young's modulus of the different crystals taken from empirical measurements done in this work[1], Sun et al.[121,125] and Elbaz et al.[122] (cf. Figure 21(ii)).

For the surface degradation case, it is found that at maximum bleaching power (1.2·10$^8$ mW/cm$^2$), the surfaces of the crystals were found to be destroyed (see Figure 55(i,b)). Since the damage bon the surface of the crystal at ambient atmosphere, it is sensitive to attack from atmospheric O$_2$ and forms PbO$_x$ (which was evident as an increase of oxygen concentration in EDS analysis). When using a fraction of that bleaching laser power for producing the damage, and even much less for imaging the PL response, it is found (Figure 55(ii)) that there is a threshold intensity beyond which the PL intensity decreases (an indication for damage). Interestingly, the intensity threshold at which such damage occurs correlates nicely with the structural bond stiffness (Figure 55(iii)). This indicates that rapid degradation paths (unlike those due to mass transfer of an AX group, e.g., heating) probe the Pb-X bond strength rather than the volatility of the AX binary. However, since the degradation occurs on the surface, it is likely to end with a depleted surface when volatile species exist. Such permanent degradation occurs mostly when A is an organic group, while for Cs it recovers over time. This emphasizes the volatile nature of the organic degradation products vs. those of the inorganic ones. Since photo-bleaching is likely to form $AX_3$ products (directly or starting from an $X_2$ +AX intermediates) and since $CsBr_3$ products seem to have higher stability (similar bond nature with a higher melting/degradation temperature – see section 4.3 (p.95), Table 4), the recyclability of the damaged $CsPbBr_3$ surface can be explained via the redox reaction: $CsBr_3 + Pb^0 \rightarrow CsPbBr_3$.

It is now clear that in an open system, where volatile species may change the stoichiometric balance, degradation is highly dependent on the binaries. The question is what happens in a closed system, which, to some extent, represents better actual solar modules that are invariably encapsulated. Recently, Khenkin et al.[243] (and partially others[244]) showed that the performance of an HaP-based PV cell, as well as the PL efficiency and peak position (see Figure 56(i)) degrade with illumination time, but recover after spending time in the dark. This shows that within day-night cycles, the degradation is effectively less than what was originally assumed (within a single cycle). It is, however, essential to find which of the elements within the tested PV cells is responsible to the degradation (ETL, light harvester, HTL, contacts) and which of the applied energies (illumination, electric field, or a combination of the two).

To remove the effect of the surface/interface and the environment, the experiment described in Figure 55(i,a) is repeated, but now *inside* the bulk (~ 110 μm from a surface) of an $APbBr_3$ single crystal (Figure 56(ii,a)) using two-photon absorption of 800 nm confocal laser light (effectively 400 nm is absorbed at the focal point). Using high enough laser power, it was shown (similarly to the single-photon absorption - Figure 55(ii)) that the PL intensity varies after photo-bleaching (see Figure 56(ii,b)). However, unlike the single-photon bleaching experiment, photo-degraded areas showed recovery for *all* compositions. For the cases where the A cation was organic (MA or FA), the recovery was very rapid and complete, while for the case of $CsPbBr_3$, the damaged area showed much slower regeneration and within 12 hr – incomplete. High time resolution plotting of the PL recovery (Figure 56(ii,c)) showed that between MA and FA, FA recovers slightly faster. More details on the experiment can be found in Ceratti et al.[8]

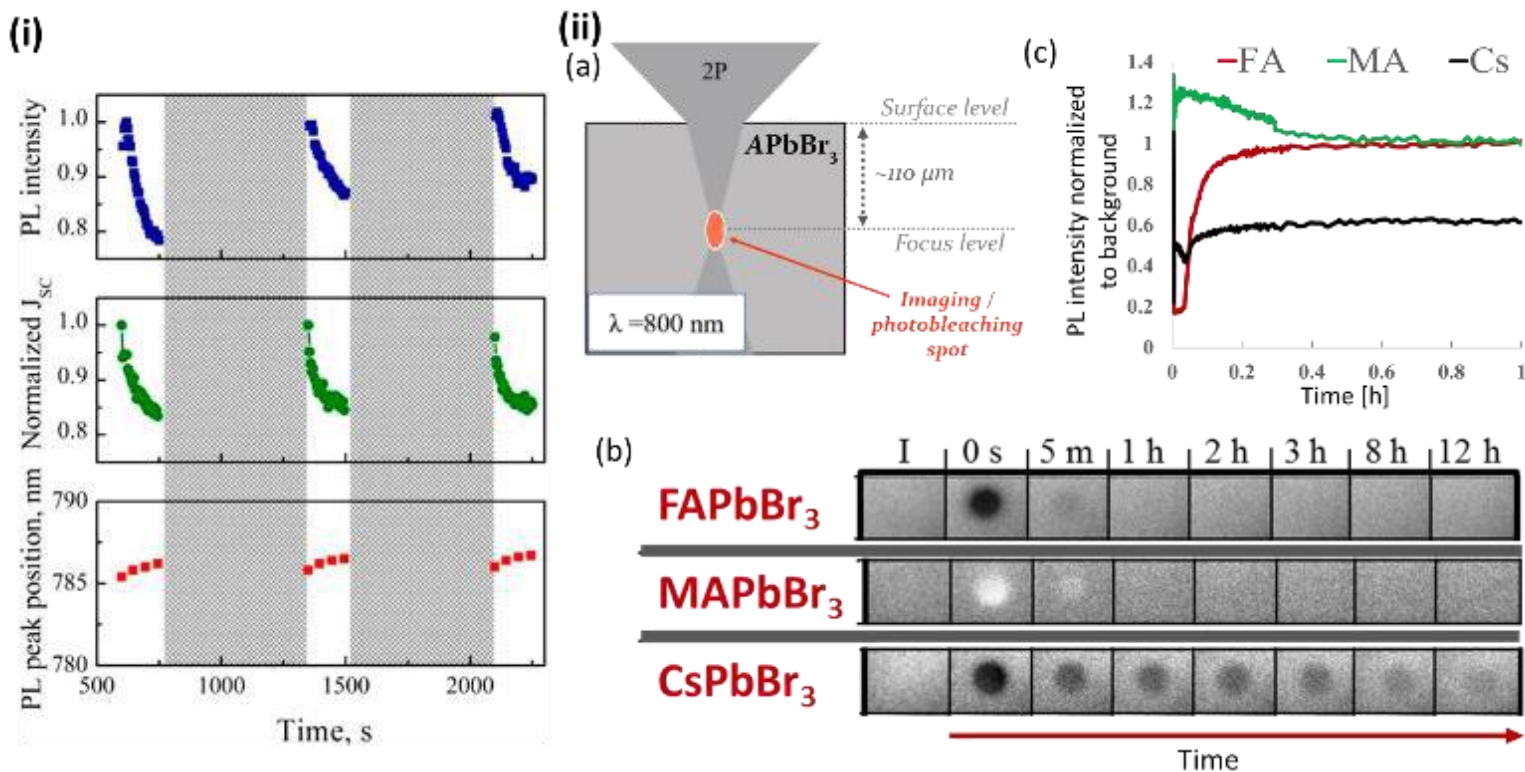

Figure 56: (i) Time evolution of the normalized wavelength integrated intensity of PL emission, normalized $J_{SC}$, and PL peak wavelength under monochromatic laser excitation ($\lambda$ = 532 nm, P = 0.6μW, d = 2 μm = ~200 suns), upon illumination−darkness cycling of a fresh mixed A and X HaP (ITO/SnO$_2$/(**Cs$_{0.05}$**(**MA**)$_{0.15}$(**FA**)$_{0.85}$)**Pb**(**I$_{2.55}$Br$_{0.45}$**)/spiro-OMeTAD/Au) solar cell (adopted from Khenkin et al.[243]). All three properties are shown to recover after a period of darkness. (ii) (a) Illustration for *two-photon* bleaching experiment done under a confocal optical microscope in which APbBr$_3$ single crystals are irradiated with a *sub-bandgap* (800 nm) pulsed laser. The bleaching in all three cases was with 2-photon generation (equivalent of 400 nm photons generated with 800 nm pulsed laser), ~110 μm inside the crystals and with the same intensity for each compound. The diameter of the bleached area is ~10 μm, much larger than the focus of the laser (~0.5 μm). (b) PL total intensity images from three different single crystals. The first left square is an image before damage (I), just after the photo-damage (0s) the remaining images are framed at different times after the photo-bleaching. The imaging intensity is much lower than the bleaching intensity and does not cause any visible damage. All three cases showed recovery of the damaged area (c) Graphical representation of the PL intensity recovery shown in (b) normalized to the background (unbleached) PL intensity. Shows that the recovery of Cs is significantly slower than that of organic A cations. Also shows that FAPbBr$_3$ recovers slightly faster than MAPbBr$_3$. The results presented in (ii) are published in ref. 8 and are a result of a collaborative work led by Dr. Ceratti.

When comparing the melting points (Table 4 - section 4.3 (p.95)) of the three ABr$_3$ solids, we find that the higher the melting temperature (CsBr$_3$>>MABr$_3$>FABr$_3$), the slower the PL recovery (CsPbBr$_3$ << MAPbBr$_3$ < FAPbBr$_3$). Such correlation strongly suggests that in an enclosed system, the nature of the AX$_3$, which is able to react back with metallic Pb resulting from photo-degradation, should play an important role in the recyclability of different HaPs. Therefore, MAPbI$_3$, which has the lowest melting point AX$_3$ intermediate (and liquid at RT – see Table 4), should show the highest recovery rate – a point that still needs to be tested.

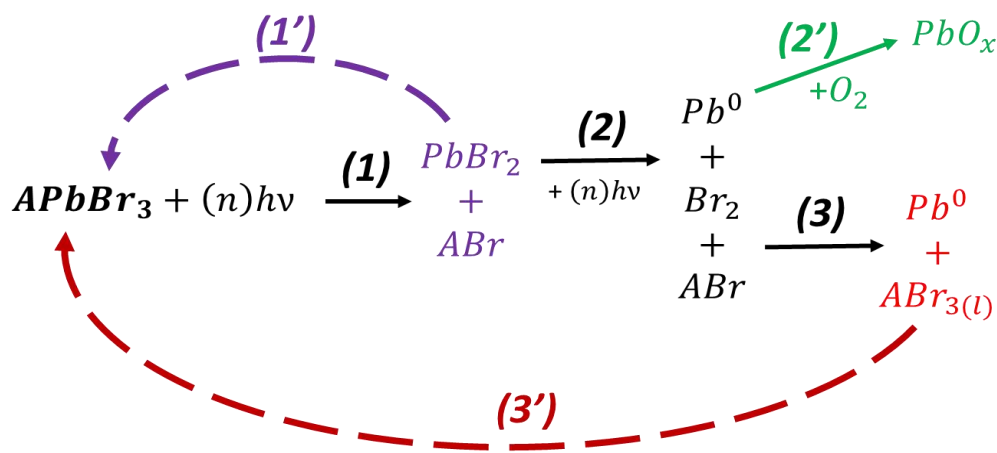

Figure 57: A scheme of photo-degradation (reactions *1, 2* and *3*), regeneration (reactions *1'* and *3'*) and oxidation (reaction *2'*) of APbBr$_3$. Reaction *2'* occurs on a surface of a crystal where it is exposed to O$_2$. Reaction *1'* is described in section 4.2. Reactions *3* and *3'* are described in section 4.3. Reaction (*1+2*) and *2'* are described in this section, Figure 53 and Figure 55(i,a), respectively.

To conclude this section and chapter, we saw that HaPs are quite reactive, especially if they are exposed to air or other environments. We found that when the A cation is inorganic, the durability of the HaP to slow decaying processes is higher. However, once the degradation process is very fast, where mass migration does not dominate the degradation process, the overall structural bond strength is probed, so the Pb-X backbone is what seems to be important. It was shown that illumination can affect local redox degradation where the B metal reduces and the X anion oxides. The oxidized X anion, which can react with an AX binary to form AX$_3$, can then readily react with the oxidized B metal to form again the HaP. It was empirically found that the lower the melting point of the AX$_3$ species, the more rapid the reformation reaction. However, in an open system, volatile species may leave the system leaving behind a depleted HaP system, which may cause a permanent degradation. Additional atmospheric species, such as O$_2$, may exploit the presence of reduced degradation products (e.g., metallic Pb$^0$) to form an oxide compound (e.g., PbO$_x$). Figure 57 summarizes the reaction paths of degradation (mostly by photons) and reformation of HaP from the degradation products. It in principle emphasizes the 'self-healing' chemistry that should reducing the defect density in HaPs.

# 5. Structural symmetry breakage and related defects

Unlike point defects, in this chapter I deal with symmetry-induced 2-D defects: *polar domain walls*. It was previously discussed that a structure with lower symmetry may lead to dislocation hardening and impeded mass-transfer during degradation processes (see section 3.2). It may lead also to some heterogeneity in charge transfer, bond stiffness and different optical (absorption/ emission) cross-sections. However, 3D perovskites are in general (pseudo-) cubic, so heterogeneity in different (opto)electronic properties are rarely observed. However, even a small deviation from centrosymmetry may lead to interesting phenomena like piezoelectricity, pyroelectricity and ferroelectricity – phenomena that are well-known in oxide-perovskites. Low-dimensional perovskites (cf. Figure 3) *do* possess strong heterogeneity (mainly 2D and 1D perovskite-related structures) as can be found in a review by D. Mitzi.[28] Since my focus is on 3D HaP, I discuss the probability to have (opto)electronically-significant symmetry breakage in 3D HaP, specifically $MAPbBr_3$ and $MAPbI_3$. This chapter summarizes two published papers: ref. 3 and ref. 5, and the reader can read them and their supplementary information for any missing background information.

## 5.1. Theoretical concepts

*Ferroelectricity* is the ability to change the spontaneous polarization in a material by an external electric field. It is a well-known dielectric phenomenon, existing in many oxide perovskites [245–247], and has been suggested to be present in HaPs, in particular $MAPbI_3$ [248,249]. If the material is indeed ferroelectric, then this suggests the possible existence of the bulk photovoltaic effect, which can arise due to existing spontaneous polarization the bulk of the material, as illustrated in Figure 58(i).[250] Ferroelectric $MAPbI_3$ could, however, have as a main benefit (as broadly discussed elsewhere) [248–251] charge transport via *domain walls* between adjacent *polar* domains (see Figure 58(ii, iii)). The existence of polar domains induces charge separation, lowers charge recombination and allows high conductivity due to local degeneracy of the SC along the domain walls [252,253]. If this is so, then it can explain the remarkably high voltage efficiency, *i.e.*, low voltage loss in HaP-based solar cells (estimated as the difference between (or ratio of) open-circuit voltage and band gap, $E_G$-$V_{OC}$ (or $V_{OC}/E_G$)).[3,254]

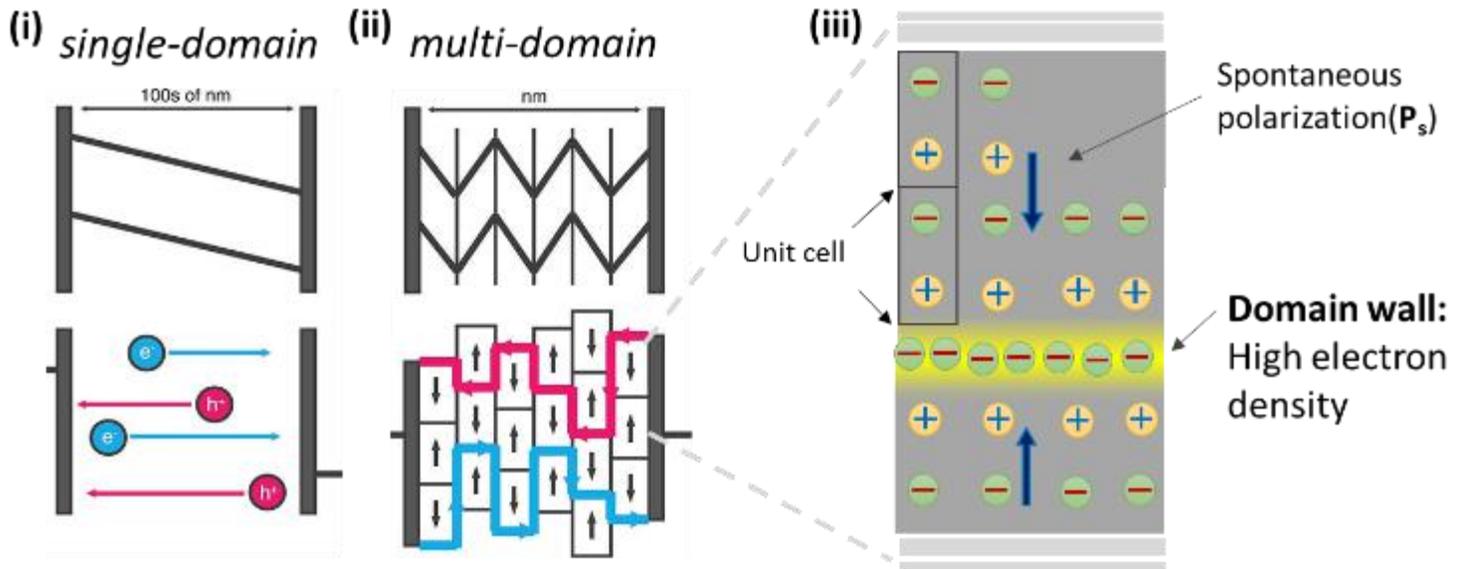

Figure 58: Illustration of (i) a *single* polar domain and (ii) *multi* polar domain structure, where the top figure represents the electric field drop in the material and the bottom figures the flow of charge in each case. The figures are adapted from Frost et al.[18]. (iii) Illustration of an area between two oppositely-oriented domains. Emphasizes the charge degeneracy at a domain wall as a result of the need to reach charge neutrality. The blue and magenta paths drawn in (ii) emphasize that due to the charge degeneracy at domain walls in a head-to-head polar multi-domain structure, the charge should flow mainly via domain walls. This concept was empirically verified in Sluka et al.[253] for $BaTiO_3$.

Whereas existence of ferroelectricity in $MAPbI_3$ was suggested *theoretically* [248,249], evidence on whether HaPs actually possess the domain structure that typifies a ferroelectric material are found to be quite contradictory [3,186,255–266]. To analyze the question if HaP are *actually* ferroelectric, we need to test for the presence/absence of several fundamental phenomena, which are tightly related to it.

Necessary conditions for bulk ferroelectricity to exist are as follows (see also Figure 59(ii)):

(1) absence of an inversion symmetry (*i.e.* a **non-centrosymmetric** material), which may lead to
(2) a unit-cell with a permanent *polarity*, where
(3) an assembly of polar unit cells facing the same direction, which will form a so-called *'polar-domain'*. An assembly of periodically ordered polar domains, facing different directions (usually 180° or 90° to each other), will form the bulk of a ferroelectric material.
(4) The polarity of these polar domains must switch when sufficiently high electric fields are applied.

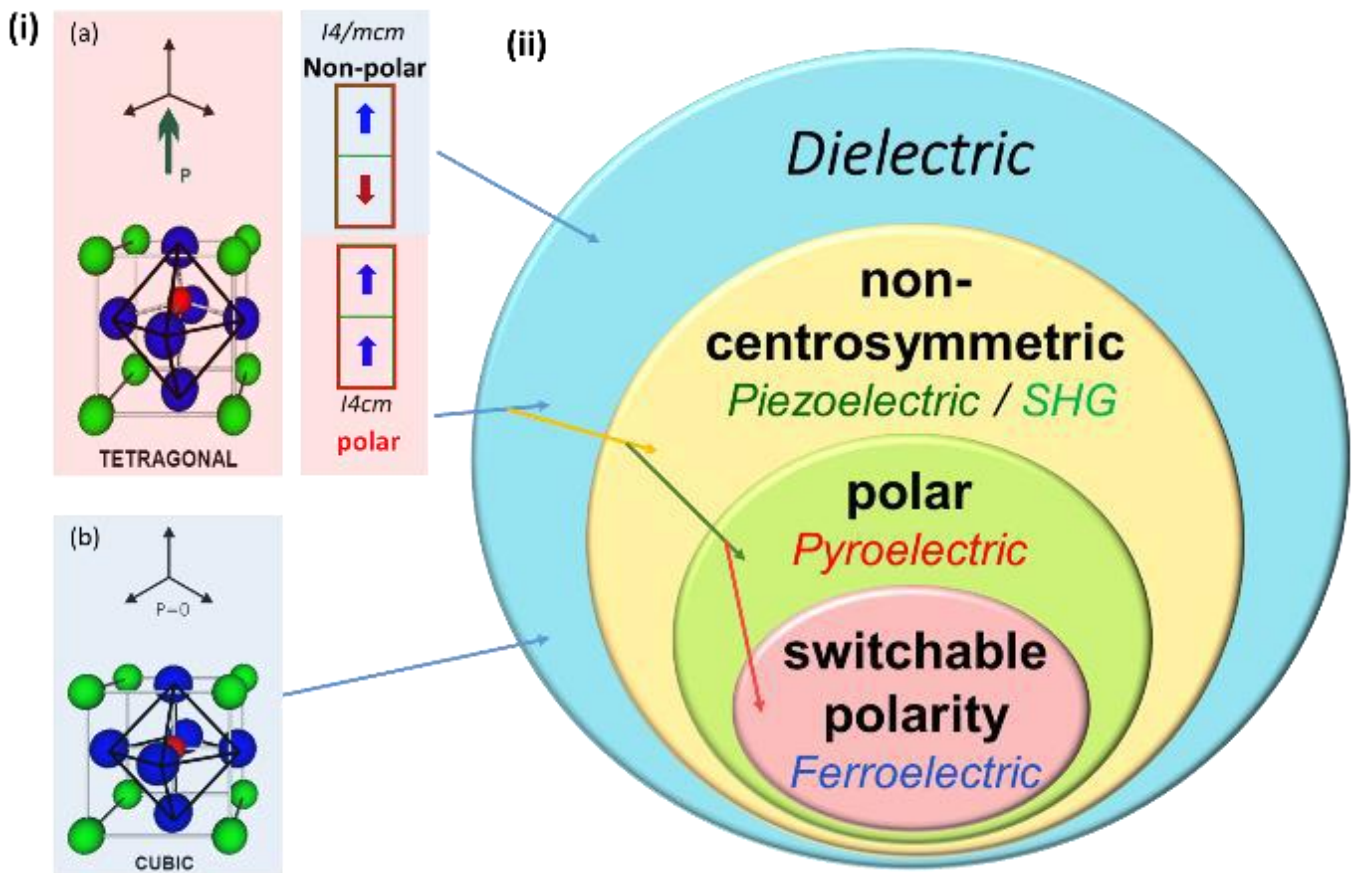

Figure 59: (i) Illustration of (a) a typical tetragonal (sub)unit cell of a perovskite that is also polar and (b) a cubic perovskite unit cell that is not polar. In (a), the polar structure may become anti-ferroelectric (non-polar), if each adjacent sub-unit cell has opposite polarity (top), or ferroelectric (polar), if the dipole of each sub-unit cell are aligned (bottom). (ii) Schematic illustration of the progression from more general to more specific dielectric properties and techniques to identify them. The arrows represent what each symmetry group presented in (i) should possess. For example, a cubic structure is a dielectric material that is centrosymmetric, and non-polar so it should *not* show SHG, pyroelectricity or ferroelectric polarization response. On the other hand, if a material is ferroelectric it *must* be polar and be non-centrosymmetric so it should show SHG and pyroelectricity.

When differences between symmetry groups in are very small, it may be challenging to distinguish experimentally (by x-ray or neutron diffraction) between cubic structures and those with small distortions from cubicity, or between a ferroelectric and an anti-ferroelectric symmetry group in a tetragonal structure (see Figure 59(i) and cf. Figure 2). In such cases, (dis-)proving the existence of the mentioned above points (1)-(4), may become pivotal evidence for determining the actual symmetry group of a semiconductor. In the following section, we will follow the scheme showed in Figure 59(ii) to (dis)prove ferroelectricity in $MAPbBr_3$ and $MAPbI_3$, which at RT are known to be (crystallographically) cubic and tetragonal, respectively.

## 5.2. Challenges in (dis-)proving ferroelectricity in MAPbBr$_3$ and MAPbI$_3$

Although MAPbBr$_3$ has been reported to be centrosymmetric, (cubic *Pm3m*)[110], very small deviations from perfect cubicity might result in a non-centrosymmetric polar structure (e.g., *R3*). The existence of a low-temperature polar phase (orthorhombic *P2$_1$cn*)[267] emphasizes the potential of this material to become polar. For MAPbI$_3$, based on x-ray- and neutron- diffraction techniques, the tetragonal phase of MAPbI$_3$ (<330 K) was ascribed to a non-polar symmetry space group, *I4/mcm*, as well as to a polar one, *I4cm* [257]. Following Aleksandrov's notations for tilt/displacement in perovskite-like systems [12], *I4/mcm* and *I4cm* have exactly the same octahedron-tilt system, where the only difference is that *I4cm* is also polar along the *c*-direction (see Figure 59(i,a)). The interpretation of data from diffraction-based techniques depends on the user's choice of the model to which the diffraction pattern should fit. Small deviations, such as between *I4/mcm* and *I4cm*, can rarely be clearly distinguished, causing disagreement in the literature in oxide perovskites, and other compounds [268,269]. Moreover, it is challenging to distinguish by diffraction methods between cubic structures and those with small distortions from cubicity, as may be the case of MAPbBr$_3$.

Since both *Pm3m* MAPbBr$_3$ and *I4/mcm* MAPbI$_3$ are centrosymmetric, using the scheme in Figure 59(ii), by testing whether there is active second harmonic generation (SHG) upon illumination, we can verify whether these are centrosymmetric or not. The concept of SHG and the experimental setup are briefly explained in Figure 60(i). By spreading µm-sized MAPbBr$_3$ and MAPbI$_3$ pieces that were grown from solution on a glass substrate, we found (see Figure 60 (ii, iii)) that SHG is absent in MAPbBr$_3$ but present in MAPbI$_3$. This immediately suggested that MAPbBr$_3$ should not be polar and, therefore, not ferroelectric. This result emphasized that ferroelectricity is not a general property of HaPs-based solar cells, as might have been thought from the literature.

It should be mentioned that the MAPbI$_3$ crystals were grown in the tetragonal phase without allowing a phase transition to the cubic phase above ~330 K. This is important since upon cooling from cubic to tetragonal phase, the system would lose its orientational memory and a homogenously-spread polar orientation may form. This could lead to a loss of SHG signal due to destructive interference (see more about that in ref. 5 or generally on SHG – read Boyd[270]). Homogenously-spread polar orientation also holds for pyroelectric and ferroelectric measurements, where the change of the *net* dipole moment is measured.

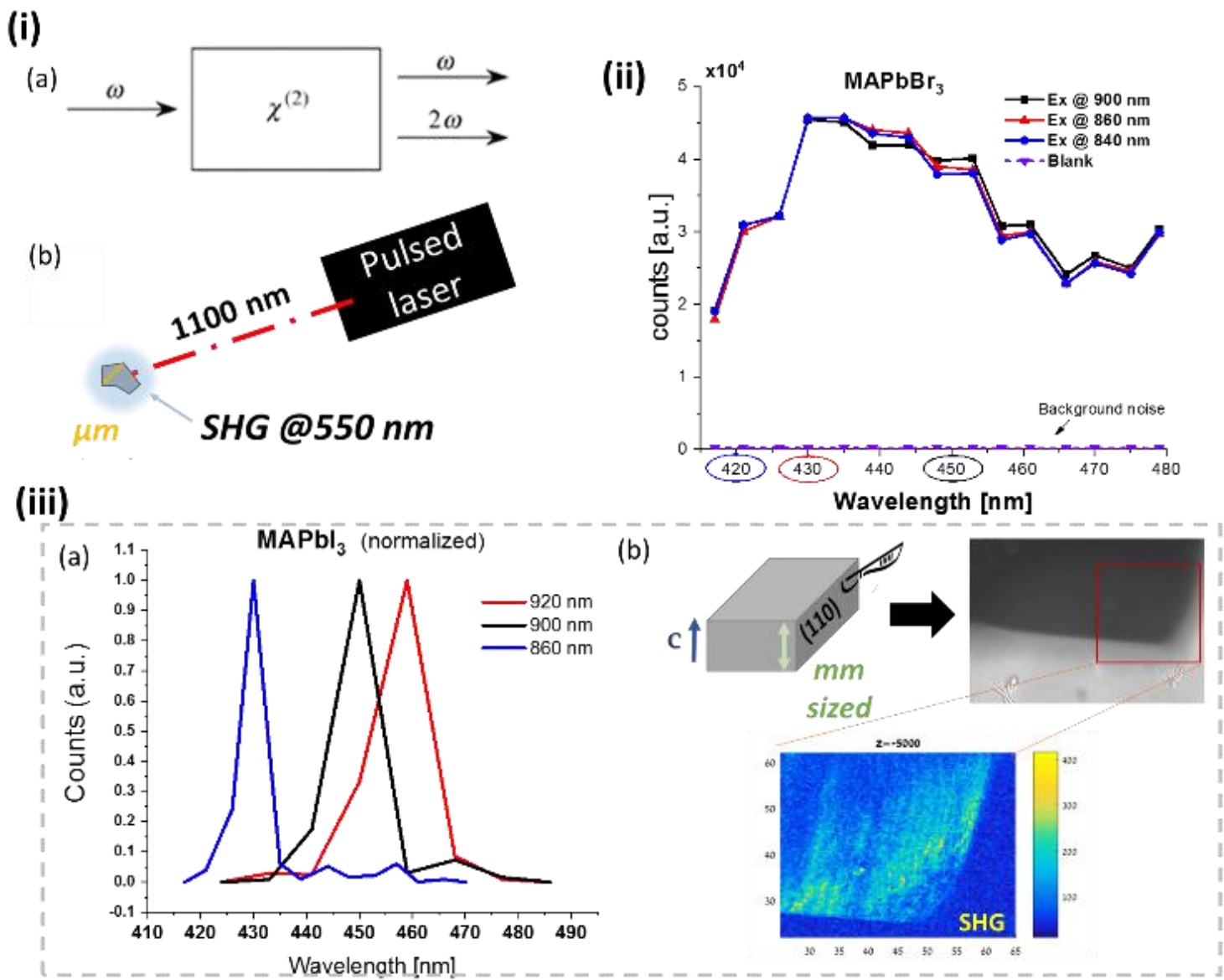

Figure 60: (i) (a) Illustration of SHG that occurs if a material is *not* centrosymmetry (meaning that its second-order susceptibility, $\chi^{(2)}$, is *not*-zero); $\omega$ is the photon frequency. (b) illustration of the experimental setup – a pulsed laser is focusing on a sample and scattered light is collected using a spectrometer. SHG should be *exactly* at half of the exciting wavelength. (ii) and (iii) are experimental results from MAPbBr$_3$ and MAPbI$_3$ μm-sized pieces, respectively, that were scratched from solution-grown crystals (VSA method). (iii,b) light microscopy image (top-right) of a MAPbI$_3$ particle that was removed from a {110} surface of MAPbI$_3$ crystal; (bottom) a SHG experiment done on a different setup, where excitation was at 1100 nm and collection at 550 nm.

Returning to Figure 59(ii), we see that a ferroelectric material has to be pyroelectric. While the opposite is not true, if a material is not pyroelectric it cannot be ferroelectric and this led us to focus on searching for pyroelectric behavior of HaPs. The reason for this approach is that, in contrast to measuring ferroelectric behavior, measuring pyroelectricity does not require an external electric field, and any possible changes in the material due to the applied electric field that are not symmetry-related (e.g., artificially induced polarity due to ion diffusion, electrochemical reaction or capacitance effects) can be ruled out.

The pyroelectricity of all samples was measured by the Chynoweth method, where the sample, in a given orientation, is exposed to a periodic temperature change[271–273] (see also Figure 61). The samples were placed on a copper plate, which was placed in a Faraday cage and connected to an oscilloscope through a current-to-voltage amplifier. Then the samples were irradiated through a small opening in the Faraday cage using a time modulated IR laser (1470 nm). The electric response was collected via an oscilloscope. The crystal was then turned over to heat the opposite side of the crystal, and the measurements were repeated.

When a sample is locally heated via illumination with an IR source, the electric current response can have several causes:

- a **thermoelectric effect**[274], i.e., an electric current as a result of a temperature gradient;
- a **flexoelectric effect**[275], which is caused by a thermally-induced stress gradient;
- **release of trapped charges**; and
- a **pyroelectric effect**, which comes from the change of the material's polarization with temperature; this effect is proportional to the temperature change.

The most significant difference between signals that come from a pyroelectric effect and those that emanate from the thermoelectric and flexoelectric effects is that only a pyroelectric electric current will reverse its sign if opposite sides of the crystal are heated. In the case of trapped charges, one can observe a response to a temperature change that is similar to that from pyroelectricity, i.e., inversion in the electric current, if the total charge on each side of the crystal is different in value or sign. Thus, because the currents, resulting from pyroelectricity and from emission of trapped charge signals could be similar, one must do the following three tests to distinguish between them[276]:

- Measure with several types of electrodes. Pyroelectricity is an intrinsic property and does not depend on the electrode type. Charge trapping may very well depend on the electrode type.
- Perform consecutive cycles of measurements or heat the sample well above the starting temperature and hold it there for a certain period of time. These should facilitate release of trapped charges.

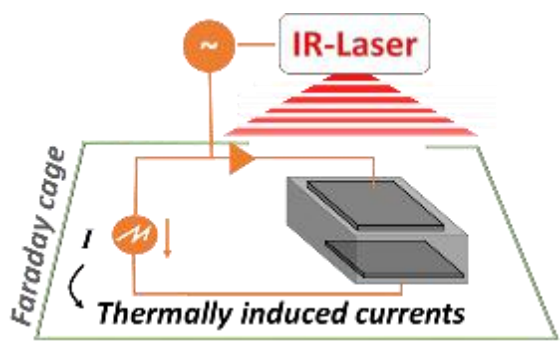

Figure 61: General scheme of pyroelectricity generation and measurement setup. The circle with the '~' symbol indicates a function generator. A circle with the '/\/\/' symbol indicates an oscilloscope. A triangle symbol indicts a current to voltage amplifier. The heat trigger is an IR pulsed laser. Pyroelectric currents, I, flow in response to a heat pulse only when there is a change in polarization occurring as a result of temperature change.

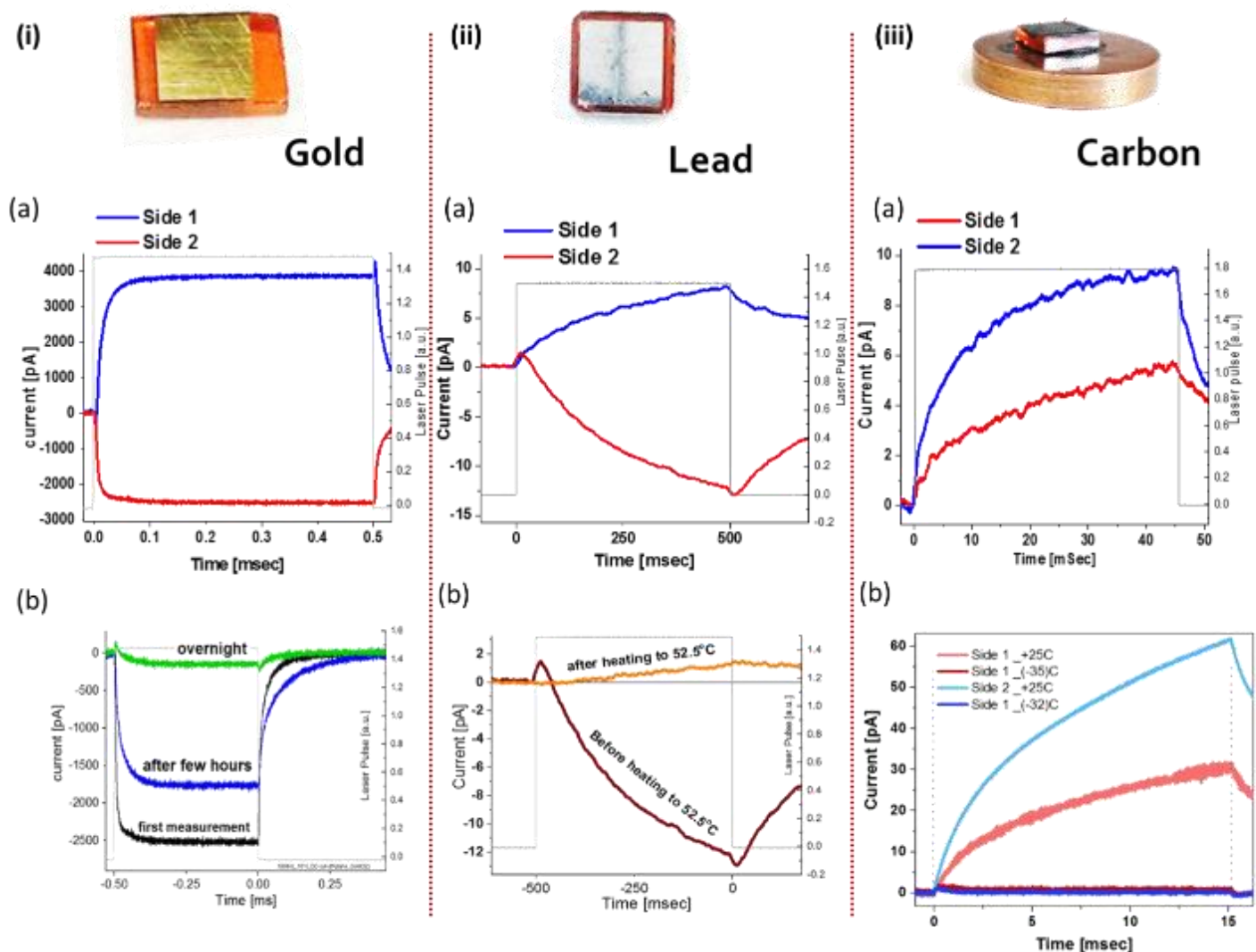

Figure 62: Electric responses of MAPbBr$_3$ crystals to heating by periodic irradiation with an IR laser. Sample with: (i) gold electrodes; (ii) lead electrodes; (iii) carbon electrodes. All sub-figures with (a) indication refer to first measurements just after electrode deposition, where 'side 1' and 'side 2' represent to two parallel sides of the crystals facing up the IR laser. Sub-figure (i,b) shows the result of many consecutive cycles of measurements of the samples with the gold electrodes, showing reduction of several orders of magnitude overnight. Sub-figure (ii.b) shows the result of before and after heating to +52.5°C, showing a non-reversible response. (i,b) and (ii,b) constitute a strong indication for trapped charges. (ii,c) shows the result of cooling the sample to below -30°C, showing vanishing of the thermoelectric currents with no indication for spontaneous polarity in the MAPbBr$_3$ crystals.

PhD Thesis – Yevgeny Rakita
September, 2018

|124|

Although MAPbBr$_3$ had been reported as cubic and no SHG was observed, we found that when we do the measurements with Au or Pb electrodes, the signal response resembles a pyroelectric response (see Figure 62(i,a) and (i,b)). When using carbon electrodes, however, the signal showed no pyroelectric-like response (Figure 62(i,c)). This led us to suspect that the currents are generated by asymmetrically-trapped charges. By performing many consecutive cycles of measurements and heating the sample well above the starting temperature to reduce the sample resistivity (Figure 62(ii,a) and (ii,b)), we indeed verified that the currents in MAPbBr$_3$ (with Au or Pb electrodes) are generated due to trapped charges and not spontaneous polarization. The possibility to cool down the pyroelectricity setup under nitrogen flow allowed us to measure MAPbBr$_3$ (with carbon electrodes) at low temperatures. This resulted in a significant reduction of the thermoelectric currents and clearly determine that there is no sign inversion in current flow upon inversion of the crystal (Figure 62(ii,c)).

Unlike MAPbBr$_3$, MAPbI$_3$ shows true pyroelectricity (Figure 63(i)). As expected, at temperatures higher than the phase transition between the tetragonal (polar) to cubic (non-polar) phases, the pyroelectric signal disappears (Figure 63(ii)). This process occurs gradually and is fully reversible and around the phase transition a local maximum is observed, indicating a rapid polarization change during the phase transition (from 'some' to 'non') as expected in a pyroelectric material (Figure 63(iii)).

Following the SHG and pyroelectricity experiments, it is clear that MAPbBr$_3$ cannot be ferroelectric, but the tetragonal MAPbI$_3$ may definitely be so. In fact, all known perovskite-structured materials that are known to be pyroelectric are also known to be ferroelectric. In materials for which the energetic cost to reverse their polarity is too high (due to lattice rigidity, as is the case for ZnO) or the required electric potential is higher than the dissociation potential of the material, observing a switchable polarization may be very challenging. However, since polarization in perovskites is induced by a relatively small translation of the atoms, polar perovskites are known to have switchable polarity.

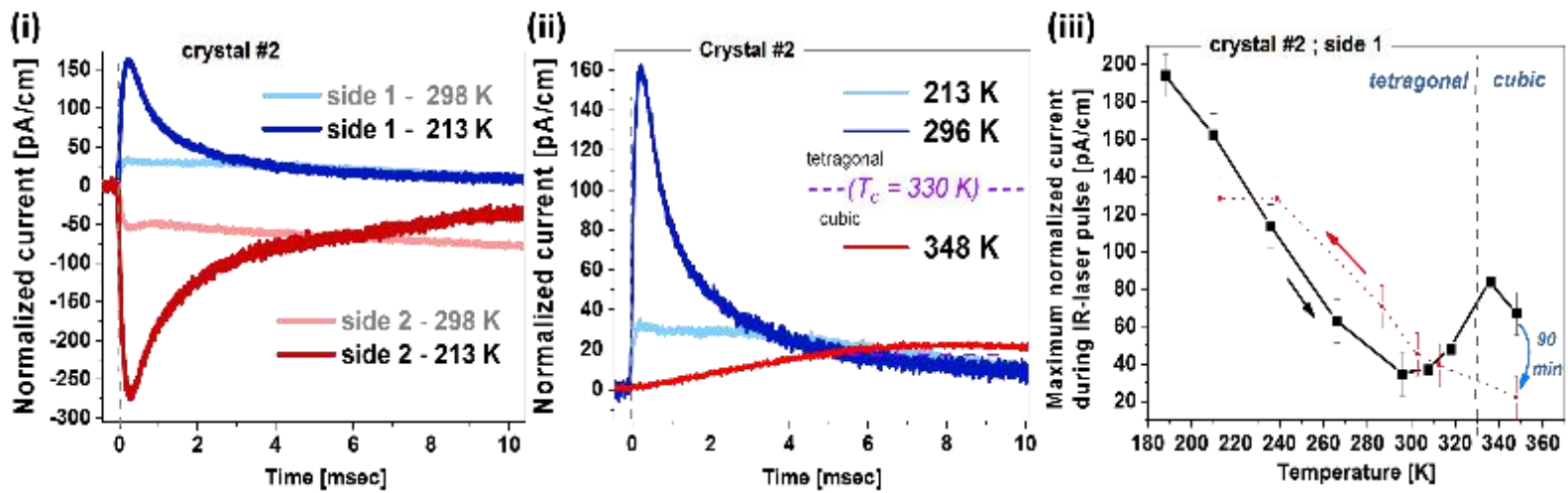

Figure 63: (i) Thermally-stimulated electric response that verifies pyroelectricity in $MAPbI_3$ crystal using carbon electrodes on the {001} planes: at RT (pale-colored plots) and 213 K, (deep-colored plots). At RT the response is convoluted with some thermoelectric currents, while at low temperatures $MAPbI_3$ shows currents that change sign when measuring from two parallel sides of the crystal (red and blue plots). (ii) results from 'side 1' showing that after the tetragonal to cubic phase transition, the signal disappears. (iii) Maximum-current value dependence on temperature that shows thermally-stimulated electrical currents are characteristic of a pyroelectric material. This is shown by the peak value at the tetragonal (polar) to cubic (non-polar) phase transition (also (ii)), and its regeneration when returning to the tetragonal phase. Unlike for the $MAPbBr_3$ case (Figure 62 (iii,b)), it shows regeneration of the pyroelectric currents when returning to the tetragonal phase.

The most straightforward way to prove a material is ferroelectric is by showing switchable polarization (*P*) under an applied electric field (*E*), which results in a commonly-presented *P-E* hysteresis loop [277,278]. Applying high electric fields (at least hundreds (100s) V×cm$^{-1}$ for soft materials[279]) result in polarization reversal of the polar domains, which is indicated as a rapid change in polarization (see Figure 64(i,c)). The probable ionic conductivity[216] with low formation energy [4,280] of HaPs increase the chances for dissociation, which makes detection of possible ferroelectric switching challenging. Indeed, previous attempts to find such evidence in $MAPbI_3$ [255,256] showed contradicting results to what was predicted from theory[248,249]. In fact, most previously-measured results are similar to those illustrated in Figure 64(iii,a). $MAPbI_3$, which has a low bandgap (1.55-1.60 eV)[281], has also a relatively (to other known ferroelectric materials) high electronic conductivity (even in the dark). High electrical conductivity and/or low stability will limit the interpretation of results using classical capacitance-based measurement systems, such as 'Sawyer Tower'-based methods [282]. To overcome this problem, one should use the *lossy* part of the dielectric response, $\varepsilon_{im}$, in addition (or instead) of the *energy storage* part, $\varepsilon_{re}$, of the dielectric response, as explained in Figure 64.

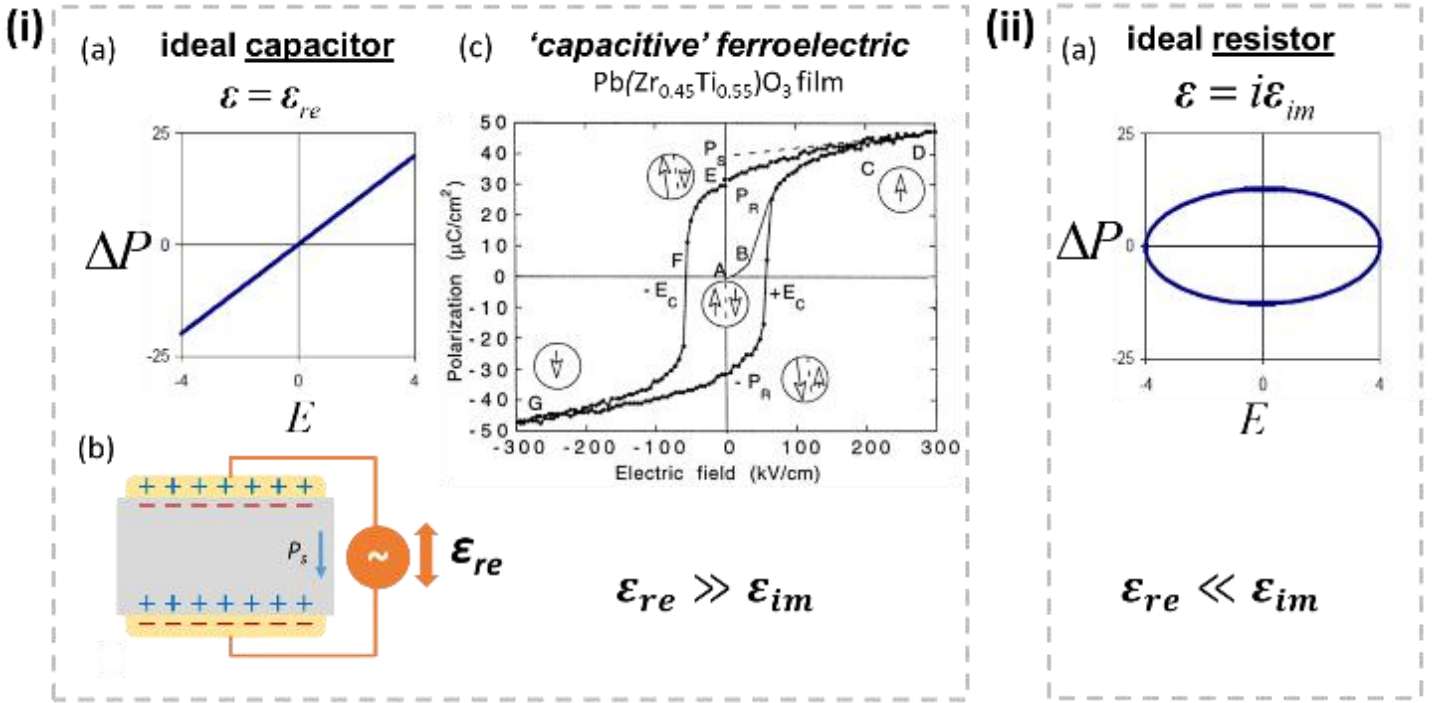
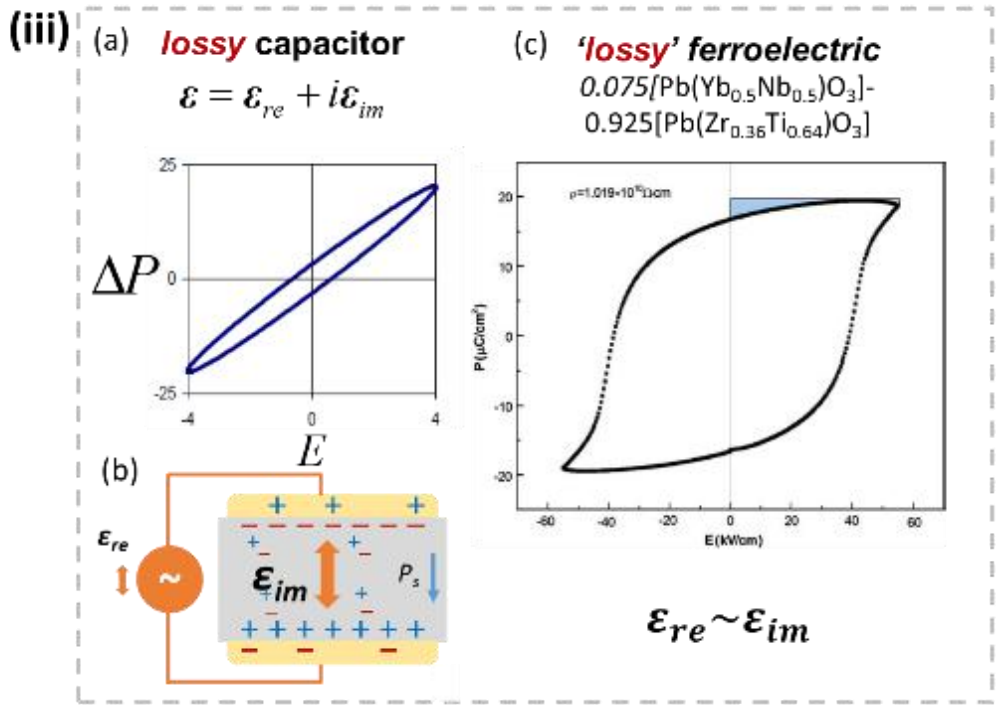

Figure 64: Representations of P-E loops of (i) an ideal capacitor (= current flow in wires), (ii) ideal resistor (= current flow through the material) and (iii) a lossy-capacitor (*some* current flows through the material). In each frame: (a) simulated polarization, $\Delta P$, vs. electric field, $E$, where $\Delta P = \varepsilon_0 \int \varepsilon \cdot \partial E$ and $\varepsilon^2 = (\varepsilon_{re} + i\varepsilon_{im})^2$; (b) an illustration of the charge flow upon an applied AC voltage, as applied during *P-E* measurements; (c) experimental P-E loops taken from (i) ref. 283 and (ii) ref. 284. To understand the different landmarks in (i,c), please see Fig. 8 in ref. 283. In all cases, $\varepsilon_{re}$ is the *energy conserving* dielectric constant and $\varepsilon_{im}$ is the energy non-conserving (= lossy) dielectric part. Note in (iii,c) the decrease in *P* at the high ends of *E*, which result from the lossy part (cf. (ii)).

PhD Thesis – Yevgeny Rakita
September, 2018
|127|

Focusing on MAPbI$_3$ as a potential ferroelectric material, we grew it from a solution using the VSA method, verified the identity of its naturally-developed facets using specular XRD, pasted carbon contacts on the {001} facets and measured its AC impedance response upon an applied background DC bias (see Figure 65). When plotting $\varepsilon_{re}$ and $\varepsilon_{im}$ as a function of the applied DC bias (see Figure 66(ii)), we found that the lossy part of the dielectric function, $\varepsilon_{im}$, dominated the overall dielectric response. This is in contrast to what is known from most ferroelectric materials where $\varepsilon_{im} \ll \varepsilon_{re}$ – even from partially organic ferroelectric materials (e.g., Rochelle salt), such as MAPbI$_3$ itself (see Figure 66(i)).

Following the relation $P = \varepsilon_0 \int \varepsilon \cdot \partial E$ to extract $P$, we integrate $\varepsilon$ (where $\varepsilon^2 = (\varepsilon_{re} + i\varepsilon_{im})^2 \approx (\varepsilon_{im})^2$) with the electric field, $E$. The result (Figure 66(iii)) shows that at lower temperatures (still in the tetragonal phase), the polarization response resembles a *lossy ferroelectric material*. At RT (Figure 66(iii,a)), the *P-E* response shows an irreversible hysteresis, which we believe indicates dissociation of the material under the high DC applied electric fields.

After proving the ferroelectric nature of MAPbI$_3$, what is still missing is direct evidence for polar domains. Polar domain walls, if dipoles inside of the domains are oriented 'head-to-head' (or 'tail-to-tail'), should be highly negatively (or positively) charged, as a screening response to the permanent dipoles (see Figure 58). The locally high charge density should vary the charge transport at domain walls with respect to its bulk – a point that was proven for BaTiO$_3$ [253]. Besides the charge dynamics, the chemical activity at the domain walls should be different between the bulk and the domain wall. Following this logic, chemical etching should uncover a periodic structuring as was demonstrated for different ferroelectric materials [285–287]. Indeed, etching the crystal with acetone for 2 minutes at RT revealed a highly periodic morphology (Figure 67(i)). To further verify that these are ferroelectric domains, we showed that the domain periodicity scales with the overall crystal size as expected (see Figure 67 (ii) and its caption for explanation). Moreover, if we heat the crystal above the tetragonal-cubic phase transition (~330 K), we find for the same crystal a very different post-etching pattern, meaning that when it was heated prior to etching, it had lost the pronounced orientation of the post-etching pattern (see Figure 67 (iii)), as expected.

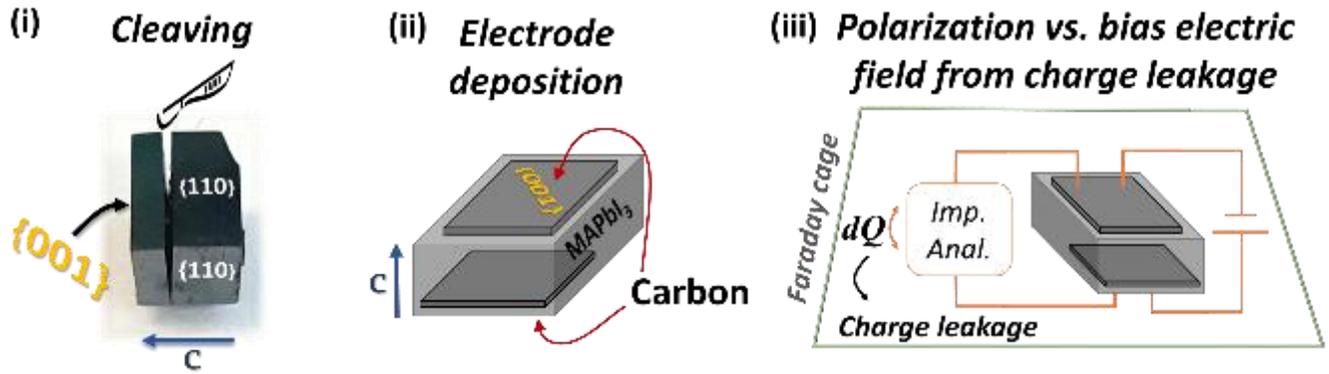

Figure 65: (i) Photograph of a MAPbI$_3$ crystal, cleaved along its {001} planes after identifying its polar direction. (ii) Illustration of the locations where the carbon electrodes were deposited for polarization measurements. (iii) General scheme of the polarization measurements under external bias, where an AC charge, dQ, is flowing *via* the impedance analyzer that is decoupled from the DC source.

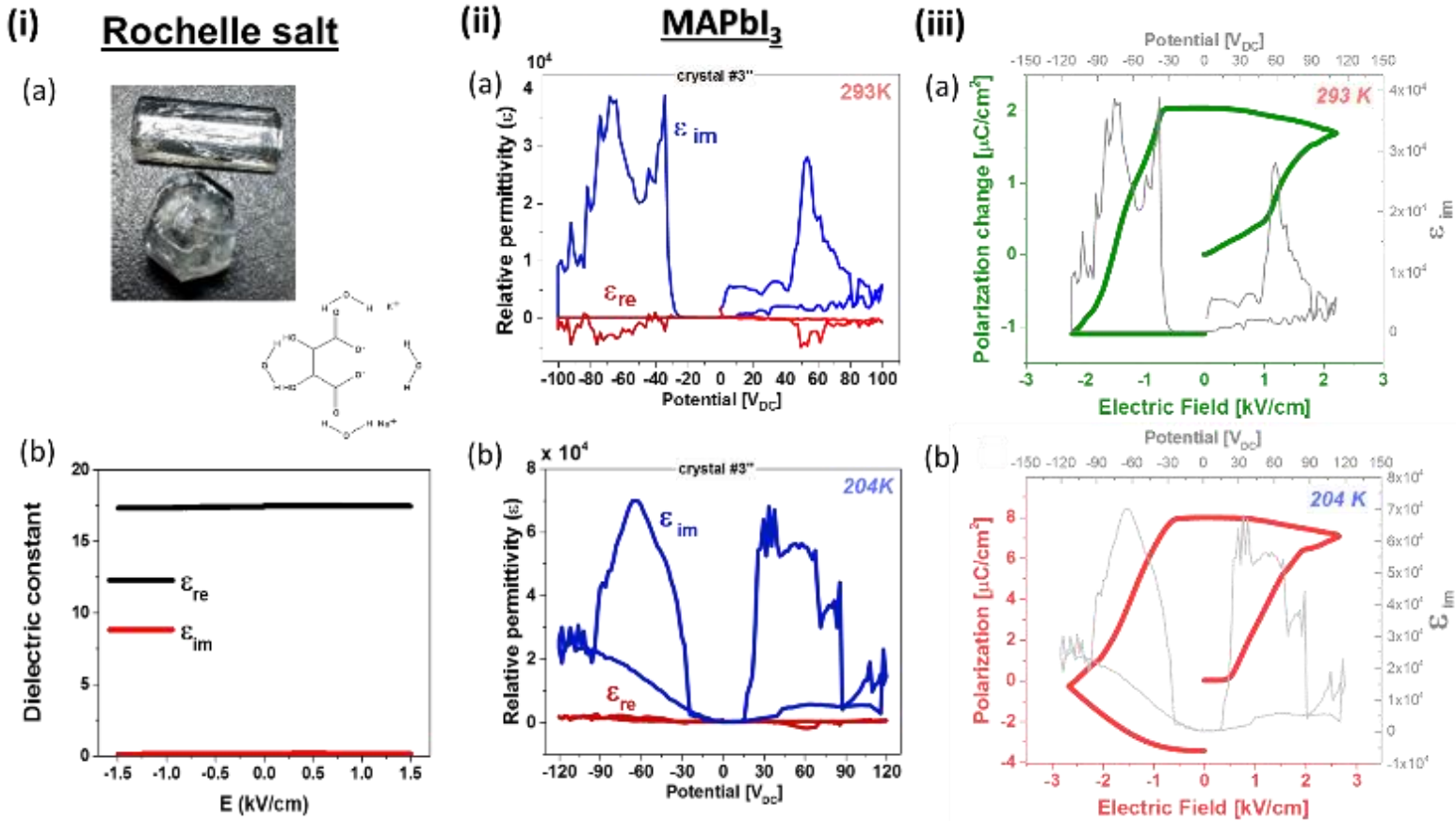

Figure 66:(i) (a) photograph and the formula of a Rochelle salt (Potassium Sodium Tartrate, KNa(OC(O)C(OH))$_2$·4H$_2$O) that is known to be ferroelectric. (b) The dielectric response of a ~1 mm thick plate that was cleaved across its polar direction. The measuring temperature was 273 K, where the polarization response to a bias electric field, $E_{DC}$ is maximal[282] (measured with an impedance analyzer (at 122 kHz and 1 V$_{AC}$). Comparison of the real, $\varepsilon_{re}$, and the imaginary (lossy) part, $\varepsilon_{im}$, of the relative permittivity clearly shows that $\varepsilon_{im}$ can be neglected for this material. (ii) Dielectric response at (a) RT and (b) 204 K of a MAPbI$_3$ crystal along its <001> direction, measured at 2,222 Hz and V$_{AC}$ =0.1 V, as a function of applied bias, $E_{DC}$ (X-axis), showing that the imaginary (dissipative) part dominates the dielectric response. (iii) P-E hysteresis loop obtained from integration of $\varepsilon_{im}$ over $E_{DC}$ following $P = \varepsilon_0 \int \varepsilon \cdot \partial E$ at (a) RT and (b) 204 K. The hysteresis loop is convoluted from a lossy bulk (i.e., resistor) response and ferroelectric polarization response, as explained in Figure 64.

How will ferroelectricity affect the operation of photovoltaic cells, based on MAPbI$_3$? To answer this question, we need to take into account the leaky nature of MAPbI$_3$, which will prevent the build-up of an internal electric field to help separate electrons and holes due to spontaneous polarization. However, within a single domain, this screening effect might possibly play a role by reducing intra-domain charge-recombination. Since recombination in other HaPs is higher, but not significantly (esp. MAPbBr$_3$), there is a quite a road ahead to find evidence in favor of such a mechanism, and to understand its significance. Our finding that etching reveals a domain-like structure, implies a high density of charge-carriers at the domain walls, which might fit the idea of 'ferroelectric highways' [18].

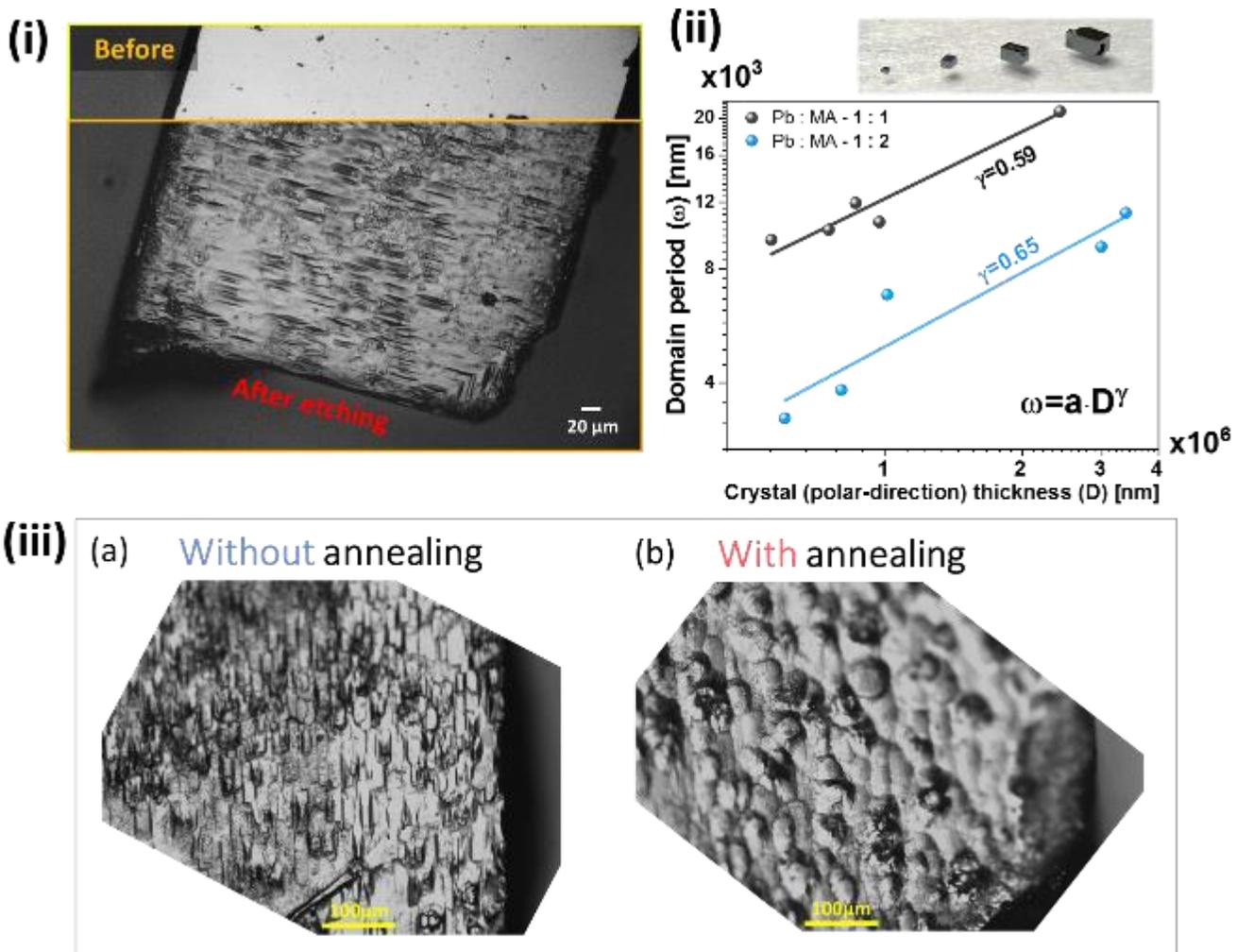

Figure 67: Evidence for polar domains and their scalability with crystal size: (i) Bright field image from a light microscope of a crystal before and after etching in acetone for 120 sec. (ii) Domain periodicity, $\omega$, with respect to the crystal thickness, D, averaged from microscope images after etching in acetone for ~ 120 sec at RT. The values of $\gamma$ are obtained by fitting to $\omega = a \cdot D^\gamma$. The theoretical value is 0.5, but in practice can show some deviations, also for other materials.[5] (iii) Bright field images from a light microscope of an etched MAPbI$_3$ crystal pieces (coming from one split crystal) where (a) had never undergone annealing while (b) was annealed at 75°C for two hours in a low-volume sealed vial to avoid degradation before etching.

It is not yet clear what the contribution of these domains is to free carriers at RT. What is missing now is evidence that these domain walls possess a very different conductivity than the bulk of the domain as predicted by theory[18] and was evident in ferroelectric oxide perovskites [253]. Although suffering some technical challenges, I am looking for such evidence with Dr. Beena Kaliski from Bari Ilan University. In the project with Dr. Kaliski we pass currents through contacts that are ~50 μm from each other on a surface of the single crystal. The currents that pass should induce small magnetic fields, whose magnitudes increase with current density. These fields can be then detected with a very sensitive magnetometer. If the conductivity is not homogenous but goes preferentially through domain walls, we should see this by the measured magnetic field contrast.

# 6. Summary and concluding remarks

In my thesis, I tried to draw conclusions about the underlying nature of the promising opto-electronic materials – halide perovskites, HaPs – and the implications of their nature. Besides the intrinsic properties of optoelectronic materials, which are a direct result of their symmetry, atomic composition and bond-nature, structural defects also play an important role. Since there are no defect-free bulk materials (quantum dots can be defect-free, if we do not count the surface as a defect), I divided my study between:

(1) the study of the averaged structural properties
(2) the potential effects defects may have on those same averaged structural properties
(3) the probability of defect existence.

Each chapter started with some theoretical understandings that helped to design hypotheses that then were tested experimentally as presented in following sections. In experiments, I was using mainly homemade solution-grown single crystals or thermodynamically-guided films as 'clean' platforms for fundamental studies. In few cases, where the formation/decomposition chemistry of HaP was studied, polycrystalline films were used.

Although some mention on Sn-based perovskites was made in section 4.3, I mostly limited my study to Pb-based HaPs. Some conclusions may change for Sn- or Ge-based halide perovskites – mostly due to their more chemically reactive $ns^2$ lone pair, which can affect their stability. Sn and Ge tend to further oxidize to +4, rather than stay +2 like Pb; therefore, their experimental investigation is challenging.

Neglecting chemical stability aspects, it was found that the influence of the A group on the B-X bond is minor. The A cation mostly influences the structural symmetry, but much less affects the electronic density and bonding between the B cation and X anions (as can be deduced from nanoindentation and SS-NMR experiments – sections 3.2 and 3.3, respectively). Since the B-X bond forms the VBM and CBM in these compounds (two important fundamental features which determine the intrinsic (opto)electronic properties of a semiconductor), the A cation is found to be *functionally* of much less significance than that of B and X. This conclusion is emphasized since the literature is full of publications that attribute the very unique (opto)electronic properties of HaPs (which allow them to perform as well as they do in PV cells, photodetectors and LEDs) to the organic nature of the A cation. Moreover, even if the A cation does affect the structural symmetry, deviation from a perfect cubic to a pseudo-cubic symmetry does not seem to affect the performance (whether good or poor) of HaPs. This is emphasized in chapter 5, where ferroelectricity was found to be not fundamental, and, when found to exist (in MAPbI$_3$), its significance at operational temperatures (i.e., ~RT) seems to be questionable. Therefore, I claim that for assessing the fundamental potential of HaPs for any technology, the focus should be on the B-X bond and less on the A group.

Nevertheless, for any technology, the long-term stability (in) and durability (to) the natural environment of a technology is important. It was found that atmospheric constituents (mostly H$_2$O as humidity and O$_2$) may chemically degrade HaPs, where those possessing an *organic* A cation seem to be more sensitive to such degradation than those with *inorganic* ones. Moreover, under conditions of constant photo-irradiation, temperature rise or even electron beam bombardment, HaPs with organic A cations are found to be more vulnerable. Some of these potential degradation paths must be considered when dealing with solar cells. In this work, I found that that the stability of the AX or AX$_3$ binaries, which may appear as degradation products, plays a central role in keeping HaPs stable. The much more dynamic (and volatile) nature of AX and AX$_3$, when A is *organic*, drives a more rapid degradation. The chemical potential window for formation of these materials is very narrow, and they can transform into their binaries or low-dimensional perovskite phases (e.g., 2D Ruddlesden-Popper phase – see Figure 3 and Figure 44). As a result, degradation should (and does) occur upon a loss of a perfect stoichiometric balance between the A, B and X subgroups. Rapid dynamics may allow a faster degradation whenever competing-chemical-reactions (e.g., with atmospheric H$_2$O or O$_2$) are thermodynamically favored. Thus, proper encapsulation may solve some of the problems as with any practical PV cell. Some degradation paths, however, like mechanical deformation or other very-high impact types of degradation tend to probe the B-X bond, almost regardless of the A cation. This is because degradation via an A cation occurs via diffusion, while high impact degradation paths are usually faster than diffusion processes, and therefore, probe the structural stiffness.

The faster dynamics of the binaries seem to have a positive side as well. It is clearly shown that upon photo-degradation in the bulk of APbBr$_3$ HaPs, recovery is faster when A is organic (Figure 56). It is important to note that even when A is inorganic self-healing seems to be *fundamental*, even it is slower than with organic A. This self-healing nature is found to be fundamental due to the following reasons:
- Possible degradation products (constituents that make up HaPs), such as metallic Pb (or Sn), PbX$_2$, AX and AX$_3$, HX or neutral A molecule are found to react with each other to form a HaP. Some of these reactions will produce a lot of heat and are, therefore, enthalpically favorable.

- Careful calorimetric experiments (done by other groups)[205] show that some reactions to form HaPs have very low negative (and favorable), or even positive (and unfavorable) enthalpies of formation with respect to their binaries (i.e., $AX_{(s)}$ and $PbX_{2(s)}$). However, entropy (probably mostly configurational-entropy) is found to be a dominating factor to stabilize $APbX_3$ compounds (even with positive enthalpies), resulting in free energy of formation of ~0.1 eV (for $MAPbI_3$) to 0.15 eV for $MAPbBr_3$ and $MAPbCl_3$.[205] Our experiments of $MAPbI_3$ formation from $PbI_2$ in MAI solution (section 4.2) yield a similar value of 0.1 eV for $MAPbI_3$. This fine balance may favor decomposition of $APbX_3$ compounds upon illumination, as shown by photodegradation experiments of HaPs and other halide-based systems (e.g., AgX) (section 4.4). It is interesting that upon illumination mixed $MAPb(Br,I)_3$ decompose to $MAPbBr_3$ and $MAPbI_3$, and, basically, looses its 'mixing' entropy. This deserves further study.

This chemically-based recyclability, however, may cause some fundamental limitations when using Sn or Ge based HaPs. Since one of the intermediates, $AX_3$, is highly oxidizing and since Sn(IV) and Ge(IV) are the most stable oxidation states (unlike Pb(II)) of Sn and Ge, the 'recycling' reaction, whenever $AX_3$ is formed, may lead to Sn(IV) compounds and form a permanent degradation of the material upon illumination. This point, however, needs to be further investigated.

The process of a spontaneous formation/decomposition (Figure 56) stresses the need for good encapsulation, since volatile degradation products or any competing reactions that may be favored by introduction of other chemicals, may lead to permanent degradation. Because we performed degradation experiments on $APbBr_3$ single crystals, and it will be interesting to compare this conclusion with results of other members of the $APbX_3$ family. In this way it is possible to probe if the nature of the X cation plays a significant role in the rate of 'self-healing' observed in HaPs, and if solid solutions (combining different A, B or X subgroups) can favor 'self-healing' due to an increase of entropy as a stabilizing factor.

This constant formation/decomposition, together with highly mobile ions or neutral binaries, seems to allow the system to reach its thermodynamic equilibrium quite rapidly. On the one hand, this leads to a problem (at least for some applications) for kinetic stabilization of a sufficient extrinsic doping density to modify carrier concentration significantly. So far, doping with $Bi^{3+}$ or $Li^+$ showed some modification in the carrier concentration, but it came along with a significant reduction in the bandgap in comparison with the hosting HaP; an effect that by-itself causes an increase in carrier concentration.[288–291]. This point is important, since in classical semiconductors (e.g., Si, GaAs), what allows doping are kinetically stabilized defects. The difficulty to form such kinetically-stabilized defects implies that HaPs are so well thermodynamically stabilized that kinetic stabilization will be challenging across a reasonable lifetime of a device made by them.

On the other hand, for PV applications, high purity is usually beneficial. The measured defect density in HaPs, almost regardless of the fabrication method, is ~$10^{10}$ cm$^{-3}$ in single crystals and up to ~$10^{16}$ cm$^{-3}$ in thin films. These values seem to be also thermodynamically driven, since $10^{10}$ cm$^{-3}$ is the point defect density that results from an activation energy for defect formation of ~1.6 eV (a reasonable value for defect formation in APbBr$_3$ and other Pb-Br based compounds). $10^{16}$ cm$^{-3}$ seems to be reasonable if defects are self-segregating at surfaces or interfaces. These conclusions are still at the level of speculation and require evidence for self-diffusion (for example by isotopic tracing), where determination of what is moving and how fast should be deduced as a following step.

It is interesting that electrostatic forces in an ionic system are known to follow Madelung forces ($E_{\text{Madelung}} \propto \frac{1}{n}$ ; n = coordination number). Increasing the coordination number in an easily polarizable system automatically implies higher configurational entropy. Higher configurational entropy means an increased stability of the system. All of this can happen only if the bonds are naturally soft enough (low *d'*), where changing configurations do not cost lots of energy. For example, the delocalization of the ns$^2$ electrons becomes more energetically favorable when one moves from fluoride ('rigid' bonds) to iodide ('soft/polarizable' bonds) (hemidirected to holodirected, respectively – see Figure 26 in section 3.3).[131]

Overall, it seems as if studying the thermodynamic state of HaPs provides the correct picture of their performance, while kinetically stabilized states seem to be highly unlikely (short time after its fabrication), so that *the role of defects (either intentional or unintentional) may become irrelevant*. Preparation chemistry and chemical environment (e.g., at interfaces), however, should have a strong impact on the material's (opto)electronic properties.

Another way of explaining the observed low defect densities is by assuming a 'defect-tolerating' system that is different from physical removal of defects by segregation. The meaning of this is that existing defects may stay in the lattice, but complex formation (such as that of AX$_3$) or an 'anti-bonding' nature of the VBM, may lead to formation of very shallow defect states with respect to the CBM and VBM position, or even within these bands. A special case was shown for AX$_3$ complexes (Figure 51), in case they form inside the HaP matrix, e.g., due to illumination, they seem to be easily tolerated, because of their very similar optical bandgap to that of their hosting HaP matrix. Similarly to the idea of shallow defect states due to an 'anti-bonding' VBM nature (see section 3.1), the bands of the *in situ*-formed AX$_3$ may lie at approximately the same energy position as the HaP. This point, that needs further verification, suggests that complex formation (AX$_3$ complexes or other chemicals) can reduce the defect energy in the bandgap of the HaP with respect to their initial (un-complexed) position in the bandgap. This means that not all defects necessarily affect (opto)electronic properties in Pb-based HaPs.

Regarding the second 'defect-tolerating' property, i.e. an 'anti-bonding' VBM, as was suggested from DFT calculations, experimental derivation of the 'deformation potential' showed a *positive* value, which means that the VBM increases upon strain faster than the CBM, indicating an 'anti-bonding' VBM. Since theory predicts shallow defect formation when VBM is 'anti-bonding', the unique situation where halide-based and Pb-based compounds (including HaPs) possess a positive VBM, suggests 'defect-tolerating' systems. Moreover, the low absolute value of the deformation potential of halide-based compounds (e.g., HaPs, AgX, TlX) suggests a system that can tolerate much higher strain fields than 'rigid' systems (e.g., next to grain boundaries, dislocations, interstitials, etc.) due to smaller distortion of the free charge energetic landscape upon strain.

For heteropolar systems, it was found that materials with a *positive* deformation potential should also be with a highly polarizable lattice, meaning with a low $\varepsilon^*$ value. Since these cases are also assumed to be with low defect density (as explained above), the dominant relevant carrier scattering mechanism is assumed to be due to polar fluctuations, or polar optical phonons (POP). Since $\mu_{POP} \sim \varepsilon^{*(+3.4)}$, there is a very strong decrease (orders of magnitude) in the optimal mobility values (i.e., low defect density) with decreasing $\varepsilon^*$. The correlation between the lattice-polarizability and the deformation potential showed in Figure 49(iv), predicts a positive deformation potential for highly polarizable systems (i.e., when $\varepsilon_s > 2\varepsilon_\infty$), making it a 'defect-tolerating' system, but also causing a fundamentally lower charge mobility.

**To summarize with some 'take-home' messages:**
- The stability of Sn(IV) together with a permanent 'deformation-formation' process – especially upon illumination, may limit the applicability of Sn-based HaPs. Although HaP-based PV cells containing Sn have been shown to be fairly stable, it is still unknown whether

P h D   T h e s i s  –  Y e v g e n y   R a k i t a
S e p t e m b e r ,   2 0 1 8

|136|

- they will 'self-heal' after photo-degradation, or slowly (but irreversibly) degrade upon illumination at normal (1-sun) PV conditions.

- Organic groups on the A site seem to reduce the durability of HaPs in the absence of effective encapsulation. At the same time, if encapsulation is sufficient and no mass migration out of the system is possible, recovery from existing damage is improved.

- Low enthalpically (un)stabilized systems with respect to their solid constituents, together with a significant entropic stabilization of a solid system, such as some HaPs, may result in a system with (1) 'self-healing' ;(2) highly mobile ions. This may result in low-defect material, even if synthesized under 'sloppy' conditions, but may limit the capability of the material to be doped and used in applications where spatially-specific doping is needed.

- Concentrated light PV cells seems to be less suitable in the case of HaPs because of the proven light-degradation. However, concentrated light after proper encapsulation may be useful for regeneration of defected areas.

- High lattice-polarizability predicts a defect-tolerating system: Figure 49(iv) predicts that if - $\varepsilon_s > 2\varepsilon_\infty$, then a material should have a positive deformation potential, and thus possess a defect-tolerating nature due to an anti-bonding VBM.

- Since the charge mobility of heteropolar systems is strongly reduced with an increase of its polarizability (assuming low scattering defect density based on the previous point), highly polarizable systems (like HaPs) are *intrinsically* expected to possess significantly lower charge mobility values than other less polarizable (more 'rigid') heteropolar systems (like GaAs).

- Mechanical softness is a very crude estimate for defect-tolerance. The deformation potential (which takes into account also the change of the bands upon strain) is a much better guideline for defect tolerance. Low absolute values of the deformation potential suggest a material that does not change its energetic landscape even with significant strain fields (e.g., next to grain boundaries or due to point defects), and therefore can tolerate uncharged defects. Charged defects may become energetically unfavorable because of the low stabilization energy of HaPs (~0.1 eV) and, therefore, become unlikely to appear above their thermodynamic limit (~$10^{10}$ cm$^{-3}$ in single crystals).

- Ferroelectricity, even when it exists in HaPs, does not seem to have much of an effect at RT. Therefore, ferroelectricity is not the 'secret' of HaP PV cells, as was sometimes claimed.